\documentclass{book}

\newcommand{\negsmallskip}{\vspace{-\smallskipamount}}
\newcommand{\negmedskip}{\vspace{-\medskipamount}}

\usepackage[fleqn]{amsmath}
\usepackage{amssymb}
\usepackage{theorem}
\usepackage{bbm} 
\usepackage{calc} 

\newcommand{\NN}{\ensuremath{\mathbb{N}}}
\newcommand{\RR}{\ensuremath{\mathbb{R}}}
\newcommand{\CC}{\ensuremath{\mathbb{C}}}

\newcommand{\ZZ}{\ensuremath{\mathbb{Z}}}
\newcommand{\PP}{\ensuremath{\mathbb{P}}}
\newcommand{\EE}{\ensuremath{\mathbb{E}}}
\newcommand{\TT}{\ensuremath{\mathbb{T}}}

\newcommand{\half}{\ensuremath{\frac{1}{2}}}

\newcommand{\upwards}{\ensuremath{\uparrow}}
\newcommand{\upwardsto}{\upwards}

\newcommand{\isom}{\ensuremath{\simeq}}
\newcommand{\Union}{\ensuremath{\bigcup}}
\newcommand{\union}{\ensuremath{\cup}}
\newcommand{\intersection}{\ensuremath{\cap}}

\newcommand{\after}{\circ}
\newcommand{\Identity}{\mathbb{Id}}

\newcommand{\indicator}{\mathbbm{1}}

\newcommand{\supnorM}[1]{\ensuremath{\|\,#1\,\|_\infty}}

\newcommand{\Iff}{\ensuremath{\quad\Longleftrightarrow\quad}}
\newcommand{\Implies}{\ensuremath{\quad\Longrightarrow\quad}}
\renewcommand{\implies}{\ensuremath{\Rightarrow}}
\newcommand{\implication}[2]%
{\ensuremath{#1\negmedspace\Rightarrow\negmedspace#2}}
\newcommand{\equivalence}[2]%
{\ensuremath{#1\negmedspace\Leftrightarrow\negmedspace#2}}

\newcommand{\In}{\hbox{ in }}

\renewcommand{\and}{\hbox{ and }}

\renewcommand{\Re}{\ensuremath{\mathrm{Re}\hspace{1pt}}}
\renewcommand{\Im}{\ensuremath{\mathrm{Im}\hspace{1pt}}}

\newcommand{\ssum}[2]%
{\textstyle\overset{#2}{\underset{#1}{\sum}}\;}

\newcommand{\tsum}%
{\textstyle\sum}

\newcommand{\dsum}%
{\displaystyle\sum}

\renewcommand{\ae}{\hbox{a.e.}}

\newcommand{\al}{\ensuremath{\alpha}}
\newcommand{\be}{\ensuremath{\beta}}

\newcommand{\de}{\ensuremath{\delta}}
\newcommand{\ep}{\ensuremath{\varepsilon}}

\newcommand{\la}{\ensuremath{\lambda}}
\newcommand{\ph}{\ensuremath{\phi}}

\newcommand{\ta}{\ensuremath{\tau}}
\newcommand{\ro}{\ensuremath{\rho}}
\newcommand{\si}{\ensuremath{\sigma}}
\newcommand{\om}{\ensuremath{\omega}}

\newcommand{\Ga}{\ensuremath{\Gamma}}

\newcommand{\Si}{\ensuremath{\Sigma}}
\newcommand{\Om}{\ensuremath{\Omega}}

\theoremstyle{break}
\newtheorem{theorem}{Theorem}[chapter]  %

\newtheorem{lemma}[theorem]{Lemma}
\newtheorem{proposition}[theorem]{Proposition}
\newtheorem{definition}[theorem]{Definition}
\newtheorem{corollary}[theorem]{Corollary}
\newtheorem{convention}[theorem]{Convention}

\theorembodyfont{\rmfamily}
\newtheorem{example}[theorem]{Example}
\newtheorem{examples}[theorem]{Examples}

\newtheorem{remark}[theorem]{Remarks}

\newenvironment{remarks}%
{\begin{remark}
 \negmedskip%
 \begin{list}{}{\setlength{\leftmargin}{0em}%
                \setlength{\itemindent}{1.2em}%
		\setlength{\labelwidth}{0pt}%
		\setlength{\labelsep}{0pt}%
		\setlength{\parsep}{0pt}%
		\setlength{\itemsep}{0.3em}%
               }
}%
{\end{list}\end{remark}%
}

\newenvironment{proof}[1][]%
{\smallskip{\sl Proof#1}\\}%
{\hfill\ensuremath{\square}\medskip}

{\smallskip{\bf Proof#1}\\}%
{\hfill\ensuremath{\square}\medskip}

\newenvironment{theorem*}[1][]%
{\medskip\noindent{\bf Theorem} #1%
\newline\slshape}%
                         {\medskip}
\newenvironment{lemma*}{\medskip\noindent{\bf Lemma}\newline\slshape}{}

\newenvironment{definition*}{\medskip\noindent%
			    {\bf Definition}\newline\slshape}{\medskip}

\newcommand{\statement}[2][0em]%
{\smallskip{\slshape#2}\vspace{0.5\smallskipamount}\vspace{#1}}

\newcommand{\qed}{\hfill\ensuremath{\square}\medskip}

\newenvironment{pf}{\begin{list}{}{\setlength{\leftmargin}{1em}
                                   \setlength{\topsep}{0pt}
				   \setlength{\parsep}{0pt}
				   \setlength{\rightmargin}{0pt}
				   \setlength{\labelwidth}{0em} 
				   \setlength{\itemindent}{-.8em}
				   \setlength{\labelsep}{0.1em} 
                                  }
                                   \item[$\vartriangleleft$]\small\ignorespaces
		    }%
                   {\unskip\hspace{1pt}$\vartriangleright$\smallskip\end{list}
                   }%

\newenvironment{pf*}{\smallskip\begin{list}{}{\setlength{\leftmargin}{1em}
                                   \setlength{\topsep}{0pt}
				   \setlength{\parsep}{0pt}
				   \setlength{\rightmargin}{0pt}
                                  }
                                  \item[]\small\ignorespaces} {\unskip
                                  \smallskip
                       \end{list}
                      }
                   {\unskip\hspace{1pt}$\vartriangleright$\smallskip\end{list}
                   }%

\usepackage{ifthen}              

\newcommand{\greek}[1]{
    \ifthenelse{\value{#1} = 1}{\ensuremath{(\alpha)}}{}%
    \ifthenelse{\value{#1} = 2}{\ensuremath{(\beta)}}{}%
    \ifthenelse{\value{#1} = 3}{\ensuremath{(\gamma)}}{}%
    \ifthenelse{\value{#1} = 4}{\ensuremath{(\delta)}}{}%
    \ifthenelse{\value{#1} = 5}{\ensuremath{(\epsilon)}}{}%
    \ifthenelse{\value{#1} = 6}{\ensuremath{(\zeta)}}{}%
    \ifthenelse{\value{#1} > 6}{\hbox{EXPAND COMMAND greek}}{}%
                       }

{\begin{list}{\greek{cnt_equivalences}}%
             {\usecounter{cnt_equivalences}%
	      \setlength{\labelwidth}{1em}\setlength{\labelsep}{.4em}%
	      \setlength{\leftmargin}{1.8em}\setlength{\itemindent}{-.3em}%
	      \setlength{\itemsep}{-.1em}%
              }
}		
{\end{list}
}

\newcommand{\refequivalence}[1]%
{   \ifthenelse{\ref{#1} = 1}{\ensuremath{(\alpha)}}{}%
    \ifthenelse{\ref{#1} = 2}{\ensuremath{(\beta)}}{}%
    \ifthenelse{\ref{#1} = 3}{\ensuremath{(\gamma)}}{}%
    \ifthenelse{\ref{#1} = 4}{\ensuremath{(\delta)}}{}%
    \ifthenelse{\ref{#1} = 5}{\ensuremath{(\epsilon)}}{}%
    \ifthenelse{\ref{#1} = 6}{\ensuremath{(\zeta)}}{}%
    \ifthenelse{\ref{#1} > 6}{\hbox{EXPAND COMMAND refequivalence}}{}%
}

\newenvironment{conditions}{%
                            \negsmallskip
                            \begin{enumerate}%
			    }%
			   {\end{enumerate}%
                            \negsmallskip}

			   {\end{enumerate}%
                             }

\newcounter{Property}

\newenvironment{Properties}[1][0em]%
{
 \vspace{-\medskipamount}
 \begin{list}{\hfill\roman{Property}.}%
             {\usecounter{Property}
	      \setlength{\labelwidth}{1.2em}\setlength{\labelsep}{.6em}%
	      \setlength{\leftmargin}{1.6em+#1}
	      \setlength{\listparindent}{0pt}\setlength{\itemsep}{-.2em}%
	      }
}%
{\end{list}\vspace{-\medskipamount}} 

\newcommand{\romanref}[1]%
{   \ifthenelse{\ref{#1} = 1}{\!\!i}{}%
    \ifthenelse{\ref{#1} = 2}{\!\!ii}{}%
    \ifthenelse{\ref{#1} = 3}{\!\!iii}{}%
    \ifthenelse{\ref{#1} = 4}{\!\!iv}{}%
    \ifthenelse{\ref{#1} = 5}{\!\!v}{}%
    \ifthenelse{\ref{#1} = 6}{\!\!vi}{}%
    \ifthenelse{\ref{#1} = 7}{\!\!vii}{}%
    \ifthenelse{\ref{#1} = 8}{\!\!viii}{}%
    \ifthenelse{\ref{#1} = 9}{\!\!ix}{}%
    \ifthenelse{\ref{#1} > 9}{\hbox{EXPAND COMMAND romanref}}{}%
}

\newcounter{Enu}
\newenvironment{Enumerate}[1][0em]%
{

 \vspace{-\medskipamount}
 \begin{list}{\hfill\arabic{Enu}. }%
             {\usecounter{Enu}
	      \setlength{\labelwidth}{2em}\setlength{\labelsep}{0pt}%
	      \setlength{\leftmargin}{2em+#1}
	      \setlength{\listparindent}{0pt}%
              \setlength{\itemsep}{-.2em}%
	      }
}%
{\end{list}\vspace{-\medskipamount}}

\newenvironment{Itemize}[2][0em]%
{\negmedskip
 \begin{list}{#2}%
       	     {\setlength{\labelwidth}{.4em}\setlength{\labelsep}{.3em}%
	      \setlength{\leftmargin}{0.9em+#1}
	      \setlength{\listparindent}{0pt}\setlength{\itemsep}{-.3em}%
               }
}%
{\end{list}
}

\newenvironment{displaymatH}%
{\vspace{-1.6\smallskipamount}%
\begin{displaymath}\textstyle%
}%
{\end{displaymath}%

\vspace{-1.6\smallskipamount}%
}%

\newenvironment{aligN}%
{\newcommand{\ats}{&\textstyle}
\vspace{-1.6\smallskipamount}%
\begin{displaymath}\begin{aligned}\textstyle%
}%
{\end{aligned}\end{displaymath}%

\vspace{-1.6\smallskipamount}%
}%

\newenvironment{equatioN}%
{\negmedskip\begin{equation}}%
{\end{equation}

\negmedskip}

\newlength{\EBDskip} 
\newlength{\EADskip} 

{\setlength{\EBDskip}{#1}\setlength{\EADskip}{#2}
\vspace{-1.6\smallskipamount}%
\vspace{\EBDskip}%
\begin{displaymath}\textstyle%
}%
{\end{displaymath}%

\vspace{-1.6\smallskipamount}%
\vspace{\EADskip}%
}%
{\begin{list}{}{%
 \settowidth{\labelwidth}{\textsf{#1}}%
 \setlength{\leftmargin}{\labelwidth+\labelsep}}%
}%
{\end{list}}

{\begin{list}{}{%
 \settowidth{\labelwidth}{\textup{#1}}%
 \setlength{\leftmargin}{\labelwidth+\labelsep}}%
}%
{\end{list}}

\newsavebox{\claimsymbol}   

{\medskip%
 \savebox{\claimsymbol}{#1}
 \begin{tabular}{p{1ex}@{}p{.85\textwidth}p{2em}}%
 &%
 \begin{minipage}{.85\textwidth}\slshape%
}%
{\end{minipage}%
 &%
 \hfill%
 \usebox{\claimsymbol}%
 \end{tabular}%
 \medskip%
}%

\newenvironment{claim+}[1]%
{\medskip%
 \savebox{\claimsymbol}{#1}
 \begin{tabular}{p{1ex}@{}p{.85\textwidth}p{2em}}%
 &
 $
 \left.
 \begin{minipage}{.85\textwidth}\slshape%
}%
{\end{minipage}%
 \;\;
 \right
 \rgroup
 $
 &
 \hfill%
 \usebox{\claimsymbol}%
 \end{tabular}%
 \medskip%
}%

\newcommand{\Dindex}[2][!*!*!]%
{%
\ifthenelse%
{\equal{#1}{!*!*!}}
{\ifthenelse{\equal{#2}{}}{}{\textsf{#2}\index{#2}}}
{
\ifthenelse{\equal{#1}{}}{}{\index{#1}}%
\ifthenelse{\equal{#2}{}}{}{\textsf{#2}}%
}%
}

\newcommand{\Rindex}[2][!*!*!]%
{%
\ifthenelse%
{\equal{#1}{!*!*!}}
{\ifthenelse{\equal{#2}{}}{}{\textnormal{#2}\index{#2}}}
{
\ifthenelse{\equal{#1}{}}{}{\index{#1}}%
\ifthenelse{\equal{#2}{}}{}{\textnormal{#2}}%
}%
}

\usepackage{amscd} 

\newcommand{\K}{\mathcal{K}}
\newcommand{\U}{\mathcal{U}}

\newcommand{\Ep}{\mathcal{E}}

\newcommand{\F}{\mathcal{F}}

\newcommand{\Var}[1]{\textsf{Var}(#1)}
\newcommand{\Cov}[2]{\textsf{Cov}(#1\,,\,#2)}

\newcommand{\ip}[3][]%
{%
 \ifthenelse{\equal{#1}{text}}
 { \ifthenelse{\equal{#2}{}}
   {\ensuremath{\langle\:|\:\rangle}}
   {\ensuremath{\langle#2|#3\rangle}}
 }%
 {}%
 \ifthenelse{\equal{#1}{big}}%
 { \ifthenelse{\equal{#2}{}}%
   {\ensuremath{\bigl\langle\:\bigm|\:\bigr\rangle}}%
   {\ensuremath{\bigl\langle#2\bigm|#3\bigr\rangle}}%
 }%
 {}%
 \ifthenelse{\equal{#1}{Big}}%
 { \ifthenelse{\equal{#2}{}}%
   {\ensuremath{\Big\langle\Bigm|\;\Bigr\rangle}}%
   {\ensuremath{\Bigl\langle#2\Bigm|#3\Bigr\rangle}}%
 }%
 {}%
 \ifthenelse{\equal{#1}{bigg}}%
 { \ifthenelse{\equal{#2}{}}%
   {\ensuremath{\bigg\langle\biggm|\;\biggr\rangle}}%
   {\ensuremath{\bigg\langle#2\biggm|#3\biggr\rangle>}}%
 }%
 {}%
 \ifthenelse{\equal{#1}{Bigg}}%
 { \ifthenelse{\equal{#2}{}}%
   {\ensuremath{\Biggl\langle \: \Biggm|\:\Biggr\rangle}}%
   {\ensuremath{\Biggl\langle #2 \Biggm|#3\Biggr\rangle}}%
 }%
 {}%
 \ifthenelse{\equal{#1}{}}%
 { \ifthenelse{\equal{#2}{}}%
   {\ensuremath{\left\langle \: \mid|\:\mid|\:\right\rangle}}%
   {\ensuremath{\left\langle#2|#3\right\rangle}}%
 }%
 {}%
}

\newcommand{\oip}[4][]%
{%
 \ifthenelse{\equal{#1}{text}}
 { \ifthenelse{\equal{#2}{}}
   {\ensuremath{\langle\:|\:|\:\rangle}}
   {\ensuremath{\langle#2|#3|#4\rangle}}
 }%
 {}%
 \ifthenelse{\equal{#1}{big}}%
 { \ifthenelse{\equal{#2}{}}%
   {\ensuremath{\bigl\langle\:\bigm|\:\bigm|\:\bigr\rangle}}%
   {\ensuremath{\bigl\langle#2\bigm|#3\bigm|#4\bigr\rangle}}%
 }%
 {}%
 \ifthenelse{\equal{#1}{Big}}%
 { \ifthenelse{\equal{#2}{}}%
   {\ensuremath{\Big\langle\Bigm|\;\Bigm|\;\Bigr\rangle}}%
   {\ensuremath{\Bigl\langle#2\Bigm|#3\Bigm|#4\Bigr\rangle}}%
 }%
 {}%
 \ifthenelse{\equal{#1}{bigg}}%
 { \ifthenelse{\equal{#2}{}}%
   {\ensuremath{\bigg\langle\biggm|\;\biggm|\;\biggr\rangle}}%
   {\ensuremath{\bigg\langle#2\biggm|#3\biggm|#4\biggr\rangle}}%
 }%
 {}%
 \ifthenelse{\equal{#1}{Bigg}}%
 { \ifthenelse{\equal{#2}{}}%
   {\ensuremath{\Biggl\langle \: \Biggm|\:\Biggm|\Biggr\rangle}}%
   {\ensuremath{\Biggl\langle #2 \Biggm|#3\Biggm|#4\Biggr\rangle}}%
 }%
 {}%
 \ifthenelse{\equal{#1}{}}%
 { \ifthenelse{\equal{#2}{}}%
   {\ensuremath{\left\langle \: \mid|\:\mid|\:\right\rangle}}%
   {\ensuremath{\left\langle#2 | #3|#4\right\rangle}}%
 }%
 {}%
}

\newcommand{\norM}[2][]%
{%
 \ifthenelse{\equal{#1}{text}}
 { \ifthenelse{\equal{#2}{}}
   {\ensuremath{\| \: \|}}
   {\ensuremath{\| #2 \|}}
 }%
 {}%
 \ifthenelse{\equal{#1}{big}}%
 { \ifthenelse{\equal{#2}{}}%
   {\ensuremath{\big\| \: \big\|}}%
   {\ensuremath{\big\| #2 \big\|}}%
 }%
 {}%
 \ifthenelse{\equal{#1}{Big}}%
 { \ifthenelse{\equal{#2}{}}%
   {\ensuremath{\Big\| \: \Big\|}}%
   {\ensuremath{\Big\| #2 \Big\|}}%
 }%
 {}%
 \ifthenelse{\equal{#1}{bigg}}%
 { \ifthenelse{\equal{#2}{}}%
   {\ensuremath{\bigg\| \: \bigg\|}}%
   {\ensuremath{\bigg\| #2 \bigg\|}}%
 }%
 {}%
 \ifthenelse{\equal{#1}{Bigg}}%
 { \ifthenelse{\equal{#2}{}}%
   {\ensuremath{\Bigg\| \: \Bigg\|}}%
   {\ensuremath{\Bigg\| #2 \Bigg\|}}%
 }%
 {}%
 \ifthenelse{\equal{#1}{}}%
 { \ifthenelse{\equal{#2}{}}%
   {\ensuremath{\left\| \: \right\|}}%
   {\ensuremath{\left\| #2 \right\|}}%
 }%
 {}%
}

\newcommand{\norm}[2][]%
{%
 \ifthenelse{\equal{#1}{text}}
 { \ifthenelse{\equal{#2}{}}
   {\ensuremath{\,\| \: \|\,}}
   {\ensuremath{\,\|  #2 \|\,}}
 }%
 {}%
 \ifthenelse{\equal{#1}{big}}%
 { \ifthenelse{\equal{#2}{}}%
   {\ensuremath{\,\big\| \: \big\|\,}}%
   {\ensuremath{\,\big\| #2 \big\|\,}}%
 }%
 {}%
 \ifthenelse{\equal{#1}{Big}}%
 { \ifthenelse{\equal{#2}{}}%
   {\ensuremath{\,\Big\| \: \Big\|\,}}%
   {\ensuremath{\,\Big\| #2 \Big\|\,}}%
 }%
 {}%
 \ifthenelse{\equal{#1}{bigg}}%
 { \ifthenelse{\equal{#2}{}}%
   {\ensuremath{\,\bigg\| \: \bigg\|\,}}%
   {\ensuremath{\,\bigg\| #2 \bigg\|\,}}%
 }%
 {}%
 \ifthenelse{\equal{#1}{Bigg}}%
 { \ifthenelse{\equal{#2}{}}%
   {\ensuremath{\,\Bigg\| \: \Bigg\|\,}}%
   {\ensuremath{\,\Bigg\| #2 \Bigg\|\,}}%
 }%
 {}%
 \ifthenelse{\equal{#1}{}}%
 { \ifthenelse{\equal{#2}{}}%
   {\ensuremath{\,\left\| \: \right\|\,}}%
   {\ensuremath{\,\left\| #2 \right\|\,}}%
 }%
 {}%
}

\newcommand{\pnorM}[2][]%
{%
 \ifthenelse{\equal{#1}{text}}
 { \ifthenelse{\equal{#2}{}}
   {\ensuremath{\| \: \|_p}}
   {\ensuremath{\| #2 \|_p}}
 }%
 {}%
 \ifthenelse{\equal{#1}{big}}%
 { \ifthenelse{\equal{#2}{}}%
   {\ensuremath{\big\| \: \big\|_p}}%
   {\ensuremath{\big\| #2 \big\|_p}}%
 }%
 {}%
 \ifthenelse{\equal{#1}{Big}}%
 { \ifthenelse{\equal{#2}{}}%
   {\ensuremath{\Big\| \: \Big\|_p}}%
   {\ensuremath{\Big\| #2 \Big\|_p}}%
 }%
 {}%
 \ifthenelse{\equal{#1}{bigg}}%
 { \ifthenelse{\equal{#2}{}}%
   {\ensuremath{\bigg\| \: \bigg\|_p}}%
   {\ensuremath{\bigg\| #2 \bigg\|_p}}%
 }%
 {}%
 \ifthenelse{\equal{#1}{Bigg}}%
 { \ifthenelse{\equal{#2}{}}%
   {\ensuremath{\Bigg\| \: \Bigg\|_p}}%
   {\ensuremath{\Bigg\| #2 \Bigg\|_p}}%
 }%
 {}%
 \ifthenelse{\equal{#1}{}}%
 { \ifthenelse{\equal{#2}{}}%
   {\ensuremath{\left\| \: \right\|_p}}%
   {\ensuremath{\left\| #2 \right\|_p}}%
 }%
 {}%
}

\newcommand{\pnorm}[2][]%
{%
 \ifthenelse{\equal{#1}{text}}
 { \ifthenelse{\equal{#2}{}}
   {\ensuremath{\,\| \: \|_p\,}}
   {\ensuremath{\,\|  #2 \|_p\,}}
 }%
 {}%
 \ifthenelse{\equal{#1}{big}}%
 { \ifthenelse{\equal{#2}{}}%
   {\ensuremath{\,\big\| \: \big\|_p\,}}%
   {\ensuremath{\,\big\| #2 \big\|_p\,}}%
 }%
 {}%
 \ifthenelse{\equal{#1}{Big}}%
 { \ifthenelse{\equal{#2}{}}%
   {\ensuremath{\,\Big\| \: \Big\|_p\,}}%
   {\ensuremath{\,\Big\| #2 \Big\|_p\,}}%
 }%
 {}%
 \ifthenelse{\equal{#1}{bigg}}%
 { \ifthenelse{\equal{#2}{}}%
   {\ensuremath{\,\bigg\| \: \bigg\|_p\,}}%
   {\ensuremath{\,\bigg\| #2 \bigg\|_p\,}}%
 }%
 {}%
 \ifthenelse{\equal{#1}{Bigg}}%
 { \ifthenelse{\equal{#2}{}}%
   {\ensuremath{\,\Bigg\| \: \Bigg\|_p\,}}%
   {\ensuremath{\,\Bigg\| #2 \Bigg\|_p\,}}%
 }%
 {}%
 \ifthenelse{\equal{#1}{}}%
 { \ifthenelse{\equal{#2}{}}%
   {\ensuremath{\,\left\| \: \right\|_p\,}}%
   {\ensuremath{\,\left\| #2 \right\|_p\,}}%
 }%
 {}%
}

\newcommand{\quasinorm}[2][]%
{%
\ifthenelse{\equal{#1}{text}}%
{\ensuremath{\,\rfloor\negthickspace\rfloor#2\lfloor\negthickspace\lfloor\,}}%
{}%
\ifthenelse{\equal{#1}{big}}%
{\ensuremath{\,\big\rfloor\negthickspace\big\rfloor#2\big\lfloor\negthickspace\big\lfloor\,}}%
{}%
\ifthenelse{\equal{#1}{Big}}%
{\ensuremath{\,\Big\rfloor\negthickspace\Big\rfloor#2\Big\lfloor\negthickspace\Big\lfloor\,}}%
{}%
\ifthenelse{\equal{#1}{bigg}}
{\ensuremath{\,\bigg\rfloor\negthickspace\bigg\rfloor#2\bigg\lfloor%
\negthickspace\bigg\lfloor\,}%
}%
{}%
\ifthenelse{\equal{#1}{Bigg}}%
{\ensuremath{\,\Bigg\rfloor\negthickspace\Bigg\rfloor#2\Bigg\lfloor\negthickspace\Bigg\lfloor\,}}%
{}%
\ifthenelse{\equal{#1}{}}%
{\ensuremath{\,\left\rfloor\negthickspace\left\rfloor#2\right\lfloor\negthickspace\right\lfloor\,}}%
{}%
}

\newcommand{\textquasinorm}[1]%
{\ensuremath{\,\rfloor\negthickspace\rfloor#1\lfloor\negthickspace\lfloor\,}}

\newcommand{\quasinorM}[2][]%
{%
\ifthenelse{\equal{#1}{text}}%
{\ensuremath{\rfloor\negthickspace\rfloor#2\lfloor\negthickspace\lfloor}}%
{}%
\ifthenelse{\equal{#1}{big}}%
{\ensuremath{\big\rfloor\negthickspace\big\rfloor#2\big\lfloor\negthickspace\big\lfloor}}%
{}%
\ifthenelse{\equal{#1}{Big}}%
{\ensuremath{\Big\rfloor\negthickspace\Big\rfloor#2\Big\lfloor\negthickspace\Big\lfloor}}%
{}%
\ifthenelse{\equal{#1}{bigg}}
{\ensuremath{\bigg\rfloor\negthickspace\bigg\rfloor#2\bigg\lfloor%
\negthickspace\bigg\lfloor}%
}%
{}%
\ifthenelse{\equal{#1}{Bigg}}%
{\ensuremath{\Bigg\rfloor\negthickspace\Bigg\rfloor#2\Bigg\lfloor\negthickspace\Bigg\lfloor}}%
{}%
\ifthenelse{\equal{#1}{}}%
{\ensuremath{\left\rfloor\negthickspace\left\rfloor#2\right\lfloor\negthickspace\right\lfloor}}%
{}%
}

\newcommand{\metric}[1]%
{\ensuremath{\,|\hspace{-.48ex}\rangle#1\langle\hspace{-.51ex}|\,}}

\newcommand{\qn}[1]%
{\ensuremath{\left\rfloor\negthickspace\left\rfloor#1\right\lfloor\negthickspace\right\lfloor}}

\newlength{\ExtraSpace}

\newcommand{\restricteD}[2]%
{\setlength{\ExtraSpace}{#1}%
\raisebox{-.6ex+\ExtraSpace}{\ensuremath{ {\bigm |}_{#2} }}%
}

\newcommand{\restricted}[1]%
{\restricteD{0pt}{#1}}

\newcommand{\expv}{\ensuremath{\ep}}
\newcommand{\sx}[2]{\ensuremath{\sigma(#1|#2)}}
\newcommand{\vN}[1]{\textit{v}\mathcal{N}}
\newcommand{\vNeu}{\textit{v}\mathcal{N}}

\newcommand{\Linf}{\ensuremath{L^\infty}}
\newcommand{\adj}[1]{\ensuremath{#1^{\dagger}}}
\newcommand{\Borel}{\texttt{Borel}}

\newcommand{\Cite}[1]{#1}

\newcommand{\B}{\mathcal{B}}

\newcommand{\ket}[1]{\left|#1\right\rangle}
\newcommand{\bra}[1]{\left\langle#1\right|}
\newcommand{\Coh}{\textrm{Coh}}

\usepackage{pslatex}

\usepackage{a4}
\usepackage{fancyhdr}
\newcommand{\Index}[1]{#1}
\newcommand{\Appref}{\ref}
\newcommand{\Exp}{\mathcal{E}}

\newcommand{\question}[1]{}
\newcommand{\questionst}[1]{}
\newcommand{\HH}{\ensuremath{\mathcal{H}}}
\begin{document}

\begin{titlepage}
 \hbox{}
 \vspace{10em}
 \begin{center}
  \Huge\textbf{Wave-particle duality\\ in the \\ damped harmonic oscillator} 
 \end{center}
  \normalsize
 \vfill
 \begin{flushright}
   \begin{minipage}{20em}
      Master thesis Theoretical Physics\\
      Radboud University Nijmegen\\
      Steven Teerenstra\\
      2006
   \end{minipage}
 \end{flushright}

\end{titlepage}



\hbox{}
\newpage
\thispagestyle{empty}
\hbox{}
\vfill
{\sf \large 
\textbf{DE MATHEMATICUS}
\\[2em]
In de tijdsevolutie van Q\\[0.2em]
Waar de Von Neumann algebra gegenereerd wordt\\[0.2em]
Commutatief\\[0.2em]
Gerepresenteerd\\[0.2em]
Op de Wiener ruimte (Gaussische rep van de CCR over L2(R))\\[0.2em]
Waar Qt voldoet aan,\\[0.2em]
Waar Bt is een Brownse beweging met variatie 2n,\\[0.2em]
waar\\[0.2em]
pure\\[0.2em]
po\"ezie\\[0.2em]
ontstaat
\\[1em]
Dit is de integrale versie van de stochastische differentiaal vergelijking
\\[0.2em]
Dus Qt kan ge\"\i dentificeerd worden met\\[0.2em]
Een Ornstein-Uhlenbeck proces\\[0.2em]
Quantum Mechanische\\[0.2em]
Gedempte\\[0.2em]
Harmonische Oscillator\\[0.2em]
Het begrip verloren\\[0.2em]
In woorden\\[0.2em]
vergaan
\\[1em]
Mt is dan te identificeren\\[0.2em]
Met een sterfteproces op\\[0.2em]
L2(Delta,P)\\[0.2em]
De Guichardet ruimte\\[0.2em]
Met Fock representatie\\[0.2em]
Van geduld\\[0.2em]
En toewijding\\[0.2em]
Loyaal in alle dimensies
\\[1em]
De mathematicus 
\\[4em]
\textsl{Rogier Teerenstra, mei 2006}
}
\vfill

\vfill


\hbox{}
\newpage
\thispagestyle{empty}

\chapter*{Guide to the reader}

When writing this text, I had a two-fold aim and audience in mind
which is roughly reflected in the division of the text into a main part
and the appendices.

\bigskip

First of all, I wanted to give a gentle
introduction to (elements of) quantum probability theory using quantization
of the damped harmonic oscillator as \textit{leitmotiv}. 
This is the focus in the main text. 
As audience I had (graduate) students in physics in mind,
people like myself when I started the research of this thesis,
although also others with similar level or interest are welcome to have a
look. 

My aim was to explain some of the physical intuition and motivation
behind the, sometimes overwhelming, math machinery of quantum
probability theory in order to convey some of its intrinsic appeal.
To this end, we start in the main text from a piece of quantum mechanics 
that is treated in almost all basic textbooks: 
the quantization of the (undamped) harmonic oscillator from the
Heisenberg and Schr\"odinger point of view. 
We show how both treatments are special instances of 
a quantum probabilistic quantization procedure: 
the second quantization functor. 
As a corrollary, quantum probability theory is a generalization 
of the quantum mechanics of Heisenberg, Dirac, Schr\"odinger etc.\ 
as we know it.
We then proceed to build up the quantum probability machinery 
needed to quantize the damped harmonic oscillator. 
I have tried to stress the ideas and intuition in the main
text and defer the details of the proofs and reasoning
to the appendices, but I have not been entirely consistent in
that. 

\bigskip

Secondly, this thesis also reflects the research I have done. 
Actually, parts of this research were used and extended in
lectures that my thesis supervisor, dr Hans Maassen, gave in a summer
school in Grenoble in 1998. 
In particular, Proposition 4.1 and Proposition 6.3 of his lecture notes
\cite{Maassen}
correspond to my theorem \ref{th_generalDefGammaFunctor} 
and lemma \ref{lem_PVMofN}. 
The details of their proofs are given 
in appendix \ref{ap_quantizationFunctor} and \ref{ap_PVMofN}, respectively.

It goes without saying that my own writing has benefited
greatly from the theoretical expos\'e that dr Maassen provided in his
lecture notes \cite{Maassen}.

\bigskip

Steven Teerenstra

\part{Harmonic Oscillator}
\chapter{Quantum Mechanical Harmonic Oscillator}

\questionst{add introduction:
We discuss: issue of representations, issue of uncertainty, give a another
quantization procedure

interpretation (see above: vibration around equilibrium);, examples

As we will see: a quantum mechanical system can be obtained from the classical
system via a translation formalism (correspondence principle)

(? As turns out Different: 
 lowest energy is non-zero,  discrete energy spectrum ?)

Observables, states, time evolution:
Time evolution: Dynamical rule: Schr\"odinger equation 

Coherent states!
Energy states are stationary in time (except for a phase factor evolution)
Coherent state resemble minimum uncertainty wave packets that have sinusoidally
motion
Coherent state for 1 h.o., for $n$ independent (uncoupled), 
for $\infty$-independent.
}

\section{Dirac's Quantization procedure}

Following Dirac's quantization rules \cite{Dirac}, 
we translate the (1-dimensional) classical harmonic oscillator 
into a quantum mechanical one:

\begin{description}
 \item[Observables] 
 The classical conjugated (generalized) coordinates 
 for position and momentum, $q$ and $p$,
 are replaced by self-adjoint operators $Q$ and $P$ 
 (acting on some Hilbert space $\mathcal{H})$.
 Any observable of the classical harmonic oscillator $a= f(q,p)$ is
 replaced by its quantum counter part $A= f(Q,P)$ 
 (which is hermitized if necessary). 
 In particular, the
 Hamiltonian of the quantum mechanical oscillator is 
 \begin{displaymatH}
  H= \frac{P^2}{2m} + \frac{m\om^2 Q^2}{2}
  \, .
 \end{displaymatH}

 \item[Correspondence principle]
 The relation between two quantum observables $A=f(Q,P)$ and
 $B=g(Q,P)$ of the oscillator system 
 is expressed by their commutator $[A,B]:=AB-BA$.
 If $a=f(q,p)$ and $b=g(q,p)$ are the corresponding classical observables
 having Poisson bracket 
 \begin{aligN}
  \{a,b\} =  
  \frac{\partial}{\partial q}f(q,p) \frac{\partial}{\partial p}g(q,p)
   - 
  \frac{\partial}{\partial q} g(q,p) \frac{\partial}{\partial p} f(q,p)
  \,
 \end{aligN}
 then the correspondence principle holds:
 \begin{equatioN}
 \label{eq_correspondence_principle}
  [A,B]= i\hbar \{a,b\}\indicator
  \,.
 \end{equatioN}
 As a special case, Heisenberg's  commutation relation emerges
 \begin{equatioN}
 \label{eq_ccr_commutator}
  [Q,P]= i\hbar \indicator \,. 
 \end{equatioN}

 \item[Time evolution]
 If $A=a(Q,P)$ is an observable that does not depend explicitly on time 
 (i.e.\ does not depend on time other than via the time evolution of 
  $Q$ en $P$), 
 then the classical description of the time evolution,
  $\frac{d a}{dt} = \{a,h\}$ where $h$ is the classical Hamiltonian,
 is translated
 via the correspondence principle into
 the \Dindex{Heisenberg equation of motion}
 \begin{equatioN}
  \frac{d A}{dt} = \frac{1}{i\hbar}[A,H]
  \,,
 \end{equatioN}
 having $A(t)= e^{itH/\hbar} A e^{-itH/\hbar}$ as formal solution.

 In particular,
  $\frac{d Q}{dt} = P/m$ and $\frac{d P}{dt} = -Q$.

 \item[States]
 The states of the system are described by (normalized) vectors 
 $\ket{\alpha}$ in the Hilbert space $\mathcal{H}$.
 \question{Hans: het lijkt me op dit moment nog niet geschikt om over
           density matrices te praten}

 \item[Measurements]
 Any single measurement of an oscillator observable $A$ yields a value $a$
 in the spectrum of $A$ (as self-adjoint operator) i.e.\ there is a
 normalized
 eigenstate $\ket{a}$ such that $A\ket{a}=a\ket{a}$.
 Suppose for simplicity that the state of the system is completely
 determined by $A$, then these eigenstates of $A$ form a 
 complete set
 in  $\mathcal{H}$ i.e.\
 every state $\ket{\zeta}$ can be decomposed as a ``superposition'' 
 of eigenstates of $A$:
 \begin{equatioN}
 \label{eq_decompInEigenstates}
  \ket{\zeta}
   =
   \begin{cases}
    \sum_a \ket{a}\ip{a}{\zeta} 
      & \text{if the spectrum of $A$ is discrete,}\\
    \int\!\! da\, \ket{a}\ip{a}{\zeta} 
      & \text{if the spectrum of $A$ is continuous.}\\
  \end{cases}
 \end{equatioN}
 Quantum theory states that when the system is in state  $\ket{\zeta}$,
 the probability of measuring the value $a$ for $A$,
 i.e.\ exactly $a$ in the discrete case or 
 within the resolution $\Delta$ of the measuring apparatus in the continuous
 case, is 
 \begin{equatioN}
 \label{eq_ProbInterpretation}
   \begin{array}{ll}
    |\ip{a}{\zeta}|^2 
    & \text{if the spectrum of $A$ is discrete,}\\[.5em] 
    \int_{a-\Delta/2}^{a+\Delta/2}  |\ip{\widetilde{a}}{\zeta}|^2  
   d\widetilde{a}
     \approx 
    \Delta |\ip{a}{\zeta}|^2 
    & \text{if the spectrum of $A$ is continuous.}
     \\
   \end{array}
 \end{equatioN}
 
 Directly after having measured the observable $A$, 
 the system is in the eigenstate  $\ket{a}$ belonging to the value $a$
 (\Dindex{collapse of the wave function}).
 The probability of measuring another value $a'\neq a $ directly after, 
 is then zero, 
 since $a\ip{a}{a'}= \ip{Aa}{a'} = \ip{a}{\adj{A}}{a'} = \ip{a}{Aa'}= a' \ip{a}{a'}$ which implies
 (as either $a\neq 0 $ or $a'\neq 0$) that $\ip{a}{a'}=0$.

 \item[Expected value]
 On average, measurements of $A$ on identical systems, all in state 
 $\ket{\zeta}$, will yield an average value of the observable $A$, 
 which is called the \Dindex{expected value} of $A$ 
 and is denoted by $\langle A \rangle_{\zeta}$.

 Quantum theory states that 
 $\langle A \rangle_{\zeta} = \oip{\zeta}{A}{\zeta}/\ip{\zeta}{\zeta}$, which can be
 justified from the probabilistic interpretation 
 (\ref{eq_ProbInterpretation}) above.
 For instance, 
 if $A$ has a complete set of eigenvectors and $\zeta$ is normalized, 
 then
 \begin{aligN}
  \oip{\zeta}{A}{\zeta}
   = 
   \sum_{a,\widetilde{a}} 
         \ip{\zeta}{a}\oip{a}{A}{\widetilde{a}}\ip{\widetilde{a}}{\zeta}
    = \sum_{a,\widetilde{a}} 
         \ip{\zeta}{a} a \de_{a,\widetilde{a}}\ip{\widetilde{a}}{\zeta}
    = \sum_a a |\ip{\zeta}{a}|^2
 \end{aligN}
 in  the discrete case, while
 \begin{aligN}
  \oip{\zeta}{A}{\zeta}
   \ats = 
   \int\!da\; d\widetilde{a}\; 
         \ip{\zeta}{a}\oip{a}{A}{\widetilde{a}}\ip{\widetilde{a}}{\zeta} 
     =
   \int\!da\; d\widetilde{a}\; 
         \ip{\zeta}{a} a\de(a-\widetilde{a})\ip{\widetilde{a}}{\zeta} 
     \\
    \ats = \int a |\ip{\zeta}{a}|^2 da 
  \end{aligN}
 in the continuous case.
 In both cases, we see that the possible outcomes of measurements $a$
 are weighted by their corresponding probabilities in state $\zeta$.

 \item[Maximal system of commuting observables: quantum numbers]
 \questionst{Necessary to discuss this? Or shorter: sometimes eigenstates
  correspond to distinct eigenvalues, then measuring determines system.
 Else adding other observables, that can be simultaneously be measured
 such that the joint eigenstates do have distinct labels}

 For the one dimensional harmonic oscillator, every eigenvalue $q$ of 
 the position operator $Q$ is non-degenerate i.e.\ there is essentially
 one eigenstate $\ket{q}$ such that $Q\ket{q}= q\ket{q}$:  
 all vectors $\psi\in\mathcal{H}$ that satisfy $Q\psi = q \psi$ 
 are scalar multiples of $\ket{q}$. 
 As a convenient consequence, there is an observable $A$ 
 such that measuring the observable means that we know
 that the system can only be in one state, viz.\ the (one and only) eigenstate
 $\ket{a}$ that corresponds to the measured eigenvalue $a$.
 In general, it is rare for a quantum system to have one observable that
 completely determines the system in this sense.
 More often than not there are more essentially different (i.e.\
 linearly independent) eigenvectors 
 for a given eigenvalue $a$ i.e.\
 the eigenspace of $a$ consisting of all vectors in $\mathcal{H}$ on
 which $A$ acts as multiplication by $a$ is not one-dimensional.
 For example, the $x$-position of a two-dimensional oscillator 
 does not uniquely determine its state.
 In such cases, one looks for observables $\{A_i\}_i$ that 
 commute pairwise: $[A_i,A_j]=0$. 
 For such observables, an eigenvector of one $A_i$
 is at the same time an eigenvector of another $A_{j}$
 \cite[p.\ 207]{Bransden+Joachain}, \cite[p. 30]{Sakurai}, and therefore a  
 simultaneous eigenvector. In particular, such observables can be measured
 simultaneously. 
 The simultaneous eigenvectors can be labeled
 by their eigenvalues $a_1, a_2, \ldots$ 
 with respect to the commuting set of observables $A_1, A_2, \ldots$ i.e.\
 $A_i\ket{a_1, a_2, a_3,\ldots}=a_i\ket{a_1, a_2, a_3,\ldots}$.
 A set of commuting observables $\{A_i\}_i$ 
 is called maximal if there is no observable $B$ such that 
 $[B,A_i]=0$ for all $i$ (and, of course, $B$ should not be a function
 of the $\{A_i\}_i$). 
 If we have a maximal system of commuting observables $\{A_i\}_i$ 
 then the joint eigenvalues $(a_i)_i$, enable a unique labeling 
 of the joint eigenstates $\ket{a_1, a_2, \ldots}$.
 These labeling eigenvalues $a_1,a_2,\ldots$ are referred to as 
 quantum numbers.
 In that case, measuring all the $\{A_i\}_i$ simultaneously 
 determines the state of 
 the system completely, namely $\ket{a_1,a_2, \ldots}$.
 E.g.\ in case of the two dimensional harmonic oscillator $\{Q_x, Q_y\}$,
 $\{P_x, P_y\}$, $\{Q_x, P_y\}$, $\{Q_y, P_x\}$ all form a maximal system
 of commuting observables.
\end{description}

\section{Representations of the Canonical Commutation Relations and 
         the problem of uniqueness}

The above procedure does not provide a clue as to which Hilbert space
one could choose and how the self-adjoint observables $Q$, $P$, and $H$ 
of the oscillator system would look like. 
Starting from a set of generalized coordinates of the system,
some general clues can be given, but as those clues have a high 
``who was first: the chicken or the egg'' content, 
first quantization remains all in all a mystery.

\paragraph{First quantization}
In general, first quantization of a system follows the following steps.
\begin{Enumerate}
 \item Take a set $X$ representing the degrees of freedom of the system i.e.\
       if the system has $n$ degrees of freedom and generalized coordinates
       $q_1,\ldots, q_n$ with $q_i \in X_i$, then for $X$ we can take
       $X=\Pi_i X_i = \{(q_1,\ldots,q_n): q_i \in X_i \text{ for all } i \}$.

       Examples:
      \begin{Itemize}{-}
       \item       
       if we are considering a free particle with 3 degrees of freedom
       and take for $(q_1,q_2,q_3)$ the $x$, $y$ and $z$ position of 
       the particle (or if we take for $(q_1,q_2,q_3)$ the $x$, $y$ and $z$ 
       composition of the momentum):
       $X=\RR^3$.
       \item
       if we are considering a spin $\half$-particle (so 1 degree 
       of freedom $q\in\{-\half,\half\}$):
       $X=\{-\half, \half\}$.
      \end{Itemize}
      
 \item Form the Hilbert space $\mathcal{H}=L^2(X)$ consisting of the complex square
       integrable functions on $X$.
 
 \item The quantization of the system is given by:
       \begin{Itemize}{-}
        \item states of the system at time $t$ 
              are (wave) functions $\psi(X,t) \in L^2(X)$ 
              with 
	      \begin{displaymatH}
	       \int_X |\psi(x,t)|^2 \,dx =1
                \,,
	      \end{displaymatH}
        \item observables of the system are self-adjoint operators 
              on $L^2(X)$.
              In particular, 
              the observable $Q_i$ attached to measurement of the $i^{th}$
              generalized coordinate is diagonal: 
              $Q_i$ is the multiplication by $q_i$ in $L^2(X)$.
       \end{Itemize}
\end{Enumerate}
\bigskip

Using the above hints, 
we now recall two familiar representations 
of the quantum mechanical harmonic oscillator: 
the Schr\"odinger and Heisenberg representation.

\subsection{Schr\"odinger representation}
As generalized coordinate, we take the position $q$ of the oscillator. 
Then $X=\RR$, $\mathcal{H}=L^2(\RR)$, and $Q=M_x$ i.e.\ the multiplication 
operator by $x$ in $L^2(X)$. 
As $Q$ is to be a self-adjoint operator, 
we must specify a domain and we could, for example, take a ``maximal'' domain:
\begin{displaymatH}
 (Q\psi)(x) = x\psi(x)
 \quad
 \text{ for } \psi\in 
 \mathcal{D}_Q :=\{\psi \in L^2(\RR) : \int_\RR |x\psi(x)|^2 dx <\infty\}
 .
\end{displaymatH}
For the conjugated coordinate $P$, one could take for example 
the derivation operator
with a factor $\hbar/i$ thrown in 
(no excusing explanation offered here) and again a ``maximal'' domain:
\begin{displaymatH}
 (P\psi)(x) = \frac{\hbar}{i}\frac{d}{dx}\psi(x) = -i\hbar\psi'(x)
 \quad
 \text{ for } \psi\in 
 \mathcal{D}_P :=\{\psi \in L^2(\RR) : \int_\RR |\psi'(x)|^2  <\infty\}
 .
\end{displaymatH}
If one feels up to it, 
then one can check that, with the above definitions,
$Q$ and $P$ are indeed self-adjoint and that the commutation relation 
$[Q,P]=i\hbar\indicator$ holds on the intersection of their domains.

The Hamiltonian $H= \frac{P^2}{2m} + \frac{m\om^2 Q^2}{2}$ boils down to
\begin{displaymatH}
   H= -\frac{\hbar^2}{2m} \frac{d^2}{d x^2} + \frac{m\om^2 x^2}{2}
   .
\end{displaymatH}

At this moment, we just want to note that there is a high degree of
arbitrariness in the choice of $Q$, $P$, $H$ and their domains.

Also note that the symmetry between $P$ and $Q$ in the Hamiltonian
$H= \frac{P^2}{2m} + \frac{m\om^2 Q^2}{2}$ is maximal when 
(units are chosen such that) $m=\om=1$. Then
$H= \frac{P^2}{2} + \frac{Q^2}{2}$.

\subsection{Heisenberg representation}
As generalized coordinate we now take the energy $e$ of the oscillator.
Then $X\subset [0,\infty)$ and $H$ is a multiplication operator on 
$L^2(X)$, but matters turn out to be not as simple as putting $X=[0,\infty)$
and $H=M_x$. 

To deduce Heisenberg's representation of the harmonic oscillator,
most textbooks on quantum mechanics (e.g.\ \cite[p.90-92]{Sakurai}) follow 
the \Dindex{factorization method} due to Dirac who rewrote
the Hamiltonian as 
\begin{displaymatH}
 H= \hbar\om(\adj{A}A + \half)
 ,
\end{displaymatH}
with
\begin{equation}
 \label{def_LadderOp}
 A:= \sqrt{\frac{m\om}{2\hbar}} \Bigl(Q+\frac{iP}{m\om}\Bigr)
 \quad\text{and }\quad
 \adj{A}= \sqrt{\frac{m\om}{2\hbar}} \Bigl(Q-\frac{iP}{m\om}\Bigr)
 .
\end{equation}
By formal calculation, 
the anti self-adjoint operators $\adj{A}$ and $A$ are then shown to 
lower and raise the eigenvalue of $H$ respectively, and are therefore
called \Dindex{annihilation} and \Dindex{creation} operators.
In particular, the set of eigenvalues of $H$ (i.e.\ the quantum energy levels) 
turns out to be discrete, consisting of 
$E_n= \hbar\om(n+\half)$, $n\in\NN:=\{0,1,\ldots\}$ 
and the action of $\adj{A}$ and $A
$ on the corresponding eigenstates $\ket{n}$,
$n\in\NN$, is given by 
\begin{equation}
\label{eq_ladder_op_ho}
 \begin{aligned}
  \adj{A}\ket{n} & =\sqrt{n+1}\ket{n+1} \quad(n=0,1,\ldots)\,,\\ 
          A\ket{n} & =\sqrt{n}\ket{n-1} \quad(n=1,2, \ldots)\,,\\
          A\ket{0} &= 0\,, \\
 \end{aligned}
\end{equation}
i.e.\ the action of $A$ on the state (of an existing oscillator system) with
no energy quanta is to destroy the system (indicated by $0$).
Although Dirac's algebraic method does not use an explicit representation
it is easy to conceive one by taking $X=\NN$, 
$\mathcal{H}=\ell^2(\NN)=\{(\zeta_0,\zeta_1, \ldots) : \sum_{k=0}^\infty |\zeta_n|^2 <\infty\}$
and 
\begin{aligN}
  \ket{0} & =(1,0,0,0,\ldots) \\
  \ket{1} & =(0,1,0,0,\ldots) \\
  \ket{2} & =(0,0,1,0,\ldots)\\
  \vdots
\end{aligN}
Using that  $Q=\sqrt{\hbar/2m\om}(\adj{A}+A) $, 
$P=\sqrt{\hbar/2m\om}(\adj{A}-A)$, 
$H=\hbar\om( \adj{A}A + \frac{\hbar}{2}\indicator)$ 
and the known action of $A$, $\adj{A}$
(\ref{eq_ladder_op_ho}) one obtains

\begin{displaymath}
\,\hspace{0.5em}
Q = 
\sqrt{\frac{\hbar}{2m\om}}
\left[\;
\begin{matrix}
 0       & 1        & 0        & 0        & 0         & \ldots \\
 1	 & 0        & \sqrt{2} & 0        & 0         & \ldots \\
 0	 & \sqrt{2} & 0        & \sqrt{3} & 0         & \ldots \\
 0	 & 0        & \sqrt{3} & 0        & \sqrt{4}  & \ldots \\
 0	 & 0        & 0        & \sqrt{4} & 0         & \ldots \\
 \vdots  & \vdots   & \vdots   & \vdots   & \vdots    & \ddots \\
\end{matrix}
    \;\right] 
\!,\hspace{1em}
\end{displaymath}
\begin{displaymath}
P = 
\sqrt{\frac{\hbar}{2m\om}}
\frac{1}{i}
\left[\;
\begin{matrix}
 0       & 1        & 0        & 0        & 0         & \ldots \\
 -1	 & 0        & \sqrt{2} & 0        & 0         & \ldots \\
 0	 &-\sqrt{2} & 0        & \sqrt{3} & 0         & \ldots \\
 0	 & 0        &-\sqrt{3} & 0        & \sqrt{4}  & \ldots \\
 0	 & 0        & 0        &-\sqrt{4} & 0         & \ldots \\
 \vdots  & \vdots   & \vdots   & \vdots   & \vdots    & \ddots \\
\end{matrix}
    \;\right]
\!\!,
\end{displaymath}
and
\begin{displaymath}
H = 
\frac{\hbar \om}{2}
\left[
\begin{matrix}
 1 & 0 & 0 & 0 & 0 & \ldots \\
 0 & 3 & 0 & 0 & 0 & \ldots \\
 0 & 0 & 5 & 0 & 0 & \ldots \\
 0 & 0 & 0 & 7 & 0 & \ldots \\
 0 & 0 & 0 & 0 & 9 & \ldots \\
 \vdots & \vdots & \vdots &  \vdots & \vdots & \ddots  \\
\end{matrix}
\right]
.
\end{displaymath}

\subsection{Uniqueness of representations \label{ssec_UniquenessRep}} 
Having now (at least) two representations of the quantum mechanical 
harmonic oscillator, the question arises whether the are ``the same''
i.e.\ whether the ``physical content'' of both representations is the same.
By the ``same'', 
we mean of course that the observable physical quantities 
calculated in either representation are equal.
In quantum theory, 
the physical observable quantities of a system 
are the expected values
$\langle A \rangle = \oip{\zeta}{A}{\zeta}$ of the system's observables $A$ 
in a given state of the system $\zeta$.
It can be shown then \cite[p.\ 210-212]{Bransden+Joachain}
that representations have the same
physical content if they are \Dindex{unitarily equivalent} i.e.\
if states $\zeta$ and observables $A$ in the one
representation are related to states $\zeta'$ and observables $A'$ in the
other by one and the same unitary map $U$:
\begin{displaymatH}
  A \mapsto A'= UAU^{-1}, \qquad \zeta\mapsto \zeta'=U\zeta
  \:.
\end{displaymatH}

Although perhaps not evident at this point, the Schr\"odinger and Heisenberg
representation of the harmonic oscillator are indeed unitarily equivalent.
We will show this later, after having developed theory which makes
both representations simple examples of  general ``tools'' to make 
representations (that are automatically equivalent).

For the moment, however, we confine ourselves to noting that it has been rather 
coincidental that we have found two pairs $(Q,P)$ that are
unitarily equivalent indeed. To illustrate this statement, 
we will provide a reasonable definition of canonical pairs and
show that such canonical pairs are not necessarily unitarily equivalent.

\subsubsection{A reasonable but failing definition of canonical pairs}
Let us start with the the following reasonable definition.

\begin{definition}[Heisenberg canonical pair]
\label{def_Heisenberg_ccr}
 A pair $(Q,P)$ of self-adjoint operators on a Hilbert space $\mathcal{H}$ 
 is called a (Heisenberg) \Dindex{canonical pair} if there exists a dense domain 
 $D\subset\mathcal{H}$ so that
 \begin{conditions}
  \item 
  $Q:D\to D$, $P:D\to D$ (i.e.\ $D$ is stable),
  \item
  $QP\phi - PQ\phi = i\phi$ for all $\phi \in D$,
  \item
  $Q$ and $P$ are essentially\footnote{i.e.\ although $Q$ and $P$ are not necessarily self-adjoint 
               on the domain $D$, 
               they do have unique self-adjoint extensions.} 
   self-adjoint on $D$.
 \end{conditions}

 Two canonical pairs $(Q,P)$ on  $\mathcal{H}$ and 
 $(Q',P')$ on  $\mathcal{H'}$ are called unitarily equivalent if there 
 exists a unitary map $U:\mathcal{H} \to \mathcal{H'}$ such that
 $Q'= U Q U^{-1}$ and $P'= U P U^{-1}$.
 (In particular, this imposes domain requirements:
  $U^{-1} \mathcal{D}_P = \mathcal{D}_{P'}$ and 
  $U^{-1} \mathcal{D}_Q = \mathcal{D}_{Q'}$).
\end{definition}

The following example, due to Nelson \cite[p.\ 275]{Reed+Simon:FA}, 
\cite[Remark 3, p.~90 ; Problem 4, p.~94]{Thirring},
shows that the above definition does not uniquely determine one pair of
self-adjoint operators.

\begin{example}[Nelson]
 Let $M$ be the Riemann surface of $\sqrt{z}$ and let $\mathcal{H}=L^2(M)$
 equipped with the Lebesgue measure (locally).
 Let 
 \begin{displaymath}
  Q:=x\,, \; P:=\frac{1}{i}\frac{\partial}{\partial x}\,,
  \qquad\text{and}\qquad
  Q':=x+ \frac{1}{i}\frac{\partial}{\partial y}\,, 
  \; P':=\frac{1}{i}\frac{\partial}{\partial x}\,,
 \end{displaymath}
 be defined  on the domain $D$ that consists of all infinitely 
 differentiable functions with compact domain not containing the origin.

 Then $(Q,P)$ and $(Q',P')$ are canonical pairs that are not 
 unitarily equivalent. In fact, the unitary groups generated by $Q'$ and $P'$ 
 respectively do not satisfy the Weyl relations \ref{eq_ccr_Weyl} further on, 
 while the unitary groups generated by $Q$ and $P$ do.
\end{example}

\subsubsection{The Weyl-Von Neumann formulation of the canonical commutation relations}

Apparently, the canonical commutation relation 
as formulated in (\ref{eq_ccr_commutator}), p.\ \pageref{eq_ccr_commutator},
does not determine the position and momentum observable of a quantum system
uniquely.
To circumvent the domain complications associated with the self-adjoint
character of $Q$ and $P$, 
Weyl's idea was to formulate the canonical commutation relation 
(\ref{eq_ccr_commutator}) of $Q$ and $P$
in terms of bounded operators (which are everywhere defined on $\mathcal{H}$). 
For this he used the one-to-one
correspondence between self-adjoint operators and one-parameter,
strongly continuous unitary groups.

\begin{theorem}[Stone's theorem]
Let $\mathcal{H}$ be a Hilbert space.
If $A$ is a self-adjoint operator on $\mathcal{H}$, then
$U_t:=e^{itA}$ is a unitary operator, $U_{t+s}=U_t U_s$ for all $s,t\in\RR$ 
(i.e.\ $(U_t)_t$ is a one-parameter group),
 and  $\phi\in\mathcal{H}$ and $t\to t_0$ implies $U_t\phi\to U_{t_0}\phi$
(i.e.\ $t\mapsto U_t$ is strong continuous).

Conversely, if $U_t$ ($t\in\RR$) is a 
strongly continuous 1-parameter unitary group
on $\mathcal{H}$, then there is a self-adjoint operator
$A$ on $\mathcal{H}$ so that 
\begin{displaymatH}
 U_t = e^{itA}
 .
\end{displaymatH}
See \cite[p.\ 265-267]{Reed+Simon:FA} for a proof.
\end{theorem}

Inspired by this correspondence, 
Weyl proposed to formulate the canonical commutation relation 
(\ref{eq_ccr_commutator}) of $Q$ and $P$
in terms of the associated unitary groups.
Letting 
\begin{displaymatH}
 T_t:= e^{-itP}, \quad\text{and}\quad S_s:=e^{isQ},
\end{displaymatH}
and formally
applying the Campbell-Baker-Hausdorff formula (\ref{BCH-formula}):
\begin{equatioN}
 \exp(A+B))= \exp(A)\exp(B)e^{-\half[A,B]}
 \quad\text{if } [A,B] \text{ is a scalar,}
\end{equatioN}
one \textsl{formally} derives the Weyl relation:
\begin{equation}
\label{eq_ccr_Weyl}
 T_t S_s = e^{-i\hbar st} S_s T_t \quad(s,t\in\RR).
\end{equation}

Taking this one step further, Von Neumann combined the two unitary groups
into a two-parameter family of \Dindex{Weyl operators}:
\begin{equation}
\label{eq_WeylOp_TtSs}
  W(t,s):= e^{\frac{i\hbar}{2}st} T_t S_s 
         = e^{-\frac{i\hbar}{2}st}  S_s T_t
 \quad(s,t\in\RR),
\end{equation}
that satisfy the Weyl relation as expressed by
\begin{equation}
 W(t,s)W(v,u) = e^{-\frac{i\hbar}{2}(tu-sv)} W(s+u,t+v)
 \quad(s,t,u,v\in\RR)
 .
\end{equation}
Note that by the Baker-Campbell-Hausdorff formula
\begin{displaymatH}
e^{ix(A+B)}=e^{ixA}e^{ixB}e^{-\half [iA,iB]}
\quad\text{(see (\ref{BCH-formula}))}
\end{displaymatH}
we have 
\begin{equation}
 W(t,s) = e^{isQ-itP}.
\end{equation}

To compact notation and prepare for generalization further on, 
we reformulate the Weyl operators by writing
$z=t+is\in\CC$
\begin{equation}
\label{eq_WeylOpSchroRep}
 W(z) = W(t+is)= W(t,s)=e^{isQ -itP} = e^{i\Im(z)Q - i\Re(z)P}
      = e^{\frac{i\hbar}{2}st} T_t S_s 
     = e^{-\frac{i\hbar}{2}st}  S_s T_t
 . 
\end{equation}
The Weyl relation then takes the form
\begin{equation}
\label{eq_WeylRelationWeylOp}
 W(z)W(w) = e^{-i\frac{\hbar}{2}\Im(\overline{z}w)} W(z+w)
 \quad 
 (z=t+is, w=v+iu \in\CC)
 .
\end{equation}

\begin{definition}[Weyl canonical pair]
When we refer to a family of  unitary operators 
$W(z)$, $z\in\CC$, as \Dindex{Weyl operators}, then we implicitly
assume that the family is strongly continuous and that 
the Weyl relation (\ref{eq_WeylRelationWeylOp}) holds.

The pair of generators $Q$ and $P$ of the strongly continuous unitary groups
$s\mapsto W(is)$ and $t\mapsto W(t)$ respectively, are called a 
Weyl canonical pair.
\end{definition}

In contrast to the Heisenberg form of the commutation relation,
the Weyl form of the commutation relation \textsl{does} uniquely determine
the two unitary one-parameter groups $T_t$, $t\in\RR$, and $S_s$, $s\in\RR$,
or, equivalently, the two-parameter family $W(z)$, $z\in\CC$
(\cite[p.\ 279]{Reed+Simon:FA}, \cite[p.\ 32-35]{Maassen}).

Furthermore, a Weyl canonical pair satisfies the conditions of
(\ref{def_Heisenberg_ccr}) \cite{Prugovecki}, \cite[p.\ 275]{Reed+Simon:FA}. 
In conclusion,
Weyl canonical pairs satisfy Heisenberg's form of the canonical
commutation relation, but are in addition unique (up to unitary equivalence).
For the cases we consider we will infer this uniqueness property from a
general purpose tool that we will need and develop later.

For the time being, let us note the following:
\begin{Itemize}{-}
 \item
 In the Schr\"odinger representation of the harmonic oscillator 
 ($\mathcal{H}=L^2(\RR)$),
 we have that $T_t:= e^{-itP}$ is a shift to the right (over an amount
 $\hbar t$), 
 $S_s:=e^{isQ}$ is the multiplication by $e^{isx}$ and the 
 Weyl operators are 
 $W(t+is): = e^{isQ-itP}= e^{\frac{i\hbar}{2}st} T_t S_s$.
 The commutation relation between $Q$ and $P$ is expressed
 by the Weyl relation $W(u)W(z)= e^{-\frac{i\hbar}{2}\Im(\overline{u}z}) W(u+z)$
 for the Weyl operators.
 \item
 Similarly, in the Heisenberg representation, 
 a family of Weyl operators $z\mapsto W(z)$ exists
 such that $Q$ is the matrix of the generator
 of $s\mapsto W(is)$ in the energy basis 
 and such that $P$ the matrix of the generator of $t\mapsto W(t)$.
 \item 
 The operators $Q'$ and $P'$ in Nelson's example above do not
 satisfy the Weyl relation.
\end{Itemize}

As a consequence of the general theory we develop later, 
the Schr\"odinger and Heisenberg representation of the harmonic oscillator
are unitarily equivalent, and the ``representation'' derived from
Nelson's example is unitarily equivalent to neither of them.

\section{Coherent states}

The non-vanishing commutator of $Q$ and $P$ prohibits to measure both
simultaneously with arbitrary precision.
Striving to maximize precision in simultaneous measurement of $(Q,P)$,
one arrives at coherent states, which first appeared 
in Schr\"odinger's treatment of the harmonic oscillator 
\cite{Schrodinger}.
Coherent states turn out to be closely related to states 
with minimal and equal uncertainty in $Q$ and $P$. 
As we will see, coherent states furnish an context in which it is easy
to establish
the equivalence of the Schr\"odinger and Heisenberg representation of the
harmonic oscillator.
Apart from that, coherent states form the link to classical states and,
more importantly in our context,  
a natural entry point to second quantization which is to be discussed in the
next section. In fact, we will be build all our theory on coherent states.

\subsubsection{The uncertainty principle for $Q$ and $P$}

Let us first recall the proof of the uncertainty relation 
$\Delta Q \cdot \Delta P \geq \frac{\hbar}{2}$.

For an observable $A$ of a system in state $\ket{\psi}$, 
the uncertainty $\Delta A$ of $A$ in state $\ket{\psi}$ is defined via
\begin{displaymatH}
 (\Delta A)^2 
   = \left\langle \bigl(\:A- \langle A \rangle\:\bigr)^2 \right\rangle
   = \oip[big]{\psi}{\bigl(\:A- \oip{\psi}{A}{\psi}\: \bigr)^2}{\psi}
   ,
\end{displaymatH}
i.e.\ the average deviation%
\footnote{more precisely, it is the root-mean square deviation} 
of $A$ from its expected value 
$\langle A \rangle= \oip{\psi}{A}{\psi}$ in state $\ket{\psi}$.

If $A$, $B$ are two observables (self-adjoint operators),
then their uncertainties $\Delta A$, $\Delta B$ are related
to their commutator. Indeed,
as the quadratic polynomial
\begin{displaymatH}
 f(\alpha):= \alpha^2 \langle A^2 \rangle + \alpha \langle i[A,B] \rangle 
           + \langle B^2 \rangle
      = \ip{(\alpha A + iB)\psi}{(\alpha A + iB)\psi} 
      = \norM{\alpha A + iB)\psi}^2 
 \quad(\al\in\CC)
  ,
\end{displaymatH}
is non-negative, 
its discriminant 
$(\langle i[A,B] \rangle)^2 - 4 \langle A^2 \rangle \langle B^2 \rangle$
must be non-positive:
\begin{displaymatH}
 \langle A^2 \rangle \langle B^2 \rangle 
   \geq 
\frac{1}{4} |\langle [A,B] \rangle|^2 
 .
\end{displaymatH}
Replacing $A$ and $B$ with $(A-\langle A \rangle)$ and 
$(B-\langle B \rangle)$ respectively, and observing that
\begin{displaymatH}
 [A-\langle A \rangle \,,\, B-\langle B \rangle]=[A,B]
 ,\qquad
  \text{(as } \langle A \rangle \text{ and } \langle B \rangle
  \text{ are scalars),}       
\end{displaymatH}
we find the uncertainty product relation:
\begin{equation}
\label{eq_product_uncertainty}
 (\Delta A)^2 (\Delta B)^2 
   \geq 
\frac{1}{4} |\langle [A,B] \rangle|^2 
 \quad \text{or equivalently,}\quad
 \Delta A \cdot \Delta B
   \geq 
 \frac{1}{2} |\langle [A,B] \rangle|
  .
\end{equation}

Note from the above argument that minimum product uncertainty arises,
when the discriminant of $f$ is equal to zero, i.e.\ when
$f$ has one zero. In that case there is an $\alpha\in\CC$ such that
 $(\alpha \widehat{A} + i \widehat{B})\ket{\psi}=0$, where
$\widehat{A}=A- \langle A \rangle$ and $\widehat{B}=B- \langle B \rangle$.

\medskip

We now apply the above knowledge to the pair $(Q,P)$.
First, the familiar product uncertainty relation 
$\Delta Q \cdot \Delta P \geq \frac{\hbar}{2}$ emerges.
Secondly, states of minimal (product) uncertainty in $Q$ and $P$ arise when 
$(\al \widehat{Q} + i\widehat{P})\ket{\psi} = 0$.

Observe that if $\ket{\psi}$ is a state with 
$\langle Q \rangle_\psi = t$ and $\langle P \rangle_\psi = s$, then 
we can translate it via $T_{-t/\hbar}=e^{itP/\hbar}$ and 
$S_{-s/\hbar}=e^{-isQ/\hbar}$
to obtain a state $\ket{\psi_0}$ with
$\langle Q \rangle_{\psi_0} = \langle P \rangle_{\psi_0} = 0$.
We then  have (see \Appref{ap_alphaQ+iP}) that $\ket{\psi}$ is a solution of
\begin{equatioN}
 (\al \widehat{Q} + i\widehat{P})\ket{\psi} 
  = \Bigl(\al(Q-t) + i(P-s)\Bigr)\ket{\psi}  = 0
  \tag{$*$}
\end{equatioN}
if and only if $\ket{\psi_0}$ is a solution of
\begin{equatioN}
 (\al Q + iP)\ket{\psi_0}  =  0.
 \tag{$**$}
\end{equatioN}
Therefore, we can find the solutions of $(*)$ by
translating solutions of $(**)$ in position and momentum.

Let us find the minimal uncertainty states with 
$\langle Q \rangle = \langle P \rangle = 0$ in the position representation
(so $\widehat{P}=P=\hbar/i\frac{d}{dq}$ and $\widehat{Q}=Q$ is diagonal with eigen states 
$Q\ket{q}=q\ket{q}$, $q\in\RR$).
The minimum product uncertainty condition then reads:
\begin{equation}
\label{eq_alphQ+iP}
  (\al Q  + iP))\psi_0(q) = \al q \psi_0(q) + \hbar \frac{d \psi_0}{dq} (q) = 0
   \quad
   (q\in\RR), 
\end{equation}
where $\psi_0(q)=\ip{q}{\psi_0}$ is the position-representation of state 
$\ket{\psi_0}$.
The solution of this differential
equation is $\exp(-\frac{\al}{2\hbar} q^2)$ 
which is normalizable if $\Re(\al) > 0$.
Including an appropriate normalization constant 
(which we take to be real-valued), 
the solution reads 
\begin{displaymatH}
 \psi_0(q)= \sqrt[4]{\frac{\Re(\al)}{\pi\hbar}} 
          \exp(-\frac{\al}{2\hbar} q^2)
 \quad
 (q\in\RR, \;\Re(\al) > 0).
\end{displaymatH}
Then
\begin{aligN}
 (\Delta Q)^2 
   \ats\overset{\phantom{(\ref{integration_formulae})}}{=} 
        \langle Q^2 \rangle
       = \oip{\psi}{Q^2}{\psi} 
       =  {\displaystyle\int_{-\infty}^{\infty}} \;q^2 |\psi_0(q)|^2 dq
       =  \sqrt{\frac{\Re(\al)}{\pi\hbar}} 
          {\displaystyle\int} q^2 \exp(-\frac{\Re(\al)}{\hbar} q^2) dq
  \\
  \ats \overset{(\ref{integration_formulae})}{=}  \frac{\hbar}{2\Re(\al)}  
  \,,
\end{aligN}
and since $\psi$ has the minimum product uncertainty 
$(\Delta Q )^2 (\Delta P )^2 = \frac{\hbar^2}{4}$,
we have 
\begin{displaymatH}
 (\Delta P )^2 = \frac{\hbar\Re(\al)}{2}
  \, .
\end{displaymatH}
Therefore, the uncertainty in $Q$ and $P$ is equal, 
whenever $\Re(\al)=1$.
The wave function of a state with $\langle Q \rangle = \langle P \rangle = 0$
and $\Delta Q=\Delta P$ 
equal to their (simultaneous) minimum $\sqrt{\hbar/2}$ is 
therefore
\begin{displaymatH}
 \psi_0(q)= \frac{1}{\sqrt[4]{\pi\hbar}} 
          \exp(-\frac{q^2}{2\hbar})\exp(-i\frac{\Im(\al) q^2}{2\hbar})
 \quad
 (q\in\RR).
\end{displaymatH}
for some $\al=1+i\Im(\al)\in\CC$.
Observe that the choice of $\al$ 
does not change the probability density $|\psi_0(q)|^2$
as it only adds a phase. 
The usual \textsl{convention} is to take $\al$ real i.e.\ $\al=1$
and thus
\begin{equatioN}
\label{conv_psi_0}
 \psi_0(q)= \frac{1}{\sqrt[4]{\pi\hbar}} 
          \exp(-\frac{q^2}{2\hbar})
 \quad
 (q\in\RR).
\end{equatioN}
If we take units such that $m\om=1$,
we see from (\ref{eq_alphQ+iP}) and (\ref{def_LadderOp}) 
that $\ket{\psi_0}$ is a solution of 
$A\ket{\psi_0} = 0$ i.e.\ $\ket{\psi_0}=\ket{0}$, the ground state of 
the oscillator which has a Gaussian form in position representation.

We now derive the form of minimal equal uncertainty states with 
non-zero $\langle Q \rangle = t$ and $\langle P \rangle = s$.
As mentioned before, these are obtained 
by translating the position (via $T_{t/\hbar}=e^{-itP/\hbar}$) 
and the momentum (via $S_{s/\hbar}=e^{isQ/\hbar}$) 
of the ground state $\ket{0}$
and they are precisely the solutions of
\begin{equation}
\label{eq_MinUncert}
 \Bigl( (Q-t) + i(P - s)\Bigr)\ket{\psi} = 0,
\end{equation}
which, after taking units such that $m\om=1$ in (\ref{def_LadderOp}),
boils down to 
\begin{equation}
\label{eq_AoverSqrt2hbar}
 A \ket{\psi} = \frac{z}{\sqrt{2\hbar}}\ket{\psi}
 \;\text{where } z= (t+ is) 
 \text{ i.e.}
 \left\lgroup
 \begin{array}{l}
  \Re(z)\text{ is the expected position, and }\\
  \Im(z)\text{ is the expected momentum.}
 \end{array}
 \right.
\end{equation}

In position representation, equation (\ref{eq_MinUncert})
takes the form
\begin{displaymatH}
 q\psi(q) + \hbar\psi'(q) = z\psi(q) 
 \quad(q\in\RR),
\end{displaymatH}
which can be directly integrated to (including a normalization constant)
\begin{displaymatH}
 \psi(q) = N e^{-\frac{(q-z)^2}{2\hbar}}
 \quad\text{with } |N|=\frac{1}{\sqrt[4]{\pi\hbar}} \frac{1}{e^{s^2/2\hbar}}
 \quad(q\in\RR) 
\end{displaymatH}
or can be obtained by shifting the 
position and momentum of the ground state to $t$ and $s$ respectively.

To motivate the choice of normalization constant $N$, we first observe
that, with $\psi_0(q)$ as in (\ref{conv_psi_0}), 
the order in which the translations in position ($e^{isQ}=S_s$) 
and momentum space ($e^{itP}=T_t$) are 
applied matters:
\begin{aligN}
 \Bigl( S_{s/\hbar} \after T_{t/\hbar} \psi_0 \Bigr)(q)
   & = e^{ist/\hbar} 
      \frac{1}{\sqrt[4]{\pi\hbar}}
      \frac{1}{e^{s^2/2\hbar}}
       e^{-\frac{(q-z)^2}{2\hbar}}
   ,\quad\text{while }
   \\
  \Bigl(T_{t/\hbar} \after S_{s/\hbar}  \psi_0 \Bigr)(q)
   & = \phantom{e^{ist/\hbar}}  
      \frac{1}{\sqrt[4]{\pi\hbar}}
       \frac{1}{e^{s^2/2\hbar}} 
       e^{-\frac{(q-z)^2}{2\hbar}}
   .
\end{aligN}
In order to retain as much symmetry between $Q$ and $P$ as possible,
the usual convention is to ``translate in position and momentum space
simultaneously'' via our old friend 
\begin{displaymatH}
e^{(isQ-itP)/\hbar}=W((t+is)/\hbar)
  \overset{(\ref{eq_WeylOp_TtSs})}{=}
           e^{\frac{i}{2\hbar}st} T_{t/\hbar} S_{s/\hbar} 
         = e^{-\frac{i}{2\hbar}st}  S_{s/\hbar} T_{t/\hbar}
.
\end{displaymatH}

\begin{convention}
The minimal (equal) uncertainty state with position $t$ and momentum $s$ 
is defined as
\begin{equatioN}
\label{eq_relationMinUncAndWeyl}
 \textsl{MinUnc}(z)
  = e^{(isQ-itP)/\hbar} \psi_0 = W((t+is)/\hbar)\psi_0
 .
\end{equatioN}
\end{convention}
In particular, in position space representation it takes 
the following form:
\begin{equatioN}
\label{eq_MinUncInQrep}
 \ip{q}{\textrm{MinUnc}(z)} = e^{ist/2\hbar}  
      \frac{1}{\sqrt[4]{\pi\hbar}} \frac{1}{e^{s^2/2\hbar}} 
       e^{-\frac{(q-z)^2}{2\hbar}}
  \quad(q\in\RR)
  , 
\end{equatioN}
which is indeed the ``middle way'' between first applying $T_{t/\hbar}$ and
then  $S_{s/\hbar}$ or vice versa:
\begin{displaymatH}
 \textrm{MinUnc}(z) 
    = e^{-ist/2\hbar} S_{s/\hbar}  T_{t/\hbar} \psi_0
    = e^{+ist/2\hbar} T_{t/\hbar} S_{s/\hbar} \psi_0 
  .\  
\end{displaymatH}

Equation (\ref{eq_AoverSqrt2hbar}) tells us that,
if in units are chosen such that $m\om=1$,
$\textrm{MinUnc}(z)$ is an eigenvector of $A$ for eigenvalue
$z/\sqrt{2\hbar}$.
Things simplify in terms of the eigenstates of $A$.

\begin{convention}
Coherent states are defined as the eigenstates of the annihilation operator 
$A$, and labeled as
\begin{displaymatH}
  A\Coh(z)= z\Coh(z)
  .
\end{displaymatH}

If (units are chosen such that) $m\om=1$, therefore in particular for
the standard Hamiltonian $H=(Q^2 + P^2)/2$,  
 $\Coh(z)$ is a minimum uncertainty state of the harmonic oscillator,
more precisely $\Coh(z)$ is 
characterized by
\begin{Properties}
 \item
 expected position $\oip{\Coh(z)}{Q}{\Coh(z)}=\sqrt{2\hbar}\Re(z)$, 
 \item
 expected momentum $\oip{\Coh(z)}{P}{\Coh(z)}=\sqrt{2\hbar}\Im(z)$,
 \item
 equal uncertainty in $Q$ and $P$ such that the product uncertainty is
 minimal:
 $\Delta Q=\Delta P = \sqrt{\hbar/2}$,
\end{Properties}
\end{convention}
since $\textrm{MinUnc}(z)=\Coh(z/\sqrt{2\hbar})$.

\subsubsection{Coherent states in Schr\"odinger representation}
In position representation, we have from (\ref{eq_MinUncInQrep})
\begin{equatioN}
\label{eq_CohStateSchrodinger}
\begin{aligned}
 \ip{q}{\Coh(z)} 
   & = \frac{e^{i\Re(z)\Im(z)-\Im(z)^2}}{\sqrt[4]{\pi\hbar}} 
       e^{-(q/\sqrt{2\hbar}-z)^2}
   \\
   & = \frac{e^{-\half(|z|^2 - z^2)}}{\sqrt[4]{\pi\hbar}} 
       e^{-(q/\sqrt{2\hbar}-z)^2}
  \quad(q\in\RR)
  .
\end{aligned}
\end{equatioN}
Also, from (\ref{eq_AoverSqrt2hbar}) and (\ref{eq_relationMinUncAndWeyl})
\begin{equatioN}
\label{eq_MinUncState}
\begin{aligned}
  \Coh(z) & =\textrm{MinUnc}(\sqrt{2\hbar}z)
   =W(\sqrt{2\hbar}z/\hbar)\ket{0}=
   \\
  & = W(\frac{z}{\sqrt{\hbar/2}}) \ket{0}
   .
\end{aligned}
\end{equatioN}
Furthermore, a calculation shows (see \Appref{ap_ipCohPosRep})
\begin{equatioN}
\label{eq_CohzFromWeylOpOnGroundStateSchrodinger}
 \langle \Coh(w)  | \Coh(z) \rangle
   =
  \int dq \ip{\Coh(w)}{q} \ip{q}{\Coh(z)}
   =
   e^{-\half|w|^2} e^{-\half|z|^2} e^{\overline{w}z}
  .
\end{equatioN}
In particular, the ground state expectation of a Weyl operator is
\begin{equatioN}
\label{eq_ExpectationWeylOpInGroundStateSchrodinger}
 \oip{0}{W(\frac{z}{\sqrt{\hbar/2}})}{0} 
  = \ip{\Coh(0)}{\Coh(z)} = e^{-\half|z|^2}
 .
\end{equatioN}
Since 
\begin{displaymatH}
 \ip{q}{\Coh(is)}
  = \frac{e^{-q^2/2\hbar}}{\sqrt[4]{\pi\hbar}} 
    e^{isq/\sqrt{\hbar/2}}
 \quad (q\in\RR)
\end{displaymatH}
we see, by the uniqueness of the Fourier transform on $L^2(\RR)$, 
that the set $\{\Coh(is): s\in\RR\}$ and thus the set of 
coherent vectors is total in $L^2(\RR)$.

From a mathematical point of view, the coherent states $\Coh(z)$ are
normalized versions of so-called exponential states:
\begin{displaymatH}
\epsilon(z):=e^{\half |z|^2} \Coh(z),
\end{displaymatH}
which read in the Schr\"odinger 
representation (use equation (\ref{eq_CohStateSchrodinger})) 
\begin{displaymatH}
 \ip{q}{\Coh(z)} 
  =
  \frac{e^{\half z^2}}{\sqrt[4]{\pi\hbar}} 
       e^{-(q/\sqrt{2\hbar}-z)^2}
  \quad(q\in\RR)
  .
\end{displaymatH}
Mathematically, exponential states are easier to use in calculations,
which for example is expressed  in their inner product:
\begin{displaymatH}
 \langle \epsilon(w)  | \epsilon(z) \rangle
   =
   e^{\overline{w}z}
\end{displaymatH}
(use (\ref{eq_CohzFromWeylOpOnGroundStateSchrodinger})).
The latter property and their completeness in fact determines 
exponential vectors up to a unitary transformation
as we will see later (see section \ref{ex_KolmogorovDilation}).

\subsubsection{Coherent states in the Heisenberg representation}
In the Heisenberg representation, the eigenvalue equation $A\ket{z}= z\ket{z}$
can be solved by expanding $\ket{z}$ in the energy basis,
\begin{displaymatH}
 \ket{z}=\sum_{n=0}^{\infty} \al_n \ket{n}
\end{displaymatH}
and rewriting the eigenvalue equation as
\begin{displaymatH}
 \sum_{n=0}^\infty z\al_n\ket{n} = z\ket{z} = A\ket{z} 
                        = \sum_{k=1}^\infty \al_k\sqrt{k}\ket{k-1}
\end{displaymatH}
so that a recursion relation for its coefficients emerges
\begin{displaymatH}
 \al_n =\frac{z}{\sqrt{n}} \al_{n-1}
\end{displaymatH}
which has 
\begin{displaymatH}
 \al_n =\frac{z^n}{\sqrt{n!}} \al_0
\end{displaymatH}
as solution. The coefficient $\al_0$ can be determined from normalization:
\begin{displaymath}
 1= \sum_{n=0}^\infty |\al_n|^2 
  = \sum_{n=0}^\infty \frac{|z^2|^{n}}{n!} |\al_0|^2 = e^{|z|^2} |\al_0|^2
 , 
\end{displaymath}
so that the coherent states read
\begin{equatioN}
\label{eq_HeisRelCohWeyl}
 \ket{z}= e^{-|z|^2/2} \sum_{n=0}^\infty \frac{z^n}{\sqrt{n!}}\ket{n}
        \overset{(\ref{eq_ladder_op_ho})}{=} e^{-|z|^2/2} e^{z\adj{A}}\ket{0}
 .
\end{equatioN}

The coherent states in Heisenberg representation are
total. Indeed, the linear span of $\{ \ket{z}: z\in\CC \}$
contains the vectors $\pi(z):=\sum_{n=0}^\infty \frac{z^n}{\sqrt{n!}}\ket{n}$
which are total as follows from the lemma below.

\begin{lemma}[Exponential vector in Heisenberg representation]
\label{lem_ExpVecHeisRep}
Define for $z\in\CC$, the exponential vector 
\begin{displaymatH}
 \pi(z):=\sum_{n=0}^\infty \frac{z^n}{\sqrt{n!}}\ket{n}.
\end{displaymatH}
Then
\begin{Properties}
 \item
  $\ip{\pi(z)}{\pi(w)}= e^{\overline{z}w} \quad (z,w\in\CC)$,
 \item
  $\Bigl[\displaystyle\frac{d^n}{d t^n} \pi(t)\Bigr]_{t=0} = \sqrt{n!} \ket{n}$.
\end{Properties}
(see (\ref{ap_ExpVecHeisRep}).
\end{lemma}

Again the coherent vectors can be generated from $\ket{0}$ by the Weyl 
operators if units are chosen such that $m\om=1$.
To see this, assume that $m\om=1$ and observe from  (\ref{def_LadderOp})
that 
\begin{displaymatH}
 isQ -itP= is \left( \sqrt{\hbar/2} (A+\adj{A})\right)
          -it \left( \sqrt{\hbar/2} (A-\adj{A})/i \right)
         = \sqrt{\hbar/2}\left( z\adj{A} - \overline{z}A\right)
 ,
\end{displaymatH}
where $z=t+is$ as usual.
By the Baker-Campbell-Hausdorff formula,
\begin{aligN}
 W(z) \ats
     = e^{isQ -itP}= e^{\sqrt{\hbar/2}\left( z\adj{A} - \overline{z}A\right)}
     = e^{\sqrt{\hbar/2}z\adj{A}} e^{-\sqrt{\hbar/2}\overline{z}A}
       e^{-\half [\sqrt{\hbar/2}z\adj{A},-\sqrt{\hbar/2}\overline{z}A]}
     \\
     \ats
     = e^{\sqrt{\hbar/2}z\adj{A}} e^{-\sqrt{\hbar/2}\overline{z}A}
       e^{-\frac{\hbar}{4}|z|^2}    
    ,
\end{aligN}
hence $W(z/\sqrt{\hbar/2})= e^{z\adj{A}} e^{-\overline{z}A} e^{-\half |z|^2}$.
Since $e^{-\overline{z}A}\ket{0}
= (1 + \sum_{n=1}^{\infty}\frac{-\overline{z}A}{n!}) \ket{0}=\ket{0} + 0$,
we see that
\begin{aligN}
 W(z/\sqrt{\hbar/2})\ket{0}
  \ats
  = e^{z\adj{A}} e^{-\overline{z}A} e^{-\half |z|^2}\ket{0}
  = e^{-\half |z|^2} e^{z\adj{A}} \ket{0}
  \\
 \ats
  \overset{(\ref{eq_HeisRelCohWeyl})}{=} \ket{z}
  .
\end{aligN}

The distribution of the number of quanta $N=\adj{A}A$ in a coherent state
is given by the probabilities $p_n$ of measuring $n$ quanta
\begin{displaymatH}
	p_n=|\ip{n}{\Coh(z)}|^2
           =|\sum_{n'}\al_{n'}\ip{n}{n'}|^2
           =|\al_n|^2 = \frac{|z^2|^n}{n!}e^{-\half|z^2|^{n}}  
        , 
\end{displaymatH}
which describes a Poisson distribution with parameter $|z^2|$.
In particular, the expected number of quanta and the variance thereof
are $|z^2|$ in a coherent state $\ket{z}$.

\bigskip

\subsection{Passage from physical to quantum probability}

In the formulas that appeared on stage to now, Planck's constant
$\hbar$ figured as $\hbar/2$, $\hbar$, or $2\hbar$, and all of these are
multiples of $\hbar/2$. Therefore, 
formulas simplify if we set $\hbar/2 = 1$ i.e.\ $\hbar=2$, and 
the latter is a convention often adopted in quantum probability theory.

In particular, we then have the following reductions.
\begin{Enumerate}
\item 
 The Weyl relation (\ref{eq_WeylRelationWeylOp}) reads:
 \begin{equation}
 \label{eq_WeylRelationOverCC}
  W(z)W(w) = e^{-i\Im(\overline{z}w)} W(z+w)
  \quad 
  (z, w \in\CC)
  .
 \end{equation} 
\item
The coherent vectors (\ref{eq_CohzFromWeylOpOnGroundStateSchrodinger})
in the Schr\"odinger representation (and as we will see later: also for cyclic 
vacuum representation of the harmonic oscillator)
are obtained by applying Weyl operators to the 
ground state:
 \begin{equation}
  \Coh(z)= W(z)\ket{0}  
  \quad 
  (z\in\CC)
  .
 \end{equation}
\item
The expectation of the Weyl operator in the ground state 
(\ref{eq_ExpectationWeylOpInGroundStateSchrodinger}) 
in Schr\"odinger's representation
is 
\begin{equatioN}
 \oip{0}{W(z)}{0} 
  =  e^{-\half|z|^2}
 .
\end{equatioN}
\end{Enumerate}
Again, we will see later that this holds 
for so-called cyclic vacuum representations of the harmonic oscillator.

\bigskip
\questionst{elaborate: Coherent states are exponential states with this
 convention and Weyl operators}

\subsection{Different kind of representations}
Summarizing we have seen three kinds of representations for the
quantum mechanical oscillator:

\begin{enumerate}
 \item The Schr\"odinger representation diagonalizes the position operator 
       and may therefore be called the configuration space representation.
       This representation emphasizes the wave like character of the
       quantum mechanical oscillator, which is why it is also called a 
       ``wave representation''.
 \item The Heisenberg (Fock) representation diagonalizes the energy operator.
       The corpuscular nature of the quantum mechanical oscillator
       is evident is this representation, which is the reason why it is 
       also called the particle representation.
 \item The Bargmann-Segal representation is less well-known than the former
       two. This representation ``diagonalizes'' the creation operators:
       when we identify the phase space with the complex plane via 
       $q+ip\leftrightarrow z$, then states can be realized as analytic
       functions on the complex plane and creation operator becomes 
       the multiplication with $z$. One could call this ``complex wave''
       representation \cite{Baez}\cite[p.~23]{Baez_thread} 
       the phase space representation.
       We will not work explicitly with this representation, but 
       this representation is always lurking in the background,
       because quantization of fields is most elegant in this
       representation. The latter will become clearer 
       when we look at the second quantization functor
       and its interpretation, see chapter \ref{ch_quantizationFunctor}.
\end{enumerate}

\section{An apology for functorial quantization}
Although Dirac correspondence principle 
is a familiar and respected procedure to quantize the harmonic oscillator,
some of its inherent difficulties could already be tasted. First of all,
there is no direct recipe how to translate classical observables into 
self-adjoint operators. Although physical reasoning and intuition often can 
provide formal expressions for the operators, one still faces the (sometimes 
formidable) problems of choosing a suitable Hilbert space and finding a domain
on which the operator can be proven to be self-adjoint. 
In addition, sometimes more than one self-adjoint operator is a reasonable 
candidate for the observable and one has to choose the ``right one'' 
\cite[p.\ 303]{Reed+Simon:FA}.
A second and for our context worse problem is 
that the Dirac procedure breaks down for damped 
(so dissipative) oscillators, because it requires a Hamiltonian to generate 
the quantum time evolution. In this sense, the Dirac correspondence
principle is only applicable to conservative (so at least energy conserving)
classical systems.

This section introduces another way of quantizing the harmonic
oscillator to deal with the above shortcomings. This approach can be
traced back to work of Simon \cite[Chapter 1, paragraph 4]{Simon}.
The basic idea is to quantize 
the time evolution in the classical phase space via a functor
(the second quantization functor $\Gamma$).
When applied to the harmonic oscillator evolution, 
it reproduces well-known results, which gives us faith that
the procedure makes sense in other situations as well. 

Besides mathematical elegance, the motivation for functorial
quantization is its ability to deal with dissipative systems
in a non ad-hoc manner.

We start with a section to develop the bits and pieces that get the functorial
machine going.

\section{A mathematical toolkit for Hilbert spaces and unitary maps}
Hilbert spaces and unitary maps are corner stones in quantum theory.
The former, because the (pure) states of a system can be described
by vectors in a suitable Hilbert space and the latter, because 
physically equivalent representations are linked by unitary maps
between the representing Hilbert spaces.
Moreover, the observables of a quantum system are generators of 
unitary one-parameter groups living on the representing Hilbert space.

This section provides general tools to generate 
Hilbert spaces and unitary maps.
As a particular application of the tools developed, 
we will see that the Schr\"odinger and
Heisenberg representation of the quantum harmonic oscillator
are equivalent.

\subsection{Extension of linear bounded maps on total sets}
More often than not, 
the unitary maps that we will consider 
are difficult or tedious to describe in terms of their action 
on arbitrary vectors.
A more suitable way then is to define the action of such maps on a 
particular set of vectors which is ``total'', i.e.\  such that it uniquely 
determines the action on an arbitrary vector.

\begin{definition}
A linear subspace $\mathcal{H}_0$ of a Hilbert space $\mathcal{H}$ 
is \Dindex{dense} in $\mathcal{H}$  
if for every $\xi\in\mathcal{H}$ there exists a sequence 
$\xi_n\in\mathcal{H}_0$ 
such that $\norM{\xi-\xi_n}^2=\ip{\xi-\xi_n}{\xi-\xi_n} \to 0$.
\end{definition}

If we have defined a map $A_0$ on a dense subspace $\mathcal{H}_0$
that is 
\begin{Properties}
 \item linear on $\mathcal{H}_0$, and 
 \item bounded (i.e.\ there is a constant $K$ such that 
       $\norM{A_0\xi_0} \leq K\norM{\xi_0}$ 
       for all $\xi_0\in\mathcal{H}_0$),
\end{Properties}
then the action of $A_0$ can uniquely be extended to the whole of
$\mathcal{H}$ by setting 
$A\xi:=\lim_n A_0\xi_n$ for any sequence $(\xi_n)_n$  in 
$\mathcal{H}_0$ with $\norM{\xi_n -\xi}\to 0$ (since the limit does not
dependent on the sequence chosen by virtue of the boundedness). 
The extension $A$ is automatically linear
and bounded with the same constant: 
$\norM{A\xi} \leq K\norM{\xi}$ ($\xi\in\mathcal{H}$).

\begin{definition}
A subset $\mathcal{X}$ of $\mathcal{H}$ is \Dindex{total} if
its linear span is dense in $\mathcal{H}$.
\end{definition}

In a Hilbert space, the closure of the linear span of a subset $\mathcal{X}$
is $\mathcal{X}^{\perp\perp}$, 
where $\mathcal{X}^{\perp}$ is the \Dindex{orthogonal complement}
of $\mathcal{X}$ which is defined as 
$\{\eta\in\mathcal{H} : \ip{\eta}{\xi} = 0 
\text{ for all } \xi\in\mathcal{X} \}$.
Consequently,

\begin{lemma}
$\mathcal{X}\subset\mathcal{H}$ is \Dindex{total} if only if
for any $\eta\in\mathcal{H}$   
 \begin{displaymatH}
   \ip{\eta}{\xi} = 0 \quad(\text{for all }\xi\in\mathcal{X})
   \Implies
   \eta=0.
 \end{displaymatH}
\end{lemma}

We will make extensively use of the following:
\begin{lemma}
\label{lem_extLinMapTotalSet}
 Let $\mathcal{X}$ be a total subset of $\mathcal{H}$
 and let $A_0$ be a map $\mathcal{X} \to \mathcal{H}$.

 Suppose that 
 \begin{conditions}
  \item 
   $A_0$ can be extended to a linear map on the linear span of $\mathcal{X}$
   i.e.\
   \begin{displaymatH}
     \sum_{i=1}^n \al_i \xi_i = \sum_{i'=1}^{n'} \al'_{i'} \xi'_{i'}
     \Implies
     \sum_{i=1}^n \al_i A_0(\xi_i) = \sum_{i'=1}^{n'} \al'_{i'} A_0(\xi'_{i'})
      ,
   \end{displaymatH}
   or, equivalently,
   $\sum_{i=1}^n \al_i\, A_0(\xi_i) = 0$ 
   whenever $\sum_{i=1}^n \al_i \xi_i = 0$;
  \item
  $A_0$ is bounded on the linear span of $\mathcal{X}$ 
  i.e.\ there is a constant
  $K> 0$ such that
  \begin{displaymatH}
    \norM{\sum_{i=1}^n \al_i A_0(\xi_i)} 
      \leq 
     K \norM{\sum_{i=1}^n \al_i \xi_i}
     \quad
    (n\in\NN,\:\al_i\in\CC,\:\xi\in\mathcal{X}).  
  \end{displaymatH}
 \end{conditions}
 Then $A_0$ can be extended to a bounded linear map on all of $\mathcal{H}$.
\end{lemma}

\begin{example}[Weyl operators and exponential vectors]
\label{ex_WeylOpExpVec}
 Let $\mathcal{K}$ be a Hilbert space and suppose that
 there exists a Hilbert space $\mathcal{F}(\mathcal{K})$
 (a so called Fock space over $\mathcal{K}$)
 that contains
 a total set $\mathcal{E}(\mathcal{K}):=\{ \Exp(f): f\in\mathcal{K} \}$ 
 such that
 \begin{equation}
  \ip{\Exp(f)}{\Exp(g)}_{\mathcal{F}(\mathcal{K})}
   = e^{\ip{f}{g}_{\mathcal{K}}}
   \quad(f,g\in\mathcal{K}).
  \tag{$*$}
 \end{equation}
 A set of vectors with property $(*)$ will be called a set of
 \Dindex{exponential vectors}. 
 We will prove the existence of such a Hilbert space 
 $\mathcal{F}(\mathcal{K})$ later.

 We now show how a collection of Weyl operators (indexed by $\mathcal{K}$)
 can be defined on such a set of exponential vectors.

 For $f\in\mathcal{K}$ set
 $W(f): \mathcal{E}(\mathcal{K}) \to \mathcal{F}(\mathcal{K})$ via 
 \begin{displaymatH}
  W(f)\Exp(g):=e^{-\ip{f}{g}-\half\norM{f}^2} \Exp(f+g)
  \quad
  (g\in\mathcal{E}(\mathcal{K})).
 \end{displaymatH}
 Then
 \begin{aligN}
   \| W (f) & [\sum_i^n \al_i\Exp(g_i)] \|^2 
     = 
   \sum_{i,j} \overline{\al}_i \al_j
    \ip{W(f)\Exp(g_i)]}{W(f)\Exp(g_j)}
    \\
    \ats
    =
   \sum_{i,j} \overline{\al}_i \al_j \:
   \overline{e^{-\ip{f}{g_i}-\half\norM{f}^2}} 
   e^{ -\ip{f}{g_j}-\half\norM{f}^2 }
   e^{\ip{f+g_i}{f+g_j}}
   \\
   \ats
    =
   \sum_{i,j} \overline{\al}_i \al_j \:
   e^{-\ip{g_i}{f}-\half\norM{f}^2}
   e^{-\ip{f}{g_j}-\half\norM{f}^2}
   e^{\norM{f}^2 + \ip{f}{g_j} + \ip{g_i}{f} + \ip{g_i}{g_j}}
   \\
   \ats
   =
   \sum_{i,j} \overline{\al}_i \al_j \: e^{\ip{g_i}{g_j}}
   =
  \| \sum_i^n \al_i\Exp(g_i) \|^2
   . 
 \end{aligN}

 Thus, 
 $W(f)$ is bounded (in fact isometric) 
 on the linear span of the exponential vectors.
 In particular, $\| \sum_i^n \al_i\Exp(g_i) \|=0 $ implies 
 $   \| W (f)  [\sum_i^n \al_i\Exp(g_i)] \| =0$.
 By lemma (\ref{lem_extLinMapTotalSet}), we therefore conclude that
 $W(f)$ can be extended to a unitary operator 
 on the whole of $\mathcal{F}(\mathcal{K})$. 
 A straightforward calculation shows that
 $W(f)W(g)=e^{-i\Im\ip{f}{g}}W(f+g)$ on exponential vectors and 
 hence on the whole of $\mathcal{F}(\mathcal{K})$ by continuity. 
\end{example}

\subsection{Kolmogorov's dilation theorem for positive kernels}
A \Dindex{positive definite kernel} on a set $S$ 
is a map $K: S\times S \to \CC$ such
that 
\begin{displaymath}
 \sum_{i=1}^n \sum_{j=1}^n \overline{\la_j}\la_i K(s_j,s_i) \geq 0
  \quad
   (n\in\NN\,,\; (\la_1,\ldots, \la_n)\in\CC^n\,,\; s_1,\ldots, s_n\in S)
   .
\end{displaymath}

For example, 
if $\mathcal{H}$ is a Hilbert space with inner product $\ip{.}{.}$, 
then $K(\xi, \eta):=\ip{\xi}{\eta}$ is a positive definite kernel.
Slightly more general: any map $V: S\to\mathcal{H}$ defines
a positive definite kernel $K(s',s):=\ip{V(s')}{V(s)}$.
Indeed,
\begin{displaymatH}
\sum_{j=1}^n  \overline{\la_j}\la_i K(s_j,s_i) 
 = \ip{\sum_{j=1}^n \la_j V(s_j)}{\sum_{i=1}^n \la_i V(s_i)} 
 = \norM{\sum_{i=1}^n \la_i V(s_i)}^2 \geq 0
 . 
\end{displaymatH}
The following theorem shows that
 every positive definite kernel can be written this way. 

\begin{theorem}
Let $K$ be a positive definite kernel on a set $S$. 
Then, up to unitary equivalence, there exists a unique Hilbert space
$\mathcal{H}_K$ and a unique embedding $V:S\to\mathcal{H}_K$ such that
\begin{align}
 \, & \ip{V(s')}{V(s)} = K(s',s) \quad (s,s'\in S) 
      \label{def_KolmogorovDilation}\\
 \, & \{ V(s) : s\in S \} \text{ is total in }\mathcal{H}_{\mathcal{K}}.  
      \label{KolmogorovTotal}
\end{align}
\end{theorem}
Given a kernel $K$ on a set $S$, 
a map $V$ from $S$ into some Hilbert space $\mathcal{H}$ 
satisfying (\ref{def_KolmogorovDilation}) is called
a \Dindex{Kolmogorov dilation} of $K$. Such a dilation is called 
\Dindex{minimal} if, in addition, (\ref{KolmogorovTotal}) holds.
For the proof we refer to \protect{\cite[section 1.5, p.\ 30]{Maassen}}

The following examples show that many (constructions with) Hilbert spaces
in quantum mechanics are Kolmogorov dilations of suitable kernels. 

\begin{examples}
\label{ex_KolmogorovDilation}
\begin{Itemize}{-}
 \item 
 Let $S$ be a set and $K(s',s):=\de_{s',s}$. 
 Then $\mathcal{H}_K=\ell^2(S)$ and $V$ maps $S$ to the standard orthonormal
 basis of $\mathcal{H}$ defined by $e_s(t):= \de_{s,t}$, $s,t\in S$.
 In particular, taking $S=\NN$ leads to $\ell^2(\NN)$.
 \item 
 Let $S$ be the Cartesian product of two Hilbert spaces $\mathcal{K}_1$ 
 and $\mathcal{K}_2$
 \begin{aligN}
  \, & \mathcal{K}_1 \times \mathcal{K}_2
     =\{(\xi_1,\xi_2) : \xi_1 \in \mathcal{K}_1\,,\; \xi_2\in\mathcal{K}_2\} 
    , \text{ equipped with}
    \\
  \, &  K\bigl( (\xi_1,\xi_2)\,,\, (\eta_1,\eta_2)\bigr):  =
   \ip{\xi_1}{\xi_2}_{\mathcal{H}_1} 
      \cdot \ip{\eta_1}{\eta_2}_{\mathcal{H}_2}
     .
 \end{aligN}
 Then $\mathcal{H}_K= \mathcal{K}_1 \otimes \mathcal{K}_2$ 
 is the tensor product
 of $\mathcal{K}_1$ and $\mathcal{K}_2$ and 
 $V(\xi_1,\xi_2) = \xi_1 \otimes \xi_2$.
 
 \item
 Let $S=\mathcal{K}$ be a Hilbert space and 
 $K(\xi,\eta):= \ip{\xi}{\eta}^2$.
 Then $\mathcal{H}_K= \mathcal{K}\otimes_{s}\mathcal{K} $ is the 
 symmetric tensor product of $K$ and $V(\xi)=\xi\otimes\xi$.

 \item 

 Let $S=\mathcal{K}$ be a Hilbert space and 
 $K(\xi,\eta):= e^{\ip{\xi}{\eta}}$.
 Then $K$ is a positive definite kernel and 
 $\mathcal{H}_K= \mathcal{F}(\mathcal{K})$ is the 
 (symmetric) \Dindex{Fock space} over $\mathcal{K}$, defined as
 \begin{displaymatH}
   \mathcal{H}_K= \mathcal{F}(\mathcal{K})
      = 
    \CC \oplus \mathcal{K} 
        \;\oplus\; (\mathcal{K}\otimes_{s}\mathcal{K})
        \;\oplus\; (\mathcal{K}\otimes_{s}\mathcal{K}\otimes_{s}\mathcal{K})
        \;\oplus\; \ldots
     ,
 \end{displaymatH}
 and $V(\xi)$ is the so-called \Dindex{exponential vector} or
 \Dindex{coherent vector} given by
 \begin{displaymatH}
   \mathrm{Exp}(\xi)= 
      1 \oplus \frac{\xi}{\sqrt{1!}}
        \oplus \frac{\xi\otimes\xi}{\sqrt{2!}}
        \oplus \frac{\xi\otimes\xi\otimes\xi}{\sqrt{3!}}
        \oplus \frac{\xi\otimes\xi\otimes\xi\otimes\xi}{\sqrt{4!}}
        \ldots 
 \end{displaymatH}

 \item
 The last example will be of use later, 
 when we want to view the damped harmonic oscillator as part of 
 an energy conserving system.

 Let $S=\RR$ and $K_{\eta,\om}(s,t):=e^{-\eta|s-t| + i\om(s-t)}$ 
 with $\eta > 0$.
 A minimal Kolmogorov dilation of this kernel is presented by
 $\mathcal{H}=L^2(\RR)$ and $V: t\mapsto v_t \in L^2(\RR)$ defined as
 \begin{displaymatH}
  v_t(x):=
   \begin{cases}
    e^{(\eta+i\om)(x-t)} & \text{if } x\leq t\,, \\
    0                  & \text{if } x>  t\,.
   \end{cases}
 \end{displaymatH}
\end{Itemize}
\end{examples}

The first kernel ($\de_{s,s'}$) above is clearly positive definite, 
and the last kernel ($K_{\eta,\om}(s,t)$)
can be checked to be positive definite by verifying its dilation.
It is perhaps less obvious that the other kernels are positive definite, but 
for this the following lemma  helps:

\begin{lemma}[see \protect{\ref{ap_PosDefKernel}}]
Let $K_1, K_2, K_3, \ldots : S\times S \to \CC $ be positive definite kernels.

Then 
\begin{Properties}
 \item
  If $r_1,r_2\in [0,\infty)$, then 
  $r_1 K_1 + r_2 K_2$ is a positive definite kernel.
 \item
  The point wise product $K_1 * K_2: (s',s) \mapsto K_1(s',s) K_2(s',s)$ 
  is a positive definite kernel.
 \item 
  If $\sum_{n=1}^\infty K_n(s',s)$ exists for all $(s',s)\in S\times S$,
  then $K: (s',s) \mapsto \sum_{n=1}^\infty K_n(s',s)$ is a positive
  definite kernel.
\end{Properties} 
In particular, 
$e^{K(s',s)}$ is a positive definite kernel is if $K$ is one.
\end{lemma}

\subsection{Cyclic vacuum representations 
of the Canonical Commutation Relations (CCR)}
As promised we would show that 
the Schr\"odinger and Heisenberg representation of
the harmonic oscillator are equivalent. 
The deeper reason for this equivalence is that both are
so-called cyclic vacuum representations of the canonical commutation 
relations over $\CC$, and are therefore automatically unitarily equivalent.
As it is no extra cost, we prove the latter in a much more general
setting: 
that of Weyl representations of the canonical commutation relations.

As introduction, 
recall from (\ref{eq_WeylRelationOverCC}) that Weyl's form of the canonical 
commutation relation of \textsl{one} harmonic oscillator can be 
represented with Weyl operators 
$W(z) ''='' \exp(i\Im(z) Q- i\Re(z)P)$ that satisfy the
Weyl relation  
\begin{displaymatH}
 W(z)W(w) = e^{-i\Im(\overline{z}w)} W(z+w)
 \quad 
 (z,w \in\CC)
 .
\end{displaymatH}
In terms of Heisenberg's commutation relation, 
two independent harmonic quantum oscillator systems
are described by two canonical pairs $(Q_1,P_1)$ and $(Q_2,P_2)$ satisfying
$[Q_i,P_j]= i\hbar \delta_{ij}$, $[Q_i,Q_j]=[P_i,P_j]=0$.

The same is accomplished by
\begin{Enumerate}
 \item 
 describing the canonical pairs $(Q_1,P_1)$ and $(Q_2,P_2)$
 by the corresponding two families of
 Weyl operators, $W_1(z_1)$,$z_1\in\CC$ , and $W_2(z_2)$, $z_2\in\CC$
 and,
 \item 
 capturing the independence ($[Q_1,Q_2]=[Q_1,P_2]=[Q_2,P_1]=0$)
 by taking the tensor product of the two Weyl operators
 $W_1(z_1)\otimes W_2(z_2)$, $z_1,z_2\in\CC$,
 which are unitary operators on $\mathcal{H}_1\otimes\mathcal{H}_2$.
 \question{(Check by formal differentiation?)}
\end{Enumerate}

These tensored Weyl operators satisfy
\begin{aligN}
 \Bigl(&W_1(z_1) \otimes W_2(z_2)\Bigr)
 \Bigl(W_1(w_1)\otimes W_2(w_2) \Bigr) 
  =  W_1(z_1) W_1(w_1)\otimes W_2(z_2) W_2(w_2) 
  \\
  & = \Bigl(e^{-i\Im(\overline{z_1}w_1)} W_1(z_1+w_1) \Bigr)
    \otimes
    \Bigl(e^{-i\Im(\overline{z_2}w_2)} W_1(z_2+w_2) \Bigr)
  \\
  & = e^{-i\Im\Big(\overline{(z_1,z_2)}(w_1,w_2)\Bigr)}
     W_1(z_1+w_1) \otimes W_2(z_2+w_2)
 .
\end{aligN}
Reformulating in terms of 
$W\Bigl( (z_1,z_2) \Bigr) :=  W_1(z_1)\otimes W_2(z_2)$,
$(z_1,z_2)\in\CC^2$
and taking the usual (coordinate wise) addition and inner product on $\CC^2$,
we have 
\begin{aligN}
 W\Bigl( (z_1,z_2) \Bigr)
 W\Bigl( (w_1,w_2) \Bigr)
  =  
 e^{-i\frac{\hbar}{2}\Im\Big(\overline{(z_1,z_2)}\cdot(w_1,w_2)\Bigr)}
    & W\Bigl( (z_1,z_2) + (w_1, w_2)\Bigr)
 \\
 \quad
 \text{for }(z_1,z_2)\,,\;(w_1,w_2) \in \CC^2
 .
\end{aligN}

This motivates to say that  a representation of 
the CCR of two independent harmonic oscillators ``is''
a single representation CCR over $\mathcal{K}:=\CC^2$ in the
following sense:

\begin{definition}
Let $\mathcal{K}$ be a complex Hilbert space. A representation of the
Canonical Commutation Relations (CCR) over $\mathcal{K}$ is a map
$W$ from $\mathcal{K}$ to the unitary operators on some Hilbert space
$\mathcal{H}$ such that 
\begin{displaymatH}
 W(f)W(g) = e^{-i\Im\ip{f}{g}}W(f+g)
 \quad
 (f,g\in\mathcal{K})
 ,
\end{displaymatH}
and
\begin{displaymatH}
 t\mapsto W(tf)\xi \;\;\text{ is continuous }
 \quad(f\in\mathcal{K}\,,\;\xi\in\mathcal{H})
 .
\end{displaymatH}
The map $W$ is called a \Dindex{vacuum representation} of the CCR 
if, in addition, there is a unit vector $\Om\in\mathcal{H}$ such that
\begin{displaymatH}
 \oip{\Om}{W(f)}{\Om} = e^{-\half\norM{f}^2} 
 \quad
 (f\in\mathcal{K}).
\end{displaymatH}
A vacuum representation of the CCR is called \Dindex{cyclic}
if $\{W(f)\Om : f\in\mathcal{K}\}$ is total in $\mathcal{H}$.
\end{definition}

Existence and uniqueness of 
cyclic vacuum representations of a CCR  is provided by:
\begin{lemma}[Existence and uniqueness of cyclic vacuum representations of the 
              CCR]
 Let $\mathcal{K}$ be complex Hilbert space.

 Then there exists a Hilbert space $\mathcal{H}$ and a map 
 $W$ from $\mathcal{K}$ into the unitary operators $\mathcal{U}(\mathcal{H})$
 on $\mathcal{H}$ such that
 \begin{conditions}
  \item 
  \textbf{(Strong continuity)}\\
  $t\mapsto W(tf)\xi$ is continuous for each 
  $f\in\mathcal{K}$ and   $\xi\in\mathcal{H}$,
  \item
  \textbf{(Weyl relation)}\\
  $W(f)W(g) = e^{-i\Im\ip{f}{g}}W(f+g)$ for all $f,g\in\mathcal{K}$,
  \item 
  \textbf{(Cyclic vacuum vector)}\\
  There exists a vector $\Om\in\mathcal{H}$ that represents a vacuum state
  i.e.\  $\oip{\Om}{W(f)}{\Om}= e^{-\half \norM{f}^2}$ and 
  that is cyclic i.e.\
  the set $\{ W(f)\Om : f\in\mathcal{K} \}$ is total in $\mathcal{H}$. 
 \end{conditions}

 Moreover, 
 if $(W',\mathcal{H}', \Om')$ is another cyclic vacuum representation
 of the CCR over $\mathcal{K}$, then there exists a unitary map
 $U:\mathcal{H} \to \mathcal{H}'$ 
 such that $W'(f)= U W(f) U^{-1}$ ($f\in\mathcal{K}$).
\end{lemma}

(For the proof see \cite[p.\ 49]{Maassen}).

\medskip

\begin{remarks}
\begin{Itemize}{-}
 \item
 As an application of the above, we see that the two-fold tensor product
 of the CCR over $\CC$ ($(W_i,\mathcal{H}_i, \Om_i)$, $i=1,2$) is
 unitarily equivalent to a representation of the CCR over $\CC^2$
 $(\mathcal{H}, W(z_1,z_2),\Om))$:
 \begin{displaymatH}
   (\mathcal{H}_1 \otimes \mathcal{H}_2, W_1(z_1)\otimes W_2(z_2),
  \Om_1\otimes\Om_2) \isom  (\mathcal{H}, W(z_1,z_2),\Om))
  \end{displaymatH}
 \item
 A slight generalization of the above yields that a representation of
 CCR of $n$ independent harmonic oscillators (which is an $n$-fold tensor product of
 the representation of the CCR over $\CC$), is unitarily equivalent 
 to a representation of the CCR over $\CC^n=\ell^2(\{1,\ldots,n\})$
 \item
 In this sense, the dimension  of the Hilbert space $\mathcal{K}$ describes 
 the number of independent oscillators. 
 In fact, if $S$ is a set  (measure space) of certain cardinality, 
 then $\mathcal{K}= L^2(S)$ describes a system of as many independent
 oscillators as the cardinality of $S$ is.
 
 For $S=\NN$ (so $\mathcal{K}=\ell^2(\NN)$), 
 we obtain a countable chain of oscillators, while for
 $S=\RR$ (so $\mathcal{K}=L^2(\RR)$) we get a continuum of oscillators.
 \item
 Representations that are not cyclic vacuum representation need not be
 unitarily equivalent, for example those associated with
 positive temperatures \cite{Bratelli+Robinson}.
\end{Itemize}
\end{remarks}

We will now discuss 
two generic methods to obtain cyclic vacuum representations:
via a Fock space functor or via a Gaussian functor.

\subsection{Cyclic vacuum representations of CCR via the Fock space functor}

Let $\mathcal{K}$ be a complex Hilbert space.
By Kolmogorov's dilation theorem, we obtain the
Fock space $\mathcal{F}(\mathcal{K})$ over $\mathcal{K}$.
The Fock space contains a total set, viz.\ that of the  
exponential vectors $\Exp(f)$, $f\in\mathcal{K}$ 
and these have the property that $\ip{\Exp(f)}{\Exp(g)}=e^{\ip{f}{g}}$,
see the examples in section (\ref{ex_KolmogorovDilation}).
Weyl operators on $\mathcal{F}(\mathcal{K})$ can be defined via
\begin{displaymatH}
  W(f)\Exp(g):=e^{-\ip{f}{g}-\half\norM{f}^2} \Exp(f+g)
  \quad
  (f, g\in\mathcal{E}(\mathcal{K})).
\end{displaymatH}
(see \ref{ex_WeylOpExpVec}).
Since $\Om:=\Exp(0)$ satisfies 
\begin{displaymatH}
 \oip{\Om}{W(f)}{\Om}=\ip{\Exp(0)}{e^{-\half\norM{f}^2}\Exp(0)}=
 e^{-\half\norM{f}^2}e^{\ip{0}{0}}=e^{-\half\norM{f}^2}
 ,
\end{displaymatH}
we can view $\Om$ as a vacuum vector.
Furthermore, as $W(f)\Om= e^{-\half\norM{f}^2} \Exp(f)$ and the exponential
vectors are dense in $\mathcal{F}(\mathcal{K})$,
$\Om$ is a cyclic vector.

A contraction $C:\mathcal{K}\to\mathcal{K}$ induces a contraction
$\mathcal{F}(C): \mathcal{F}(\mathcal{K})\to\mathcal{F}(\mathcal{K})$
that maps $\Exp(f)$ to $\Exp(Cf)$ for every $f\in\mathcal{K}$ 
(Lemma \ref{lem_extLinMapTotalSet}). 
Viewing the Fock space as 
\begin{displaymatH}
\mathcal{F}(\mathcal{K}) 
    = 
    \bigoplus_{n=0}^\infty \mathcal{K}^{\otimes_s^n}
    ,
\end{displaymatH}
this map can be written as
\begin{displaymatH}
\mathcal{F}(C)
    = 
    \bigoplus_{n=0}^\infty C^{\otimes^n}
    .
\end{displaymatH}
Given an orthogonal projection $P$ on $\mathcal{K}$, we can define 
(\ref{ap_countOperator})
a self-adjoint operator  $d\mathcal{F}(P)$ on $\mathcal{F}(\mathcal{K})$
via
\begin{equatioN}
\label{eq_countOperator}
 e^{i\lambda d\mathcal{F}(P)} : = \mathcal{F}(e^{i\lambda P})
 \quad
 (\lambda\in\RR)
  .
\end{equatioN}
This operator  $d\mathcal{F}(P)$ is interpreted as the quantum random
variable that counts how many particles the ``question'' $P$ is answered
``yes''. For example, the total number of particles $N$ equals 
$d\mathcal{F}(\Identity)$.

The above procedure derives its strength from its generality, but remains
rather abstract. 
Some concrete representations of Fock spaces provide 
some flesh and blood to it.

\subsubsection{The Hilbert space $\ell^2(\NN)$ as a representation of the Fock space over $\CC$}

As $\CC$ is one-dimensional with orthonormal base $\{ 1\}$,
an orthonormal base of $\CC \otimes \CC$ is given by $\{ 1\otimes 1\}$ 
and as this basis (= one vector) is already symmetrical, it is a basis
of $\CC \otimes_s \CC$ too.
In other words, $\CC \otimes \CC$ and $\CC \otimes_s \CC$
are one-dimensional and hence unitary equivalent to $\CC$. 

Therefore, 
\begin{aligN}
 \mathcal{F}(\CC)
 \ats = \CC \oplus \CC \oplus \CC^{\otimes^2_s} \oplus
\CC^{\otimes^3_s} \oplus  \ldots \isom \CC \oplus \CC \oplus \CC \oplus \ldots
 \\
 \ats
  = \{ (z_n)_{n\in\NN} : z_n\in\CC\,, \; \sum_1^\infty |z_n|^2 <\infty \}
 \\
 \ats
 = \ell^2(\NN)
 .
\end{aligN}

Otherwise stated,
the map $\pi: \CC \to \ell^2(\NN)$, 
\begin{displaymatH}
 z\mapsto \pi(z):=(1,z,\frac{z^2}{\sqrt{2!}},\frac{z^3}{\sqrt{3!}},\ldots)  
 ,
\end{displaymatH}
defines a minimal Kolmogorov dilation of the kernel 
$K(z,w):=e^{\overline{z}w}$,
since 
$\ip{\pi(z)}{\pi(w)}_{\ell^2(\NN)}=e^{\overline{z}w}$
and the set $\{ \pi(z) : z\in\CC \}$ is dense in $\ell^2(\NN)$
(lemma \ref{lem_ExpVecHeisRep}).

In particular, the Heisenberg representation of the harmonic oscillator
is (a Fock representation of) the canonical commutation relations over $\CC$.

\subsubsection{The $L^2$ over Guichardet's space $(\Delta(\RR),\mu)$ as 
                a representation of the Fock space over $L^2(\RR)$}
The Fock space over $L^2(\RR)$ is formally
\begin{displaymatH}
 \mathcal{F}(L^2(\RR))
   = \CC \oplus L^2(\RR) 
         \oplus \:\Bigl( L^2(\RR) \otimes_s L^2(\RR) \Bigr)\:
         \oplus \: \Bigl( L^2(\RR) \otimes_s L^2(\RR) \otimes_s L^2(\RR) 
                   \Bigr)\:
         \oplus \ldots
\end{displaymatH}

Note that $L^2(\RR) \otimes_s L^2(\RR)  \isom L^2_{symm}(\RR^2)$ i.e.\
there is a natural correspondence between $L^2$-functions on $\RR^2$ 
that are symmetrical:
\begin{displaymatH}
 f(t_1,t_2) = f(t_2,t_1) \quad (\text{for almost all } t_1,t_2\in\RR)
\end{displaymatH}
and symmetrical tensor products of $L^2$-functions on $\RR$:
\begin{displaymatH}\displaystyle
 \Bigl( \frac{f \otimes g  + g\otimes f}{2} \Bigr)(t_1,t_2)
  = \frac{f(t_1) g(t_2)  + g(t_1) f(t_2)}{2}
 ,
\end{displaymatH}
since the ``elementary'' symmetrical functions 
$\half(f\otimes g + g\otimes f)$, $f,g\in L^2(\RR)$,
are dense in the Hilbert space of symmetrical $L^2$-functions on
$\RR^2$ (\ref{ap_SymTensorDenseInSymmL2}).
In the same vein, 
\begin{displaymatH}
 L^2(\RR)^{\otimes_s^n}
 \isom 
  L^2_{symm}(\RR^n)
  = \Bigl\{ f\in L^2(\RR^n) : 
        f(t_1,\ldots, t_n) = f(t_{\tau(1)},\ldots, t_{\tau(n)})
        \text{ for all } \tau\in S_n
    \Bigr\}
  .
\end{displaymatH}

Due to its symmetry, a symmetrical function on $\RR^n$ is determined by it values
on $\{ (t_1,\ldots, t_n) \in\RR^n: t_1 < t_2 < \ldots <t_n \}$.
Hence, we can identify a symmetrical function on $\RR^n$ with a 
function on $\Delta_n(\RR):= \{ \si \subset \RR : \#(\si) = n \}$ 
i.e.\ $\Delta_n(\RR)$ is the collection of all subsets of $\RR$ that
contain $n$ (distinct) points.

Using the correspondence
\begin{displaymatH}
 (t_1,\ldots, t_n) \text{ with }  t_1 < t_2 < \ldots <t_n 
 \quad\longleftrightarrow\quad
 \text{the set of distinct points } \{t_1, \ldots, t_n \} 
\end{displaymatH}
we can transfer the Lebesgue measure $\lambda_n$ 
on $\RR^n$ to $\Delta_n$.

Let us introduce the Guichardet space over $\RR$ 
\begin{aligN}
 \Delta(\RR) 
   & = \{ \si \in \RR : \#(\si) < \infty \} 
   \\
   & = \Union_{n\in\NN} \Delta_n(\RR)  
\end{aligN}
where
\begin{aligN}
  \Delta_0(\RR) & := \{ \emptyset \} \\
  \Delta_n(\RR) & := \{ \si \subset \RR : \#(\si) = n \}    \quad (n \geq 1)
  ,      
\end{aligN}
i.e.\ $\Delta(\RR)$ is the collection of all finite subsets of $\RR$,
and equip $\Delta(\RR)$ with a  measure $\mu$ given by
\begin{aligN}
  \mu(\{\emptyset\}) & = 1 \\
  \mu\mid_{\Delta_n(\RR)} & = \lambda_n\mid_{\Delta_n(\RR)}
  . 
\end{aligN}

By the above, $L^2_{\text{symm}}(\RR^n) \isom L^2(\Delta_n(\RR))$, 
and therefore:
\begin{aligN}
 \mathcal{F}(L^2(\RR)) 
    & = \oplus_{n=0}^\infty \; L^2(\RR)^{\otimes_s^n} 
    \\
    & \isom \oplus_{n=0}^\infty \; L^2_{\text{symm}}(\RR^n)
    \\
    & \isom \oplus_{n=0}^\infty \; L^2(\Delta_n(\RR))
    \\
    & \isom  L^2( \union_{n=0}^\infty \Delta_n(\RR))
    \\
    & = L^2(\Delta(\RR))
    .
\end{aligN}

Under the identification 
$ \oplus_{n=0}^\infty \; L^2(\RR)^{\otimes_s^n}  \isom  L^2(\Delta(\RR))$, 
the exponential vector 
$\oplus_{n=0}^\infty f^{\otimes^n} / \sqrt{n!}$
corresponds to the ``product'' vector  $\pi(f): \Delta(\RR) \to \CC$
given by 
\begin{displaymatH}
  \pi(f)(\si): = \Pi_{s\in\si} f(s)
  \quad
  \text{ with}\quad \pi(f)(\emptyset) : = 1
   .
\end{displaymatH}
Then
\begin{aligN}
 \ip{\pi(f)}{\pi(g)} 
   & = \int_{\Delta(\RR)} \overline{\pi(f)(\si)}\pi(g)(\si)\: \mu(d\si) 
   \\
   & = \sum_{n=0}^\infty \int_{\Delta_n(\RR)} 
              \overline{\pi(f)(\si)}\pi(g)(\si)\:\mu\mid_{\Delta_n(\RR)}(d\si)
   \\
   & = \sum_{n=0}^\infty \int_{s_1 < s_2 < \ldots < s_n}
          (\overline{f}g)(s_1) \cdots (\overline{f}g)(s_n) ds_1\cdots ds_n
   \\  
   & = \sum_{n=0}^\infty \frac{1}{n!}\int_{\RR^n}
          (\overline{f}g)(s_1) \cdots (\overline{f}g)(s_n) ds_1\cdots ds_n
   \\
   & = \sum_{n=0}^\infty \frac{1}{n!}
          \biggl( \int_{-\infty}^{\infty} \overline{f(s)}g(s) ds\biggr)^n
   \\
   & = e^{\ip{f}{g}}
   .
\end{aligN}
In particular, 
the vacuum state $\Om$ 
is represented as the function that is $1$ on $\{\emptyset\}$
and $0$ elsewhere.

The interpretation of an element of $\Delta_n(\RR))$ 
is a configuration of positions $\{s_1, \ldots, s_n\}$ of 
$n$ indistinguishable particles.

\subsection{Cyclic vacuum representations of the CCR via Gaussian functors}
We will provide the Gaussian representation of the harmonic oscillator, 
but will seize the opportunity to present it as a simple (sometimes degenerate)
case of a more general construction which will be discussed in the next
section. 
This general construction will be of
use in the dilation of the harmonic oscillator which will be needed further on.
We illustrate the finite dimensional case with $E=\RR$ and the 
infinite dimensional case with $E=L_\RR^2(\RR)$.

\subsubsection{Finite dimensional case illustrated with $E=\RR$} 
To emphasize the similarity in the construction for the finite dimensional case
with the infinite dimensional case, we will treat the case $E=\RR$ along the
same lines as the infinite case, which may look at first sight a bit 
pedantic.

\paragraph{Step 1: the basic ingredients}
We start with $E=\RR$ as a (one dimensional) real vector space and its dual
$E'$, which is identified with $\RR$ again: each linear continuous functional
$\om:\RR\to \RR $ is of the form $\om_u (r) = r \cdot u$ for some  $u\in \RR$. 
In the general case (discussed later on), there is a Hilbert space $H$ 
in between $E$ and $E'$, $E\subset H \subset E'$. In this case,
$H= L^2_\RR(\{1\})\isom \RR $ and now coincides with $E=\RR$ as well as
with  $E'\isom\RR$.
In particular, $\norM{r}_H^2= r^2$ ($r\in E\subset H$).
The bilinear form linking $E$ with $E'$ is denoted by 
$\ip{\cdot}{\cdot}: E \times E'\to \RR$, $\ip{r}{u}= r\cdot u$.
Second, we need a measure $\PP$ on the dual $E'\isom\RR$ such that
$\int_{E'} e^{i\ip{r}{u}} \:d\PP(u) = e^{-\half\norM{r}_H^2}$, $(r\in E)$.
From (\ref{eq_Gaussian_integral}), it follows that the Gaussian measure $\PP$ 
on $E'=\RR$ with density 
\begin{equatioN}
\label{eq_DensityGamma}
 \gamma(x)= e^{-\half x^2}/\sqrt{2\pi} 
\end{equatioN}
with respect to the
Borel-Lebesgue measure does the job.

Summarizing we have
  \begin{equation}
  \label{eq_BochnerMinlos_R}
   \int_{\RR} e^{iru} e^{-\half u^2}\frac{du}{\sqrt{2\pi}}
    = e^{-\half r^2}
   \quad
   (r\in \RR)
   \,,
  \end{equation}
where it should be born in mind that
\begin{Enumerate}
 \item
  we are dealing with a triple of spaces $E\subset H \subset E'$ 
  which are here all identifiable with $\RR$;
 \item
  the integration is over $u\in E'\isom \RR$; 
  the $\si$-algebra on $E'$ is that which makes all point evaluations
  \begin{displaymatH}
   \be(r):  u \mapsto \ip{r}{u} = r\cdot u
   \quad
    (u\in E')\,,
  \end{displaymatH}
  $r\in E$,
  measurable functions from
  $E' \to \RR$ i.e.\ it is the
  Borel(-Lebesgue) $\si$-algebra on $E'\isom\RR$;
  the measure $\PP$ on $E'$ is 
   $\PP(B) = \int_B e^{-\half u^2} \frac{du}{\sqrt{2\pi}}$ 
   ($B$ a Borel-Lebesgue set in $E'\isom \RR$);
 \item
  the term $e^{iru}$ in the integrand is actually $e^{i\ip{r}{u}}$, where
  $\ip{\cdot}{\cdot}$ is the bilinear form linking $E$ with $E'$.
 \item
  the exponential  $e^{-\half r^2}$ is actually $e^{-\half \norM{r}_\RR^2}$
  where $\norM{r}^2_\RR :=r^2$ is the Hilbert norm on $H=\RR$.

\end{Enumerate}

\paragraph{Step 2: an isometry $\be: H \to L^2_\RR(E',\PP)$}
By taking $r=0$ in (\ref{eq_BochnerMinlos_R}), it is clear that $\PP$ is a
Gaussian probability measure on $E'$, 
and (\ref{eq_BochnerMinlos_R}) can be read as:
\begin{displaymath}
   \int_{E'} e^{i\be(r)} \:d\PP = e^{-\half r^2}
   \quad
   (r\in \RR)
   \,,
\end{displaymath}
i.e.\ $\be(r)$  is a Gaussian random variable on $E'$ with zero expectation 
and variance $r^2$. 
In other words, 
$\norM{\be(r)}_{L^2_\RR(E')}= \textsl{Var}(\be(f))= r^2 = \norM{r}^2_\RR$,
and $\be: E = H \to L^2_\RR(E')$ is an isometry. 

\paragraph{Step 3: complexification of $\be$ to $\be_\CC$}
The complexification $\be_\CC$ of $\be$ is simply defined as
$\be_\CC(r+is) = \be(r) + i\be(s)$.
The isometric character of $\be$ passes on to $\be_\CC$:
\begin{aligN}
 \ip{\be_\CC(r+is)}{\be_\CC(r+is)}_{L^2_\CC(E')}
   \ats =
 \ip{\be(r) + i\be(s)}{\be_\CC(r) +i \be(s)}_{L^2_\CC(E')}
   \\
  \ats
   =
 \ip{\be(r)}{\be(r)}_{L^2_\RR(E')} + \ip{\be(s)}{\be(s)}_{L^2_\RR(E')}  
  = 
  r^2 + s^2 
  \\
  \ats
  = \ip{r+is}{r+is}_\CC
 \end{aligN}
From now on, we will refer to $\be_\CC$ with $\be$.

\paragraph{Step 4: exponential vectors}
For $z=r+is\in\CC$, we define the exponential vector 
$\expv(z) : = e^{\be(z) - \half z^2}$. 
To calculate the inner product of two such vectors, we consider
\begin{aligN}
 \ip{e^{\be(z)}}{e^{\be(w)}}_{L^2_\CC(E')}
  \ats
  = 
 \ip{e^{\be(r+is)}}{e^{\be(p+iq)}}_{L^2_\CC(E')}
  \\
  \ats
  =
 \ip{e^{\be(r) +i\be(s)}}{e^{\be(p)+i\be(q)}}_{L^2_\CC(E')} 
  = 
 \EE\Bigl(e^{\be(r+p) + i \be(q-s)}\Bigr)
 \,.
\end{aligN}
Using that $\be$ is an isometry $\RR\to L^2_\RR(E')$ and 
using (\ref{eq_expectation_iX+Y}), the latter term is seen to be
\begin{aligN}
 \exp\Bigl[\ats
          \half\bigl\{ (r+p) + i(q-s) \bigr\}^2
     \Bigr] 
   =
 \exp\Bigl[\half\bigl\{ (\overline{r+is}) + (p+iq) \bigr\}^2\Bigr] 
  \\
  \ats
   =
 \exp\Bigl[\half\bigl\{ \overline{z} + w \bigr\}^2\Bigr] 
   =
 \exp\bigl[\half\overline{z}^2 
           + \half w^2 
           + \overline{z}w
     \bigr] 
 \,,
\end{aligN}
i.e.\ $\ip{\expv(z)}{\expv(w)}_{L^2_\CC(E')}= e^{\overline{z}w}$.

\paragraph{Step 5: the exponential vectors form a total subset of $L^2_\CC(E')$}
Actually, by the uniqueness of the Fourier-transform, the subset of all
$e^{i\be(r)}: u\mapsto e^{iru}$, $r\in E=\RR$, is already total in 
$L^2_\CC(E')$.
\paragraph{The Gaussian representation of the CCR over $\CC$}
The representation obtained above is called the Gaussian representation of
the CCR over $\CC$.

In this representation, the action of the Weyl operator 
takes a rather simple form as
\begin{aligN}\displaystyle
 \Bigl[W(t) \expv(z) \Bigr](x)
   & =\Bigl[ e^{-\ip{t}{z}} e^{-\half | t|^2} \expv{t + z} \Bigr](x)
      = e^{-2tz} e^{-t^2} e^{-\half z^2} e^{(t+z)x}
   \\
   & = \expv(x-2t)\Bigl[\frac{\gamma(x-2t)}{\gamma(x)}\Bigr]^{1/2}    
   \quad(x\in\RR),
\end{aligN}
where $\gamma$ is the density of the Gaussian measure $\PP$ on $\RR$, see
equation (\ref{eq_DensityGamma}).
Further,
\begin{aligN}\displaystyle
 \Bigl[W(is) \expv(z) \Bigr](x)
   & =\Bigl[ e^{-\ip{is}{z}} e^{-\half | is|^2} \expv(is + z) \Bigr](x)
      = e^{zx -\half z^2} e^{isx}
  \quad(x\in\RR)
  .
\end{aligN}
Concluding: apart from a mixing in of a factor $\sqrt{\gamma(x-2t)/\gamma(x)}$,
$W(t)$ is a shift to the right, and $W(is)$ is a multiplication 
with $e^{isx}$.
Things take on a familiar form when dividing away a 
factor $\sqrt{\gamma(x)}$:

\paragraph{Equivalence of the Schr\"odinger and Heisenberg representation 
of the CCR over $\CC$}

The linear map $U_{GS}: L^2_\CC(\RR, \PP)\to L^2_\CC(\RR)$ defined by
\begin{displaymatH}
  (U_{GS} f)(x):= \sqrt{\gamma(x)}f(x) 
   \quad(x\in\RR)
  ,
\end{displaymatH}
is isometric and surjective, hence unitary.
Using this map to transfer the Weyl operators to $L^2_\CC(\RR)$,
we see them in the following form:
\begin{displaymatH}
 \left.
 \begin{array}{l}
  \bigl[W(t) f\bigr](x) = f(x-2t)\\[3pt]
  \bigl[W(is) f\bigr](x) = e^{isx}f(x)\\
 \end{array}
 \right\rgroup
 \quad
 (x\in\RR)
 ,
\end{displaymatH}
i.e.\ we obtain the Schr\"odinger representation.

In conclusion, we have seen that the Heisenberg representation coincides with
the Fock representation of the CCR over $\CC$ which is unitarily equivalent
with the Gaussian representation of the CCR over $\CC$, as both are
cyclic vacuum representations. Further the Schr\"odinger representation
is unitarily equivalent with the Gaussian, and so with the Heisenberg's.


\subsubsection{Infinite dimensional case illustrated with $E=L^2_\RR(\RR)$}
\paragraph{Step 1: basic ingredients}
We illustrate the infinite dimensional case of the Gaussian functor
with $E=L^2_\RR(\RR)$. To start with, we  shortly stating the ingredients, 
and, after that, we will provide their definition, comment and illustrate.
The basic ingredients are:
\begin{enumerate}
 \item 
  a triple of spaces $E\subset L^2_\RR(T)=:H \subset E'$, and 
 \item 
  a measure $\PP$ on $E'$ such that
  \begin{equation}
  \label{eq_BochnerMinlos}
    \int_{E'} e^{i\ip{f}{\om}}d\,\PP(\om) = e^{-\half\norM{f}^2_{H}}
     \quad
     (f\in E)
     \;.
  \end{equation}
\end{enumerate}

We now give the details:
\begin{Itemize}{-}
 \item
 the middle space is a real Hilbert space $L^2_\RR(T)$ 
 where $T$ is a countable set ($T\subset\NN$) equipped with the counting
 measure (so that $L^2_\RR(T)=\ell^2(T)$)
 or a continuum ($T\subset\RR$ finite or infinite interval)
 equipped with the Borel-Lebesgue measure; 
 \item $T$ can be interpreted as time;
 \item
 $E$ is a so-called countably Hilbert nuclear space. 
 The details of the defining properties of this can be found in
  \cite[Appendix]{Hida}, but for our concern, the following examples 
  (to be used further on) will suffice:
 \begin{Itemize}{$*$}
  \item 
  $E=H=E'=\RR=L^2_\RR(\{1\})$ which we saw before;
  \item 
  $E=\mathcal{S}$, the real-valued test functions of Schwartz on $\RR$, 
  $H=L^2_\RR(\RR)$, and $E'=\mathcal{S}'$ is the space of tempered real-valued 
  distribution (also known as generalized functions);
 \end{Itemize}
 \item
 The existence of a measure as mentioned in (\ref{eq_BochnerMinlos})
 (under the appropriate conditions)
 is asserted by the Bochner-Minlos theorem (\Cite{Hida});
 \item
 The space $E$ is dense in $L^2_\RR(T)$ with respect to $\norM{}_{L^2_\RR(T)}$;
 \item
 $E'$ is the dual of $E$
 \item 
 the bilinear form $E\times E'\to \RR$ that sends $(f,\om)$ to $ \om(f)$
 is denoted by $\ip{\cdot}{\cdot}$ i.e.\ $\ip{f}{\om}:=\om(f)$
 (The reason for using this order of $E'$ and $E$ in the bilinear
  form will become clear later);
 \item
 the injections $E\to H$ and $H\to E'$ are continuous; 
 (\cite[Appendix]{Hida}).
 \item Note that, since $f$ is real-valued,
        $\norM{f}_H^2=\int_T f(t)^2 dt$.
\end{Itemize}

The elements of $E$ induce point-evaluations on $E'$:
$\be(f)$, $f\in E$, where $\be(f)$ is the map 
$E' \ni \om  \mapsto \om(f):=\ip{f}{\om} \in \RR$.

On $E'$, we have the weak-$*$ topology, $\sigma(E',E)$, 
i.e.\ the smallest topology for which all $\be(f)$, $f\in E$, 
are continuous.
Let $\mathcal{F}$ be the $\sigma$-algebra generated by the weak-$*$ topology
$\sigma(E',E)$, 
otherwise stated 
$\mathcal{F}$ is the $\sigma$-algebra generated by the the cylinder sets
\begin{displaymatH}
 A_{f_1,\ldots, f_n,B} 
   := 
  \{\om\in E': 
     (\ip{f_1}{\om}, \ldots, \ip{f_n}{\om}) \in B \}
   \,,
\end{displaymatH}
where $n\geq 1$, $f_1,\ldots, f_n \in E$ and $B$ is a Borel set in $\RR^n$.

\paragraph{Step 2: an isometry $\be: L^2_\RR(T) \to L^2_\RR(E',\PP)$}
By virtue of (\ref{eq_BochnerMinlos}), $\PP(E')= 1$ (take $f=0$), so
$(\Om:=E', \mathcal{F})$ is a probability space.
For $f\in E$, 
$\be(f): \Om= E'\to\RR$ can therefore 
be viewed as a random variable on this probability space.
Moreover, (\ref{eq_BochnerMinlos}) implies that the characteristic function
of $\be(f)$ is
\begin{displaymatH}
    \int_{E'} e^{it\be(f)}d\,\PP(\om) = e^{-\half t^2 \norM{f}^2_{H}}
     \quad
     (t\in \RR)
     \,,
\end{displaymatH}
i.e.\ $\be(f)$ is a centered ($\EE(\be(f))=0)$ 
Gaussian random variable with variance $\norM{f}_H= \int_T f(t)^2 dt$. 
Reformulating:
\begin{aligN}
 \norM{\be(f)}^2_{L^2_\RR(E',\,\PP)} 
   & = \int_{E'} \be(f)^2 d\PP
     = \EE(\be(f)^2)= \EE(\be(f)^2) - [\EE(\be(f))]^2
   \\
   & = \textsf{Var}(\be(f))
     = \norM{f}^2_{L^2_\RR(T)}
   \,,
\end{aligN}
which implies that $\be$ is a linear isometry $E\to L^2_\RR(E')$.
Since $E$ is a dense subset of $H=L^2_\RR(T)$, $\be$ can be extended
to a linear isometry between the Hilbert spaces 
$H=L^2_\RR(T)$ and $L^2_\RR(E')$, also denoted by $\be$.
As such, $\be$ automatically preserves the inner product.
In particular
\begin{aligN}
 \textsf{Cov}(\be(f),\be(g))
  &=
 \EE(\be(f)\be(g))- \EE(\be(f))\EE(\be(g))
  =
 \EE(\be(f)\be(g))
  \\
  & =
 \ip{\be(f)}{\be(g)}_{L^2_\RR(E')}
  =
 \ip{f}{g}_{L^2_\RR(T)}
\end{aligN}

\paragraph{Step 3: complexification of the map $\be$ to 
$\be_\CC: L^2_\CC(T) \to L^2_\CC(E')$}

We define the complexification of $\be$ as:
\begin{displaymatH}	
 \be_\CC(f + ig) = \be(f) +i \be(g)
 \quad (f,g\in L^2_\RR(T))
 \,,
\end{displaymatH}

That $\be_\CC: L^2_\CC(T) \to L^2_\CC(E')$ is an isometry,
can be seen from
\begin{aligN}
 \ip{\be_\CC(f+ig)}{\be_\CC(f+ig)}_{L^2_\CC(E')}
  \ats
   =
 \ip{\be(f)+i\be(g)}{\be(f)+i\be(g)}_{L^2_\CC(E')}  
  \\
  \ats
  =
 \ip{\be(f)}{\be(f)} + \overline{i}i\ip{\be(g)}{\be(g)}
  -i\ip{\be(g)}{\be(f)} + i\ip{\be(f)}{\be(g)}
  \\
  \ats
  =
 \int_T f^2 + \int_T g^2 -i \int_T gf + i \int_T fg
  =
 \int_T f^2 + \int_T g^2 
  \\
 \ats
  =
 \ip{f+ig}{f+ig}_{L^2_\CC(T)}
 \,.
\end{aligN}

From now on we drop also the subscript $\CC$ from $\be_\CC$.

\paragraph{Step 4: Exponential vectors}

For $f_c \in L^2_\CC(T)$, we define the exponential vector 
$\expv(f_c)$ as
\begin{equation}
 \expv(f_c) = e^{\be(f_c) - \half\int_T f_c^2}
 \,,
\end{equation}
and we calculate the inner product of two such vectors.

Suppose $f_c :=f_1 +if_2\in$  and $g_c :=g_1 +ig_2\in L^2_\CC(T)$ 
with $f_1,f_2,g_1,g_2 \in L^2_\RR(T)$.
Then  
\begin{aligN}
  \ip{e^{\be(f_c)}}{e^{\be(g_c)}}
   \ats
    =
  \ip{e^{\be(f_1+if_2)}}{e^{\be(g_1+ig_2)}}_{L^2_\CC(E')}
  \\
  \ats
   =
 \ip{e^{\be(f_1)+i\be(f_2)}}{e^{\be(g_1)+i\be(g_2)}}_{L^2_\CC(E')}  
   =
 \EE e^{\be(f_1+g_1)+ i\be(g_2-f_2)}
\end{aligN}
which is of the form $\EE e^{iX + Y}$ with
$X:= \be(g_2 - f_2)$ and $Y= \be(f_1 +g_1)$ centered real Gaussian random
variables. Since $\be$ is an isometry $L^2_\RR(T)\to L^2_\RR(E')$, 
we have that 
\begin{aligN}
 -\Var{X} \ats + \Var{Y} + 2i\Cov{X}{Y} 
  \\
  \ats 
   = 
  -[\int_T (g_2 - f_2)^2 dt] + [\int_T (f_1 + g_1)^2 dt] 
   + 2i[ \int_T (g_2 - f_2)\cdot (f_1 + g_1) dt]
  \\
  \ats
  = \int_T [ (f_1+ g_1) + i(g_2 -f_2)]^2 dt
  = \int_T [ (\overline{f_1 + if_2}) + (g_1 + ig_2) ]^2 dt
  \\
  \ats 
  = \int_T [ \overline{f_c} + g_c]^2 dt
  = \int_T \overline{f_c}^2 dt  + \int_Tg_c^2 dt + 2\int_T\overline{f_c}g_c dt
\end{aligN}
so that by (\ref{eq_expectation_iX+Y})
\begin{displaymatH}
  \EE  e^{i\be(g_2-f_2)+ \be(f_1+g_1) }
    = \exp\left[\half\int_T \overline{f_c}^2  
            + \half\int_Tg_c^2 
           +  \int_T\overline{f_c}g_c
          \right]
 \,.
\end{displaymatH}
Combining the above yields,
\begin{equation}
 \ip{\expv(f_c)}{\expv(g_c)}_{L^2_\CC(E')} = e^{\ip{f_c}{g_c}_{L^2_\CC(T)}}
 \quad
  (f_c, g_c \in L^2_\CC(T)\,)
  \,,
\end{equation}
whence the name exponential vectors.

\paragraph{Step 5: Exponential vectors form a total subset of $L^2_\CC(E')$}
For $f_c=f_1 + i f_2 \in L^2_\CC(T)$, 
the random variable $|e^{\be(f_c)}|^2 = e^{2\be(f_1)}$ is of the form 
$e^Y$ where
$Y=\be(2f_1)$ is a centered Gaussian with variance $4\int_T f_1(t)^2 dt$.
From (\ref{eq_expectation_iX+Y}), it therefore follows that 
\begin{displaymatH}
\expv(f_c) = e^{\be(f_c) - \half\int_T f_c(t)^2 dt}  \in L^2_\CC(E')
.  
\end{displaymatH}
Next, the set of exponential vectors is also total in $L^2_\CC(E')$, i.e.\
the linear span of the exponential vectors is a dense linear subspace of
$L^2_\CC(E')$.
Actually, it suffices to consider real-valued test functions:

\begin{lemma}
The set $\{ e^{i\be(g)}: g\in E \}$ 
is a total subset in $L^2_\CC(E')$.
\end{lemma}

\begin{pf}
(See \cite[p.\ 116]{Hida}).
Note that $E$ is a dense linear subspace of $L^2_\RR(T)$.
Now, 
let $\mathbb{A}$ be the complex linear subspace spanned by the $e^{i\be(g)}$, 
$g\in E$. 
Since $e^{i\be(g_1)} \cdot e^{i\be(g_2)}= e^{i\be(g_1 + g_2)}$, 
$\mathbb{A}$ is actually an algebra.

To prove that $\mathbb{A}$ is dense in $L^2_\CC(E')$ is suffices to prove that
if $\ph\in L^2_\CC(E')$ is orthogonal to all random variables of the form
\begin{displaymatH}
 \exp\bigl[ i\be(\sum_1^n t_i g_i) \bigr]\,,
 \quad
 (t_i\in\RR\,,\; g_i \in E\,)
 \,, 
\end{displaymatH}
then $\ph = 0$ $\PP$-almost everywhere ($\PP$-almost surely).

\statement{Step 1}

Fix $n\in\NN$, $g_1, \ldots, g_n \in E$, 
and let $\mathcal{F}_n \subset \mathcal{F}$ be the $\si$-algebra generated by
$X_1:=\be(g_1)$, \ldots, $X_n:=\be(g_n)$. 
Suppose we have for every $(t_1,\ldots, t_n) \in \RR^n$:
\begin{displaymatH}
  \int_{E'} \exp\bigl[\sum_1^n it_k X_k\bigr] \overline{\ph}\: d\PP= 0  
  \,.
\end{displaymatH}
Since $E' \in \mathcal{F}_n$, the second equality in the following chain
is valid:
\begin{aligN}
  0 \ats 
   =  \int_{E'} \exp\bigl[\sum_1^n it_k X_k\bigr] \overline{\ph} \: d\PP
   = \int_{E'}  \EE\Bigl(
                 \exp\bigl[\sum_1^n it_k X_k\bigr] \overline{\ph}
                  \Bigm| \mathcal{F}_n
                   \Bigr)
                \: d\PP
  \\
  \ats 
    = \int_{E'}  \exp\bigl[\sum_1^n it_k X_k\bigr] 
                 \EE\bigl(\overline{\ph} \bigm| \mathcal{F}_n \bigr)
                 \:d\PP
  \,.
\end{aligN}
where we have used in the third equality that 
$\exp\bigl[\sum_1^n it_k X_k\bigr]$ is $\mathcal{F}_n$-measurable.
Since a function is $\mathcal{F}_n$-measurable if and only if it is a function
of $(X_1,\ldots,X_n)$ we can write 
$\EE\bigl(\overline{\ph} \bigm| \mathcal{F}_n \bigr) = f(X_1, \ldots, X_n)$
for some measurable $f:\RR^n\to\RR$ and conclude that
\begin{displaymatH}
 0 = \int_{E'}  e^{it_1 X_1 + \ldots + it_n X_n} f(X_1,\ldots,X_n) d\PP
   = \int_{E'}  e^{it_1 x_1 + \ldots + it_n x_n} f(x_1,\ldots,x_n) 
     dP_{(X_1,\ldots,X_n)}(x_1, \ldots, x_n)   
  \,,
\end{displaymatH}
i.e.\ the $n$-dimensional Fourier-transform of $f(x_1,\ldots,x_n)$ with respect
to the joint-probability distribution of $(X_1,\ldots,X_n)$ 
vanishes on $\RR^n$. 
This means that $f(x_1,\ldots,x_n)= 0$ for $P_{(X_1,\ldots,X_n)}$-almost all
$(x_1,\ldots,x_n) \in \RR^n$
i.e.\
$\PP[ f(X_1,\ldots,X_n) \neq 0 ] = P_{(X_1,\ldots,X_n)}[ f\neq 0 ]  = 0$.

Conclusion: $\EE(\overline{\ph} | \mathcal{F}_n) = 0 $ $\PP$-a.e.

\statement{Step 2}

Choose a system $\{g_1,g_2,\ldots\}$ in $E$ that is complete in $L^2_\RR(T)$. 
Then the corresponding $\mathcal{G}_n:=\sigma\{g_1,\ldots,g_n\}$ increase to $\mathcal{F}$
(i.e. $\mathcal{G}_n \subset \mathcal{G}_{n+1} \subset \mathcal{F}$ and 
$\mathcal{F}$ is the smallest $\si$-algebra 
containing $\union_n \mathcal{G}_n$).
 \begin{pf}
 Recall that $\mathcal{F}$ is the smallest $\si$-algebra such that all
 $\be(f)$, $f\in E$, are measurable. 
 
 Let $f \in E$. We prove that $f$ is measurable 
 with respect to the $\si$-algebra generated by $\union_n \mathcal{G}_n$,
 so that $\mathcal{F}$ is also generated by $\union_n \mathcal{G}_n$.

 Indeed, 
 since $E\subset L^2_\RR(T)$ and the linear span of 
 $\{g_n : n\in\NN\}$ is dense in $L^2_\RR(T)$, there exists a sequence
 $f_n = \sum_i^{N_n} \al_{n i} g_{n i}$ of finite linear combinations 
 of $\{g_n : n\in\NN\}$ such that $\norM{f-f_n}_2 \to 0$. 
 In particular, $f_n(t) \to f(t)$ 
 for almost all $t\in T$, and $f$ is measurable with respect to the 
 $\si$-algebra generated by $\{f_n: n\in\NN\}$. 
 Since each $f_n$ is $\mathcal{G}_{N_n}$ measurable, $f$ is 
 measurable with respect to the $\si$-algebra generated by $\union_n \mathcal{G}_n$.
 \end{pf}

Let $\mathcal{G}:=\{ A\in \mathcal{F} : \int_A \overline{\ph} \:d\PP = 0\}$.
Then $\mathcal{G}$ is a $\si$-algebra that contains each $\mathcal{G}_n$ 
(by step 1 above), 
whence $\mathcal{G}\supset\mathcal{F}$ i.e.\
$\int_A \overline{\ph} \:d\PP = 0 $ for all $A\in\mathcal{F}$. 
This implies that $\ph= 0 $ $\PP$-a.e. 
\end{pf}

\section{Functorial quantization of the harmonic oscillator}
Instead of Dirac's quantization procedure, we can now outline a
functorial quantization procedure that acts on the classical evolution
in phase space. We take the simple classical harmonic oscillator as 
illustrating example.

\paragraph{Step 1: Identify the classical phase space with a Hilbert space}
The simple harmonic oscillator has a position $q$ and momentum $p$. 
Its phase space $\RR^2$ can be identified with $\CC$ via
$(p,q) \leftrightarrow z=p+iq$. If we have $n$ oscillators, the phase space
corresponds with $\CC^n=\ell^2(\{1,\ldots,n\})$. For a chain of oscillators,
we have as phase space $\ell^2(\{1,2,\ldots\})=\ell^2(\NN)$, and a continuum
of oscillators labeled by $x\in\RR$ may correspond to $L^2(\RR)$.

\paragraph{Step 2: Identify the evolution in phase space with an evolution
                   in the corresponding Hilbert space}
As the explicit time evolution of the harmonic oscillator is given by
\begin{aligN}
 p(t) & = - q(0) \sin(\om t) + p(0) \cos(\om t)\,,\\
 q(t) & =  \phantom{-}q(0) \cos(\om t) + p(0) \sin(\om t)\,,\\
\end{aligN}
the evolution in terms of $z$ reads
\begin{displaymatH}
 z(t) = e^{i\om t}z(0)
 ,
\end{displaymatH}
i.e.\ the time evolution operator $T_t$ is the multiplication by $e^{i\om t}$.

In the same vein, $n$ harmonic oscillators, each with frequency $\om_i$, 
can be described in time by the evolution 
$(z_1,\ldots,z_n)\mapsto (e^{i\om_1 t}z_1,\ldots, e^{i\om_n t}z_n)$
in $\CC^n$.

\paragraph{Step 3: Quantization of the phase space}
Next we need to specify how to obtain the canonical observables $(Q,P)$, and 
in particular the Hilbert space they act on. 

This is done by taking a cyclic vacuum representation of the CCR over $\CC$. 
For example, one could take as Hilbert space the Fock space over $\CC$ and 
end up with the Heisenberg representation of the harmonic oscillator.
Another, unitarily equivalent, choice would be to take a Gaussian 
representation which is unitarily equivalent with the Schr\"odinger 
representation.

For $n$ oscillators, one would take the CCR over $\CC^n=\ell^2(\{1,\ldots,n\})$
and for a chain of oscillators the CCR over $\ell^2(\NN)$ is appropriate.

\paragraph{Step 4: Quantization of the time evolution}
The classical time evolution $z\to e^{i\om t}z$, can now be quantized by 
describing a time evolution of the corresponding Weyl operators 
$W(z) \mapsto W(e^{i\om t}z)$.
In terms of the canonical observables $Q$ and $P$, defined as the generators
of the strongly continuous unitary maps $s\mapsto W(is)$ and $t\mapsto W(t)$
respectively,
the evolution comes down to
\begin{displaymatH}
\left.
\begin{aligned}
 P(t) & = - Q(0) \sin(\om t) + P(0) \cos(\om t)\,,\\
 Q(t) & =  \phantom{-}Q(0) \cos(\om t) + P(0) \sin(\om t)\,,\\
\end{aligned}
\right\rgroup
\quad
(t\in\RR),
\end{displaymatH}
which corresponds to what we would expect  based on the 
Heisenberg equation of motion for $Q,P$ 
\cite[2.2.19, p.~83; 2.3.4, p.~95]{Sakurai}.

For $n$ oscillators the evolution in $\CC^n$ described by 
$W(z_1,\ldots,z_n)\mapsto W(e^{i\om_1 t}z_1,\ldots, e^{i\om_n t}z_n)$
reads in terms of corresponding infinitesimal generators
\begin{displaymatH}
\left.
\begin{aligned}
 P_i(t) & = - Q_i(0) \sin(\om_i t) + P_i(0) \cos(\om_i t)\,,\\
 Q_i(t) & =  \phantom{-}Q_i(0) \cos(\om_i t) + P_i(0) \sin(\om_i t)\,,\\
\end{aligned}
\right\rgroup
\quad
(t\in\RR\,,\;i=1,\ldots,n)
.
\end{displaymatH}

In the remainder of this section we will discuss observables, states and 
wave-particle duality in the functorial quantization.

\subsection{Wave-particle duality in the harmonic oscillator}
Applying the functorial procedure outlined above yields the same 
quantum analogue of the classical harmonic oscillator as the usual
Dirac procedure. 
Indeed, by choosing appropriate cyclic vacuum representations of
the CCR, the familiar Schr\"odinger and Heisenberg representations are obtained 
respectively. States in the Schr\"odinger representation are described by
(complex-valued) functions on $\RR$ (= configuration space of position)  
that express the wave-like properties when measuring
the position the harmonic oscillator, whence their name wave functions.
Also, momentum states are described by wave functions.

In the Heisenberg representation, states are (complex-valued) functions on 
$\NN$ (= configuration space of energy) that relate to the particle-like
properties found when measuring the the energy of the oscillator.

Identifying the exponential states in Schr\"odinger representations with those
in Heisenberg's representation, both representations are easily seen to be
unitarily equivalent.

\subsection{Algebra of observables}
For a classical harmonic oscillator, the position and momentum observable 
$q$ and $p$ completely determine the state of the system and 
therefore each observable $o$ of the system can be viewed as a function 
$o(q,p)$ of $q$ and $p$.

After quantization, $q$ and $p$ are represented as unbounded self-adjoint 
operators $Q$ and $P$ on a Hilbert space. A reasonable guess at the quantum 
observable $O$ corresponding to $o=o(p,q)$ can be made if $o(p,q)$ admits a 
Taylor series expansion in $q$ and $p$. 
In that case, a formal substituting $q\mapsto Q$
and $p\mapsto P$ in the Taylor series of $o$ can be effectuated and 
after that,
a domain for $O$ on which $O$ is well-defined and 
in addition self-adjoint can be looked for.

From a mathematical viewpoint, the above way to obtain quantum analogues
of the classical observables is full of obstacles and pitfalls, since it
is often hard to find a domain on which the Taylor series expansion can be 
proven to be self-adjoint.

Therefore, Weyl took another approach using bounded operators, as 
the latter are well-defined on the whole Hilbert space and therefore
mathematically easier to deal with. 
The main ideas are the following:
\begin{Properties}
 \item
 Restrict to bounded observables of the system instead of unbounded ones.
 \item
 If $o$ is a bounded observable and therefore a function $o(q,p)$ of $q$ and 
 $p$, then use its Fourier transform $\widehat{o}(s,t)$ that satisfies
 $o(q,p)=\int\int \widehat{o}(s,t) e^{isq-itp} ds dt$ to define
 the quantum observable corresponding to the classical observable $o$ as
 \begin{displaymatH}\displaystyle
   O=O(Q,P)= \int_{-\infty}^\infty \int_{-\infty}^\infty
                 \widehat{o}(s,t) 
                  e^{isQ-itP} 
             ds dt
  .
 \end{displaymatH}
\end{Properties}

We comment on the above ideas.

The first idea is no restriction as 
the properties of unbounded self-adjoint
operators can be encoded in bounded operators.
For example, the event that a measurement of an observable $O$ yields a 
value in $(a,b)$ (i.e. the observed value $o$ of $O$ has $a\leq o \leq b$)
is described by a projection operator $P_{(a,b)}$.
One can therefore imagine that the family of bounded self-adjoint operators 
$\{ P_{(a,b)}: a<b \text{ in } \RR \}$ describes the observable $O$ completely.

A mathematical description of that is given by Von Neumann's spectral theorem.
Before formulating that, let us describe its ingredients:
\begin{definition}
\label{def_spectralMeasure}
 A \Dindex{projection-valued measure} (better known as \Dindex{spectral measure}
 in the physics literature) $P$ is a map from the Borel $\si$-algebra
 $\Borel(\RR)$  to the algebra of bounded operators  
 on a Hilbert space $\mathcal{H}$, $\mathcal{B}(\mathcal{H})$, such that
 \begin{Properties}
  \item
   $P(B)$ is a projection for each Borel set $B\subset \RR$,
  \item
   $P(B_1\intersection B_2)= P(A)\cdot P(B)$ for Borel sets  
   $B_1, B_2 \subset \RR$
  \item
   if $B_1,B_2,\ldots$ are disjoint Borel sets in $\RR$, then
   \begin{displaymatH}
      \lim_{N\to\infty} \sum_1^N P(B_n)\xi = P(\union_n B_n) \xi
      \quad
      (\xi\in\mathcal{H}).
   \end{displaymatH}
 \end{Properties}
\end{definition}
For each $\xi \in \mathcal{H}$, the map $P_\xi: B\mapsto \oip{\xi}{P(B)}{\xi}$
is a measure on the Borel sets on $\RR$.
If $\int_\RR |x|^2 P_\xi(dx) < \infty$
then the integral $\int_\RR x P_\xi(dx)=:\int_\RR x \oip{\xi}{P(dx)}{\xi}$
exists (in $\RR$), 
and the set $\{\xi : \int_\RR |x|^2 \oip{\xi}{P(dx)}{\xi} < \infty\}$ is a linear
subspace of $\mathcal{H}$.

\begin{theorem}[Von Neumann]
There is a one-to-one correspondence between self-adjoint operators $O$
on a Hilbert space $\mathcal{H}$ and projection-valued measures
$P:\Borel(\RR)\to\mathcal{B}(\mathcal{H})$.

\medskip

Specifically, if $O$ is a self-adjoint operator on $\mathcal{H}$,
then there is a projection valued measure $P$ on $\mathcal{H}$
such that
\begin{Properties}
 \item 
  the linear subspace 
   $\mathcal{H}_P=
    \{ \xi\in\mathcal{H} : \int_\RR |x|^2 \oip{\xi}{P(dx)}{\xi} < \infty\; \} $
  is dense in $\mathcal{H}$;
 \item
  $O$ is well-defined on $\mathcal{H}_P$ and
  $\oip{\xi}{O}{\xi}= \int_\RR x P_\xi(dx)$ for all $\xi\in\mathcal{H}_P$.

  In particular, 
  $\norM{O\xi}^2=\ip{O\xi}{O\xi}=\int_\RR |x|^2 \oip{\xi}{P(dx)}{\xi}$.
\end{Properties}
Symbolically, we denote this as
\begin{displaymatH}
 O = \int_\RR x P(dx)
 .
\end{displaymatH}

\medskip

Conversely, given a projection-valued measure $P$ on $\mathcal{H}$ and
$\eta\in \mathcal{H}$, 
we have that the sequence $\eta_n:=P(-n,n) \eta$ converges to $\xi$,
and is contained in the linear subspace
$\mathcal{H}_P:=
   \{ \xi\in\mathcal{H} : \int_\RR |x|^2 \oip{\xi}{P}{\xi} < \infty\; \}$
(\ref{ap_PVM}).
Hence, the latter subspace is dense in $\mathcal{H}$. A self-adjoint
operator $O$ can be defined on $\mathcal{H}_P$ via the formula
$\oip{\xi}{O}{\xi} := \int_\RR  x \oip{\xi}{P(dx)}{\xi}$  (and 
$\oip{\xi}{O}{\eta}$ is from there defined by polarization)
(see \ref{ap_PVM}).
\end{theorem}

\begin{example}
 In the Schr\"odinger cyclic vacuum representation of the CCR over $\CC$,
 $W(is)=e^{isQ}$ is the multiplication  operator $[W(is) f]= e^{isq} f(q)$ 
  in $L^2(\RR)$
 (see the end remarks of section \ref{ssec_UniquenessRep}).
 The form of the projection-valued measure of $Q$ in the Schr\"odinger
 representation is $P_Q(a,b)= \indicator_{(a,b)}$. 
 This corresponds to the fact that in a normalized state $\psi$, 
 the probability  of measuring a value of $Q$ that lies between $a$ and $b$ is 
 $\int_\RR \indicator_{(a,b)}(q) |\psi(q)|^2 dq=\int_a^b |\psi(q)|^2 dq $.
 \cite[p.~101]{Sakurai}.
\end{example}

\medskip

Weyl's second idea, viz.\ defining quantum analogues via the Fourier transform,
works if the bounded observable $o=o(q,p)$ has a Fourier transform which
is guaranteed if $o$ is integrable over the phase space 
($\int \int | o(q,p)| dq dp < \infty$).
Identifying as usual the phase space with $\CC$  
(via $(q,p) \leftrightarrow p+iq = z$) and 
$e^{isQ+itP} \leftrightarrow W(t+is)$, 
we want the classical observable 
$o(q,p)=\int\int \widehat{o}(s,t) e^{isq+itp} ds dt$ 
to correspond to 
$O(Q,P)= \int\int o(s,t) W(t+is) dt ds$.

The latter integral symbolizes that some limit of (superpositions of) 
Weyl operators is taken, and we have to elucidate what kind of convergence
is involved.
It turns out that so-called strong convergence enables us to obtain
interesting operators from (superpositions of)  Weyl operators.

\begin{definition}[\protect{\Dindex{Strong convergence}}]
  A sequence of operators $(T_n)_{n=1}^\infty$ in $\mathcal{B}(\mathcal{H})$
  converges strongly 
  to an operator $T\in\mathcal{B}(\mathcal{H})$ if 
  $T_n \xi \to T\xi$ ($n\to\infty$) for each $\xi\in\mathcal{H}$ 
  (i.e.\ $\lim_n\norM{T_n\xi- T\xi}= 0$ for all $\xi\in\mathcal{H}$).
\end{definition}

Strong convergence implies weak convergence:
\begin{definition}[\protect{\Dindex{Weak convergence}}]
  A sequence of operators $(T_n)_{n=1}^\infty$ in $\mathcal{B}(\mathcal{H})$
  converges weakly
  to an operator $T\in\mathcal{B}(\mathcal{H})$ if 
  $\oip{\xi}{T_n}{\eta} \to \oip{\xi}{T}{\eta}$ ($n\to\infty$) for each 
  $\xi,\eta \in\mathcal{H}$.
\end{definition}

To obtain interesting (bounded) operators from the Weyl operators 
and to do all the 
operations that are commonly 
used in quantum mechanics, 
we need to take linear combinations (superpositions), 
compositions, adjoints and strong limits.
 
Therefore, the collection of all bounded 
observables of the quantum harmonic oscillator
system is mathematically best described by a Von Neumann algebra:

\begin{definition}
 A \Dindex{von Neumann} algebra on a Hilbert space $\mathcal{H}$
 is a collection $\mathcal{N}$ 
 of bounded operators on $\mathcal{H}$ that is closed under taking
 \begin{Properties}
   \item superpositions (linear combinations): 
     $A,B\in\mathcal{N},\: \al,\be\in\CC  \implies 
        \al A + \beta B\in\mathcal{N}$,
   \item composition: 
     $A,B\in\mathcal{N} \implies AB \in\mathcal{N}$,
   \item adjoints: 
     $A\in\mathcal{N} \implies \adj{A}\in\mathcal{N}$,
   \item strong limits:
     $\mathcal{N} \ni A_n \to  A\in\mathcal{B}(\mathcal{H}) \text{ strongly } 
      \implies A\in\mathcal{N}$.
 \end{Properties}
 For simplicity we will always assume that a von Neumann algebra in
 $\mathcal{B}(\mathcal{H})$ contains the identity of 
 $\mathcal{B}(\mathcal{H})$: $ \Identity\in\mathcal{N}$.

 \medskip

 Let $\mathcal{B}_0$ be a subset of $\mathcal{B}(\mathcal{H})$.
 Then the  von Neumann algebra generated by the elements of $\mathcal{B}_0$,
 i.e.\ the intersection of all von Neumann algebra that contain 
  $\mathcal{B}_0$, is denoted by $\vN\{\mathcal{B}_0\}$.
\end{definition}

The interesting operators 
that we can obtain via strong convergence 
from the Weyl operators of the harmonic oscillator 
include all bounded measurable functions of $Q$, $P$ 
and of $N$,
and therefore in particular the spectral measures 
(projection valued measures) associated with $Q$, $P$, and $N$.

The mathematical formulation of this is as follows.

\begin{definition}[ bounded measurable functions of a self-adjoint operator]
 Let $O$ be a self-adjoint operator with projection-valued measure $P_O$,
 and let $f:\RR\to\RR $ be a bounded Borel function.
 Then for $\psi\in\mathcal{H}$, $B\mapsto \oip{\psi}{P_O(B)}{\psi}$ is
 a finite measure and 
 $\int_{-\infty}^\infty f(x) \oip{\psi}{P_O(dx)}{\psi} < \infty$. 
 Therefore, an bounded operator $f(O)$ can be defined via 
 \begin{displaymatH}
  \oip{\psi}{f(O)}{\psi}= \int_{-\infty}^\infty f(x) \oip{\psi}{P_0(dx)}{\psi}
  \quad
  (\psi \in\mathcal{H})
  ,
 \end{displaymatH}
 since matrix elements $\oip{\psi_1}{f(O)}{\psi_2}$ can be defined via 
 polarization (\ref{ap_PolarizationFormula}).
 If $f$ is real-valued then $f(O)$ is Hermitean.
 In particular, if $f=\indicator_B$ for some Borel set $B$, 
 then $f(O)= P_O(B)$ i.e.\ the projection-valued measure of $O$ is obtained.
\end{definition}

With the notion of functional calculus above, we can now state

\begin{theorem}
\label{th_vNWeylOpCCRoverC}
Let $(\mathcal{H},\Om)$ be a cyclic vacuum representation of the
CCR over $\CC$ 
(so a cyclic vacuum representation of the quantum harmonic oscillator). 
\begin{Properties}
 \item
 \label{th_vNWeylOpCCRoverCi}
 Let $Q$ be the position operator and $P_Q$ its spectral measure. Then
 \begin{aligN}
   \vNeu\{e^{isQ}: s\in\RR\} 
    & = \vNeu\{ P_Q(a,b): a<b\In\RR\}
    = \vNeu\{ f(Q) : f \text{ step function}\}
    \\
    &
    = \vNeu\{ f(Q) : f \in L^\infty \}. 
 \end{aligN}
 \item 
 \label{th_vNWeylOpCCRoverCii} 
 Let $P$ be the momentum operator and $P_P$ its spectral measure. Then 
 \begin{aligN}
   \vNeu\{e^{-itP}: t\in\RR\} 
     & = \vNeu\{ P_P(a,b): a<b\In\RR\}
     = \vNeu\{ f(P) : f \text{ step function}\}
     \\
     & = \vNeu\{ f(P) : f \in L^\infty\} 
 \end{aligN}
 \item
 \label{th_vNWeylOpCCRoverCiii} 
 Let $H=\hbar(N+ \half)$ be the Hamiltonian and let $P_N$ be the spectral 
 measure of the number operator $N$. Then
 \begin{aligN}
  \vNeu\{s^N: s\in (-1,1)\} & = \vNeu\{P_N(n): n\in\NN\} = 
   \vNeu\{f(N): f\in \ell^\infty\}
   \\
   & \subset
  \vNeu\{W(z) : z\in\CC \}
 \end{aligN}
\end{Properties}

In particular, the von Neumann algebra generated by the Weyl operators,
$\mathcal{W}= \vNeu\{W(z): z\in\CC \}$, 
 contains all bounded functions of $Q$, $P$, and $N$.
\end{theorem}

A key lemma to prove the above, is Weyl's previously mentioned idea of 
writing an operator $O(Q,P)= \int\int o(s,t) W(t+is) dt ds$.

\begin{theorem}[Proof in \protect{\ref{ap_IntegralWeylOp}}]
\label{th_integralWeyl}
 Suppose:
 \begin{conditions}
  \item $\ro:\CC\to \CC$ is continuous, 
  \item $W:\CC\to\B(\mathcal{H})$ is a strongly continuous family in $\B(\mathcal{H})$, 
  \item the map $z\to |\ro(z)|\norM{W(z)}$ from $\CC$ into $[0,\infty)$ is 
        integrable with respect to the Lebesgue measure on $\CC$.
 \end{conditions}

 Then,
 $T\in \vNeu\{W(z): z\in\CC \} \subset B(\mathcal{H})$, 
 and $\norM{T}_{\B(\widehat{H})} \leq \int_{\CC} |\ro(z)|\norM{W(z)} \la_2(dz)$.
\end{theorem}
(A version in which continuity of $\rho$ is dropped can be found in
\cite[chapter 3]{Thirring}).

\medskip
Using the above lemma, we can prove the follow lemma, which is a step up 
to the first two statements of theorem \ref{th_vNWeylOpCCRoverC}.

\begin{lemma}[Proof in \ref{ap_PVMofQandP}]
\label{lem_PVMofQandP}
Let $(a,b)\subset\RR$.
Then there exists a sequence of functions $f_k$ such that each $f_k$ is
\begin{Properties}
 \item infinitely differentiable;
 \item rapidly decreasing in all its derivatives 
       i.e.\ $x\mapsto x^nf_k^{(m)}$ is bounded for all $n,m\in\NN$;
 \item $0\leq f_k \leq \indicator$;
 \item $f_k$ coincides with $\indicator_{(a,b)}$ 
       on $\RR \setminus ( [a-\frac{1}{k},a] \union [b,b+\frac{1}{k}])$
\end{Properties}
(Note that the first two properties tell that the $f_k$ a test functions
in the sense of Schwartz).

\medskip

Then the Fourier transform of each $f_k$ is (continuous and) integrable,
and 
\begin{displaymatH}
 \int_{-\infty}^\infty \widehat{f_k}(s) e^{-isx} ds = f_k(x)
 \quad (x\in\RR)
 .
\end{displaymatH}

By theorem \ref{th_integralWeyl}, 
the operators 
\begin{aligN}
 P_Q^{k}(a,b) 
  & = \int_{-\infty}^\infty \widehat{f_k}(s) W(is) ds, \text{ and} 
  \\
 P_P^{k}(a,b) 
  & = \int_{-\infty}^\infty \widehat{f_k}(t) W(-t) ds
  \quad(\text{the minus sign occurs due to the minus sign 
               in $e^{isQ-itP}$})
\end{aligN}
are in $\vNeu\{W(z): z\in\CC\}$ 
and so are their strong limits
\begin{aligN}
 P_Q(a,b) & = s\text{-}\lim_{k\to\infty}  P_Q^{k}(a,b)\,, \\  
 P_P(a,b) & = s\text{-}\lim_{k\to\infty}  P_P^{k}(a,b)  ,
\end{aligN}
where $P_Q$ and $P_P$ are the spectral measures of $Q$ and $P$ respectively
(see \ref{th_vNWeylOpCCRoverC}).
\end{lemma}

Using the above result we can complete the first two statements of theorem
\ref{th_vNWeylOpCCRoverC}.

\begin{proof}[ of theorem \ref{th_vNWeylOpCCRoverC} \romanref{th_vNWeylOpCCRoverCi},\romanref{th_vNWeylOpCCRoverCii}]
We will prove the following sequence of inclusions for the operator $Q$ 
(the proof for $P$ is analogously):
\begin{aligN}
 \vNeu\{f(Q): f \in L^\infty(\RR)\} 
   &\supset
 \vNeu\{e^{isQ}: s\in\RR \}
   \supset
 \vNeu\{P_Q(a,b): a < b \text{ in } \RR \} 
   \\
   & \supset
  \vNeu\{f(Q): f \text{ step function }\} 
   \supset
 \vNeu\{f(Q): f \in L^\infty(\RR)\} 
   .
\end{aligN}

The first inclusion is trivial, 
since $\{f(Q): f \in L^\infty(\RR)\} \supset \{e^{isQ}: s\in\RR \}$.
The second is the purport of lemma \ref{lem_PVMofQandP} above.
The third inclusion follows from a standard measure theoretic argument.
\begin{pf}
 Let 
 \begin{displaymatH}
  \mathcal{G}
   = 
  \{ A \in \Borel(\RR): \indicator_A(Q) 
                        \in 
                        \vNeu\{P_Q(a,b): a < b \text{ in } \RR \}
  \}
  . 
 \end{displaymatH}
 We will prove that $\mathcal{G}=\Borel(\RR)$ meaning that
 $\indicator_B(Q)$, for each Borel set $B$, and hence also linear combinations
 of them, i.e.\ step functions are in 
 $\vNeu\{P_Q(a,b): a < b \text{ in } \RR \}$.

  First observe that the intervals $(a,b)\subset \RR$ are contained in 
  $\mathcal{G}$ as $\indicator_{(a,b)}(Q)= P_Q(a,b)$ 
  in the functional calculus for self-adjoint operators:
  \question{need elucidation? introduce functional calculus after introducing
            the spectral measure?}
  \begin{aligN}
    \oip{\psi}{\indicator_{(a,b)}(Q)}{\psi}
     : & =\int_{-\infty}^{\infty} \indicator_{(a,b)}(q) 
                                   \oip{\psi}{P_Q(dq)}{\psi}
         =\int_{a}^{b} \oip{\psi}{P_Q(dq)}{\psi}
    \\
    &=\oip{\psi}{P_Q(a,b)}{\psi}
    \quad
    (\psi\in\mathcal{H})
     .
  \end{aligN}
  Next, observe that $\mathcal{G}$ is $\si$-algebra:
  
  \begin{Properties}
   \item
     $\RR\in \mathcal{G}$, since every von Neumann algebra contains 
     $\Identity\in\mathcal{B}(\mathcal{H})$.
   \item   
     $A\in\mathcal{G}\implies\RR\setminus A \in\mathcal{G}$ , since
     $\indicator_{\RR\setminus A}(Q)= \indicator_\RR(Q) - \indicator_A(Q)
      = \Identity -\indicator_A(Q)$ is a linear combination of elements
     in $\vNeu\{P_Q(a,b): a < b \text{ in } \RR \}$. 
    \item
     $A_1, A_2, \ldots \in \mathcal{G} \text{ disjoint }
        \implies 
      \union_n A_n \in \mathcal{G}$, since 
     in $L^\infty(\RR)$ the functions 
     $f_N:= \sum_{n=1}^N \indicator_{A_n}$ converge
     point wise (almost everywhere) to $f=\indicator_{\union_n A_n}$.
     Therefore the multiplication operators 
     $M_{f_n}$ converge strongly to $M_f$ in $\mathcal{B}(L^2(\RR))$: by 
     Lebesgue's dominated convergence theorem,
     $\norM{f_n g - f g}_2^2 \to 0$ for every $g\in L^2(\RR)$.
     Thus in the Schr\"odinger (cyclic vacuum) representation of the 
     CCR we see that 
     $\indicator_{\union_n A_n} 
       = f
       = s-\lim_N f_N \in \vNeu\{P_Q(a,b): a < b \text{ in } \RR \}$. 
  \end{Properties}
  Concluding, $\mathcal{G}$ is a $\si$-algebra that contains the open intervals
  of $\RR$ and must therefore contain $\Borel(\RR)$.
\end{pf}
The last inclusion follows from the fact that every $f\in L^\infty(\RR)$ can
be uniformly approximated by step functions: 
$f_N:=\sum_{n=-N}^N n/N  
         \supnorM{f}
           \cdot 
         \indicator_{[ f\in [n/N\supnorM{f},(n+1)/N\supnorM{f}) ]}
$,
so that $f_n \to f$ strongly in the Schr\"odinger (cyclic vacuum) 
representation of the CCR  (Lebesgue's dominated convergence theorem).
\end{proof}

The following lemma underpins the last statement of
theorem \ref{th_vNWeylOpCCRoverC}.

\begin{lemma}
\label{lem_PVMofN}
We can write $s^N$ as a strong integral with respect to Weyl operators:
 \begin{displaymath}
  s^N =  \frac{1}{\pi}
             \int_\CC 
              \frac{e^{-\half\frac{1+s}{1-s}|z|^2}}{1-s}
              W(z) 
            \la_2(dz)
   \quad(s\in (-1,1) )\;. 
 \end{displaymath}
 
Furthermore,
\begin{displaymatH}\displaystyle
 P_n = (n!)^{-1} \biggl[\frac{d^n}{ds^n} s^N\biggr]_{s=0}
 ,
\end{displaymatH}
where the derivative can be taken in the strong operator topology.
\end{lemma}

\begin{proof}[ of lemma \ref{lem_PVMofN} and theorem 
\ref{th_vNWeylOpCCRoverC} iii.]
By theorem \ref{th_integralWeyl}, the integral over the Weyl operators exists
and a verification shows that it coincides with $s^N$ 
(see \ref{ap_PVMofN}).
We can write $s^N = \sum_{k=0}^\infty s^k P(k)$ ($s\in (-1,1)$, 
where $P(k)$ is the projection
onto the linear subspace spanned by the eigenvector $e_k$ of $N$ 
with eigenvalue
$k$ (i.e.\ the $k$-particles subspace).
Since the series $s^k$ is absolutely summable on (say) $[-\half, \half]$ 
and the collection operators
$\{P_k: k\in\NN\}$ is bounded (in the uniform operator topology),
we can prove by induction 
that $s^N$ is infinitely differentiable 
(in the uniform operator topology \ref{ap_uniformDerivative}, and
hence) in the strong operator topology and has as $n$-th derivative
\begin{displaymatH}
 {\displaystyle\frac{d^n}{d s^n} s^N}
  =
 \sum_{k=n}^\infty k\cdot (k-1)\cdots (k-n+1)\cdot s^{k-n}\cdot P(k)
 \quad (s\in (-\half,\half) )
 .
\end{displaymatH}
This shows that $\vNeu\{s^N: s\in (0,1)\} $ contains all $\{P_N(k)$, $k\in\NN$,
and, a fortiori all functions in $\ell^\infty(\NN)$ with finite support:
$f=\sum_{k=1}^K \al_k P_N(k)$. By taking strong limits of the latter, we obtain
all functions in $\ell^\infty(\NN)$.
\end{proof}

Resuming: the quantum harmonic oscillator is completely described by
its  canonical observables $(Q,P)$, 
or equivalently, given the family Weyl operators $W(z)=e^{isQ-itP}$, 
$z=t+is \in\CC$.
All bounded observables of the quantum harmonic oscillator are 
bounded functions of $Q,P$ and we can obtain these by taking the 
von Neumann algebra generated by $W(z)=e^{isQ-itP}$, $z=t+is\in\CC$.

\subsection{States of the harmonic oscillator}
In quantum mechanics states of a quantum system are often 
given by a normalized vector 
$\psi$ in the Hilbert space $\mathcal{H}$ i.e.\ 
quantum theory postulates that all  physical information
of the system is encoded in $\psi$ as follows:
the expected value of a harmonic oscillator observable $A$ is given by
\begin{displaymatH}
 \EE_\psi(A) = \oip{\psi}{A}{\psi}
 ,
\end{displaymatH}
and, more informatively, the probability of measuring a value of $A$ 
in the interval $(a_1,a_2)$ is given by
\begin{equation}
\label{eq_ProbInterprationPureState}
 \PP_\psi((a_1,a_2) =\oip{\psi}{P_A(a_1,a_2)}{\psi}= \EE_\psi(P_A(a_1,a_2)) 
\end{equation}
where $P_A$ is the projection-valued measure associated with $A$ (which lies
in the von Neumann algebra generated by $A$, whence in 
$\mathcal{W}=\vNeu\{Q,P\}$.

For $\PP_\psi$ to be interpretable as a probability measure
(as postulated by quantum theory), we need that
\begin{Enumerate}
 \item $\PP_\psi(\RR) = 1$,
 \item $\PP_\psi(S)\geq 0 $ for all Borel sets $S$ in $\RR$,
 \item $\PP_\psi(\RR\setminus S) = 1- \PP_\psi(S)$ for all Borel set $S$,
 \item If $S_1, S_2, S_3, \ldots$ are disjoint 
       Borel sets with union $\union_n S_n = S$,
       then $\PP_\psi(S)= \sum_1^\infty \PP_\psi(S_n)$.
\end{Enumerate}

This is taken care of, since ($\PP_\psi= \EE_\psi \after P_A$, $P_A$ is a 
projection-valued measure, see definition (\ref{def_spectralMeasure}), and)
$\EE_\psi$ has the following properties:
\begin{Enumerate}
 \item $\EE_\psi(A)$ is unity-preserving: $ \EE_\psi(\indicator) = 1$,
 \item $\EE_\psi(A)$ is positive: 
       $\EE_\psi(\adj{A}A) \geq 0$ for all bounded observables $A$,
 \item $\EE_\psi$ is additive: $\EE_\psi(A + B) = \EE_\psi(A)  + \EE_\psi(B)$,
 \item $\EE_\psi(A)$ is normal i.e.\
       if  $A_1\leq A_2 \leq ...\leq A$ and $A_n \to A$ strongly, 
       then $\EE_\psi(A_n) \upwards \EE_\psi(A)$.
\end{Enumerate}
Furthermore, $\EE_\psi$ is not only additive, but in fact linear, 
since we have $\EE_\psi(\la A)=\la \EE_\psi(A)$ ($\la \in\CC$).

It is clear that $\EE_\psi$ can be defined on the whole of 
$\mathcal{B}(\mathcal{H})$, but for describing the state of 
the oscillator observables, we might as well restrict it to 
the von Neumann algebra of all observables
of the harmonic oscillator, i.e.\ $\mathcal{W}=\vNeu\{Q,P\}$. 

The above example of a state, induced by a normalized vector in $\mathcal{H}$,
is called a \Dindex{pure state}. Pures states often arise after a 
``filtration'' (selective measurement)
i.e.\ an experiment in which the system is exhaustively measured and
is described by a (simultaneous) eigenstate of a complete 
(system of) observable(s).
Normally, the state of the system is not completely known and must be regarded
as a convex combination of pure states: $\psi= \sum_n r_n \psi_n$ 
where $r_n$ are positive weights reflecting the relative amount of 
state $\psi_n$: $\sum_n r_n =1$ (\cite[14.1]{Bransden+Joachain}). 
Such a state is called a \Dindex{mixture} 
and described by a density matrix \cite[p. 177]{Sakurai}\cite[14.1]{Bransden+Joachain} i.e.\ there is a
positive operator $\rho$ such that $\oip{\psi}{A}{\psi}=Tr(\rho A)$, where
$Tr$ denotes taking the trace.
In any case, the states of the form $\EE_\psi(A)=Tr(\rho A)$ also 
satisfy the conditions 1.-4. above. 
\question{Hans: karakteriseren ze ook de normale toestanden?}

Thus, if we choose to describe the observables of a quantum system with a 
von Neumann algebra, then we are led to the following definition of a state:
\begin{definition}
A \Dindex{state} on a von Neumann algebra $\mathcal{N}$ a positive, linear, 
identity-preserving functional $\phi:\mathcal{A}\to\CC$. It is called
\Dindex{normal} if $0\leq A_n \upwards A $ and $A_n\to A$ strongly implies 
that $\ph(A_n) \upwardsto \ph(A)$.
\end{definition}

The normal states are those that associate probability measures with 
projection valued measures of observables.

\markboth{Chapter 2. The quantization functor $\Gamma$ in general setting}{}
\chapter{The quantization functor $\Gamma$ for an arbitrary number of 
oscillators and dissipative time evolutions\label{ch_quantizationFunctor}}
\markboth{Chapter 2. The quantization functor $\Gamma$ in general setting}{}

The previous chapter has shown that the functorial quantization procedure
above reproduces well-known results 
when used to quantize the harmonic oscillator.
This gives us faith to apply it in a wider context, in particular to the
damped harmonic oscillator.
To this end, this section will discuss the general framework of the 
quantization functor $\Gamma$.

\section{Classical input: a phase space with a time evolution described as a Hilbert space with a contractive map}

In principle the general quantization procedure will not be applicable to all 
conceivable classical systems, but only to those that can be described by
a system of oscillators (possibly after a suitably choice of generalized 
coordinates). This restriction is not that stringent as it may seem at first 
sight, since many systems near stable equilibrium can be well approximated by
such an oscillator system.
A further demand is that the evolution in phase space 
has to correspond to a linear contraction,
in the sense that will become more clear below.

\paragraph{Phase space $\mathcal{K}$}
To prepare an oscillator system for quantization, 
its phase space must be framed 
as a (complex) Hilbert space $\mathcal{K}$ with as dimension 
the number of independent oscillators 
i.e.\ the degrees of freedom of the system. 
Thus if the system has $n$ degrees of freedom (possibly $n=\infty$) then
the standard choice for the Hilbert space is $\ell^2(\{1,\ldots,n\}$ 
(i.e.\ $\CC^n$ if $n<\infty$ and $\ell^2(\NN)$ if $n$ is countably infinite).
The independent oscillators then correspond with orthogonal one-dimensional
subspaces in this Hilbert space i.e. the phase space $(q_k,p_k)$ of the
$k^{th}$ oscillator is $\CC\cdot e_k$ where the vectors $\{e_k\}_k$ form an
orthogonal base of $\mathcal{K}$. 
Observables are linear (continuous) functionals $\mathcal{K}\to \CC$ and 
states are vectors in $\mathcal{K}$. E.g.\ an harmonic oscillator evolution in
oscillator $k$ can be described by the time-dependent state 
$\zeta_k(t)= e^{-i\om t}e_k$. On the other hand, the position observable of the
$k^{th}$ oscillator is $\zeta\mapsto \Re(\ip{e_k}{\zeta})$.

\paragraph{Energy $\norM{}^2$}
The square of the norm of the Hilbert space $\mathcal{K}$ is interpreted 
as the energy of the oscillator system 
(possibly after a suitable transformation of the Hamiltonian). 
This is most easily seen in the standard choice of Hilbert space: 
every oscillator $(q_k,p_k)$ contributes $\half( q_k^2 + p_k^2)$ 
to the total energy of the system, 
whence the square of the Hilbert norm $\half\sum_k(q_k^2 + p_k^2)$ 
corresponds to the total energy of the system.

\paragraph{Time-evolution $C_t$}
The most ideal time-evolution map in phase space is that of a 
energy conservative system,
and this corresponds in the above framework $(\mathcal{K},\norM{})$ with
the most ideal map in a Hilbert space: a unitary (i.e.\ norm-preserving) map.
As we will see further on, 
an advantage of the quantization functor $\Gamma$ is that it can
deal with contractive maps in $\mathcal{K}$ as well
which opens the door to quantize dissipative oscillator systems. 

Summarizing, a time-evolution in the 
above sense is a family of contractions $C_t:\mathcal{K}\to\mathcal{K}$.
Here $t$ runs through an index set $\TT$ which denotes the times; 
$\TT$ can either be discrete  ($\NN$, $\ZZ$) or continuous 
($[0,\infty)$, $\RR$).

\medskip

The above framework is summarized in the following definition
\begin{definition}
\label{def_clasDynSyst}
 A \Dindex{classical dynamical system} consists of a Hilbert space 
 $\mathcal{K}$ equipped with a family of contractions $C_t$, $t\in\TT$.
\end{definition}

\markright{\thesection.\ Quantum output: a Von Neumann algebra with completely positive maps}
\section{Quantum output: a quantum system described 
            by a von Neumann algebra of observables and a time evolution
            of completely positive maps \label{ssec_outputGamma}
         }
\markright{\thesection.\ Quantum output: a Von Neumann algebra with completely positive maps}

The quantization functor $\Gamma$ is specified by describing what it does 
to a Hilbert space $\mathcal{K}$ and to a contraction $C$.

\paragraph{$\Gamma(\mathcal{K})$ is a von Neumann algebra $\mathcal{W}_\mathcal{K}$ with vacuum state $\phi_\mathcal{K}$}
First we take a representation of the Canonical Commutation Relation over 
$\mathcal{K}$. This gives us a representation of the Weyl operators
$W(f)$, $f\in\mathcal{K}$, over some Hilbert space $\mathcal{H}$. 
The Weyl operators encode for the position and momentum observables 
of the system of oscillators, 
but to obtain the algebra of all observables of the system 
(at least those that are determined by the position and momentum operators 
of the system such as the spectral measures of the positions and momenta of the
oscillators), we take the strong closure and obtain a von Neumann algebra
$\mathcal{W}_\mathcal{K}$.
States of the system are given by normalized positive normal functionals 
$\ph: \mathcal{W}_\mathcal{K} \to \CC$ that are unity-preserving i.e.\ 
$\ph(\indicator) = 1$.

For simplicity (and because it suffices for the situation we want to describe),
we will assume that our representation of the CCR($\mathcal{H}$)
is a cyclic vacuum representation, i.e.\ there is a normalized 
vector $\Om_\mathcal{K}$ in the representation Hilbert space $\mathcal{H}$
such that 
\begin{displaymatH}
 \ph_\mathcal{K}(W(f))
  :=
 \oip{\Om_\mathcal{K}}{W(f)}{\Om_\mathcal{K}}= e^{-\half\norM{f}_\mathcal{K}^2}
 \quad
 (f\in\mathcal{K})
 .
\end{displaymatH}
Being a pure state, $\phi_\mathcal{K}$ is induced by a vector $\Om_{\mathcal{K}}\in \mathcal{H}$, therefore naturally defined on the whole of 
$\mathcal{B}(\mathcal{H}) \supset \mathcal{W}_\mathcal{K}$.
Also note that all cyclic vacuum representations of 
$\Gamma(\mathcal{K})= (\mathcal{W}_\mathcal{K}, \ph_\mathcal{K})$ are
unitarily equivalent (\ref{ap_repGamma(K)Equivalent}). 

A physical interpretation of choosing a vacuum representation is that all
states of the system can be obtained by successively applying ladder operators
(annihilation and creation operators) associated with $f\in\mathcal{K}$ 
to the vacuum state. In other words, a general state of the system is made of
excitations of each of the constituting oscillators (which are labeled by
$f\in\mathcal{K}$).

\paragraph{$\Gamma(C)$ is a linear strongly continuous, completely positive, 
           state- and unity-preserving 
           map between von Neumann algebras of type
           $(\mathcal{W}_{\mathcal{K}_i},\phi_{\mathcal{K}_i})$}

Up-to-now, the contraction $C$ has had the interpretation of a
time-evolution in the Hilbert phase space $\mathcal{K}$. 
This time-evolution could either correspond to that of a energy-conservative
system or a energy-dissipative system.
Later on, we will like to consider the damped harmonic oscillator, 
which is dissipative quantum system, as a subsystem of a larger, enveloping, 
energy-conservative system and investigate relations between the two systems.
This means that we must be able to shuttle ``up'' from a subsystem
to its envelop (via an isometric inclusion) and shuttle ``down''
from the envelope to the subsystem (via an orthogonal projection). 
To accommodate this, we need  the general formalism of quantizing a
contraction between two different Hilbert phase spaces: 
$C:\mathcal{K}_1\to\mathcal{K}_2$.

We will first motivate the mathematical properties of $\Gamma(C)$, 
for $C: \mathcal{K}_1 \to \mathcal{K}_2$ a contraction 
between two Hilbert spaces. In the
next section we will define its action on the Weyl operators, 
leaving the proof of its extension to the appendix.
To relieve the burden of parentheses in notation, 
we will write $\Gamma_C:=\Gamma(C)$ below. 

\smallskip

As we want that linear relations between observables 
and taking limits of them 
in the von Neumann algebra of observables of 
system 1 ($\mathcal{W}_{\mathcal{K}_1}$)
are reflected  in that of system 2 ($\mathcal{W}_{\mathcal{K}_2}$),
we require \textsf{linearity} and \textsf{strong continuity}:
\begin{aligN}
 \, & 
 A_3 = \alpha_1 A_1 + \alpha_2 A_2 
   \Implies
  \Gamma_C(A_3)=  \alpha_1 \Gamma_C(A_1) + \alpha_2 \Gamma_C (A_2) 
  ,
  \\
 \,& 
 A_n \to A \text{ strongly in } \mathcal{W}_{\mathcal{K}_1}
   \Implies 
 \Gamma_C(A_n) \to \Gamma_C(A)
         \text{ strongly in } \mathcal{W}_{\mathcal{K}_2}
  .
\end{aligN}

\smallskip

$\Gamma_C$ maps observables of system 1 ($\mathcal{W}_{\mathcal{K}_1}$) to 
observables of system 2 ($\mathcal{W}_{\mathcal{K}_2}$) and induces 
therefore also a map from the linear functionals on system 2 
to the linear functionals on system 1: 
if $\ph_2$ is a linear functional on system 2, 
then the induced linear functional 
on system 1 is $\Gamma_C^*(\ph_2)(A)= \ph_2\bigl(\Gamma_C(A)\bigr)$.

The duality between $\Gamma_C$ and $\Gamma_C^*$ can be interpreted as 
the duality of the Heisenberg picture and Schr\"odinger picture in
quantum mechanics.
$\Gamma_C$ reflects the Heisenberg picture of quantum mechanics: 
it describes how observables change. It is then natural to have the 
Schr\"odinger picture of quantum mechanics implemented by $\Gamma_C^*$ 
and this is possible provided that this map describes how states change.
At the very least, it should then map states to states,
in other words, we require that 
$\Gamma_C(\ph_2)= \ph_2 \after \Gamma_C$ is  positive, identity preserving
whenever $\ph_2$ is a state. 
Sufficient for that is that $\Gamma_C$ is \textsf{positive} and 
\textsf{identity-preserving}:
\begin{aligN}
 \, & A\geq 0 \implies \Gamma_C(A) \geq 0, \\
 \, & \Gamma_C(\Identity)= \Identity .
\end{aligN}
Moreover, we would like that normality of states is preserved, so that
the resulting state can be interpreted as a probability measure.
Sufficient for that turns out to be that $\Gamma_C$ maps 
vacuum states to vacuum states:
\begin{displaymatH}
 \Gamma_C^*(\ph_{\mathcal{W}_2})
   :=\ph_{\mathcal{W}_2} \after \Gamma_C 
    = \ph_{\mathcal{W}_1}
 .
\end{displaymatH}
The latter is not an unreasonable requirement, 
since the time-evolutions we consider (either that of energy-conservative or
energy-dissipative systems) and the embeddings (in larger quantum systems)
or projections (onto quantum subsystems) do not increase
the energy of the system. From the interpretation of a vacuum state as the
state of lowest energy level, 
we do not expect that $\Gamma_C^*$ yields a state of higher energy than the
vacuum state when starting from a vacuum state.

There is one complication with the above, however: 
positivity alone is not sufficient to make the functorial machine 
$\Gamma_C$ work well. The oil needed is called \Dindex{complete positivity}.
\begin{definition}
 A linear map $T$ between von Neumann algebras $\mathcal{A}$ and $\mathcal{B}$
 is called \Dindex{$n$-positive} ($n\in\NN$) if the map
 $T\otimes \Identity$ maps positive elements of $\mathcal{A}\otimes M_n$ 
 to positive elements of $\mathcal{B}\otimes M_n$,
 where $M_n$ is the algebra of $n\times n$-matrices with complex coefficients.
 
 $T$ is called \Dindex{completely positive} 
 if it is $n$-positive for all $n\in\NN$.
\end{definition}

The rationale for this is the following. The map $T^*$ describes some 
time-evolution of (or, more generally, a physical operation on) a state of 
the quantum system described by $\mathcal{B}$ yielding a quantum system
$\mathcal{A}$ in some state. If a system $\mathcal{R}$ is not affected by 
this operation then naturally the map $T^*$ is extended to a map sending 
states of $\mathcal{B}\otimes \mathcal{R}$ to states 
of $\mathcal{A}\otimes \mathcal{R}$. In particular, the corresponding (dual) 
map $T\otimes \Identity_\mathcal{R}:\mathcal{A}\otimes \mathcal{R}
 \to\mathcal{B}\otimes \mathcal{R}$ must be positive. It turns out that it
is sufficient to consider finite-dimensional $\mathcal{R}$ only, i.e.\
von Neumann algebras over finite Hilbert spaces ($\CC^n$), which are
the matrix algebras $M_n$.

This leads to the following definition:
\begin{definition}
\label{def_operation}
Let $\mathcal{A}$ and $\mathcal{B}$ be von Neumann algebras with normal states
$\ph_\mathcal{A}$ and $\ph_\mathcal{B}$ respectively.

A physical \Dindex{quantum operation} 
on a quantum system (also referred to as a \Dindex{quantum operation})
described by
$\mathcal{B}$ in state $\ph_\mathcal{B}$ that yields a 
state $\ph_\mathcal{A}$ of some quantum system described by
$\mathcal{A}$ 
is a linear map 
$T: (\mathcal{A},\ph_\mathcal{A}) 
     \to 
    (\mathcal{B},\ph_\mathcal{B})
$
that is
\begin{Properties}
 \item
  strongly continuous,
 \item
  completely positive,
 \item
  identity preserving,
 \item 
  state preserving.
\end{Properties}
\end{definition}

We can now formulate the quantum version of a dynamical system.

\begin{definition}
\label{def_quantumDynSyst}
A \Dindex{quantum dynamical system} consists of a quantum system described by
a von Neumann algebra and normal state $(\mathcal{A},\phi_\mathcal{A})$
equipped with a family of quantum operations 
$T_t:(\mathcal{A},\phi_\mathcal{A}) \to (\mathcal{A},\phi_\mathcal{A})$,
$t\in\TT$.
\end{definition}

\bigskip

\section{Mathematical formulation of $\Gamma$}
Although the input and the output of the quantization functor $\Gamma$ 
seem very different at first sight, they bear resemblance on an abstract level:
both consist of ``objects'' (Hilbert spaces on the one hand and 
von Neumann algebras equipped with a normal state on the other) and there are
``morphisms'' between those objects (contractions from one Hilbert space
to the other versus completely positive, strongly continuous, 
state- and identity-preserving maps between the von Neumann algebras).
Such a collection of objects and morphisms (together with some properties, 
which we do not spell out explicitly \cite{Lawvere}
is called a category by mathematicians.
Thus, $\Gamma$ maps the category whose objects are Hilbert spaces and whose
morphisms are contractions to the category 
whose objects are von Neumann algebras equipped with normal states and whose
morphisms are completely positive, strongly continuous, state- and 
identity-preserving maps.

As we will see below, $\Gamma$ preserves compositions of morphisms and 
maps special types of morphisms in the one
category to special types of morphisms in the other, which mathematically 
justifies naming it 
a \Dindex{functor} \cite{Lawvere}

First, we will define the action of $\Gamma$ in general terms i.e.\
on Von Neumann algebras generated by arbitrary cyclic vacuum representations.

\begin{theorem}
\label{th_generalDefGammaFunctor}
 Let $\mathcal{K}_i$, $i=1,2$, be Hilbert spaces and 
 let $W_i: \mathcal{K}_i \to \mathcal{B}(\mathcal{H}_i)$, $f\mapsto W_i(f)$ be
 cyclic vacuum representations of the CCR over $\mathcal{K}_i$ (with cyclic
 vacuum vector $\Om_{\mathcal{K}_i}$ in the representing Hilbert spaces 
 $\mathcal{H}_i$ respectively).
 Let $\mathcal{W}_0(\mathcal{K}_i)$ 
 be the linear span of the Weyl operators over  $\mathcal{K}_i$.

 Assume that $C:\mathcal{K}_1 \to \mathcal{K}_2$ is a contraction and
 define 
 $\Gamma_0(C): \mathcal{W}_0(\mathcal{K}_1) \to \mathcal{W}_0(\mathcal{K}_2)$
 by
 \begin{equatioN}
 \label{eq_Gamma0}
  \Gamma_0(C)\Bigl( W_1(f) \Bigr) 
   := 
  e^{-\half(\norM{f}^2 - \norM{Cf}^2)} 
  W_2(Cf)
  .
 \end{equatioN}
 Then $\Gamma_0(C)$ has a unique  strongly continuous extension to an 
 operation $\Gamma(C)$ from 
 $\Gamma(\mathcal{K}_1)=(\mathcal{W}_{\mathcal{K}_1}, \ph_{\mathcal{K}_1})$
 to
 $\Gamma(\mathcal{K}_2)=(\mathcal{W}_{\mathcal{K}_2}, \ph_{\mathcal{K}_2})$.

 \medskip
 More precisely stated,
 \begin{Properties}
  \item formula (\ref{eq_Gamma0}) determines by linearity 
        unambiguously a map 
        $\mathcal{W}_0(\mathcal{K}_1) \to \mathcal{W}_0(\mathcal{K}_2)$ and
  \item this map is strongly continuous, hence admits a strongly continuous
        extension $\Gamma(C)$ from 
        $\vNeu\{\mathcal{W}_0(\mathcal{K}_1)\}= \mathcal{W}(\mathcal{K}_1)$
        to
        $\vNeu\{\mathcal{W}_0(\mathcal{K}_2)\}= \mathcal{W}(\mathcal{K}_2)$,
  \item The required properties for $\Gamma(C)$ to be an operation 
        (see \ref{def_operation})
	are easily checked on $\mathcal{W}_0(\mathcal{K}_1)$  
	and then pass on to $\mathcal{W}(\mathcal{K}_1)$
	by strongly continuous extension.
 \end{Properties}
\end{theorem}

The proof of theorem \ref{th_generalDefGammaFunctor} takes several steps and is
deferred to the appendix \ref{ap_quantizationFunctor}.

The following theorem underlines the functorial character of $\Gamma$.
One aspect of that needs some elucidation: 
the property of $\Gamma$ to map special morphisms in the category of 
Hilbert spaces with contractions to special morphisms in the category
of Von Neumann algebras with operations.
The special morphisms in the category of Hilbert spaces with contractions
that we consider are isometries, projections, and unitary maps.
An isometry $J$ is special in the sense that it allows 
to identify one Hilbert space ($\mathcal{H}_1$)
with a \textsl{sub}space of another ($\mathcal{H}$) in such a way that
all the Hilbert space structure of $\mathcal{H}_1$ (the linear structure and the inner products) is preserved. In other words: an isometry embeds 
a Hilbert space into an enveloping Hilbert space.
The analog of this in the category of Von Neumann algebras with operations
is an injective map that preserves all Von Neumann algebra structure 
(the algebraic structure and the adjoints), a so-called
injective $*$-homomorphism.

A projection $P_1$ is special in the sense that it decomposes each vector
in a Hilbert space $\mathcal{H}$ in two components: 
one lying in a Hilbert subspace $\mathcal{H}_1$ and the other in 
its orthogonal complement $\mathcal{H}_1^{\perp}$. In other words,
it allows us to pass from a Hilbert space onto a Hilbert subspace.
If $J=P_1^*:\mathcal{H}_1\to \mathcal{H}$ is the corresponding isometry 
embedding $\mathcal{H}_1$ into $\mathcal{H}$, then $P$ is 
characterized by $P_1\after J=\Identity_{\mathcal{H}_1}$.
The analog of a projection in the category of Von Neumann algebras 
with operations is called a conditional expectation
and is defined with respect to an injective $*$-homomorphism $j$
as the unique morphism $P$ such that $P\after j=\Identity$.

Unitary maps are special because they allow to identify Hilbert spaces
with each other. In particular, if unitary maps of a Hilbert space
to itself are considered, they can describe reversible time-evolutions.
The analog of unitary maps in the category of Von Neumann algebras and
operations are called $*$-isomorphism which are bijective $*$-homomorphisms.
$*$-Isomorphisms that map a Von Neumann algebra to itself are called
$*$-automorphisms. The latter can be used to describe time evolutions 
in a Von Neumann algebra.

Summarizing:
\begin{definition}
\label{def_homomorphism}
Let $(\mathcal{A},\ph_{\mathcal{A}})$, $(\mathcal{B},\ph_{\mathcal{B}})$ 
be Von Neumann algebras equipped with normal states.

An operation 
$j: (\mathcal{A},\ph_{\mathcal{A}}) \to (\mathcal{B},\ph_{\mathcal{B}})$ 
that is linear such 
that $j(A_1A_2)=j(A_1)j(A_2)$ and ${j(A)}^*=j(A^*)$ is called 
a  \Dindex[homomorphism!$*$-homomorphism]{$*$-homomorphism}
(here $\phantom{\cdot}^*$ 
refers to taking the adjoint). 

If 
$j: (\mathcal{A},\ph_{\mathcal{A}}) \to (\mathcal{B},\ph_{\mathcal{B}})$ 
is an 
\Dindex[homomorphism!injective $*$-homomorphism]{injective $*$-homomorphism}, 
then the \Dindex{conditional expectation} with respect to $j$ 
is the unique operation 
$P: (\mathcal{B},\ph_{\mathcal{B}})\to(\mathcal{A},\ph_{\mathcal{A}})$ 
such that $P\after j= \Identity_{\mathcal{A}}$ (if it exists).

An operation 
$T: (\mathcal{A},\ph_{\mathcal{A}}) \to (\mathcal{B},\ph_{\mathcal{B}})$ 
is called an \Dindex[isomorphism!$*$-isomorphism]{$*$-isomorphism} 
if $T$ is an bijective $*$-homomorphism
and an \Dindex[automorphism!$*$-automorphism]{$*$-automorphism} if in addition 
$(\mathcal{B},\ph_{\mathcal{B}})= (\mathcal{A},\ph_{\mathcal{A}})$.
\end{definition}

Thus, an injective $*$-homomorphism allows us to embed a quantum dynamical
system into a larger one; a conditional expectations allows us 
to project from a larger quantum dynamical system to one of its subsystems,
and a $*$-automorphism can be used to describe a transformation (e.g.\
by time evolution) of a quantum dynamical system.

\begin{theorem}
Assume the setting of theorem \ref{th_generalDefGammaFunctor} above.

\smallskip

Then $\Gamma$ preserves composition of morphisms: 
if $C_1: \mathcal{K}_1\to\mathcal{K}_2$ 
and $C_2:\mathcal{K}_2\to\mathcal{K}_3$
are contractions 
between Hilbert spaces $\mathcal{H}_1,\mathcal{H}_2,\mathcal{H}_3$,
then $\Gamma(C_1 C_2)=\Gamma(C_1)\Gamma(C_2)$.

\smallskip

Further, $\Gamma$ maps special morphisms in the category of 
Hilbert spaces with contractions to special morphisms in the category
of Von Neumann algebras with operations:
\begin{Properties}
 \item if $J: \mathcal{H}_1 \to\mathcal{H}_2$ is an isometry,
       then $\Gamma(J):\Gamma(\mathcal{H}_1) \to \Gamma(\mathcal{H}_2)$
       is an injective $*$-homomorphism.
 \item if $J^*: \mathcal{H}_2 \to\mathcal{H}_1$ is a projection,
       then $\Gamma(J^*):\Gamma(\mathcal{H}_2) \to \Gamma(\mathcal{H}_1)$
       is a conditional expectation.
 \item if $U: \mathcal{H}_1 \to\mathcal{H}_1$ is a unitary map,
       then $\Gamma(U):\Gamma(\mathcal{H}_1) \to \Gamma(\mathcal{H}_1)$
       is a $*$-automorphism.
\end{Properties}
\end{theorem}


\section{The usual approach to second quantization 
         \label{sec_usualSecondQuantization}}

We conclude this part by relating the \textsl{functorial} 
approach to second quantization described above with
the usual approach taken in physics literature.

First of all, second quantization is used in physics 
when quantizing fields. Recall that a field is a map (observable) 
whose domain is a space-time continuum 	$\mathbb{S}\times\TT$ and 
whose range $V$ could be any set. 
Typically, $\mathbb{S}= \RR$ (one-dimensional Euclidean space) or 
$\mathbb{S}= \RR^3$ (three-dimensional Euclidean space) 
and $\TT$ is a possibly infinite time interval 
($\TT=[a,\infty)$, $\TT=(-\infty,a]$, $\TT=[a,b]$, or $\TT=\RR$).
Further, if $V\subset\RR$, then we speak of a scalar field and
when $V\subset\RR^n$ with $n>1$, then we speak of a vector field).
An example of a scalar field is the displacement $\eta(z,t)$ 
from the equilibrium of a string, which is a function of the position
along the string $z$ and time $t$.
As fields are systems with an infinite number of degrees of freedom,
their quantization is more involved.

\smallskip

The approach usually taken in physics literature can generally be 
outlined as follows 
\cite[p.~107]{Itzykson+Zuber}:
\begin{Enumerate}
 \item Let $\ph=\ph(\mathbf{x},t)$ 
       be the field whose dynamics we are interested in.
 \item Due to the continuous character of the space 
       the classical Lagrangian $L$ is a integral over a Lagrangian density
       $\mathcal{L}$
       \begin{displaymatH}
         L(t) = \int d^3x\; \mathcal{L}(\ph,\partial \ph)
         \,,
       \end{displaymatH}
       where $\partial\ph$ is a vector with 4 components, 
       $\displaystyle\frac{\partial\ph}{\partial t}(t,\mathbf{x})$, and
       $\nabla\ph(t,\mathbf{x})$ if we have a vector field on $\RR^3$.  
 \item Derive the conjugate field (field impulse) 
       \begin{displaymatH}\displaystyle
         \pi(t,\mathbf{x})
     = \frac{\delta L(t)}%
            {\delta\bigl[\frac{\partial\ph(t,\mathbf{x})}{\partial t}\bigr]}
     = \frac{\partial \mathcal{L}}%
            {\partial\bigl[\frac{\partial\ph(t,\mathbf{x})}{\partial t}\bigr]}
         \;,
       \end{displaymatH}
      where $\frac{\delta \ldots}{\delta\ldots}$ denotes a functional 
      derivative.
 \item Invert the above to obtain 
       $\frac{\partial\ph}{\partial t}(t,\mathbf{x})$ 
       in terms of $\pi(t,\mathbf{x})$ and $\ph(t,\mathbf{x})$.
 \item Construct the Hamiltonian $H$ as a function of the field 
       $\ph(t,\mathbf{x})$ and field impulse $\pi(t,\mathbf{x})$
       \begin{displaymatH}
         H(t)  
           = 
           \int d^3 x \; 
           \underbrace%
            {%
             \Bigl[ 
                \pi(t,\mathbf{x})
                \frac{\partial\ph(t,\mathbf{x})}{\partial t}
                -
                \mathcal{L}(\ph,\partial \ph)
             \Bigr]%
            }%
            _{\mathcal{H}
                \Bigl(t,\mathbf{x}, \ph(t,\mathbf{x}), \pi(t,\mathbf{x})
                \Bigr)%
             }
           \;,
       \end{displaymatH}
       Note that in the above $\frac{\partial\ph}{\partial t}(t,\mathbf{x})$
       is a function of $\ph(t,\mathbf{x})$ and $\pi(t,\mathbf{x})$ (and of
       $t$ and $\mathbf{x}$).
       Further, $\partial \ph$ in the Lagrangian density is to be read as
       the vector with as first component
       \begin{displaymatH}\displaystyle
         \frac{\partial\ph}{\partial t}
         \Big(t,\mathbf{x},\;\ph(t,\mathbf{x}),\; \pi(t,\mathbf{x})\Bigr)
          \;,
       \end{displaymatH}
       and as second till fourth component $\nabla\ph(t,\mathbf{x})$.
       Finally, $\mathcal{H}$ is a Hamiltonian density. 
         
 \item The Hamiltonian operator is obtained by viewing $\ph(t,\mathbf{x})$ 
       and $\ph(t,\mathbf{x})$ as a continuum of operators satisfying the
       equal time commutators
       \begin{displaymatH}
         [\ph(t,\mathbf{x}),\;\pi(t,\mathbf{x'})] 
            = i\delta^3(\mathbf{x}-\mathbf{x'})
         \;, 
       \end{displaymatH}
       while 
       $[\ph(t,\mathbf{x}),\ph(t',\mathbf{x'})]
         =[\pi(t,\mathbf{x}),\pi(t',\mathbf{x'})]=0$.
\end{Enumerate}
Variants of the above sketched second quantization procedure involve 
going backwards-forwards between continuous and discrete versions
of the system at steps 2 and 5 respectively.

\medskip 
     
Often, obstacles arise when applying this quantization procedure, which 
physics traditionally bypasses ``along the road'':
\begin{enumerate}
 \item Naive substitution of operators for $\ph$ and $\pi$ may lead to
       a non self-adjoint Hamiltonian. Problems of these kind have to 
       be solved ad-hoc by hermitization procedures.
 \item Nothing has been said about the Hilbert space on which the fields
       $\ph$ and $\pi$ act, but in the particular case of the free scalar field
       one can fortunately do without 
       (see below and \ref{ap_usualSecondQuantization}).
 \item Multiplication of operator fields at the same point may introduce
       new problems, such as infinite ground state energy levels. 
       Problems of the latter kind have to be tackled ad-hoc by 
       renormalization procedures
       (putting the system in a finite box with periodic boundary conditions,
        and see if the ground state energy level is proportional to the 
        size of the box. If so, then one simply subtracts the infinity caused
        by the size of the system). 
\end{enumerate}

\section{Quantization of the free scalar field: the ``usual''  way}
As a example to make the comparison between the ``usual'' approach 
to second quantization and the functorial approach easier,
we will discuss the quantization of a (quadratic) 
real scalar field $\ph$
which yields a self-adjoint (quantum) field. Our approach follows that
of \cite[p.~107-114]{Itzykson+Zuber} to a great extent, but
we will only mention the main facts, leaving the details to the
appendix (see \ref{ap_quantizationFreeScalarField}).
 
The Lagrangian density of the quadratic free scalar field $\ph(x,t)$ reads
\begin{displaymatH}
  \mathcal{L}= \half(\partial \ph)^2 - \half m^2\ph^2
  ,
\end{displaymatH}
leading to the Hamilton density 
\begin{displaymatH}
  H=\int d x \,
         \Bigl\{ \half\bigl[{\pi(x)}^2 
                  + \left(\frac{d \phi}{d x}(x)\right)^2 \bigr] 
                  +  m^2{\ph(x)}^2 
         \Bigr\}
  ,
\end{displaymatH}
which reads after Fourier transformation:
\begin{displaymatH}
  H = \frac{1}{4\pi} \int\!\!dk
       \biggl\{ 
     \adj{\widehat{\pi}}(k) \widehat{\pi}(k) 
           +  
     \underbrace{[m^2 +k^2]}_{\om(k)} \adj{\widehat{\ph}}(k) \widehat{\ph}(k)  
       \biggr\}
     \;.
\end{displaymatH}
After defining
\begin{aligN}
  a(k)       \ats := \frac{1}{\sqrt{4\pi\om(k)}}
               \left[
                     \om(k) \widehat{\ph}(k) + i\widehat{\pi}(k)
               \right]  \\
  \adj{a}(k) \ats  = \frac{1}{\sqrt{4\pi\om(k)}}
               \left[
                     \om(k) \adj{\widehat{\ph}}(k) - i\adj{\widehat{\pi}}(k)
               \right] , 
\end{aligN}
we obtain
\begin{displaymatH}
 H = \half\int_{-\infty}^\infty  dk\quad
                \om(k) \Bigl[ \adj{a}(k)a(k)+ a(k)\adj{a}(k) \Bigr]
   .
\end{displaymatH}
The vacuum state $\ket{0}$ has infinite energy and has to be renormalized by
applying a Wick ordering of the Hamiltonian:
\begin{displaymatH}
  H=  \int_{-\infty}^\infty  dk\quad \om(k)\: \adj{a}(k)a(k)
  .
\end{displaymatH}
Excited states where only the oscillator corresponding with ``mode'' $k$ 
contains an energy quantum
can formally be created as $a(k)\ket{0}$, but have infinite energy.
Therefore, it makes more sense to build ``wave packets''
\begin{displaymath}
 \adj{a}_f \ket{0}:= \int_\infty^\infty dk\; f(k) \adj{a}(k)\ket{0}
\end{displaymath}
which are normalizable provided that
\begin{aligN}
  \infty 
   & >  \oip{0}{a_f \adj{a}_f}{0} 
     =
  \int_{-\infty}^\infty\hspace{-1em}dk \int_{-\infty}^\infty\hspace{-1em}dk'\quad 
   \overline{f(k')}f(k) \; 
    \langle 0 |\hspace{-1.3em}
     \underbrace{a(k')\adj{a}(k)}_{\adj{a}(k)a(k') + \delta(k-k')}
              \hspace{-1.3em} | 0 \rangle 
     \\
   & =
     \int_{-\infty}^\infty\hspace{-1em}dk \; |f(k)|^2
    \,,
\end{aligN}
i.e.\ the ``density kernel'' $f$ has to be square integrable: $f\in L^2(\RR)$.
Coherent states are  obtained as:
\begin{displaymatH}
  \Coh(f): = K(f) \exp( \int\!\! dk\;  f(k) \adj{a}(k) )\ket{0}
  \,,
\end{displaymatH}
where $K(f)= e^{-\half\norM{f}_{L^2}^2}$ is a normalization factor.

Thus, the Weyl operators are proportional to
\begin{displaymatH}
 W(f)= \exp( \adj{a_f})
 \quad
 (f\in L^2(\RR))
 .
\end{displaymatH}

\questionst{Look up what are the appropriate constants 
in the exponential for $W(f)$}
Thus, the usual quantization of the free scalar field 
corresponds to functorial quantization (via $\Gamma$) 
of the phase space $L^2(\RR)$ 
of an infinite number of oscillators, indexed by $k\in\RR$.

\part{Damped Harmonic Oscillator}
\chapter{The problem of irreversibility}
Some fundamental laws in physics, 
the Schr\"odinger equation,
are invariant under time reversal.
\question{Hans: zijn er andere voorbeelden?}
This has the following implication for a physical system that 
obeys these laws:
we cannot distinguish  (by no physical quantity whatsoever)
whether time is running `forward' or `backward'.
Another way of saying this is that 
in principle the course of events 
can be reversed: none of the changes that the system has gone through
is permanent.
Owing to the reversibility of its time evolution, 
such a system called a \Dindex{reversible dynamical system}.

An example of a reversible dynamical system is a classical harmonic oscillator.
An harmonic oscillator
is (fully) characterized by its frequency $\om$, amplitude $A$ (and mass $m$)
and it has the following time evolution  
\begin{displaymatH}
 \left.
 \begin{array}{l}
  q(t)  = A \sin \om t \; , \\
  p(t)  = \frac{d q(t)}{dt}= m A\om \cos \om t,  
 \end{array}
 \right\rgroup
 \quad(t\in\RR)
  .
\end{displaymatH}
If we replace in the above equations $t$ by $-t$ 
(which can be realized in practice by giving at time $0$ 
the oscillator the same speed initial speed 
but now in the converse direction), 
we again obtain a harmonic oscillator 
with the same frequency and mass. 
From a physical point of view,
we obtain the {\it same} harmonic oscillator: 
the fact that the direction of the oscillation
has flipped from `positive' to `negative' has no physical meaning because
if we have two frames of references that can be obtained from each other
by a rotation, then there is no physical way of distinguishing 
the one from the other.
Note that the average energy of the harmonic oscillator over one period 
which is proportional to the amplitude, 
is conserved quantity in this case.
and this conservation of energy is a feature of reversible dynamical systems.
We may think of this as follows: 
the flow of energy stays within the system and cannot leave it.
Therefore, the reversible systems are also called {\it closed} systems.

\smallskip

There is, however, a major problem with time reversibility.
In nature most courses of events are not reversible: who has ever
seen a damped harmonic oscillator that started from rest position and began
to oscillate with increasing amplitude 
(without applying external forces)?

Since the reversible theories are too good to discard, 
physicists (from Boltzmann on wards) have taken up the challenge to explain
irreversible behavior from the reversible theories.

One point of view (the open system interpretation) is the following.
First observe that an irreversible dynamical system 
looses energy to its surroundings and that 
we may think of this as follows: the flow of energy is not confined to 
the irreversible system, but leaves it and enters the system's environment. 
For this reason a irreversible system is called an {\it open} system.

An example of an irreversible dynamical system is 
the classical damped harmonic oscillator:
\begin{displaymatH}
 \left.
 \begin{array}{l}
  q(t)  = e^{-\gamma t}A\sin\om t , \\
  p(t)  = m \frac{d q(t)}{d t}, \\
 \end{array}
 \right
 \rgroup
 \quad
 \bigl(t\in [0,\infty)\,\bigr)
 .
\end{displaymatH}
The average energy of the oscillator over one period 
(roughly proportional to its amplitude $e^{-\gamma t}A$) 
decreases and the lost energy is found back as heat 
caused by friction and deposited to the oscillator's surroundings.

If we take the environment large enough, 
then the flow of energy of the total system consisting 
of irreversible system plus environment,
stays within the total system.
Then the total system is a closed system again and it is expected to have 
reversible time evolution. 

In case of the damped harmonic oscillator, we could imagine that it is a
point on a string on which a wave packet is traveling. 
As the (center of the) wave packet moves away from the point, the oscillation of
the point is fading away with it.

Summarizing, the \Dindex{open system interpretation} states that   
a irreversible (open) system is  a subsystem of a reversible (closed) system
and that the time evolution and the observables of the irreversible system 
are `projections' of the time evolution and the observables 
respectively of the reversible system.

Let us now formulate the open system interpretation in a mathematical language
for both classical and quantum dynamical systems.

\section{Dilation theory of dynamical systems}

Dilation derives from the Latin ``dilatare'', which means ``to make broader''.
Dilation theory aims to give a mathematical description of embedding an
irreversible dynamical system 
$\Bigl(\mathcal{S}, (T_{s,t})_{s, t\in [0,\infty)}\Bigr)$
into a larger reversible dynamical system 
$\Bigl(\widehat{\mathcal{S}}, (\widehat{T}_{s,t})_{s,t\in (-\infty,\infty)}\Bigr)$.

We can depict this symbolically as follows:
\begin{displaymatH}
 \begin{CD}
   \mathcal{S}           @>T_{t_1\! , \: t_2}>>            \mathcal{S}\\
    @Vj_{t_1}VV                                          @AA P_{t_2}A \\
   \widehat{\mathcal{S}} @>\widehat{T}_{t_1\! ,\: t_2}>>  \widehat{\mathcal{S}}
 \end{CD}
\end{displaymatH}
\questionst{Elucidate the t and s in the time evolution}
Here $\mathcal{S}$ and $\widehat{\mathcal{S}}$ describe the observables
of the subsystem and the supersystem respectively.
$T_{s,t}:\mathcal{S}\to \mathcal{S}$ is the irreversible time-evolution of 
$\mathcal{S}$ (hence
$s,t\in [0,\infty)$) and 
$\widehat{T}_{s,t}:\widehat{\mathcal{S}}\to\widehat{\mathcal{S}}$ 
is the reversible time-evolution of $\widehat{\mathcal{S}}$ (hence
$t_1,t_2\in\RR$).
Note that a description in terms of states can be obtained by taking 
the dual spaces $\mathcal{S}^*$ and $\widehat{\mathcal{S}}^*$ and 
translating the time-evolutions correspondingly.

The embedding of irreversible system into the reversible one at time $t_1$
is described by a one-to-one map 
$j:\mathcal{S}\to\widehat{\mathcal{S}}$
that identifies $S \isom j_{t_1}(\mathcal{S})$ 
as a subsystem of $\widehat{\mathcal{S}}$. 
A corresponding map $P_{t_2}:\widehat{\mathcal{S}}\to\mathcal{S}$ has to exists
that projects $\widehat{\mathcal{S}}$ to $\mathcal{S}$ such that 
$P_t\after j_t = \Identity_{\mathcal{S}}$ for all $t\in\RR$.
Furthermore, it must be possible to view the evolution $T_{t}$ in the 
irreversible system as a projection of an evolution in the
reversible system: 
$T_{t_1\!,\:t_2}= P_{t_2}\after \widehat{T}_{t_1\!,\: t_2} \after j_{t_1}$.

In the following we will restrict ourselves to the situation that the
time-evolutions are 
\Dindex[time-evolution!translation-invariant]{translation-invariant}:
$T_{s,s+t}=T_{0,t}=:T_t$ for all $s,t\in[0,\infty)$,
$\widehat{T}_{s,s+t}=\widehat{T}_{0,t}=:\widehat{T}_t$ 
for all $s,t\in\RR$, and moreover $j_t=j$, $P_t=P$ i.e.\

\begin{displaymatH}
 \begin{CD}
   \mathcal{S}           @>T_{t}>>            \mathcal{S}\\
    @VjVV                                          @AA P A \\
   \widehat{\mathcal{S}} @>\widehat{T}_{t}>>  \widehat{\mathcal{S}}
 \end{CD}
\end{displaymatH}

\subsection{Dilations of classical dynamical systems}
Let $(\mathcal{K}, C_t)$ be a classical dynamical system 
(see \ref{def_clasDynSyst}), so a member of the category 
of Hilbert spaces with contractions. 
As isometries, unitary maps, and orthogonal projections are the 
embeddings, isomorphisms, and projections respectively in this category
(see the motivation before \ref{def_homomorphism}),
a \Dindex{dilation of a classical dynamical system} $(\mathcal{K}, C_t)$ is a 
quadruple
$\Bigl(J, \widehat{\mathcal{K}}, \widehat{C}_t,  J^* \Bigr))$  
where $J$ is an isometry  $\mathcal{K}\to\widehat{\mathcal{K}}$,
$J^*:\widehat{\mathcal{K}}\to\mathcal{K}$ an orthogonal projection, and
$\widehat{C}_t$, $t\in\RR$ a one-parameter unitary group.
Summarizing:
\begin{displaymatH}
 \begin{CD}
   \mathcal{K}           @>C_{t}>>            \mathcal{K}\\
    @VJVV                                          @AA J^* A \\
   \widehat{\mathcal{K}} @>\widehat{C}_{t}>>  \widehat{\mathcal{K}}
 \end{CD}
\end{displaymatH}

Given an irreversible dynamical system, the question arises 
whether there exists dilations and if so, if there are more than 
one. 
\question{Hans: moet het bestaan van dilataties per geval bekeken worden?}
For special classes of irreversible systems, general constructions for
dilating them have been found. 
In general, if there exists one dilation, there are more which are
not necessarily unitarily equivalent. 
To force uniqueness, so that the properties of the dilating system
$(\widehat{\mathcal{S}},\widehat{T}_t)$ 
are determined by the irreversible subsystem 
$(\mathcal{S}, T_t)$, the concept of a minimal dilation is useful.

A dilation is called \Dindex[dilation!minimal]{minimal}
if $J(\mathcal{K})$ is dense in $\widehat{\mathcal{K}}$.

\medskip

As example, we present a dilation of a ``symmetrically''
damped harmonic oscillator which we will use later.

\begin{theorem}[Sz.~Nagy, Foias 1953; special case]
\label{th_dilationCHO}
Let $\mathcal{K}=\CC$ be the phase space of one oscillator and
consider the damped time evolution
\begin{displaymatH}
 C_t:\CC\to\CC, \quad C_t(z)= e^{(-\gamma+i\omega)t}z ,
  \qquad(t\ge0).
\end{displaymatH}

Unique up to unitary equivalence there exists a Hilbert space
$\widehat{\mathcal{K}}$
with a unit vector $v$ and a one-parameter group of unitary transformations
$\widehat{T}_t=:U_t$ on $\widehat{\mathcal{K}}$ 
such that the span of the vectors $U_t v$, $t\in\RR$, is
dense in $\widehat{\mathcal{K}}$ and
\begin{displaymatH}
 \ip{v}{U_tv}=e^{(-\gamma+i\omega)t},
  \quad(t\ge0)
  .
\end{displaymatH}

Therefore, we have the following minimal dilation of $(\CC,C_t)$:
\begin{displaymatH}
 \begin{CD}
   \CC @>C_t=e^{(-\gamma+i\omega)x}>> \CC\\
   @VJ:z\mapsto zv VV                @AAJ^*: f\mapsto\ip{v}{f}A \\
   \widehat{\mathcal{K}} @> U_t >>     \widehat{\mathcal{K}}
 \end{CD}
\end{displaymatH}
\end{theorem}

The existence follows e.g.\ from taking
$\widehat{\mathcal{K}}:=L^2(\RR,2\gamma\,dx)$, 
$U_t:=R_t$ the shift to the right, and 
\begin{displaymatH}
 v(x):=\begin{cases}
         0 & \text{if } x >0,\\
         e^{(\eta-i\omega)x} &\text{if }x\le 0,
       \end{cases}
\end{displaymatH}
while the uniqueness follows from Kolmogorovs dilation theorem
(see the last example of \ref{ex_KolmogorovDilation}). The minimality
follows from the bijectivity of the Fourier transform. 
For details see \ref{ap_minClassicDilationDampedHarmonicOsc}.

\smallskip

In practice several (unitarily equivalent) minimal unitary dilations
of $(C_t)_t$ can be useful. If $\mathcal{K}=L^2(\RR)$ and $U_t$ is the shift,
then we speak of \Dindex{translation dilations} of $(C_t)_t$. They differ
only in the shape of $v\in L^2(\RR)$ which must satisfy (in terms of their
Fourier transforms)
\begin{displaymatH}
 |\widehat{v}(\lambda)|^2 = \frac{1}{(\lambda-\om)^2 + \eta^2},
 \quad
 (\lambda \in \RR).
\end{displaymatH}
Particular solutions are 
$\widehat{v}_{\pm}(\lambda)=1/(\lambda-\om \pm i\eta)$. Here
$v_{+}$, which occurred above, leads to the \Dindex{incoming translation
dilation} and $v_{-}$ to the \Dindex{outgoing translation dilation}:
\begin{displaymatH}
 v_{-}(x):=\begin{cases}
         e^{-(\eta+i\omega)x} &\text{if }x\ge 0,\\
         0 & \text{if } x <0,
       \end{cases}.
\end{displaymatH}
The former is more useful for the study of incoming fields and particles,
the latter for outgoing ones.
The unitary equivalence of these two unitary dilations, asserted by
theorem \ref{th_dilationCHO}, is implemented by the scattering operator
$S$ which in terms of the Fourier transform $F$ can be written as
\begin{displaymatH}
 S:=FM_sF^{-1}, 
 \quad 
 s(\lambda):=\frac{\lambda -\om + i\eta}{\lambda -\om - i\eta}
 .
\end{displaymatH}

\subsection{Dilations of quantum dynamical systems}
The embeddings, isomorphisms, and projections in the category of
quantum dynamical systems (\ref{def_quantumDynSyst})
are injective $*$-homomorphisms, 
$*$-isomorphisms, and conditional expectations respectively, as has
been elucidated in the the motivation before \ref{def_homomorphism}.
Therefore a dilation of a quantum dynamical system 
$\Bigl( (\mathcal{A},\ph), T_t \Bigl)$ is a 
quadruple 
$\Bigl(j, (\widehat{\mathcal{A}},\widehat{\ph}), \widehat{T}_t, P \Bigl)$ 
such that $T_t= P\after \widehat{T}_t \after j$,
where $j: (\mathcal{A},\ph) \to (\widehat{\mathcal{A}},\widehat{\ph})$
is an injective $*$-homomorphism, $P$ the associated conditional expectation,
and 
$\widehat{T}_t: 
(\widehat{\mathcal{A}},\widehat{\ph})\to (\widehat{\mathcal{A}},\widehat{\ph})
$, $t\in\RR$, is a family of $*$-automorphisms.
Symbollically, we can depict this as follows
\begin{displaymatH}
 \begin{CD}
   (\mathcal{A},\phi)           @>T_{t}>>            (\mathcal{A},\phi)\\
    @VjVV                                          @AA P A \\
  (\widehat{\mathcal{A}},\widehat{\ph}) @>\widehat{T}_{t}>>  (\widehat{\mathcal{A}},\widehat{\ph})
 \end{CD}
\end{displaymatH}

Although not mentioned explicitly, it is understood implicitly 
that all morphisms are quantum operations, as we work in the
category of quantum dynamical systems with quantum operations.
In particular, $j$, $P$, and $\widehat{T}_t$ 
are strongly continuous.

We will meet a quantum dynamical dilation in just a moment,
when we quantize the damped harmonic oscillator.

\chapter{Quantum Mechanical Damped Harmonic Oscillator}

A technical problem when quantizing the classical damped harmonic oscillator
is that the r\^oles of the position $q$ and momentum $p$ are not symmetric,
because their time-evolutions differ more than a simple
phase shift. We will quantize a symmetrical spiraling motion in phase
space instead:
\begin{equatioN}
\label{eq_symmDampedHarmOsc}
 \left. 
 \begin{aligned}
   q(t)& = e^{-\eta t} \left(q(0)\cos \om t + p(0) \sin \om t\right), \\
   p(t)& = e^{-\eta t} \left(-q(0)\sin \om t + p(0)\cos\om t \right), \\ 
 \end{aligned}
 \right
 \rgroup
 \quad(t\in [0,\infty)),
\end{equatioN}
i.e.\ we take $m=1$ and discard the term proportional to 
$\eta e^{-\eta t}$ in $p(t)= m\frac{dq}{dt}$ (weak damping).
Although this may seem a crude approximation at first sight, it 
retains the essential features of a damped evolution and it can be 
viewed as a ``Van Hove'' limit of the damped harmonic oscillator
\question{Hans: heb je hier verwijzingen of meer toelichting bij?}

In view of the above and the Heisenberg equations of motion, 
we formally define a quantum damped harmonic oscillator as 
a canonical pair $(Q,P)$ obeying the time-evolution

\begin{equatioN}
\label{eq_damposc}
 \left.
 \begin{aligned}
    T_t&(Q) =e^{-\eta t}\left(Q\cos\om t+P\sin\om t\right) , \\
    T_t&(P) =e^{-\eta t}\left(-Q\sin\om t+P\cos\om t\right),     
 \end{aligned}
 \right
 \rgroup
 \quad (t\in [0,\infty) )
 .
\end{equatioN}

\section{Quantization of the damped harmonic oscillator}
Identifying the phase space $\{(q,p): q,p\in\RR\}$ with
$\CC$, we can describe 
the damped spiraling evolution (\ref{eq_symmDampedHarmOsc})
as
\begin{displaymatH}
 C_t: \CC \to\CC, \quad z\to e^{(-\eta + i\omega)t} z,
  \qquad (t\in[0,\infty)) 
  ,
\end{displaymatH}
and quantize it using the second quantization functor $\Gamma$ 
to obtain
\begin{displaymatH}
 T_t:=\Gamma(C_t): \Gamma(\CC)\to\Gamma(\CC)
 \qquad
 (t\in [0,\infty)).
\end{displaymatH}

Then 
\begin{equatioN}
\label{eq_formulaTt}
  T_t(W(z))=e^{\half(e^{-2\eta t}-1)|z|^2}W(e^{(-\eta+i\omega)t}z)
  , 
\end{equatioN}

and $T_t$ is defined on $\Gamma(\CC)$, 
the whole von Neumann algebra generated 
by the Weyl operators associated to $Q$ and $P$.

If we {\it define} for $x,y\in\RR$, 
$$xP-yQ$$ 
as the Stone-generator of 
the strongly continuous one-parameter unitary group \\
$\bigl( W(\la(x+iy)) \bigr)_{\la\in\RR} $ i.e.
	$$ 
  W(\la(x+iy))= e^{-i\la(xP-yQ)} \quad ( \la\in\RR) \;,
	$$
then we have a natural way to let $T_t$ {\it also}
act on the unbounded self adjoint 
operators 
\begin{displaymatH}
 xP-yQ.
\end{displaymatH}

Indeed, for $x,y\in\RR$ let 
\begin{equatioN}
\begin{aligned}
\label{eq_differentationTt} 
  T_t \bigl( xP-yQ  \bigr) 
   &= T_t \Bigl(
             i\frac{d}{d\la} 
               \Bigl[ 
                  W\bigl( \la(x+iy)\bigr) 
               \Bigr]_{\la=0}  
          \Bigr) 
  \\
  :& =i\frac{d}{d\la} 
           \Bigl[
                 T_t \bigl( W( \la(x+iy) )~ \bigr) 
           \Bigr]_{\la=0} 
        = \quad \hbox{using (\ref{eq_formulaTt})}\quad       
  \\
   &=i\frac{d}{d\la} \Bigl[
                     e^{\half\la^2(e^{-2\eta t}-1)(x^2+y^2)} 
                     e^{-i\la   \bigl[ 
                             \Re( e^{i\om t}(x+iy) )P - \Im( e^{i\om t}(x+iy))Q 
                               \bigr]
                        }  
                   \Bigr]_{\la=0}                                               \\
   &
   = x\,[ -\sin\om t \; Q + \cos\om t \; P] 
      + y\,[ \cos\om t \; Q + \sin\om t \; P ]          
\end{aligned}
\end{equatioN}

Formally setting 
\begin{equatioN}
\label{eq_formallinearTt}
  T_t \bigl( xP-yQ  \bigr) = x\, T_t(P) -y\, T_t(Q) 
\end{equatioN}

and  equating the coefficients of $x$ and $y$ 
in (\ref{eq_differentationTt}) and (\ref{eq_formallinearTt}) , 
we recover the equations (\ref{eq_damposc}).

In this sense, $T_t$ is a formulation 
of the damped harmonic oscillator evolution 
in terms of Weyl operators that is equal to that 
given by (\ref{eq_damposc}) 
in terms of $Q$ and $P$.

\begin{remark}
Note that in a vacuum representation $W\!:\; \CC \to \mathcal{U}(\mathcal{H})$
of the CCR over $\CC$ (thus the Gaussian, Schr\"odinger, and Heisenberg
representations),
the von Neumann algebra $\Gamma(\CC)$ generated by all Weyl operators 
coincides with all of $ \mathcal{B}(\mathcal{H}) $.
\question{can we provide a proof?}
\end{remark}

\section{Quantum dilation of the damped harmonic oscillator}
\label{sec_quantumDilationDHO}
We now focus on dilating the damped oscillator evolution to an evolution 
determined by a Schr\"odinger equation.
Taking the minimal dilation of $(C_t)_{t\geq 0}$ as in theorem 
\ref{th_dilationCHO}
and applying the second quantization functor $\Gamma$ to it,
we obtain the following dilation of $T_t$.
\begin{displaymatH}
 \begin{CD}
   \Gamma(\CC)           @>T_{t}>> \Gamma(\CC) \\
    @VjVV                                          @AA P A \\
   \Gamma\bigl(L^2(\RR)\bigr)  
            @>\widehat{T}_{t}>> \Gamma\bigl(L^2(\RR)\bigr)
 \end{CD}
\end{displaymatH}

Here $T_t=\Gamma(C_t)$, $j=\Gamma(J)$ with $J:z\mapsto zv$ and 
$P=\Gamma(J^*): f\mapsto \ip{v}{f}$.
Since $\widehat{T}_t:=\Gamma(R_t)$ is a 
point wise continuous one-parameter group of
*-automorphisms of the algebra $\Gamma(L^2(\RR))=\mathcal{B}(L^2(\RR))$,
it is implemented by a unitary strongly continuous one-parameter
group $( U_t=e^{itH} )_{t\in\RR}$ as 
\begin{displaymatH}
  \widehat{T}_t(A)= U_t A U_t^* .
\end{displaymatH}
\question{Where can you find a proof that all $*$-automorphisms of 
$\mathcal{B}(L^2(\RR))$ are of the form $A\mapsto U_tAU_t^{-1}$?
See p. 35 QP (Hans)}

So the evolution $( \widehat{T}_t )_t$ obeys a Schr\"odinger equation 
and the irreversible evolution 
\begin{displaymatH}
 ( T_t = P \circ \widehat{T}_t \circ j)_t
\end{displaymatH}
is a `shadow' of the reversible evolution $( \widehat{T}_t )_t$.

It is now a good moment to lend some physical flesh and blood to the 
mathematical skeleton depicted in the diagram above.

\section{Interpretation as quantum dynamical system: \\
         open system with (minimal) Bose dilation}

The semi-group $( T_t )_t$ 
expresses the time evolution of the damped harmonic oscillator,
although not in terms of the unbounded observables $Q$ and $P$, but
in terms of their associated Weyl operators. 
$\Gamma(\CC)$, which is the von Neumann algebra generated by the 
Weyl operators associated to $Q$ and $P$, is the algebra of 
all (bounded) observables of the damped harmonic oscillator.
Implicitly, the state $\ph_\CC$ on $\Gamma(\CC)$ is assumed, which describes
that the damped harmonic oscillator is in its vacuum state.

We may wonder how we could  physically interpret this dilation of the
quantum harmonic oscillator. 
From the discussion in section \ref{sec_usualSecondQuantization}, it is
clear that we have emerged (or coupled) the damped harmonic oscillator in
a field of oscillators labeled by $f\in L^2(\RR)$. Due to the field's 
commutation relations, this is a Bose field.
Keeping in mind that the damped harmonic oscillator should model 
an atom loosing its energy to its surroundings by emission of light,
the interpretation of $\Gamma\bigl(L^2(\RR)\bigr)$ as 
a Bose field of light quanta is readily made.
The time evolution of $\Gamma\bigl(L^2(\RR)\bigr)$ is given by 
$*$-automorphisms, and therefore reversible. These $*$-automorphisms are
a generalized way of saying that the time-evolution is given by a 
Schr\"odinger equation.

Note, however, that we could have taken the Bose field too `large'
in the sense that some parts of the field do not interact with the
oscillator. 
Removing the redundant parts of the Bose field, we are left 
with a quantum mechanical system called the 
\Dindex{minimal Bose dilation of the damped harmonic oscillator}. 

The minimal Bose dilation is unique in
the sense that any quantum mechanical system that evolves 
according to a Schr\"odinger equation and contains the damped harmonic
oscillator in the sense that 
\begin{Properties}
\item  every oscillator observable can be recovered as a `projection'
            of an observable of the total system onto the oscillator
            subsystem, 
\item  the time evolution of the damped oscillator is a `projection'
            of the overall time evolution onto the oscillator subsystem and
\item the interaction of the oscillator with the surrounding system
            is via a Bose field,
\end{Properties}
also contains an isomorphic copy of the minimal Bose dilation.

Now $\Gamma\bigl(L^2(\RR)\bigr)$ and 
$\bigl( \widehat{T}_t=\Gamma(R_t) \bigr)_t$ are 
the observable algebra and time evolution respectively 
of the minimal Bose dilation of the damped harmonic oscillator.

On  $\Gamma\bigl(L^2(\RR)\bigr)$, 
the state $\ph_{L^2(\RR)}$ is assumed.

As mentioned in ii., every observable of the damped harmonic oscillator 
is a projection of an observable of the total system (=oscillator and
minimal Bose surroundings).
Now $j$ associates to an oscillator observable the corresponding 
observable of the minimal Bose dilation, while $P$ projects observables of the
minimal Bose dilation onto oscillator observables.

\section{Interpretation as quantum stochastic process}
No one who has been exposed to quantum mechanics seriously has
failed to notice that the 
physical content of quantum mechanics is of stochastic nature.
We cannot predict the outcome of a physical quantity in a single 
experiment, but we can predict 
the probability distribution of an observable. 
The latter is the relative frequency 
with which each of the possible outcomes of measuring
that observable occurs during an (infinitely) repeated series of 
measurements on identical systems. Such a collection of identical
systems is called an ensemble, and arises e.g.\ 
because of an identical preparation procedure.
The merit of quantum theory is that it postulates
rules how to obtain and calculate with those probabilities,
and that these rules turn out to aptly predict 
how small scale physical systems behave.

Formulating these rules more comprehensively on an abstract level 
leads to quantum probability theory in which
the concepts of quantum probability spaces
and quantum stochastic processes enter stage.

\paragraph{Quantum probability space}
Recall that a quantum dynamical system $(\mathcal{A}, \phi, T_t)$
consists of two parts: 
a Von Neumann algebra $\mathcal{A}$ with normal state $\ph$, 
and a family of quantum operations $(T_t)_t$. 
It is the first part that can be interpreted as a quantum probability 
space as soon as we have identified what the events and 
their probabilities are.
 
\begin{definition}
A stochastic experiment is said to be \textsl{modeled} by a 
quantum probability space $(\mathcal{A},\ph)$ if the following is the case.
The experiment can be repeated arbitrarily often and 
sufficiently many of the orthogonal projections $E$ in $\mathcal{A}$ have an
interpretation as an event i.e.\ 
a statement about the occurrence of certain outcomes 
of the experiment that can be tested by observation. 
The relative frequency with which 
the event encoded by $E$ occurs is given by $\ph(E)$.
\end{definition}

The projection $0$ is the impossible event and $\Identity$ is the sure event.
Many events relate to the outcomes of observables of the system.
If $A$ is an observable and $P_A(\cdot)$ its projection-valued measure,
then $P_A\bigl((-\infty,s]\bigr)$ is the question whether a measurement of $A$
yielded an outcome that was less than or equal to $s$.

In the picturesque language of Mackey \cite{Mackey} events may be considered as
\textsl{questions} that one can ask the system. 

In contrast to classical stochastic experiments,
as envisaged by Kolmogorov,
we allow adjustment of the observation equipment between the trials,
that is, we are allowed to ask new questions.
When asking different questions, a quantum peculiarity arises.
If two questions $E, F$ are compatible i.e.\ $[E,F]=0$, then
those questions can be asked together in the same trial, and
$EF$ denotes the event that both $E$ and $F$ occur,
and $E\vee F:=E+F-EF$ is the event that either $E$ or $F$ occurs.

Incompatible questions can never be asked together.
They can be asked in different trials by readjusting the apparatus in between.
Inside one single trial it is sometimes possible to ask
incompatible questions one \textsl{after} the other,
but then the fact that one question was posed,
may influence the answer to subsequent questions.

If the questions $E_1, E_2, \ldots, E_k$ are asked in each trial 
and in this order,
the asymptotic fraction of the trials in which they are all answered `yes'
is
   $$\ph(E_1E_2\cdots E_{k-1}E_kE_{k-1}\cdots E_2E_1).$$

The generality of the concept of a quantum probability space is 
illuminated by the fact that a (classical) probability space is included
as special case (which we shall not prove):

\begin{theorem}
Let $(\Om,\Si,\PP)$ be a probability space and let $M_f$ be the 
multiplication by $f\in\Linf(\Om,\Si,\PP)$.
Then the algebra
   $$\mathcal{A}:=\{M_f : f\in\Linf(\Om,\Si,\PP)\}$$
is a (commutative) von Neumann algebra of operators on $L^2(\Om,\Si,\PP)$,
and $\ph:M_f\mapsto\int f d\PP$ is a faithful normal state on $\mathcal{A}$.
Conversely, every \textsl{commutative} von Neumann algebra $\mathcal{A}$ with a faithful
normal state $\ph$ is of the above form for some classical probability space.
\end{theorem}

Taking $f=\indicator_{S}$ for $S\in\Si$, it is clear that $\phi(f)=\PP(S)$ as
should be. The question arises how random variables are encoded in terms of
(commutative) Von Neumann algebras.

\paragraph{Quantum random variable}
Classically,
a random variable with values in $(\RR,\Borel(\RR))$
is a measurable function $X:(\Om,\Si) \to (\RR, \Borel(\RR))$.
We can alternatively characterize $X$ by the $\si$-algebra embedding
   $$X^{-1}:\RR\to\Si:S\mapsto X^{-1}(S):=\{\om\in\Om : X(\om)\in S\}.$$
The probability distribution $\PP_X$ of $X$ is then the induced measure
on $\RR$:
   $$\PP_X:=\PP\circ X^{-1}.$$
In fact, giving
$X$ is equivalent to giving the injective *-homomorphism
   $$j_X:\Linf(\RR,\Borel(\RR),\PP_X)\to\Linf(\Om,\Si,\PP)
        : f\mapsto f\circ X,$$
because $\ph= \widehat{\ph} \circ j_X$ where $\ph$ and $\widehat{\ph}$
are be the (standard) choices for the states on 
$\Linf(\RR,\Borel(\RR),\PP_X)$ and $\Linf(\Om,\Si,\PP)$ i.e.\ the
expectation with respect to $\PP_X$ and $\PP$.

\begin{definition}
A \textsl{quantum random variable} on a probability space $(\mathcal{A},\ph)$
is an injective *-homomorphism $j$ 
from some other von Neumann algebra $\mathcal{B}$
to $\mathcal{A}$ which maps $\Identity_\B$ to $\Identity_\mathcal{A}$.
The state $\psi:=\ph\circ j$ on $\B$ is called the
{\it probability distribution} of $j$.
\end{definition}

We say that $j$ \textsl{takes values in $(\Om',\Si')$} if 
$\mathcal{B}=\Linf(\Om',\Si')$.
The random variable $j$ is then completely determined by the projection-valued
measure
   $$E:\Si'\to\mathcal{A}:S\mapsto j(1_S).$$

In particular, we see that quantum probability generalizes
the events and random variables as known in classical probability theory.

\paragraph{Quantum stochastic processes}

\begin{definition}
A \Dindex{quantum stochastic process} is a family $(j_t)_{t\in\TT}$
of random variables 
$(\mathcal{A},\ph)\to (\widehat{\mathcal{A}},\widehat{\ph})$ indexed by time.
Here, $\TT$ is a linearly ordered set such as
$\ZZ$, $\RR$, $\NN$ or $\RR_{+}$.
If $\TT=\RR$ or $\RR_{+}$ we require that for all $A\in\mathcal{A}$ the curve
$t\mapsto j_t(A)$ is strongly continuous.

\smallskip

If $\TT$ is a group, say $\ZZ$ or $\RR$,
then the process is called \Dindex{stationary}
provided $j_t=j_0\circ\widehat{T}_t$ for some representation 
$t\mapsto\widehat{T}_t$ of $\TT$ into the automorphisms of $
(\widehat{A},\widehat{\ph})$.
\end{definition}

\begin{remarks}
If a classical process is stationary, $X_t$ and $X_s$ differ by an automorphism
$\tau_t$ of the underlying probability space: $X_t= X_0\circ \ta_t$.
These remarks generalize to quantum probability, as
quantum stochastic processes only differ by an automorphism of Von Neumann
algebras (=quantum probability spaces).
\end{remarks}

The idea behind dilation (open system interpretation) is that 
we are observing a subsystem with observable algebra $\mathcal{A}$
of a larger environment with algebra $\widehat{\mathcal{A}}$ 
that we cannot see.
According to the Heisenberg picture, the smaller algebra is
moving inside the larger one.
A question $E\in\widehat{A}$ can be asked at time $t$ only if 
$E\in j_t(\mathcal{A})$.
If $t_0\le t_1\le\cdots\le t_n$ is a sequence of times,
and $E_1, \ldots, E_n$ a sequence of events in $\mathcal{A}$, then
\begin{displaymatH}
 \widehat{\ph}
   \bigl(j_{t_1}(E_1) j_{t_2}(E_2) 
           \cdots j_{t_{n-1}}(E_{n-1})j_{t_n}(E_n) j_{t_{n-1}} \cdots
         j_{t_2}(E_2) j_{t_1}(E_1)
    \bigr)
\end{displaymatH}
is the probability that $E_1$ occurs at time $t_1$, $E_2$ at time $t_2$,
$\ldots$, and $E_n$ at time $t_n$.

Can we calculate this probability if we only observe the subsystem
(i.e.\ if we cannot evaluate $\widehat{\ph}$ and $j_t$)?

\smallskip

\textsl{Two-time-probabilities}
Suppose that for all $s\in\TT$ there exists a conditional expectation $P_s$ 
with respect to $j_s$, that is, not only do there exist embeddings $j_t$ of
the subsystem into the environment, but also the corresponding ``projections'' 
that associate observables of the total system with observables of the 
subsystem.

Then the probability for $F$ to occur at time s and $E$ at time $t>s$
can be written as
   $$\widehat{\ph}(j_s(F)j_t(E)j_s(F))=\ph(FP_s(j_t(E))F)
                             =\ph(FT_{s,t}(E)F),$$
where $T_{s,t}=P_s\circ j_t$ is an endomorphism of $(\mathcal{A},\ph)$,
the \Dindex{transition operator} from time $s$ to time $t$.
In other words, the two-time probability can be ``calculated'' from
information that is available within the subsystem.

\smallskip

\textsl{Multi-time-probabilities}
The above reduction to the subsystem succeeds for more than two time
points if there also exist conditional expectations onto the algebras
   $$\mathcal{A}_{(-\infty,t]}:=\vNeu\{j_s(\mathcal{A}) : s\le t\}.$$
and moreover the \Dindex{Markov property} holds:
   \begin{equatioN}
   \label{eq_quantumMarkovProperty}
   s\le t\Implies P_{(-\infty,s]}(j_t(\mathcal{A}))\subset j_s(\mathcal{A}).
   \end{equatioN}

\begin{theorem}
Let 
$\bigl(j_t:(\mathcal{A},\ph)
  \to
 (\widehat{\mathcal{A}},\widehat{\ph})\bigr)_{t\in\TT}$
be a stationary Markov process with conditional expectations $P_t$.
Then the transition operators form a monoid:
   $$0\le s\le t\le u\Implies T_{s,t}T_{t,u}=T_{s,u}.$$
In particular, if the process is stationary,
then $T_t:=T_{0,t}=T_{s,s+t}$ satisfies
   $$T_sT_t=T_{s+t}\;\qquad(s,t\ge0).$$
\end{theorem}

In the latter case, $(T_t)_{t\ge0}$ is known as the
\Dindex{dynamical semigroup} induced by the stationary Markov process.
Conversely, the process $(j_t)_{t\in\TT}$ is called the 
\Dindex{quantum Markov dilation}
of the dynamical semigroup $(T_t)_{t\in\TT}$.

The situation is symbolized by the commutative diagram
\begin{displaymatH}
  \begin{CD}
  (\mathcal{A},\ph) @>T_t>>(\mathcal{A},\ph) \\
   @VjVV                         @AAPA       \\
  (\widehat{\mathcal{A}},\widehat{\ph}) @>\widehat{T}_t>> 
      (\widehat{\mathcal{A}},\widehat{\ph}) 
  \end{CD}
\end{displaymatH}

\paragraph{Embedded classical processes}
In case  $\mathcal{A}_0 \subset \mathcal{A}$, and 
$\widehat{\mathcal{A}}_0 \subset \widehat{\mathcal{A}}$
are \textsl{commutative} Von Neumann \textsl{sub}algebras so that
the following diagram 
(with the appropriate restrictions of the states and operations)
\begin{displaymatH}
  \begin{CD}
  (\mathcal{A}_0,\ph) @>T_t>>(\mathcal{A}_0,\ph) \\
   @VjVV                         @AAPA       \\
  (\widehat{\mathcal{A}}_0,\widehat{\ph}) @>\widehat{T}_t>> 
      (\widehat{\mathcal{A}}_0,\widehat{\ph}) 
  \end{CD}
\end{displaymatH}

commutes, 
then there exist classical probability spaces $(R, \mathcal{B}, \mu)$
and $(\Om,\Si,\PP)$, such that 
\begin{displaymatH}
  \begin{CD}
  L^\infty(R,\mathcal{B},\mu) @>T_t>> L^\infty(R,\mathcal{B},\mu) \\
   @VjVV                         @AAPA       \\
  L^\infty(\Om,\Si,\PP) @>\widehat{T}_t>> L^\infty(\Om,\Si,\PP)
  .
  \end{CD}
\end{displaymatH}
In this case, 
we can often identify a stationary classical Markov process or chain $(X_t)_t$,
as follows:

\begin{displaymatH}
\begin{CD}
  L^\infty(R,\mathcal{B},\mu) 
     @>T_t: f\mapsto \bigl(x\mapsto \int f(y) P_t(x,dy)\bigr) >> 
          L^\infty(R,\mathcal{B},\mu) 
  \\
  @Vj: f\mapsto f(X_0) VV                         
      @AAP: Y\mapsto \EE(Y|X_0)A       
  \\
  L^\infty(\Om,\Si,\PP) 
     @>\widehat{T}_t: Y\mapsto Y\after \tau_t >> 
     L^\infty(\Om,\Si,\PP)
  .
  \end{CD}
\end{displaymatH}
where
\begin{Properties}
 \item
 $\widehat{T}_t: $ corresponds to an automorphism 
 $\tau_t$ of the underlying measure space $(\Om, \Si, \PP)$:
 $\widehat{T}_t(Y)= Y\after \tau_t$; 
 in particular:  $X_t= X_0 \circ \tau_t$ is stationary random process
 on $(\Om,\Si)$ with values in $(R,\mathcal{B})$;
 \item 
 the embedding $j$ maps $f\in L^\infty(R,\mathcal{B},\mu)$ to
 $f(X_0) \in L^\infty(\Om,\Si,\PP)$; 
 \item 
 the conditional expectation $P$ is closely related to 
 the classical conditional expectation
 \begin{displaymatH}
 Y\mapsto \EE(Y|X_0)
 \end{displaymatH}
 i.e.\ the conditional expectation given the process at time $0$;
 more precisely: 
 \begin{displaymatH}
   P(Y)=g \in L^\infty(R,\mathcal{B},\mu) \Iff \EE(Y|X_0)= g(X_0); 
 \end{displaymatH}
 \item
 the maps $(T_t)_t$ form a semi-group of transition probabilities;
 if $P_t(x,d y)$ is the probability that the process $(X_t)_t$ 
 is in the area $dy$ at time $t$ given that it started from $x$ at time 
 $0$, then
 \begin{displaymatH}
  (T_t f)(x) = \int f(y)P_t(x, dy) 
   \quad \Bigl( x\in R, \; f\in L^\infty(R,\mathcal{B},\mu) \Bigr) 
   ;
 \end{displaymatH}
  \item
  the equality in the upper right corner of the diagram:
   $\EE(f(X_t)|X_0) =   (T_t f)(X_0) = \int f(y)P_t(X_0, dy)$
  is also intuitively obvious, 
  as the expected value of $f(X_t)$ given that the process started at $X_0$
  is the  average of all possible $f(y)$ weighted by the probability
  that $X_t$ would arrive at $y$ at time $t$, when it started from $X_0$.
\end{Properties}

\section{Stochastic behavior of the oscillator}
Let us return to the quantum dilation of the damped harmonic
oscillator (section \ref{sec_quantumDilationDHO}), 
where $T_t=\Gamma(C_t)$, $\widehat{T}_t:=\Gamma(U_t)$, 
$j=\Gamma(J)$ with $J:z\mapsto zv$ and $P=\Gamma(J^*)$.
Due to the functorial character of $\Gamma$ (\ref{ap_propertiesGammaFunctor}),
this is automatically a quantum Markov dilation.
Furthermore, the functorial character of $\Gamma$ also implies that 
we can split $T_t$ as
\begin{displaymatH}
  T_t=\Gamma(e^{(-\eta+i\om)t})=\Gamma(e^{i\om t})\Gamma(e^{-\eta t}).
\end{displaymatH}

The operator $\Gamma(e^{-i\om t})$ is the automorphism of the quantum
harmonic oscillator studied previously.
Let us now look at the `dissipative' part $\Gamma(e^{-\eta t})$.

\begin{proposition}
For $0\le c<1$ the operator $\Gamma(c)$ leaves the abelian subalgebras
generated by $\Identity$ and any of the operators $xP-yQ$ invariant.
In particular its action on the algebra
\begin{displaymatH}
 \mathcal{Q}:=\{f(Q): f\in\Linf(\RR) \}
\end{displaymatH}
is given by
\begin{equatioN}
\label{eq_actionGammaQ}
  \Gamma(c)(f(Q))
     =\frac{1}{\sqrt{2\pi(1-c^2)}}
      \int e^{-\frac{x^2}{2(1-c^2)}}\;f(cQ+x)\,dx
     .
\end{equatioN}
\end{proposition}

\begin{proof}
Obviously, $\Gamma(c)$ leaves the linear span of 
$\{W(\la z): \la\in\RR\}$
invariant, and then also its strong closure.
Putting $f(Q)=e^{iyQ}$ we see that the right hand side of 
(\ref{eq_actionGammaQ}) equals
\begin{displaymatH}
 \Bigl(\frac{1}{\sqrt{2\pi(1-c^2)}}
       \int_{-\infty}^\infty e^{-\frac{x^2}{2(1-c^2)}}e^{ixy}dx
 \Bigr)
 e^{icyQ}
  =
 e^{-\half(1-c^2)y^2W(icy)}
 ,
\end{displaymatH}
which is equal to the left hand side by the definition of $\Gamma$.
The theorem follows from the strong continuity of $\Gamma(c)$. 
\end{proof}

We recognize the semigroup $T_t$ of transition operators 
restricted to $\mathcal{Q}$ as the transition operators of a diffusion
on $\RR$ with a drift towards the origin proportional to the distance
to the origin.

\subsection{The driving field}

It is interesting to look at the dilation of the semigroup $T_t$
for $\om=0$ in the \textsl{incoming} dilation.

Then we obtain an abelian subalgebra of $\Gamma(L^2(\RR,2\eta))$ by putting
\begin{displaymatH}
 \widehat{Q}:=\vNeu\{\widehat{T}_t\circ j\left(e^{isQ}\right) : t,s\in\RR\}
 .
\end{displaymatH}
In the Gaussian representation of the algebra $\widehat{A}=\Gamma(L^2(\RR))$
this is just
\begin{displaymatH}
 \widehat{Q}_t:=\widehat{T}_t\circ j(Q)=\int_{-\infty}^t e^{\eta(s-t)}dB_s
 ,
\end{displaymatH}
where $B_s$ is a Wiener process with variance $2\eta$.

Indeed, let $\Phi(f)$ for $f\in L^2(\RR, 2\eta)$ be defined as the 
generator of $W(\lambda f)$ i.e.\ $e^{i\lambda \Phi(f)}=W(\lambda f)$.
Observe that $\Phi(f)$ is a random variable 
that has a normal distribution with mean $0$ and
variance $\norM{f}^2$ in the vacuum state (because its characteristic
function is 
$\lambda \mapsto \ph_{\mathcal{K}}(W(\lambda f)) 
        = e^{-\half\lambda^2\norM{f}^2}$).
Furthermore, the Weyl relations imply that 
$\Phi(f)$ and $\Phi(g)$ are compatible if $\ip{f}{g}$ is real, and
independent if $\ip{f}{g}=0$.
In particular, if $\mathcal{K}=L^2(\RR, 2\eta dx)$, then
\begin{displaymath}
 B_t := 
   \begin{cases}
     \Phi(\indicator_{[0,t]} & \text{ if } t\geq 0 ,\\
     \Phi(-\indicator_{[t,0]} & \text{ if } t< 0 ,\\
   \end{cases}
\end{displaymath}
is a stochastic process with compatible normally distributed
independent increments having variance 
\begin{displaymatH}
 \ph_{\mathcal{K}}\Bigl((B_t -B_s)^2\bigr)= 2\eta| t-s|,
\end{displaymatH}
i.e.\ $(B_t)_t$ is a Wiener process.

\smallskip

\begin{theorem}
The process $\widehat{Q}_t:=\Phi(v_t)$ satisfies the integral equation
\begin{equatioN}
\label{eq_OU}
 \widehat{Q}_t -  \widehat{Q}_s= -\eta \int_s^t  \widehat{Q}_u du +
  B_t - B_s
  \quad (s\leq t),
\end{equatioN}
which is the integral version of the stochastic differential
equation 
\begin{displaymatH}
 d \widehat{Q}_t = -\eta  \widehat{Q}_t dt + dB_t.
\end{displaymatH}
\end{theorem}

So we find an embedded Ornstein-Uhlenbeck process in our Markov
dilation.

\begin{proof}
 The following equality between functions in $\mathcal{K}=L^2(\RR, 2\eta)$
 holds
 \begin{displaymatH}
  U_t v_{-} -   U_s v_{-} = -\eta\int_s^t U_u v_- du + \indicator_{[s,t]}
  \quad
  (s\leq t).
 \end{displaymatH}
 Acting with $\Phi$ on both sides of the equation yields (\ref{eq_OU}).
\end{proof}

\subsection{Quanta}
We now concentrate on another abelian subalgebra of 
$\Gamma(\CC)=\mathcal{B}(\HH)$,
namely
\begin{displaymatH}
 \mathcal{N}:=\{\sum_0^\infty \al_n E_n : \al\in \ell^\infty(\NN) \}
  ,
\end{displaymatH}
where $E_n$ is the orthogonal projection onto $e_n$ in the Heisenberg
representation $\mathcal{H}_{Heis}=\ell^2(\NN,\frac{1}{n!})$.
So $\mathcal{N}$ is the algebra of all diagonal matrices in Heisenberg's matrix
mechanics.

This time we do not put $\om=0$.
Let $\ph_n$ denote the state $A\mapsto\textrm{tr}(E_nA)$.
Write $H=2N+\Identity$.

\begin{proposition}
The algebra $\mathcal{N}$ is invariant for $T_t$ and
\begin{displaymatH}
  \ph_n\left(T_t\left(s^N\right)\right)=\bigl(1-e^{-2\eta t}(1-s)\bigr)^n
  .
\end{displaymatH}
This is the probability generating function
of a pure death process \cite{Grimmit+Stirzaker} with the generator
\begin{displaymatH}
  L=2\eta
     \left[
     \begin{matrix}
       0&0&0&0&\cdots\\
       1&-1&0&0&\cdots\\
       0&2&-2&0&\cdots\\
       0&0&3&-3&\cdots\\
       \vdots&\vdots&\vdots&\vdots&\ddots\\
     \end{matrix}
     \right]
     .
\end{displaymatH}
\end{proposition}

\begin{proof}
We can write $s^N$ as a weak integral over the operators $W(z)$
(lemma \ref{lem_PVMofN}, appendix \ref{ap_PVMofN})
\begin{equatioN}
\label{eq_sNint}
 s^N=\frac{1}{\pi}\int_\CC e^{-\half\frac{1+s}{1-s}|z|^2} W(z)\la_2(dz)
 ,
\end{equatioN}
where $\la_2$ denotes the two-dimensional Lebesgue measure on $\CC$.
This relation can be explicitly checked by taking matrix elements
with respect to coherent vectors.
Application of $T_t$ to both sides of (\ref{eq_sNint}) 
yields for all $u,v\in\CC$,
\begin{displaymatH}
 \ip{\pi(u)}{T_t\left(s^N\right)\pi(v)}=e^{\bar uv(1-e^{-\eta t}(1-s))}
 .
\end{displaymatH}
Since this expression is not sensitive to the relative phase of $u$ and $v$,
the operator $T_t\left(s^N\right)$ lies in $\mathcal{N}$. \question{???}
The statement is proved by taking coefficients of $(\overline{uv})^n$ on
both sides.
\end{proof}

\subsection{Emitted quanta}
Finally,
let us have a look at what happens outside the oscillator while
it is cascading down its energy spectrum.
Since we are interested in outgoing quanta at positive times, 
we consider the \textsl{outgoing} translation dilation of $(T_t)_{t\geq 0}$.
The Weyl algebra of $\Gamma(L^2(\RR))$ has a suggestive interpretation
in its Fock representation $L^2(\Delta(\RR)$.
Let $N_t$ be the number operator $d\mathcal{F}(P_{v_t})$ 
that counts the excitations of the oscillator at time $t$.

As $\mathcal{N} \isom \ell^\infty$ is invariant under the action of
$T_t=\Gamma(e^{(-\eta+i\om)t})$, we can obtain a dilation of 
$T_t: \mathcal{N}\to\mathcal{N}$ by restriction:
\begin{displaymatH}
\begin{CD}
 \mathcal{N} @>T_t>> \mathcal{N}\\
 @VjVV                  @APAA   \\
 \Gamma(\mathcal{K})@>\Gamma(U_t)>> \Gamma(\mathcal{K}).
\end{CD}
\end{displaymatH}
However, since for different times $s$ and $t$ the functions
$v_s$ and $v_t$ are neither orthogonal nor parallel (i.e.\ multiples of each
other), the one dimensional projections $P_{v_s}$ and $P_{v_t}$ do not 
commute. A fortiori the associated number operators 
$N_t:=d\mathcal{F}(P_{v_t})$ do not commute
either, since
\begin{displaymatH}
 e^{i\la N_t} e^{i\mu N_s} 
  = \mathcal{F}(e^{i\la P_{v_t}}) \mathcal{F}(e^{i\mu P_{v_s}})
  = \mathcal{F}(e^{i\la P_{v_t}}e^{i\mu P_{v_s}})
\neq \mathcal{F}(e^{i\mu P_{v_s}}e^{i\la P_{v_t}})
  =  e^{i\mu N_s} e^{i\la N_t} 
  .
\end{displaymatH}

Nevertheless, we can identify a classical process as follows.
For each $t\in\RR$, we can consider three number operators:
\begin{aligN}
 N_t \ats := d\mathcal{F}(P_{v_t}) 
     & \text{the number of quanta in the oscillator,} \hfill
 \\
 M_t \ats := d\mathcal{F}(M_{\indicator_{[t,\infty)}}) 
     & \text{the number of quanta that have not yet left the oscillator,}
 \\
 K_t \ats := d\mathcal{F}(M_{\indicator_{(-\infty,t]}}) 
     & \text{the number of outgoing quanta that have left the oscillator.}
\end{aligN}
Note that in this choice of representation for the dilation 
(the outgoing dilation), $M_t - N_t$ i.e. the number of incoming quanta, 
is not given by a multiplication operator, but the number of outgoing quanta
is (hence the name outgoing representation). Note furthermore that the 
operators $M_t$ and $K_s$ do commute for all $t,s\in\RR$, hence generate
commutative von Neumann subalgebras of $\Gamma(\mathcal{K})$.

For positive times the number of incoming quanta $M_t - N_t$ has
expectation 0 in the states of the form $\theta \after P$ 
($\theta \in \ell^1(\NN)$) that we consider. So we may expect that 
replacing $N_t$ by $M_t$ would lead to an embedded classical Markov chain.

\begin{proposition}
For $t\geq 0$, we have the following commuting diagram involving commutative
von Neumann algebras:
\begin{displaymatH}
\begin{CD}
 \mathcal{M} @>T_t>> \mathcal{M} \\
 @Vj:f\mapsto f(M_0)VV    @AAP=\Gamma(J^*)A\\
 L^\infty(\Delta,\mu_\eta) @>\Gamma(U_t)>> L^\infty(\Delta,\mu_\eta),
\end{CD}
\end{displaymatH}
where $\mathcal{M}$ is the von Neumann algebra generated by 
$T_t(s^{M})$ ($t\geq 0$, $s\in (-1,1)$). 
Furthermore, $\mathcal{M}=\mathcal{N}$.
\end{proposition}

\begin{proof}
For all $z,u\in\CC$, $s\in [0,1]$
we have that
\begin{aligN}
\oip{\pi(u)}{P\after \Gamma(U_t)\after(s^M)}{\pi(z)}
  \ats = \oip{\pi(u)}{s^{M_t}}{\pi(z)}
  \\
  \ats = \sum_{n=0}^\infty 
         \int_{\Delta_n(\RR)} 
               s^{M_t(\si)} |v^{\otimes n}(\si)|^2 
          \mu_\eta(d\si)
  \\
  \ats = \sum_{n=0}^\infty (2\eta \overline{u}z)^n \cdot \frac{1}{n!} 
         \int_{\Delta_n(\RR)} 
               \Bigl(
                 \Pi_{j=1}^n 
                 s^{\indicator_{[t,\infty)}(r_j)} 
                 |v(r_j)|^2 
               \Bigr)
          dr_1 \ldots dr_n
  \\
  \ats = \exp\Bigl( 2\eta\overline{u}z
             \int_0^\infty e^{-2\eta r} s^{\indicator_{[t,\infty)}}(r) dr
      \Bigr)
  \\
  \ats =\exp\Bigl(\overline{u}z
           \bigl([-e^{-2\eta r}]_0^t -[se^{-2\eta r}]_t^\infty \bigr)
      \Bigr)
  \\
  \ats = \exp\bigl(\overline{u}z(1-e^{-2\eta t}(1-s))\bigr)
  ,
\end{aligN}
and the latter equals $\oip{\pi(u)}{T_t(s^N)}{\pi(z)}$.

The equality of von Neumann algebras $\mathcal{M}$ and $\mathcal{N}$ follows
from the fact that $M_{[t,\infty)}$ is a strongly continuous limit of 
linear combinations of $P_{v_s}$, $s\leq t$ \question{Is this true?}
\end{proof}

Finally, since $K_t + M_t$ is equal to the total number of quanta, which
is the same constant for all times (given a state), we conclude that
a quantum is emitted precisely at the moment that the oscillator
makes a downward jump. Moreover these jumps are made at independent
exponentially distributed random times.
These phenomena turn out to be natural consequences of damped harmonic motion
in a non-commutative description.

\questionst{Elaborate:
However, the algebra generated by the embedded count operators $N_t$
i.e.\ $\{ j_t (\mathcal{N}) : t\in\RR\}$ is 
is not an abelian subalgebra of $\Gamma(\mathcal{K})$, 
and therefore this dilation cannot directly be interpreted as a 
classical stochastic process.

However, if we define
   $$F_t:=M_{f_t},\qquad\hbox{where}\qquad f_t(\s):=\#(\s\cup[0,\infty))\;,$$
then then the $F_t$ do have this property. Moreover:

Proposition:
Let $\phh_n:=\ph_n\circ P$ be the state that has $n$ excitations in the
oscillator at time 0 and vacuum outside.
Then for $t\ge0$ we have $N_t=F_t$ almost surely with respect to $\phh_n$,
and the quanta are released precisely at the times that number $N_t$ jumps down
by one.
}

\questionst{Summarize the spiraling motion and the cascading down as 
announced earlier.}


\chapter*{Epilogue: other models for \\the damped harmonic oscillator \markboth{Epilogue: other models for the damped harmonic oscillator}%
{Epilogue: other models for the damped harmonic oscillator}}

It took some time after the conception of
quantum theory before it became clear that dissipation was not only a
macroscopic phenomenon, but that there were also microscopic systems
with dissipative behavior.
For example, the quantum mechanical analysis of radiation in a 
microwave cavity needs to incorporate dissipation, 
since cavity losses are not negligible (Senitzky). Other examples
occur in nuclear physics: 
nuclear fission, giant resonances, and heavy ion collisions 
(see p.~7 of Dekker for references).

Since then the interest in developing quantum
mechanical models capable of describing dissipation emerged.
The damped harmonic oscillator (dho) functions as a standard example 
for quantum mechanics of open dissipative systems, much like
the harmonic oscillator is the standard example for quantum
mechanics of closed systems evolving via unitary time-evolution
operator.
A steady flow of articles on quantization of the dho
began in the 1940's and has not faded away today. The 1980 review of
Dekker mentioned already 232 articles on the subject (see Harris), 
but concluded there was still no generally accepted
way of quantization. Some consensus seems to
have grown, owing to the use of the word "canonical" in some more
recent articles. Based on a non-random, non-systemic literature
search, we give an impression of some of the major approaches to
quantize the dho, without claiming completeness in any sense, however.
(For a completer historical survey up to 1980, see Dekker).

\medskip

One of the first approaches could be summarized as based on
\textsf{quantizing the classical equation of motion}.  
The basic idea
in this approach is to start with the classical equation of motion for
a damped harmonic oscillator, find the Hamiltonian or Lagrangian which
leads to these equation of motion and then quantize by conventional
formal methods.

One exponent of this approach is the so-called
(Bateman)-Caldirola-Kanai model (Bateman, Caldirola, Kanai, Huang and Wu) 
which derives quantum mechanics from a "dissipative
Hamiltonian".  This Hamiltonian was actually proposed earlier by
Bateman, but in a classical context (Bateman 1938).
This approach has the attractiveness of providing an exact solution,
in essence because the classical equation of motion has an exact solution
and formal quantization merely has the effect of converting the
classical variables into operators. The solution is therefore
formally the same, but must now be interpreted as an expression in
operators. A critique on these models is that the commutator of $Q$ and $P$ 
fades away i.e.\ Heisenberg's uncertainty principle is violated.

A slightly different way to quantize directly the classical equation
starts from a "dissipative lagrangian'' that give raise to the
classical equation of motion of the dho (see note [25] of Huang \& Wu and 
see Papadopoulos).  The idea here is that the propagator of the
Schr\"odinger equation contains all quantum mechanical information of
the system (like the Hamiltonian does) and that this propagator can be
constructed from the Lagrangian. For a quadratic Lagrangian of a
particle, the propagator can be given explicitly. 
Papadopoulos extends this idea to Lagrangians with not
only time-dependent total energy, but also a time-dependent mass
and developed a method to evaluate the corresponding propagator $K$ 
by solving a certain differential equation for a matrix
related to the Lagrangian.  The dho fits in this framework as its
Lagrangian contains a time dependent coefficient for the velocity which
can be interpreted as a time-dependent mass.

\medskip

A second approach uses an \textsf{interaction Hamiltonian} and applies
perturbation theory. This approach tries to understand damping as the
result of an interaction of two types of systems. One is a rather
simple system (the undamped ho) that we want to describe completely,
and the other is an extremely complicated system (the loss mechanism)
which we want to describe very incompletely. We are interested in the
loss mechanism only to the extent that it affects the behavior of the
oscillator and even in this respect the interest is limited: we
concern ourselves only with the lowest order interactions which
produce loss, since description of higher order interactions requires
a more detailed knowledge of the loss mechanism than that provided by
a single dissipation constant.  Underlying the usual concept of
dissipation is the understanding that the loss mechanism is affected
only slightly by its interaction with the oscillator, whereas the
oscillator may undergo large changes due to the loss mechanism. Thus
the loss mechanism may be treated by a perturbation approach, but the
oscillator may not.

\medskip

Common to the above two approaches is that the description of the
environment to which the oscillator looses its energy is restricted to a 
dissipation constant (and possibly temperature).

Approaches that construct an environment of the dho also exist. These
in fact close the system, i.e. make a realistic or artificial
embedding in a larger system that preserves energy. This way,
Hamiltonians that describe a total, conserved energy can be obtained.

An example of this line of thought is the so-called \textsf{doubling
the degrees of freedom} approach.  In fact also this idea can be
traced back to a Hamiltonian that was coined by Bateman, 
the so-called dual Hamiltonian (see Bateman). The idea is that the damped
oscillator is coupled to its time-reversed image oscillator which
absorbs the energy lost, so that the energy of the whole system is
conserved (i.e. the system is closed). In fact, since the phase space of the
whole system describes the dho and its image, the degrees of freedom
are effectively doubled. Another way of looking at this is that the
adding a time-reversed oscillator restores the breaking of the 
time-reversal symmetry.  

In the earlier attemps to elaborate this idea, difficulties arose such
as that the time evolution leads out of the Hilbert space of states,
but later a satisfactory quantization could be achieved within the
framework of quantum field theory (Feshbach \textsl{et al}, Celeghini
\textsl{et al}, Blasone \textsl{et al}).  A more recent quantization
based on the doubling the degrees of freedom approach was given by
Banerjee \textsl{et al}.  He obtained separate Lagrangians for the
oscillator and its time-reversed image, although these Lagrangians were
bound to have complex-valued parameters. In fact, these Lagrangians are
complex-conjugated. This implies that resulting Hamiltonians are
so-called "pseudo-hermitian", which nevertheless is sufficient to
diagonalize the individual Hamiltonians via a generalization of the
Dirac-Heisenberg treatment of the harmonic oscillator i.e. by forming
ladder operators, a number operator, so that all eigenstates of the
Hamiltonian can be obtained by applying a suitable ladder operator to
a ground state.

\medskip

The doubling of degrees of freedom approach has the conceptual
disadvantage that the environment to which the dho is coupled
(i.e. the image of the oscillator in which time is running backwards)
is rather artificial. More realistic environments are constructed when
forming \textsf{unitary dilations} such as in this thesis.  In fact, this
thesis presented a Bose quantization of finite number of oscillators
via an explicit dilation of contraction evolution $T_t$.

A Fermi quantization of a infinite dimensional phase space of the dho
(i.e.\ an infinite number of oscillators) is discussed in Latimer. The
latter can be regarded as a logical extension of the doubling
procedure.  In fact, the contraction evolution
$T_t$ of a single dissipative oscillator space can be "embedded" in a
orthogonal evolution on the doubled phase space (see Latimer) 
and by embedding this space in an even larger space (of
$L^2(\RR)$ type) a unitary dilation can be constructed.  This has an advantage
over the doubling the degrees of freedom approach in case of 
an infinite dimensional phase space. The doubling the degrees of freedom
procedures only closes the system at each point in time and, if the
phase space is infinite dimensional, at each point in time a rep of
the canonical anti-commutations relations must be made, 
while Latimer's method closes the system at once
for all future times. Therefore, only one representing Fock space is needed
and the vacuum vector is stationary.

\bigskip

\textbf{Literature}

H.~Bateman, Phys. Rev. 38 (1931) 815. 

R.~Banerjee and P.~Mukherjee, arXiv:quant-ph/0108055.

M.~Blasone, E.~Graziano, O.K.~Pashaev and G.~Vitiello,
Ann. Phys. (N.Y.) 252 (1996) 115.

P.~Caldirola, Nuovo Cim. 18 (1941) 393.

E.~Celeghini, M.~Rasetti and G.~Vitiello, Ann. Phys. (N.Y) 215 (1992) 156.

H.~Dekker, Phys Rep, 80 (1981) 1.

H.~Feshbach and Y.~Tikochinsky, N.Y. Acad. Sci. 38 (Ser.II) (1977) 44.

E.G.~ Harris, Phys.~Rev.~A, 42 (1990) 42.

M.-C.~Huang and M.-C.~Wu, Chinese journal of physics, 36 (1998) 566.

E.~Kanai, Progr. Theor. Phys 3 (1941) 440.

D.C.~Latimer, J Phys A: Math. Gen. 38 (2005) 2021.

I.R.~Senitzky, Phys.~Rev. 119 (1960) 670.

\part{Appendix}
\appendix 
\renewcommand{\chaptername}{Appendix}

\chapter{General}

\section{partial integration}
\begin{definition}[Indefinite integral] 

Suppose $ F $ and $ f  $ are functions whose domain contains  
   a (closed bounded)  interval $ [a,b] \subset \RR $. 

Then $ F $ is called an \Index[\bf]{indefinite integral} of  $ f $ 
on $ [a,b]$ iff 
\begin{displaymatH}
\int_x^y f(t)  d t  = F(y) - F(x)   
 \quad (x,y \in [a,b] \mathrm{with } x \leq y ) 
\end{displaymatH}
\end{definition}

\begin{lemma}[Partial Integration]

 Let $ f $ and $ g $  be integrable over a closed bounded interval 
 $ [a,b] \subset \RR$.
 Suppose $ F $ and $ G $ are indefinite integrals on $ [a,b] $ of 
 $ f $ and $ g $ respectively.
 Then:
 
\[ F(b)G(b) - F(a)G(a) = \int_a^b Fg + \int_a^b fG   \]

\end{lemma}

\begin{proof}
The proof is a straightforward application of Fubini's theorem. 

\begin{aligN}
\int_a^b Fg  
   \ats := \int ( Fg)_{[a,b]}(x) dx =
      = \int F_{[a,b]}(x) g_{[a,b]}(x) dx 
   \\
  \ats
 = \int (F(a) + \int f_{[a,x]}(t) dt) g_{[a,b]}(x) dx
 = F(a) ( G(b) - G(a) ) +  \int ( \int f_{[a,x]}(t) g_{[a,b]}(x) dt ) dx
\end{aligN}
 Now, use that $ f_{[a,x]}(t) g_{[a,b]}(x) = 
                 f_{[a,b]}(t) g_{[a,b]}(x) \indicator_{[a,x]}(t) =
                  f_{[a,b]}(t) g_{[a,b]}(x) \indicator_{[t,b]}(x)=   
                  f_{[a,b]}(t)  g_{[t,b]}(x)$
 and apply Fubini (allowed since 
 $ \int |f_{[a,b]}(t) g_{[a,b]}(x) \indicator_{[t,b]}(x)| dt dx \leq 
   \int |f_{[a,b]}(t) g_{[a,b]}(x)| dt dx = \int_a^b |f| \int_a^b |g| < \infty$
 )
 
 to obtain:
\begin{aligN}
\int_a^b Fg  
   \ats := \int ( Fg)_{[a,b]}(x) dx =
   = F(a) ( G(b) - G(a) ) +  \int f_{[a,b]}(t)(\int g_{[t,b]}(x) dx ) dt
   \\
  \ats
    = F(a) ( G(b) - G(a) ) +  \int f_{[a,b]}(t)( G(b) - G(t) ) dt
   \\
  \ats
   = F(a) ( G(b) - G(a) ) +  G(b) ( F(b) - F(a) ) -  \int f_{[a,b]}(t)G(t)dt
   \\
   \ats
    =:  F(b)G(b) - F(a)G(a) - \int_a^b fG
\end{aligN}
\end{proof}

\section{Integration formulae\label{integration_formulae}}
From contour integration, one obtains
\begin{displaymatH}\displaystyle
  \int_{-\infty}^{\infty} e^{-a x^2 + zx} dx = \sqrt{\frac{\pi}{a}} e^{z^2/4a}
  \quad
  (a\in (0,\infty) \;,\; z\in\CC )
  \,.
\end{displaymatH}

Using partial integration
\begin{aligN}
 \int x^2 e^{-ax^2} dx \ats = \int -(x/2a) d e^{-ax^2} 
  \\
  \ats =
 [-(x/2a) e^{-ax^2}] - \int - e^{-ax^2}/(2a)dx
\end{aligN}
or by differentiating the above with respect to $a$, one obtains
\begin{displaymatH}
  \int_{-\infty}^{\infty} x^2 e^{-a x^2} dx = \frac{\sqrt{\pi}}{2a\sqrt{a}}
  \quad
  (a\in (0,\infty)\,)
  \,.
\end{displaymatH}

\section{Joint centered Gaussian distributions\label{eq_expectation_iX+Y}}

Let $X$ and $Y$ be joint centered (i.e.\ with $\EE(X)=\EE(Y)=0$)
Gaussian random variables i.e.
\begin{aligN}
 X & \sim N(0,a^2)\,, \\
 Y & \sim N(0,b^2)\,, \\
 \text{Cov}(X,Y)&= \rho a b\,, 
\end{aligN}
where $\rho$ is the correlation between $X$ and $Y$, 
and $a$ and $b$ are the standard deviations of $X$ and $Y$ respectively.

The covariance matrix $V$ of $(X,Y)$ reads:
\begin{displaymatH}
 V=\left[
   \begin{array}{cc}
     a^2      & \rho a b \\
     \rho a b & b^2 \\
   \end{array}
   \right]
   \,,\;
   \text{ with } 
   \det( V )= a^2 b^2(1-\rho^2)
   \;.
\end{displaymatH}
If $\det(V) \neq 0$ (i.e.\ $\rho\neq \pm 1$), so that
\begin{displaymatH}\displaystyle
 V^{-1} =\frac{1}{\det(V)}
         \left[
         \begin{array}{cc}
          b^2       & -\rho a b \\
          -\rho a b & a^2 \\
   \end{array}
   \right]
   \,,
\end{displaymatH}
the joint probability density of $(X,Y)$ is given by 
\begin{equation}
\begin{aligned}
 p(x,y) & = \frac{1}{2\pi \sqrt{\det(V)}} 
          \exp\left[-\half (x,y) V^{-1} (x,y)^t\right]
        \\
        & = \frac{1}{2\pi \sqrt{\det(V)}} 
           \exp\left[
                    -\half
                    \frac{1}{\det(V)}
                    \bigl(b^2 x^2 + a^2 y^2 - 2 \rho a b xy \bigr)
               \right]
        \\
        & = \frac{1}{2\pi \sqrt{\det(V)}} 
           \exp\left[
                    -\bigl(\be x^2 + \al y^2 - 2 \gamma xy \bigr)
               \right]
         \,,       
\end{aligned}
\end{equation}
where 
\begin{displaymath}
 \al := \frac{a^2}{2\det(V)}\,,\quad
 \be := \frac{b^2}{2\det(V)}\,,\quad
 \gamma  := \frac{\rho a b }{2\det(V)}\,.
\end{displaymath}
Using that
\begin{equation}
\label{eq_Gaussian_integral}
 \int_{-\infty}^{\infty} e^{-r x^2 + z x} = \sqrt{\frac{\pi}{r}} e^{z^2/4r}
  \quad (r \in (0,\infty)\,,\; z\in \CC)
  \,,
\end{equation}
we calculate
\begin{displaymath}
\begin{aligned}
 \EE(e^{iX + Y}) 
  & =
    \int_{-\infty}^{\infty} \int_{-\infty}^{\infty}
      e^{ix + y}
      p(x,y)
      d\,x d\,y
 \\
  & = \frac{1}{2\pi \sqrt{\det(V)}} 
      \int_{-\infty}^{\infty} 
        e^y e^{-\al y^2}
          \left\{
               \int_{-\infty}^{\infty}
                    e^{-\be x^2  + (2\gamma y + i) x}
               d\,x
          \right\}
      d\,y
 \\
  & = \frac{1}{2\pi \sqrt{\det(V)}} \sqrt{\frac{\pi}{\be}} e^{-\frac{1}{4\be}}
      \cdot
      \int_{-\infty}^{\infty} 
         e^{-(\al- \gamma^2/\be) y^2}  e^{(1+ i\gamma/\be )y}
      d\,y  
 \\
  & = \frac{1}{2\pi \sqrt{\det(V)}} \sqrt{\frac{\pi}{\be}} e^{-\frac{1}{4\be}}
      \cdot     
      \sqrt{\frac{\pi}{\al-\gamma^2/\beta}} 
      e^{(1+ i\gamma/\be )^2/4(\al-\gamma^2/\beta)}
 \\
  & = \frac{1}{2 \sqrt{\det(V)}} \frac{1}{\sqrt{\al\be-\gamma^2}} 
      \cdot
      \exp\left[
                \frac{\be-\al + 2i\gamma}{4(\al\be-\gamma^2)}
          \right]
  \\
  & = \exp\left[
                \frac{\be-\al + 2i\gamma}{4(\al\be-\gamma^2)}
           \right]
  \\
  & = \exp\left[
                \frac{b^2-a^2 + i  (2\rho a b)}{2}
          \right]
  \quad (\rho \neq \pm 1)
  \,,
\end{aligned}
\end{displaymath}
where we have used in the fourth equality that $\al - \gamma^2/\be > 0$ 
(since this is equivalent with $1 > \rho^2$), 
and the one but last and last equality 
that $\al\be-\gamma^2= 1/(4\det(V))$.

Observing that 
$b^2-a^2 + i(2\rho a b)= \Var{Y} -\Var{X} + i (2\Cov{X}{Y})$, 
we can summarize this as
\begin{equation}
 \EE(e^{iX + Y}) 
   =  \exp\left[
                \frac{-\Var{X} + \Var{Y} + 2i\Cov{X}{Y}}{2}
          \right]
   \,.
\end{equation}

\section{Polarization formula\label{ap_PolarizationFormula}}
We want to show that the inner product is actually determined by the norm
and, more generally, that a linear operator on a domain is actually determined
by its ``matrix-elements'' on that domain.

\begin{lemma}[Polarization formula]
\label{lem_polarizationFormula}
Let  $\mathcal{D}$ be a linear space and 
let $B:\mathcal{D}\times \mathcal{D}\to \CC$ be a map such that
\begin{Properties}
 \item for each $\eta\in\mathcal{D}$, 
       the map $\xi \mapsto B(\xi,\eta)$ ($\xi\in\mathcal{D}$)
       is \textsl{complex}-linear,
 \item for each $\xi\in\mathcal{D}$, 
       the map $\eta \mapsto B(\xi,\eta)$ ($\eta\in\mathcal{D}$)
       is linear,
\end{Properties}
Then 
$B(\xi,\eta)=\frac{1}{4}\sum_{k=0}^3 (-i)^k B(\xi+ i^k \eta, \xi+ i^k \eta)$
for all $\xi,\eta \in\mathcal{D}$.
\end{lemma}

\begin{examples}
\begin{Itemize}{-}
 \item If $\ip{\cdot}{\cdot}$ is an inner product, then
       $\ip{\xi}{\eta}=\frac{1}{4} \sum_{k=0}^3 (-i)^k \norM{\xi+ i^k \eta}^2$.
 \item If $A$ is a linear map, then
      $\oip{\xi}{A}{\eta}
     = \frac{1}{4}\sum_{k=0}^3 (-i)^k \oip{\xi+ i^k \eta}{A}{\xi+ i^k \eta}$.
\end{Itemize}
\end{examples}

\begin{proof}
 Since $B$ is additive in both arguments, we have
 \begin{displaymatH}
  B(\xi + \eta,\xi + \eta) - B(\xi - \eta,\xi - \eta)
   = 2 B(\xi,\eta) + 2 B(\eta,\xi).
 \end{displaymatH}
 Likewise, but now using in addition that $B(\eta,i\xi)= iB(\eta,\xi)$,
 while $B(i\eta,\xi)= -iB(\eta,\xi)$
 \begin{displaymatH}
  B(\xi + i\eta,\xi + i\eta) - B(\xi - i\eta,\xi - i\eta)
   = 2 i B(\xi,\eta) - 2i B(\eta,\xi).
 \end{displaymatH}
 Therefore, 
 $B(\xi,\eta)=\frac{1}{4}\sum_{k=0}^3 (-i)^k B(\xi+ i^k \eta, \xi+ i^k \eta)$.
\end{proof}

\section{The operator expansion formula\label{formula_operator_expansion}}
Let $A$ be a self-adjoint and $B$ be a bounded operator. Then
\begin{aligN}
 \exp\ats (ix A)B \exp(-ix A) 
     \\
     & =
      B + ix[A,B] + \frac{(ix)^2}{2!}[A,[A,B]]
      + \ldots 
      + \frac{(ix)^n}{n!} 
         \underbrace{[A,[A,[A, \ldots [A}_{n \text{ times}},B
                     \underbrace{]\ldots ]]]}_{n \text{ times}} 
      + \ldots
\end{aligN}

Let $f(x):= \exp(ix A)B \exp(-ix A)$. Then 
\begin{displaymatH}
 f'(x)= \frac{df(x)}{dx}= iA \cdot \exp(ix A)B \exp(-ix A) 
                   + \exp(ix A)B \exp(-ix A) \cdot (-iA)
                 = [ iA, f(x)]
         .
\end{displaymatH}
From this, we derive next 
\begin{displaymatH}
 f''(x)= \frac{df'(x)}{dx}= \frac{d [ iA, f(x)]}{dx} 
          = [iA, \frac{df(x)}{dx}] = [iA, [ iA, f(x)]]
         ,
\end{displaymatH}
and, with induction,
\begin{displaymatH}
 f^{(n)}(x)=  [iA, \frac{df(x)}{dx}] 
           = \underbrace{[iA,[iA,[iA, \ldots [iA}_{n \text{ times}},f(x)
                    \underbrace{]\ldots ]]]}_{n \text{ times}} 
	   .
\end{displaymatH}
This means that in a (formal) Taylor series around $x=0$:
\begin{aligN}
 f(x)& = f(0) + f'(0) x  + f''(0) \frac{x^2}{2!} +  f^{(3)}(0)  \frac{x^3}{3!} 
      + \ldots
      \\
     & = B + ix [A,B] + \frac{(ix)^2}{2!}[A,[A,B]]
       + \frac{(ix)^3}{3!}[A,[A,[A,B]]]+ \ldots
\end{aligN}

\medskip

In particular, if $[A,B]\in\CC$, then $e^{iA} B e^{-iA}=B + i[A,B]$ or
$[e^{iA},B]=[A,B]e^{iA}$.

\section{The Campbell-Baker-Hausdorff lemma\label{BCH-formula}}
Let $A$ and $B$ be two self-adjoint operators 
whose commutator $[A,B]$ commutes with
both $A$ and $B$ (this is e.g.\ the case if $[A,B]$ is a scalar). Then
\begin{equatioN}
 \exp\left(ix[A+B]\right) =  \exp(ixA)\exp(ixB) e^{-\half x^2[iA,iB]}
  .
\end{equatioN}

Indeed, it suffices to show that
the right hand side satisfies the differential equation that
defines the left hand side (Stone's theorem). 
In other words, we need to check that
\begin{displaymatH}
 \frac{d}{dt}\left( e^{ixA}e^{ixB}e^{-\half x^2 [iA,iB]}\right)
 =
 (iA+iB) e^{ixA}e^{ixB}e^{-\half x^2 [iA,iB]}
 .
\end{displaymatH}

The latter is verified by direct computation.
By the product rule,
\begin{aligN}
 \frac{d}{dt}\left( e^{ixA}e^{ixB}e^{-\half x^2 [iA,iB]}\right)
  & =
 (iA) e^{ixA}e^{ixB}e^{-\half x^2 [iA,iB]}
  +
  e^{ixA}(iB) e^{ixB}e^{-\half x^2 [iA,iB]}
  \\
  & \qquad
  +
  e^{ixA} e^{ixB}(-x[iA,iB]) e^{-\half x^2 [iA,iB]}
 .
\end{aligN}
Since $[A,B]$ commutes with both $A,B$ and thus $e^{ixA}, e^{ixB}$,
the latter term is 
\begin{displaymatH}
(-x[iA,iB]) e^{ixA} e^{ixB} e^{-\half x^2 [iA,iB]}.
\end{displaymatH}
For the middle term use that the operator expansion theorem of above
yields that
\begin{displaymatH}
e^{ixA}(iB) e^{-ixA}= iB + ix[A,iB] + 0 + 0 + \ldots 
\end{displaymatH}
i.e.
\begin{displaymatH}
e^{ixA}(iB) = iB e^{ixA} + ix[A,iB]e^{ixA}.
\end{displaymatH}
Summarizing all terms then yields
$(iA + iB)e^{ixA}e^{ixB}e^{-\half x^2 [iA,iB]}$, which was what to be verified.

(Alternatively, let $g(x):=\exp(xA)\exp(xB)$, 
differentiate both sides with respect to $x$ and, 
using the operator expansion formula above, show that 
$\frac{dg(x)}{dx}=(A+B+ x[A,B])g(x)$. Integrate with respect to $x$ like
an ordinary differential equation.)

\section{Relation of the representation of $\ket{\psi}$ in position and momentum representation}

$P/\hbar$ generates translations in position, 
$-Q/\hbar$ translations in momentum (see e.g.\ \cite{Sakurai}).

\section{Coherent states are eigenstates of $A$\label{ap_alphaQ+iP}}
Recall that 
he operator $e^{-itP/\hbar}$ implements a position translation over $t$ to 
the left ($e^{-itP/\hbar} \psi(q) = \psi(q+t)$) and  
$e^{isQ/\hbar}$ implements a momentum translation over $s$ to the left:
$e^{isQ/\hbar}\psi(p) = \psi(p+s)$.

In order to show that 
\begin{displaymatH}
 \Bigl(\al(Q-t) + i(P-s)\Bigr)\ket{\psi}  = 0 
 \Implies
 [\al Q + iP ] \underbrace{e^{-isQ+itP} \ket{\psi}}_{\ket{\psi_0}} = 0
\end{displaymatH}
and, vice versa, that
\begin{displaymatH}
 [\al Q + iP ] \ket{\psi_0} = 0
 \Implies
 \Bigl(\al(Q-t) + i(P-s)\Bigr)\underbrace{e^{isQ-itP}\ket{\psi_0}}_{\ket{\psi}}  = 0 
\end{displaymatH}
it suffices to show (for the first statement) that 
\begin{aligN}
 \,\ats  \ats [\al Q + iP ] e^{-isQ/\hbar +itP/\hbar} 
   = e^{-isQ/\hbar +itP/\hbar} \Bigl(\al(Q-t) + i(P-s)\Bigr) 
  \\
  \ats \Leftrightarrow \ats
  e^{-(-isQ+itP)/\hbar} [\al Q + iP ] e^{(-isQ+itP)/\hbar} 
   = \al(Q-t) + i(P-s) 
  \,,
\end{aligN}
which can be verified by applying the operator equality $e^A B e^{-A} = B + [A,B]$ that holds for $A,B$ with $[A,B]\in\CC$:
\begin{displaymatH}
 [(isQ -itP)/\hbar, \al Q + iP] 
  = is\al [Q,P]/\hbar   + (-it)i[P,Q]/\hbar = -\al s + -it
\end{displaymatH}

\section{Inner product of coherent vectors\label{ap_ipCohPosRep}}

\begin{align*}
 \langle \Coh(w) & | \Coh(z) \rangle
   =
  \int dq \ip{\Coh(w)}{q} \ip{q}{\Coh(z)}
   \\
 & =
  \int 
  \overline{
  \biggl(\frac{e^{-\half(|w|^2 - w^2)}}{\sqrt[4]{\pi\hbar}} 
       \; e^{-(q/\sqrt{2\hbar}-w)^2}
  \biggr)
  }
  \cdot
    \frac{e^{-\half(|z|^2 - z^2)}}{\sqrt[4]{\pi\hbar}} 
    \; e^{-(q/\sqrt{2\hbar}-z)^2}
   dq
 \\
 & =
 \frac{e^{-\half(|\overline{w}|^2 - \overline{w}^2)}
       e^{-\half(|z|^2 - z^2)}}{\sqrt{\pi\hbar}} 
 \cdot
 \int 
   e^{-(q/\sqrt{2\hbar}-\overline{w})^2}
   e^{-(q/\sqrt{2\hbar}-z)^2}
  dq
 \\
 & =
 \frac{e^{-\half(|\overline{w}|^2 - \overline{w}^2)}
       e^{-\half(|z|^2 - z^2)}}{\sqrt{\pi\hbar}} 
 \cdot
 \int e^{-\frac{q^2}{\hbar} 
         + \frac{2q(z+\overline{w})}{\sqrt{2\hbar}}
         -(z^2 +\overline{w}^2)}
  dq
 \\
  & =
 \frac{e^{-\half(|\overline{w}|^2 - \overline{w}^2)}
       e^{-\half(|z|^2 - z^2)}}{\sqrt{\pi\hbar}} 
 \cdot
 e^{-(z^2 +\overline{w}^2)}
 \sqrt{\frac{\pi}{1/\hbar}}
 \exp\biggl( \Bigl(\textstyle\frac{2(z+\overline{w})}{\sqrt{2\hbar}}\Bigr)^2 / (4/\hbar) 
     \biggr)
 \\
 & =
 e^{-\half(|\overline{w}|^2 - \overline{w}^2)}
 e^{-\half(|z|^2 - z^2)}
 \cdot e^{-(z^2 +\overline{w}^2)}
 e^{\half(z+\overline{w})^2} 
 \\
 & =
 e^{-\half(|\overline{w}|^2 - \overline{w}^2)}
 e^{-\half(|z|^2 - z^2)}
 \cdot 
 e^{-\half z^2} e^{-\half\overline{w}^2)}
 e^{z\overline{w}}
 \\
 & = 
 e^{-\half|w|^2} e^{-\half|z|^2} e^{z\overline{w}}
\end{align*}

\section{Coherent states in the Schr\"odinger representation}
We determine the form of coherent states in Schr\"odinger representation
by solving the eigenvalue equation $A\ket{\psi}= z\ket{\psi}$ in 
position representation and then evaluate the action of the
Weyl operators on them.
This action takes a particular simple form if
$\hbar/2=m\om=1$ i.e.\ $\sqrt{m\om/2\hbar}=1/2$, 
see (\ref{eq_WeylOpOnCohSchro}), 
(\ref{eq_z_sup_bullet}), (\ref{eq_VacExpWeylSchr}), . 

\subsection{Coherent vectors in position representation}

Since
\begin{displaymath}
 A = \sqrt{\frac{m\om}{2\hbar}} \Bigl(Q+\frac{iP}{m\om}\Bigr)
   = \sqrt{\frac{m\om}{2\hbar}} Q +  \frac{1}{\sqrt{2\hbar m\om}}i P
 ,
\end{displaymath}
the differential equation for $\psi(q)=\ip{q}{\psi}$ in position representation
reads:
\begin{displaymatH}
 \al q \psi(q) + \frac{1}{2\al}\psi'(q) = z\psi(q)
 \;\;\text{ with }
 \al:=\sqrt{\frac{m\om}{2\hbar}}
  ,
\end{displaymatH}
i.e.\
\begin{displaymatH}
 \psi'(q) = - (2\al^2 q -2\al z) \psi(q)
 .
\end{displaymatH}
Integrating this equation, we obtain
\begin{displaymatH}
 \psi(q) =  N \exp\Bigl( -(\al^2 q^2 - 2\al z q)\Bigr)
  \quad
  (q\in\RR)
  ,
\end{displaymatH}
where the normalization constant $N$ satisfies
\begin{aligN}
 |N|^{-2} & = \int | \exp\Bigl( -(\al^2 q^2 - 2\al z q)\Bigr) |^2 dq
            = \int  \exp\; \Re \Bigl( -(2\al^2 q^2 - 4\al z q)\Bigr) dq 
          \\
          & = \int  \exp \Bigl( -(2\al^2 q^2 - 4\al \Re(z) q)\Bigr) dq 
            = \sqrt{\frac{\pi}{2\al^2}}
              \exp\Bigl( \frac{[-4\al \Re(z)]^2}{4[2\al^2]}\Bigr) 
          \\
          & = \sqrt{\frac{\pi}{2\al^2}} \exp\Bigl( 2\Re(z)^2\Bigr)
          .            
\end{aligN}
For the convenience of $N$ taking a a simple form,
we admit a phase factor $\exp\Bigl(i\Im(z)\Re(z)\Bigr)$, so that 
$N=\sqrt[4]{\frac{2\al^2}{\pi}} \exp\Bigl(- \half(  |z|^2+ z^2)\Bigr)$.

All in all, we then have
\begin{equation}
\label{eq_CohVecSchroRep}
 \Coh(z) 
  =
  \sqrt[4]{\frac{2\al^2}{\pi}} \exp\Bigl(- \half(  |z|^2+ z^2)\Bigr)
  \exp\Bigl( -(\al^2 q^2 - 2\al z q)\Bigr)
  .
\end{equation}

\subsection{The action of $W(z)$ on the vacuum state}

From (\ref{eq_CohVecSchroRep}),
\begin{displaymatH}
 \Coh(0)
   =
  \sqrt[4]{\frac{2\al^2}{\pi}} \exp\Bigl( -\al^2 q^2\Bigr)
   ,
\end{displaymatH}
and from (\ref{eq_WeylOpSchroRep}), 
\begin{displaymatH}
 W(z) = e^{-\frac{i\hbar}{2}st}  S_s T_t
\end{displaymatH}
so that 
\begin{aligN}
 W(z)\Coh(0) \ats  = \exp(-\frac{i\hbar}{2}st)  S_s T_t 
               \biggl( 
               \sqrt[4]{\frac{2\al^2}{\pi}} \cdot \exp\Bigl( -\al^2 q^2\Bigr)
               \biggr)
             \\
             \ats = \sqrt[4]{\frac{2\al^2}{\pi}}
                     \cdot
                  \exp(-\frac{i\hbar}{2}st)\exp(isq)  
                  \exp\Bigl( -\al^2 (q-t\hbar)^2\Bigr)
             \\
             \ats = \sqrt[4]{\frac{2\al^2}{\pi}}
                     \cdot
                  \exp\Bigl(
                       -\frac{i\hbar}{2}st + isq +  -\al^2 q^2 
                       + 2\al^2 t\hbar q  - \al^2 t^2 \hbar^2
                      \Bigr) 
             .
\end{aligN}
On the other hand, 
if we define 
\begin{equation}
\label{eq_z_sub_bullet}
 z_{\bullet}:= \sqrt{\frac{\hbar}{2}}
               \Bigl( t \sqrt{m\om}  +  \frac{is}{\sqrt{m\om}}\Bigr)    
             = t \al\hbar +  \frac{is}{2\al} 
  \quad
 (z=t + is \in\CC)
 ,
\end{equation}
then from (\ref{eq_CohVecSchroRep})
\begin{aligN}
 \Coh( z_{\bullet}) & 
   =  \bigl[ \Coh(t+is)  \bigr]_{t\mapsto t\al\hbar, s\mapsto s/(2\al)} 
  \\
  & =
 \Biggl[
     \sqrt[4]{\frac{2\al^2}{\pi}} \exp\Bigl(- ( t^2 + ist)\Bigr)
    \exp\Bigl( -\al^2 q^2 + 2\al t q + 2i\al s q)\Bigr)
 \Biggr]_{t\mapsto t\al\hbar, s\mapsto s/(2\al)}
  \\
  &  =
 \Biggl[
    \sqrt[4]{\frac{2\al^2}{\pi}} 
    \exp\Bigl(- ( t^2\al^2\hbar^2 + is\frac{\hbar}{2})\Bigr)
    \exp\Bigl( -\al^2 q^2 + 2\al^2\hbar t q + i s q)\Bigr)
 \Biggr]
 . 
\end{aligN}
Concluding:
\begin{displaymath}
  W(z)\Coh(0)  = \Coh(z_\bullet)
  .
\end{displaymath}

\subsection{The action of $W(z)$ on coherent vectors}

By the Weyl relation $W(u)W(z)= e^{-i\frac{\hbar}{2}\Im(\overline{u}z)}W(u+z)$,
the action of $W(u)$ on $\Coh(z)$ can be evaluated as
\begin{equation}
\label{eq_WeylOpOnCohSchro}
\begin{aligned}
 W(u)\Coh(z)
   & = W(u)W(z^\bullet)\Coh(0) 
     = e^{-i\frac{\hbar}{2}\Im(\overline{u}z^\bullet)}W(u+z^\bullet)\Coh(0)
   \\
   &
     = e^{-i\frac{\hbar}{2}\Im(\overline{u}z^\bullet)}\Coh(u_\bullet +z)
     .
\end{aligned}
\end{equation}
where 
\begin{equation}
\label{eq_z_sup_bullet}
  z^\bullet
  = \frac{t}{\al\hbar} + i 2 \al s
  = \sqrt{\frac{2}{\hbar}}\Bigl( \frac{t}{\sqrt{m\om}} + is\sqrt{m\om}\Bigr)
 \quad(z=t+is\in\CC)
\end{equation}
i.e.\
$z\mapsto z_\bullet$ and $z\mapsto z^\bullet$ are each others inverses.

Concluding, the Weyl operators reshuffle the coherent vectors and add 
a phase factor. A particular simple form arises if $m\om=\hbar/2=1$:
$W(u)\Coh(z)=e^{-i\frac{\hbar}{2}\Im(\overline{u}z)}\Coh(u+z)$.

\subsection{Coherent vector algebra}
The vacuum state expectation of the Weyl operators are given by
\begin{equation}
\label{eq_VacExpWeylSchr}
\begin{aligned}
 \langle  \Coh(0)|& W(z) | \Coh(0)\rangle 
    =
 \ip{\Coh(0)}{\Coh(z_\bullet)}
   \\
  & =
  \int \Bigl(
       \overline{  {\textstyle\sqrt[4]{\frac{2\al^2}{\pi}}} e^{-\al^2 q^2}}
       \Bigr)
       \Bigl(
       {\textstyle\sqrt[4]{\frac{2\al^2}{\pi}}} 
       e^{- \half(  |z_\bullet|^2+ z_\bullet^2)}
       e^{-(\al^2 q^2 - 2\al z_\bullet q)}
       \Bigr)
   dq
  \\
  & =
  {\textstyle\sqrt{\frac{2\al^2}{\pi}}}
  e^{- \half(  |z_\bullet|^2+ z_\bullet^2)}        
  \int
       e^{-(2\al^2 q^2 - 2\al z_\bullet q)}
  dq
  \\
  &
   = 
  {\textstyle\sqrt{\frac{2\al^2}{\pi}}}
  e^{- \half(  |z_\bullet|^2+ z_\bullet^2)}        
  \cdot
  {\textstyle\sqrt{\frac{\pi}{2\al^2}}}
  \exp\Bigl( (2\al z_\bullet)^2 / (4\cdot 2\al^2) \Bigr)
  \\
  & 
  = \exp(-\half|z_\bullet|^2)
  \quad
  (z\in\CC),
\end{aligned}
\end{equation}
which enables us to calculate the inner product between two coherent vectors
as
\begin{equation}
\label{eq_InProdCohVecSchr}
\begin{aligned}
 \langle & \Coh(u_\bullet) |\Coh(z_\bullet)\rangle
     =
  \ip{W(u)\Coh(0)}{W(z)\Coh(0)}S
   \\
   &  =
  \oip{\Coh(0)}{\adj{W(u)}W(z)}{\Coh(0)}
    =
  \oip{\Coh(0)}{W(-u)W(z)}{\Coh(0)}  
   \\
   & =
   e^{-i\frac{\hbar}{2}\Im(-\overline{u}z)}
  \oip{\Coh(0)}{W(z-u)}{\Coh(0)}
   =  
   e^{i\frac{\hbar}{2}\Im(\overline{u}z)}
   e^{-\half|z_\bullet-u_\bullet|^2}
  \\
  &=
   e^{i\frac{\hbar}{2}\Im(\overline{u}z)}
   e^{-\half|z_\bullet|^2 - |u_\bullet|^2 
      + \Re(\overline{u}_\bullet z_\bullet)}
  =
   e^{-\half|z_\bullet|^2} e^{ - |u_\bullet|^2} 
   e^{\:\overline{u}_\bullet z_\bullet}
  ,
\end{aligned}
\end{equation}
since if we write $u=y+ix$ and $z=t+is$, then
\begin{aligN}
 \overline{u}_\bullet z_\bullet
 \ats 
  = \sqrt{\frac{\hbar}{2}} \Bigl(y\sqrt{m\om} - \frac{ix}{\sqrt{m\om}}\Bigr)
    \sqrt{\frac{\hbar}{2}} \Bigl(t\sqrt{m\om} + \frac{is}{\sqrt{m\om}}\Bigr)
  = \frac{\hbar}{2}
    \Bigl( [yt (m\om) + \frac{xs}{m\om}] + i[ys - tx]\Bigr)
  ,
\end{aligN}
i.e.\ 
$\frac{\hbar}{2}\Im(\overline{u}z) = \Im( \overline{u}_\bullet z_\bullet)$.

\section{Exponential vectors}

Exponential vectors are no more or less than ``unnormalized'' coherent vectors
i.e.\
\begin{equation}
\label{eq_ExpVecSchr}
\begin{aligned}
\varepsilon(z)
  &
  = e^{\half|z|^2}\Coh(z)
  \overset{(\ref{eq_CohVecSchroRep})}{=}
  \sqrt[4]{\frac{2\al^2}{\pi}} \exp(-\half z^2)
  \exp\Bigl( -(\al^2 q^2 - 2\al z q)\Bigr)
  \\
  & =
 \sqrt[4]{\frac{2\al^2}{\pi}} 
  \exp\Bigl(2\al z q -\half z^2 -\al^2 q^2)\Bigr)
  .
\end{aligned}
\end{equation}

From the above, the inner product between two exponential vectors is then
$\ip{\varepsilon(u)}{\varepsilon(z)}=e^{\overline{u}z}$.

In particular, if $m\om$ and $\hbar/2=1$ are taken to be $1$ 
(so that $\al=\half$), then
\begin{displaymatH}
 \varepsilon(z)=  \frac{1}{\sqrt[4]{2\pi}}
  \exp\Bigl(z q -\half z^2 -\frac{1}{4} q^2)\Bigr).
\end{displaymatH}

\subsection{The coherent (exponential) vector are total in 
                $\mathcal{H}_{\text{Schr\"o}}=L^2(\RR)$}

If $f\in L^2(\RR)$ satisfies $\ip{\varepsilon(z)}{f}=0 $ for all $z\in\CC$, 
then 
\begin{aligN}
  0 \ats = \ip{\varepsilon(-\frac{it}{2\al})}{f}
      = \sqrt[4]{\frac{2\al^2}{\pi}} 
        e^{\frac{t^2}{8\al^2}}    
         {\displaystyle\int_{-\infty}^{\infty}}  
	   \Bigl(e^{-\al^2 q^2} f(q)\Bigr)
           e^{itq} 
         d q
  \quad(t\in\RR).
\end{aligN}
In particular the Fourier transform of 
$q\mapsto e^{-\al^2 q^2} f(q)$ vanishes and, by the injectivity of the
Fourier-Plancherel transform on $L^2(\RR)$, therefore 
\begin{displaymatH}
 e^{-\al^2 q^2} f(q) = 0 \;\In L^2(\RR) \text{ i.e. } f=0 \;\In L^2(\RR)\,.
\end{displaymatH}

\section{Coherent states in the Heisenberg representation}
To calculate the action of $W(z)$ on coherent vectors,
we write the coherent vector $\Coh(u)$ in terms of $\exp(u\adj{A})$ 
and we write $W(z)$ in terms of $A$ and $\adj{A}$.

\statement{Step 1: Writing $W(z)=W(t+is)=\exp(isQ-itP)$ 
           in terms of $A$ and $\adj{A}$}

Using the relations (\ref{def_LadderOp}) between $A$ and $\adj{A}$ 
with $P$ and $Q$,
\begin{displaymath}
 Q= \sqrt{\frac{\hbar}{2m\om}} ( A+ \adj{A}) 
  = \frac{1}{2\al}( A+ \adj{A}) 
 \;\;\text{ and }\;
 P= \sqrt{\frac{\hbar m \om}{2}} ( A - \adj{A})/i 
   = \al\hbar ( A - \adj{A})/i
   ,
\end{displaymath}
we have that
\begin{displaymatH}
 isQ - itP 
 = \frac{is}{2\al}( A+ \adj{A})  - i t  \al\hbar ( A - \adj{A})/i
 = z_\bullet \adj{A} -  \overline{z}_\bullet A 
\end{displaymatH}
with $z_\bullet = t \al\hbar +  \frac{is}{2\al}$ as above.

Therefore,
\begin{equation}
 W(z) = \exp\Bigl( z_\bullet \adj{A} - \overline{z}_\bullet A \Bigr)
 \quad 
 (\textstyle z_{\bullet}:= \sqrt{\frac{\hbar}{2}}
               \Bigl( t \sqrt{m\om}  +  \frac{is}{\sqrt{m\om}}\Bigr)    
             = t \al\hbar +  \frac{is}{2\al}) 
\end{equation}

\statement{Step 2: Coherent vectors}

From the relation $\adj{A}\ket{n}=\sqrt{n+1}\ket{n+1}$, 
it follows that 
\begin{displaymath}
 \varepsilon(z):=
  \exp(u\adj{A})\ket{0}= \sum_{n=0}^\infty \frac{u^n}{\sqrt{n!}}\ket{n}
   .
\end{displaymath}
The ``exponential vectors'' $\varepsilon(z)$ are closely related to
the coherent vectors $\Coh(z)$, since $\Coh(z)=e^{-|z|/2}\varepsilon(z)$.

\statement{Step 3: The action of $W(z)$ on the vacuum state $\ket{0}$}

If $A,B$ are  operators for which $[A,B]$ commutes with $A$ and with $B$,
then the Baker-Campbell-Hausdorff formula $e^A e^B = e^{A+B} e^{\half[A,B]}$
holds. Using that, 
we derive
\begin{aligN}
 W(z) \exp(u\adj{A}) 
  & =
 \exp\Bigl( z_\bullet \adj{A} - \overline{z}_\bullet A \Bigr) \exp(u\adj{A}) 
   =
 e^{-\overline{z}_\bullet u)/2}
 \exp\Bigl( (z_\bullet+ u) \adj{A} - \overline{z}_\bullet A \Bigr)
   \\
   & =
 e^{-\overline{z}_\bullet u)/2}e^{-(z_\bullet + u)\overline{z}_\bullet}
 \exp\Bigl( (z_\bullet+ u) \adj{A} \Bigr) 
  \exp \Bigl(- \overline{z}_\bullet A \Bigr)
   \\
   &  
  =
 e^{-\overline{z}_\bullet u - \half|z_\bullet|^2}
 \exp\Bigl( (z_\bullet+ u) \adj{A} \Bigr) 
  \exp \Bigl(- \overline{z}_\bullet A \Bigr)
 .  
\end{aligN}
Since $  \exp \Bigl(- \overline{z}_\bullet A \Bigr) 
         = 1 + \sum_1^\infty (- \overline{z}_\bullet)^n A^n )$
and $A$ (and its powers) annihilate $\ket{0}$,
we have $\exp \Bigl(- \overline{z}_\bullet A \Bigr)\ket{0} = \ket{0}$.
Using the Baker-Campbell-Hausdorff lemma,
\begin{aligN}
 W(z) \varepsilon(u)
  & =
  \exp\Bigl( z_\bullet \adj{A} - \overline{z}_\bullet A \Bigr)
  \exp(u\adj{A})\ket{0}
   =  
  e^{-\overline{z}_\bullet u - \half|z_\bullet|^2}
 \exp\Bigl( (z_\bullet+ u) \adj{A} \Bigr) 
  \exp \Bigl(- \overline{z}_\bullet A \Bigr)
 \ket{0}
  \\
  & =
  e^{-\overline{z}_\bullet u - \half |z_\bullet|^2}\varepsilon(z_\bullet+ u)
  .
\end{aligN}
Therefore,
\begin{aligN}
 W(z) \Coh(0) &= W(z)\varepsilon(0)
  =
 e^{-\half |z|^2}\varepsilon(z_\bullet)
  = \Coh(z_\bullet)
  .
\end{aligN}

\statement{Step 4: The action of $W(z)$ on coherent vectors}

Using the Weyl relation, it follows as in (\ref{eq_WeylOpOnCohSchro}) that
\begin{displaymatH}
 W(u)\Coh(z) 
  = 
 e^{-i\frac{\hbar}{2}\Im(\overline{u}z)}
 \Coh(u_\bullet + z)
  .
\end{displaymatH}
The vacuum state expectation of the Weyl operators can be calculated as
\begin{equation}
\label{eq_VacExpWeylHeis}
\begin{aligned}
 \langle  \Coh(0)|& W(z) | \Coh(0)\rangle 
    =
 \ip{\Coh(0)}{\Coh(z_\bullet)}
   =
 \ip{\varepsilon(0)}{e^{-\half|z_\bullet|^2}\varepsilon(z_\bullet)}
   \\
  & =
  e^{-\half|z_\bullet|^2}
  \sum_{n=0}^\infty \frac{(z_\bullet)^n}{\sqrt{n!}} 
  \underbrace{\ip{0}{n}}_{\de_{0,n}}
  \\
  &
  =
  e^{-\half|z_\bullet|^2} 
  \qquad (z\in\CC),
\end{aligned}
\end{equation}
and as in (\ref{eq_InProdCohVecSchr}),
the inner product between two coherent vectors is then
verified to be
\begin{displaymatH}
 \langle \Coh(u_\bullet) |\Coh(z_\bullet)\rangle
  =
   e^{-\half|z_\bullet|^2} e^{ - |u_\bullet|^2} 
   e^{\:\overline{u}_\bullet z_\bullet}
  .  
\end{displaymatH}

\subsection{Coherent and exponential vectors 
               are total in $\mathcal{H}_{\text{Heis}}=\ell^2(\NN)$
               \label{ap_ExpVecHeisRep}}
As expected by formal differentiation, we have
\begin{displaymatH}\displaystyle
 \Bigl[ \frac{d^k}{d t^k}  \varepsilon(t) \Bigr]_{t=0}
  =
 \frac{d^k}{d t^k}\Bigr |_{t=0}
   \Bigl(\sum_{n=0}^{\infty} \frac{t^n}{\sqrt{n}} \ket{n}\Bigr)
  = \sqrt{n!}\ket{n}.
\end{displaymatH}
and this can be verified as follows:  inductively define for $t\in\RR$
\begin{aligN}
  \varepsilon^{(0)}(t) 
   & :=\varepsilon(t) 
      = \sum_{n=0}^\infty \frac{t^n}{\sqrt{n!}}\ket{n}
      = (1, \frac{t}{\sqrt{1!}}, \frac{t^2}{\sqrt{2!}}, \frac{t^3}{\sqrt{3!}}, \ldots)
   ,
   \\
   \varepsilon^{(n+1)}(t) 
   &
   :=\lim_{s\to t} \frac{\varepsilon^{(n)}(s) - \varepsilon^{(n)}(t)}{s-t} 
   .
\end{aligN}
and prove by induction that
$\norM[text]{\varepsilon^{(n)}(t) - \sqrt{n!}\ket{n}}^2 \to 0 $ if $t\to 0$
for all $n\in\NN$.



\section{$L^2(\RR)^{\otimes_s^n} \isom L^2_{\text{symm}}(\RR^n)$
            \label{ap_SymTensorDenseInSymmL2}}

(See \cite{Prugovecki})
We treat the case $n=2$; 
the case $n>2$ is a straightforward generalization thereof.

\statement{Step 1: $L^2(\RR) \otimes L^2(\RR) \isom L^2(\RR^2)$}

Let $\Phi: L^2(\RR) \otimes L^2(\RR) \to L^2(\RR^2)$ be the linear map that
sends $f\otimes g$ to the function $(t_1,t_2) \mapsto f(t_1)\cdot g(t_2)$.
Let $\mathcal{H}_0$ be the linear subspace of $L^2(\RR)$ consisting
of the linear combinations of indicators of measurable sets, 
\begin{displaymatH}
 \mathcal{H}_0
     = 
  \{ \sum_1^n \al_i \indicator_{B_i} 
     : n\in\NN,\: \al_i\in\CC, \: B_i \text{ measurable }
  \}.
\end{displaymatH}
Then $\mathcal{H}_0$ is dense in $L^2(\RR)$, and as consequence
$\mathcal{H}_0 \otimes \mathcal{H}_0$ is dense in $L^2(\RR)\otimes L^2(\RR)$
while its range under $\Phi$ is
\begin{displaymatH}
 \mathcal{H}_1 
  :=
   \{ \sum_1^N \al_i \indicator_{A_i\times B_i } 
     : N\in\NN,\: \al_i\in\CC, \: A_i,\, B_i \text{ measurable }
  \}
\end{displaymatH}
which is dense in $L^2(\RR^2)$.

Moreover, $\Phi$ is isometric as map 
from $\mathcal{H}_0 \otimes \mathcal{H}_0$ onto
$\mathcal{H}_1$, since
\begin{displaymatH}
 \norM{\Phi( \indicator_{B_1} \otimes \indicator_{B_2})}^2
   =
 \int_0^\infty \int_0^\infty 
    \indicator_{B_1}^2 (t_1) \cdot \indicator_{B_2}^2(t_2) dt_1 dt_2
  =
 \lambda(B_1)\cdot \lambda(B_2)
  =
 \norM{\indicator_{B_1}}^2 \norM{\indicator_{B_2}}^2
  .
\end{displaymatH}

Therefore, $\Phi$ extends to an isometric map from 
$L^2(\RR)\otimes L^2(\RR)$ onto $L^2(\RR^2)$.

\statement{Step 2: $L^2(\RR) \otimes_s L^2(\RR) 
                     \isom L^2_{\text{symm}}(\RR^2)$}

The map $P: L^2(\RR^2) \to L^2_{\text{symm}}(\RR^2)$ defined by
\begin{displaymatH}
 (P F)(t_1,t_2):=\frac{F(t_1,t_2) + F(t_2,t_1)}{2}
  \quad
  (F\in L^2(\RR^2)\,,\:(t_1,t_2)\in\RR^2)
   , 
\end{displaymatH}
is surjective i.e.\ every symmetrical $L^2$-function on $\RR^2$ can be obtained
in this way.

Concluding,
$L^2(\RR)^{\otimes_s^2} \isom L^2_{\text{symm}}(\RR^2)$.

\section{Positive definite kernels \label{ap_PosDefKernel}}

The verification that the (infinite) sum of positive definite kernels
is a positive definite kernel is straightforward from the definition.

We now show that the point wise product of two positive definite kernels
is positive definite.

\medskip

Let $K_1$ and $K_2$ be positive definite kernels on $S$, take $n\in\NN$,
$s_1, \ldots, s_n \in S$, and $\la_1,\ldots, \la_n\in\CC$. 
We need to show that 
$\sum_{i,j} \overline{\la_j} K_1(s_j,s_i)K_2(s_j,s_i)\la_i \geq 0$.

First, observe that the $n\times n$ matrix with elements 
$A_{ji}:=K_1(s_j,s_i)$ is positive
definite, since $K_1$ is positive definite ($\la^{t}A \la \geq 0$ for all
$\la\in\CC^n$). As a consequence, there exists a matrix $B$ such that
$A=B^2$ and $B^{t}=\overline{B}$. In particular,
$A_{ji}=\sum_{k=1}^n B_{jk}B_{ki}=\sum_{k=1}^n \overline{B_{kj}} B_{ki}$.

Therefore,
\begin{align*}
 \sum_{i,j}  \overline{\la_j} K_1(s_j,s_i)K_2(s_j,s_i)\la_i
   & 
   =
 \sum_{i,j}  \overline{\la_j} A_{ji} K_2(s_j,s_i)\la_i
   =
 \sum_{i,j,k}  (\overline{\la_j B_{kj}})\cdot  K_2(s_j,s_i) \cdot (B_{ki}\la_i)
   \\
    &
   = \sum_k \Bigl\{ 
           \overline{\la_j^{(k)}} K_2(s_j,s_i) \la_i^{(k)} 
           \Bigr\}
   \geq
   0
   ,
\end{align*}
since $K_2$ is a positive definite kernel, and
$(\la_i^{(k)}:= B_{ki}\la_i)_{i=1}^n$ are $n$ sequences of scalars.

\section{Projection valued measures\label{ap_PVM}}

\paragraph{$\mathcal{H}_P$ is a dense subset }
Given a projection-valued measure $P$ 
on $\mathcal{H}$ and $\eta\in\mathcal{H}$,
we have that $\eta_n:=P(-n,n)\eta $ converges to $\eta$ 
since $\union_n (-n,n) = \RR$ and $P(\RR)=I$.
Furthermore, for fixed $n$ and Borel set $B$:
\begin{aligN}
 P_{\eta_n}(B) & =  \oip{\eta_n}{P(B)}{\eta_n} 
     = \oip{P(-n,n)\eta}{P(B)}{P(-n,n)\eta} 
     = \oip{\eta}{P(-n,n)P(B)P(-n,n)}{\eta} \\
   & = \oip{\eta}{P(B \intersection (-n,n)}{\eta} 
     = P_\eta \mid_{(-n,n)}(B)
   .
\end{aligN}
This means that 
\begin{aligN}
 \int_\RR |x|^2 P_{\eta_n}(dx) 
   & = \int_{-n}^n |x|^2 P_{\eta}(dx) \leq n^2 P_{\eta}( (-n,n)) 
     =  n^2 \oip{\eta}{P(-n,n)}{\eta} \leq n^2 \norM{\eta}^2
   < \infty.
\end{aligN}
Hence 
$\mathcal{H}_P
 := \{ \xi \in \mathcal{H} :  
      \int_\RR |x|^2 \oip{\xi}{P(dx)}{\xi} < \infty 
    \}$ 
is dense.

\paragraph{$\mathcal{H}_P$ is a linear subspace}
Furthermore, for fixed $B$, $P:=P(B)$ is a projection and  
the map $N(\xi,\eta):= \oip{\xi}{P}{\eta}$ is an inner product 
except for the fact that $N(\xi,\xi)=0$ implies that $\xi\in P\mathcal{H}$
(i.e. the null space of $P$). 
This however is still enough to show with the standard argument that 
$N$ satisfies the Schwartz inequality: 
$|N(\xi,\eta)| \leq \sqrt{N(\xi,\xi)}\sqrt{N(\eta,\eta)}$.
In other words, 
$|\oip{\xi}{P}{\eta}| \leq \sqrt{\oip{\xi}{P}{\xi}}\sqrt{\oip{\eta}{P}{\eta}}$.
The product of the latter is of course less than the product of the largest
i.e.
\begin{displaymatH}
|\oip{\xi}{P}{\eta}| 
   \leq 
 \max\{\oip{\xi}{P}{\xi}, \oip{\eta}{P}{\eta} \}
   \leq 
 \oip{\xi}{P}{\xi} + \oip{\eta}{P}{\eta}
  .
\end{displaymatH}
As a result, we have that 
\begin{displaymatH}
 \oip{\xi+\eta}{P(B)}{\xi+\eta} 
    \leq 
 2 \Bigl( \oip{\xi}{P(B)}{\xi} + \oip{\eta}{P(B)}{\eta} \Bigr)
  \quad (B\in\Borel(\RR)).
\end{displaymatH}
so if $\xi,\eta\in\mathcal{H}_P$ then $\xi +\eta\in\mathcal{H}_P$ and
$\mathcal{H}_P$  is a linear subspace.

\paragraph{The formula $\oip{\xi}{O}{\xi}= \int x \oip{\xi}{P(dx)}{\xi}$ 
defines a linear operator on $\mathcal{H}_P$}

For $\xi\in\mathcal{H}_P$, we have that 
\begin{displaymatH}
 \int_\RR \int |x| \oip{\xi}{P(dx)}{\xi} 
  \leq \int_{-1}^{1} \oip{\xi}{P(-1,1)}{\xi} 
   + \int_{x: |x| > 1} |x|^2 \oip{\xi}{P(dx)}{\xi} 
< \infty,
\end{displaymatH}
hence the right hand side is well defined on $\mathcal{H}_P$.

For $\xi,\eta\in \mathcal{H}_P$,  $\oip{\xi}{O}{\eta}$ is defined by 
polarization:
\begin{displaymatH}\displaystyle
 \oip{\xi}{O}{\eta}
  :=
  \frac{1}{4}\sum_{k=0}^3 (-i)^k   \oip{\xi+ i^k \eta}{O}{\xi + i^k\eta}
  =
  \frac{1}{4}\sum_{k=0}^3 (-i)^k  
            \int_\RR  x \; P_{\xi+ i^k\eta}(dx)
  .
\end{displaymatH}
To prove that 
$\oip{\xi}{O}{\eta+\zeta} =  \oip{\xi}{O}{\eta} +\oip{\xi}{O}{\zeta}$
it therefore suffices to show that the measure 
\begin{displaymatH}
 \frac{1}{4}\sum_{k=0}^3 (-i)^k  P_{\xi+ i^k(\eta+\zeta)}
\end{displaymatH}
coincides with the sum of two other measures, viz.
\begin{displaymatH}
 \frac{1}{4}\sum_{k=0}^3 (-i)^k  P_{\xi+ i^k\eta}
  \:,\; \text{ and }\;
 \frac{1}{4}\sum_{k=0}^3 (-i)^k  P_{\xi+ i^k\zeta}
  .
\end{displaymatH}
That however is clear since for each Borel set $B$
\begin{aligN}
  \frac{1}{4}\sum_{k=0}^3 \ats (-i)^k  P_{\xi+ i^k(\eta+\zeta)}(B)
   =
 \frac{1}{4}\sum_{k=0}^3 ((-i)^k  
    \oip{\xi+ i^k(\eta+\zeta)}{P(B)}{\xi+ i^k(\eta+\zeta)} 
   \\
   & =
  \oip{\xi}{P(B)}{(\eta+\zeta)}
\end{aligN}
by polarization, and likewise
\begin{displaymatH}
 \frac{1}{4}\sum_{k=0}^3 (-i)^k  P_{\xi+ i^k\eta}(B)=   \oip{\xi}{P(B)}{\eta}
 \,\;\text{ and } \;
 \frac{1}{4}\sum_{k=0}^3 (-i)^k  P_{\xi+ i^k\zeta}(B)=   \oip{\xi}{P(B)}{\zeta}
 .
\end{displaymatH}

\paragraph{$O$ is self-adjoint}
The argumentation proceeds smoother if we realize that
\begin{displaymatH}\displaystyle
 \oip{\xi}{O}{\eta}
  :=
  \frac{1}{4}\sum_{k=0}^3 (-i)^k  
            \int_\RR  x \; P_{\xi+ i^k\eta}(dx)
  .
\end{displaymatH}
coincides with the notion of integration of $x$ with respect to the
\textsl{complex-valued} measure $B\mapsto \oip{\xi}{P(B)}{\eta}$ i.e.\
$\oip{\xi}{O}{\eta}= \int x \oip{\xi}{P(dx)}{\eta}$.

This way, we see that for $\xi,\eta\in \mathcal{H}_P = \mathcal{D}_O$

\begin{aligN}
 \overline{\oip{\eta}{O}{\xi}} 
  &= \overline{\int x \oip{\eta}{P(dx)}{\xi}} 
  = \int x \overline{\oip{\eta}{P(dx)}{\xi}}
  = \int x \ip{P(dx)\xi}{\eta}
  = \int x \oip{\xi}{P(dx)}{\eta}
  \\
 & = \oip{\xi}{O}{\eta}
 ,
\end{aligN}
i.e.\ $\oip{\xi}{O}{\eta}= \ip{O\xi}{\eta}$
which shows that $\mathcal{D}_{O^*}\supset \mathcal{D}_O$ and 
that the restriction of $O^*$ to $\mathcal{D}_O$
coincides with $O$.
We now show that actually, $\mathcal{D}_{O^*}= \mathcal{D}_O$.
To this end, suppose that $\xi\in \mathcal{D}_{O^*} \setminus \mathcal{D}_{O}$ i.e.
\begin{displaymatH}
 \oip{\xi}{O}{\eta}= \ip{O^*\xi}{\eta} 
 \quad \text{ for all }\eta\in\mathcal{H}_P,
\end{displaymatH}
but $\int_{-\infty}^\infty x^2 \oip{\xi}{P(dx)}{\xi} = \infty$.
In particular, $\xi\neq 0$.
Since  $\eta_n:=P(-n,n)O^*\xi\in\mathcal{H}_P=\mathcal{D}_O$ (see the argument
why $\mathcal{H}_P$ is a dense subset), we have by taking
$\eta=\eta_n$ in the above equation
\begin{displaymatH}
  \oip{\xi}{O}{\eta_n}= \ip{O^*\xi}{\eta_n} \quad (\text{all } n\in\NN). 
\end{displaymatH}
For $n\to\infty$ the right hand side converges to $\ip{O^*\xi}{O^*\xi}$
(since $O^*\xi\in\mathcal{H}$ and $\eta_n\to O^*\xi$) and is therefore bounded.
The left hand side evaluates to
\begin{aligN}
 \oip{\xi}{O}{\eta_n} 
   & =
 \int_{-\infty}^{\infty} x \oip{\xi}{P(dx)}{\eta_n} 
     = 
 \int_{-\infty}^{\infty} x \oip{\xi}{P(dx)P(-n,n)}{O^*\xi} 
     =
 \int_{-n}^{n} x \ip{\xi}{P(dx)O^*\xi} 
   \\
   &
  = \int_{-n}^{n} x \ip{O\: P(dx)\xi}{\xi}
  =\overline{ \int_{-n}^{n} x \oip{\xi}{O}{P(dx) \xi} }
  =\overline{ 
     \int_{-n}^{n} x 
       \int_{\infty}^{\infty} x'\oip{\xi}{P(dx')}{P(dx) \xi}
             }
  \\
  & 
  = \overline{ 
     \int_{-n}^{n} x^2 \oip{\xi}{P(dx)}{\xi} 
    }
  =    \int_{-n}^{n} x^2 \oip{\xi}{P(dx)}{\xi}
  ,
\end{aligN}
and tends to $\infty$ if $n\to\infty$.
Contradiction. Thus, there cannot be a 
$\xi \in \mathcal{D}_{O^*}\setminus\mathcal{D}_{O}$.

\section{Proof of lemma \ref{th_integralWeyl} \label{ap_IntegralWeylOp}}

For $n\in\NN$, 
let $\Box_n$ be the square in $\CC \isom \RR^2$ with corners at the
points $(\pm 2^{n},\pm 2^{n})$.  
Thus, the length of $\Box_n$ is $2\cdot 2^n$ and 
$\Box_0 \subset\Box_1 \subset \Box_2 \subset \ldots $.
We can think of $\Box_n$ as filled up with small squares of length 
$(\half)^{n}$,
which have corners at the points
$(\half)^n\cdot (k,l)$ with $k,l=0, \pm 1, \pm 2, \ldots, \pm 2^{2n}$.
\footnote{%
The centers of these squares are
$(\half)^{n+1} \cdot \bigl(\pm (2k+1), \pm  (2l+1)\bigr)  $, 
with $k,l= 0, 1,\ldots, 2^{2n}-1$. This are in total $2^{2n} \cdot 2^{2n} = 2^{4n}$ squares in each quadrant, so
$2^{4n+2}$ squares of volume $2^{-2n}$ in total (which equals the volume
$2^{2n+2}$ of $\Box_n$).
}

Let $\mathcal{P}_n$ be the collection of these squares.

What we will use of this construction is the following:
\begin{Properties}
 \item
  $\Box_n$ is an increasing sequence
  of squares which fill up $\RR\isom\CC$,
 \item
  $\Box_n$ is the union of all squares $R$ in $\mathcal{P}_n$,
 \item
  for each $R \in \mathcal{P}_n$, the distance of two points within $R$
  is less than $\sqrt{2} (\half)^{n}$, 
 \item
  if $n\geq m$ then each 
  square $R\in\mathcal{P}_m$ is the union of squares in $\mathcal{P}_n$
  that lie within it.
\end{Properties}

 For $n\in\NN$ define the step functions  
\begin{displaymatH}
  \displaystyle
  T_n := 
      \sum_{R\in\mathcal{P}_n}  \la_2(R)\cdot \ro(z_R) \cdot W(z_R)
      = 
     \int_\CC \Bigl[ 
                 \sum_{R\in\mathcal{P}_n} \ro(z_R) W(z_R)\indicator_R(z) 
                 \Bigr]\la_2(dz)
      \;,
\end{displaymatH}
where $z_R \in R$ are chosen such that
\begin{equatioN}
  |\ro(z_R)| \norM{W(z_R)} \la_2(R) \leq \int_R |\ro(z)| \norM{W(z)} \la_2(dz)
   \quad (R\in\mathcal{P}_n) 
   \;,
  \tag{\&}
\end{equatioN}
(for $z_R$ we could, for example, choose the point in $R$ 
where the continuous function 
$z\mapsto   |\ro(z)| \norM{W(z)} $ attains its minimum over $R$).

\medskip

 Then $T_n\in\B(\HH)$, and we prove that 
 \begin{Properties}
  \item for all $\xi\in\HH$, $(T_n\xi)_{n\in\NN}$ 
        is a Cauchy sequence in $\HH$,
        therefore its limit  $T\xi:=\lim_n T_n \xi$  exists;
  \item for all $\xi,\eta\in\HH$,
        $\ip{\eta}{T\xi}=\int_{\CC} \ip{\eta}{W(z)\xi}\ro(z) \la_2(dz)$.
 \end{Properties}

 \smallskip

\statement{Step 1: for each $\xi\in\HH$, the map $z\mapsto \ro(z) W(z)\xi$ 
                   from  $\CC$ into $\HH$ is continuous}
 
 Simply, because $z\mapsto W(z)\xi$ as well as $z\mapsto\ro(z)$ are continuous.

\statement{Step 2: for each $\xi\in\HH$, the sequence $(T_n\xi)_n$ in $\HH$ is 
                   Cauchy}

 Let $\xi\in\HH$ and $m\geq n\geq N$ in $\NN$.
 We split up $T_m\xi - T_n \xi$ in three terms:
 \begin{aligN}
  T_m\xi - T_n \xi 
    \ats
     = 
  \displaystyle
  \sum_{\widetilde{R}\in\mathcal{P}_m} 
    \la_2(\widetilde{R}) \ro(z_{\widetilde{R}})  W(z_{\widetilde{R}})\xi 
    -
  \sum_{R\in\mathcal{P}_n} \la_2(R)\ro(z_R) W(z_R)\xi
  \\[5pt]
   &
   = 
    \sum_{\begin{subarray}{l}
           \widetilde{R}\in\mathcal{P}_m \\ 
            \widetilde{R}\subset \CC\setminus\Box_N
         \end{subarray}
         } 
    \la_2(\widetilde{R})\ro(z_{\widetilde{R}})  W(z_{\widetilde{R}}) \xi
   \;
   +
    \sum_{\begin{subarray}{l}
           R\in\mathcal{P}_n \\ 
           R\subset \CC\setminus\Box_N
         \end{subarray}
          } 
    \la_2(R)\ro(z_R)  W(z_{R}) \xi
  \\
  \ats
 \quad
   +
  \Biggl(
   \displaystyle
    \sum_{\begin{subarray}{l}
           \widetilde{R}\in\mathcal{P}_m\\
           \widetilde{R}\subset \Box_N 
         \end{subarray}}
   \la_2(\widetilde{R})\ro(z_{\widetilde{R}}) W(z_{\widetilde{R}})
    -
   \sum_{\begin{subarray}{l}
           R\in\mathcal{P}_n \\
           R\subset \Box_N 
         \end{subarray}}
   \la_2(R)\ro(z_R) W(z_R)
  \Biggr)\xi
 \end{aligN}
 and we show that the norm of each of these terms vanish 
 if $m\geq n \to\infty$.
 Actually the first two terms will vanish because of the integrability 
 of $z\mapsto |\ro(z)|\norM{W(z)}$, and the last term will vanish because of 
 the uniform continuity of $z\mapsto \ro(z)W(z)\xi$ on compacta. 

 To make this precise, let $\ep>0$.

 First we take $N$ so large that 
 \begin{equatioN}
   \int_{\CC\setminus\Box_N} |\ro(z)| \| W(z)\| \la_2(dz) < \ep/(3\norM{\xi})
   \tag{\#1}
 \end{equatioN}
 (Lebesgue's dominated convergence theorem).
  
 With this choice of $N$ we have for the norm of the first term:
 \begin{aligN}
    \biggl\|
    \dsum_{\begin{subarray}{l}
                 \widetilde{R}\in\mathcal{P}_m \\
                 \widetilde{R}\subset \CC\setminus\Box_N
          \end{subarray}
         }
    \ats
    \la_2(\widetilde{R}) \ro(z_{\widetilde{R}})  W(z_{\widetilde{R}})\xi 
    \biggr\|
         \leq 
   \norM{\xi} 
   \dsum_{\begin{subarray}{l}
                 \widetilde{R}\in\mathcal{P}_m \\
                 \widetilde{R}\subset \CC\setminus\Box_N
          \end{subarray}
         }
    \la_2(\widetilde{R})|\ro(z_{\widetilde{R}})| \| W(z_{\widetilde{R}})\| 
    \\[3pt]
    &
    \leq
   \norM{\xi}\int_{\Box_m\setminus\Box_N}
    |\ro(z)| \| W(z)\| \la_2(dz)
    \leq
   \norM{\xi}\int_{\CC\setminus\Box_N}
    \ro(z) \| W(z)\| \la_2(dz)
   \\
   \ats
   < \ep/3
   \;,
 \end{aligN}
 where the second inequality uses the choice of $z_{\widetilde{R}}$,
 see $(\&)$.
 
 Likewise the norm of the second term is $< \ep/3$.
 
 Now, we consider the third term.
 Since $z\mapsto\ro(z)W(z)\xi$ is continuous, 
 it is uniformly continuous on the compact set $\Box_{N}$, 
 and so there exists an 
 $M\geq N$ in $\NN$  such that for all $n\geq M$
 \begin{equatioN}
  \left.
  \begin{array}{l}
   z,z'\in \Box_{N} \;, \\
   z,z' \in R \text{ with} \\
   \qquad R\in\mathcal{P}_n 
  \end{array}
   \right\rgroup
   \Implies
  \norM{\ro(z)W(z)\xi - \ro(z')W(z')\xi}
  \leq
  \displaystyle  \frac{\ep}{3\la_2(\Box_{N})}
  \tag{!}
 \end{equatioN}

 Suppose now that both $m$ and $n$ are larger than $M$. 

 Since $m\geq n$, 
 each square $R\in \mathcal{P}_n$ is a union of disjoint 
 squares $\widetilde{R}\in\mathcal{P}_m$. 
 Further, since $m\geq n \geq M\geq N$,
 $\Box_N$ is the union of all $R\in\mathcal{P}_n$ with $R\subset\Box_N$, 
 and likewise 
 $\Box_N$ is the union of all $\widetilde{R}\in\mathcal{P}_m$ 
 with $\widetilde{R}\subset\Box_N$.
 Thus,
 \begin{aligN}
  \Bigl\| 
    & 
    \displaystyle
    \sum_{\begin{subarray}{l}
           \widetilde{R}\in\mathcal{P}_m\\
           \widetilde{R}\subset \Box_N 
         \end{subarray}
         }
   \la_2(\widetilde{R}) \ro(z_{\widetilde{R}}) W(z_{\widetilde{R}}) \xi
    -
   \sum_{\begin{subarray}{l}
           R\in\mathcal{P}_n \\
           R\subset \Box_N 
         \end{subarray}
        }
   \la_2(R)\ro(z_R) W(z_R) \xi
 \Bigr\|
  \\
  \ats
  =
  \Bigl\|
    \displaystyle
    \sum_{\begin{subarray}{l}
            R\in\mathcal{P}_n \\
            R\subset \Box_N 
	  \end{subarray}
         }
    \sum_{\begin{subarray}{l}
           \widetilde{R}\in\mathcal{P}_m\\
           \widetilde{R}\subset R 
         \end{subarray}
         }
   \la_2(\widetilde{R}) \ro(z_{\widetilde{R}}) W(z_{\widetilde{R}})\xi
    -
   \sum_{\begin{subarray}{l}
           R\in\mathcal{P}_n \\
           R\subset \Box_N 
         \end{subarray}
        }
    \sum_{\begin{subarray}{l}
           \widetilde{R}\in\mathcal{P}_m\\
           \widetilde{R}\subset R 
         \end{subarray}
         }
   \la_2(\widetilde{R})\ro(z_R) W(z_R)\xi
 \Bigr\|
  \\
   &
   \leq 
  \sum_{\begin{subarray}{l}
           R\in\mathcal{P}_n \\
           R\subset \Box_N 
         \end{subarray}
        }
    \sum_{\begin{subarray}{l}
           \widetilde{R}\in\mathcal{P}_m\\
           \widetilde{R}\subset R 
         \end{subarray}
         }
    \la_2(\widetilde{R})
    \norM{\ro(z_{\widetilde{R}}) W(z_{\widetilde{R}}\xi) 
           -
          \ro(z_{R}) W(z_{R})\xi 
         }
   \\
   &
   \overset{(!)}{\leq} 
  \sum_{\begin{subarray}{l}
           R\in\mathcal{P}_n \\
           R\subset \Box_N 
         \end{subarray}
        }
    \sum_{\begin{subarray}{l}
           \widetilde{R}\in\mathcal{P}_m\\
           \widetilde{R}\subset R 
         \end{subarray}
         }
    \la_2(\widetilde{R})
    \frac{\ep}{3\la_2(\Box_N)}
   =
   \sum_{\begin{subarray}{l}
           R\in\mathcal{P}_n \\
           R\subset \Box_N 
         \end{subarray}
        }
    \frac{\ep}{3\la_2(\Box_N)}
    \la_2(R)
   =
  \la_2(\Box_N)
  \frac{\ep}{3\la_2(\Box_N)}
  \\
  &
  \leq \ep/3
 \end{aligN}

\statement{Step 3: for all $\xi, \eta\in\HH$: 
       $\lim_m\ip{\eta}{T_m\xi} = \int_\CC \ro(z)\ip{\eta}{W(z)\xi}\la_2(dz)$}

Let $\xi,\eta\in\HH$. 
First we mention that $z\to\ro(z)\ip{\eta}{W(z)\xi}$ is integrable,
since it is a continuous function (by the strong continuity of $z\to W(z)$)
dominated by the integrable function
$z\to\ro(z)\norM{W(z)}\norM{\xi}\norM{\eta}$.

Next, we split 
$ \int_\CC \ro(z)\ip{\eta}{W(z)\xi} \la_2(dz) - \ip{\eta}{T_n\xi}$ 
in three terms:
\begin{aligN}
 \int_\CC \;&\ro(z)\ip{\eta}{W(z)\xi} \la_2(dz) - \ip{\eta}{T_n\xi}
  \\
  &
  = 
 \int_{\CC\setminus\Box_N} \ro(z)\ip{\eta}{W(z)\xi} \la_2(dz)
 \; -
 \sum_{\begin{subarray}{l}
             R\in\mathcal{P}_n\\
             R\subset\CC\setminus\Box_N 
       \end{subarray}
      }
      \la_2(R)\ro(z_R) \ip{\eta}{W(z_R)\xi} 
 \\
  &
  +
   \int_{\Box_N} 
     \Biggl\lgroup\;
       \ro(z) \ip{\eta}{W(z)\xi}
       \;\;-
        \sum_{\begin{subarray}{l}
                 R\in\mathcal{P}_n\\
                 R\subset\Box_N 
              \end{subarray}
             }
        \ro(z_R) \ip{\eta}{W(z_R)\xi}\indicator_R(z)
     \;
    \Biggr\rgroup
    \la_2(dz)
    \;,
\end{aligN}
and show that each of them vanishes if $m\to\infty$.

To this end, let $\ep>0$.
 First take $N$ so large that 
 $\int_{\CC\setminus\Box_N} |\ro(z)|\norM{W(z)} \la_2(dz) 
   < \ep/(3\norM{\xi}\norM{\eta}$. 
 Then the first term is in absolute value $<\ep/3$.
 The same holds for the second term, since the absolute value of each summand
 is majorized by 
 $\norM{\xi}\:\norM{\eta}\:\int_R |\ro(z)|\norM{W(z)}\la(dz)$ 
 by definition of the $z_R$'s.
 To handle the third term, we use (as in step 3) the uniform continuity of 
 $z\to\ro(z)\norM{W(z)}$ to find an $M\geq N$ so large that for 
 $n\geq M$:
 \begin{equatioN}
  \left.
  \begin{array}{l}
   z,z'\in \Box_{N} \;, \\
   z,z' \in R \text{ with} \\
   \qquad R\in\mathcal{P}_n 
  \end{array}
   \right\rgroup
   \Implies
  \norM{\ro(z)W(z)\xi - \ro(z')W(z')\xi}
  \leq
  \displaystyle  \frac{\ep}{3\norM{\eta}\la_2(\Box_{N})}
  \;.
  \tag{!2}
 \end{equatioN}
 Next, we rewrite the integrand of the third term as:
 \begin{aligN}
       \ro(z) & \ip{\eta}{W(z)\xi} \indicator_{\Box_N}(z)
       \;\;-
        \sum_{\begin{subarray}{l}
                 R\in\mathcal{P}_n\\
                 R\subset\Box_N 
              \end{subarray}
             }
        \ro(z_R) \ip{\eta}{W(z_R)\xi}\indicator_R(z)
     \;
    \\
    & 
    =
   \sum_{\begin{subarray}{l}
                 R\in\mathcal{P}_n\\
                 R\subset\Box_N 
         \end{subarray}
         }
       \Bigl\lgroup
         \ro(z) \ip{\eta}{W(z)\xi}
         \;\;-
        \ro(z_R) \ip{\eta}{W(z_R)\xi}
       \Bigr\rgroup
       \indicator_R(z)
   \\
    &
   =
 \sum_{\begin{subarray}{l}
               R\in\mathcal{P}_n\\
               R\subset\Box_N 
       \end{subarray}
      }
   \ip{\eta}{
       \bigl[ \ro(z)W(z)\xi - \ro(z_R) W(z_R)\xi \bigr]
            }
   \cdot \indicator_R(z)
   \;,
 \end{aligN}
 which implies that for $n\geq M$ 
 the absolute value of the integrand is bounded
 by $\ep/(3\la_2(\Box_N))$ (use $(!2)$), and hence the third term is 
 bounded by $\ep/3$.

 Concluding, $\ip{\eta}{T\xi}=\lim_n \ip{\eta}{T_n\xi} 
  =\int_\CC \ro(z)\ip{\eta}{W(z)\xi}$. 

\section{Proof of lemma \protect{\ref{lem_PVMofQandP}}\label{ap_PVMofQandP}}

We only discuss the claim for $Q$. The claim for $P$ is proved analogously.
The existence of a sequence $f_k$ with the desired properties follows from
\cite[p.~93-94]{vR_Fourier}.
As the Fourier transform maps the space of Schwartz test functions
 $\mathcal{S}$ into itself bijectively \cite{vR_Fourier}, 
\question{look up pages} 
$P^k_{(a,b)}= \int_\CC \widehat{f_k}(s) W(is) ds $ exists strongly, 
lies in the von Neumann $\vN{W(is): s\in\RR}$, and 
for any state $\psi$:
\begin{displaymatH}
 \oip{\psi}{P^k_{(a,b)}}{\psi}
   = \int_{-\infty}^{\infty} \widehat{f_k}(s) \oip{\psi}{e^{isQ}}{\psi} ds
\end{displaymatH}

Taking the Schr\"odinger representation
as cyclic vacuum representation of the CCR, 
we have that $Q$ is the multiplication by $q$ 
(so that $W(is)=e^{isQ}$ is the multiplication by $e^{isq}$)
and $\ip{q}{\psi}=\psi(q)$, so that
\begin{aligN}
 \oip{\psi}{P^k_{(a,b)}}{\psi}
  \ats
  =
 \int_{-\infty}^{\infty} \widehat{f_k}(s) \oip{\psi}{e^{isQ}}{\psi} ds
  =
 \int_{-\infty}^{\infty}
 \int_{-\infty}^{\infty} \widehat{f_k}(s) e^{isq} |\psi(q)|^2  ds dq
  \\
  \ats
   =
 \int_{-\infty}^{\infty} f_k(q) |\psi(q)|^2  dq
 \quad\overset{k\to\infty}{\longrightarrow}\quad
 \int_{a}^{b}  |\psi(q)|^2  dq
\end{aligN}
which equals $\oip{\psi}{P(a,b)}{\psi}$ i.e.\ the probability that a 
measurement of $Q$ gives a value in $(a,b)$ when the system is in state $\psi$.
Hence $P_Q(a,b)$ lies in $\vN{W(is): s\in\RR}$ too.

If $f=\sum_{i=1}^n \al_i \indicator_{(a_i,b_i)}$, then 
$f(Q)=\sum_{i=1}^n \al_i \indicator_{(a_i,b_i)}(Q)=
\sum_{i=1}^n \al_i P_Q(a_i,b_i)$ lies in $\vN{W(is): s\in\RR}$.

Finally, if $f\in L^\infty(\RR)$, then there exist a sequence $f_k$ of
simple functions such that $\norm{f-f_k}_\infty \to 0$ (see \cite{vR_Fourier}).
As a consequence of the dominated convergence theorem, 
the corresponding sequence (of multiplication operators) will
converge to (the multiplication by) $f$,
and hence
$\{f(Q) : f\in L^\infty\} \subset \vN{W(is): s\in\RR}$.

\section{Proof of lemma \ref{lem_PVMofN}\label{ap_PVMofN}}
We show that
\begin{displaymatH}
 s^N = 
   \frac{1}{\pi}
      \int_\CC
        \frac{e^{-\half\frac{1+s}{1-s}|z|^2}}{1-s} W(z) 
  \la_2(dz)
  .
\end{displaymatH}
 Take $s\in (-1,1)$ and let 
 \begin{displaymatH}
   \ro(z):= \frac{1}{\pi(1-s)} \exp(-\half\frac{1+s}{1-s}|z|^2)
   \;,
 \end{displaymatH}
 and let $W(z)$, $z\in\CC$, be the Weyl operators on $\CC$.
 Then that the conditions of theorem \ref{th_integralWeyl}, are satisfied,
 so the strong integral on the right hand side makes sense.
 To show that it coincides with $s^N$, we evaluate its ``matrix elements''
 with respect to the total subset of exponential vectors.

 Take $u,v\in\CC$. Then by theorem \ref{th_integralWeyl}
 \begin{aligN}\displaystyle
  \langle \expv(u)& 
   | \Bigl\lgroup\int_\CC \ro(z) W(z) \la_2(dz)\Bigr\rgroup | 
           \expv(v)\rangle  
      =
  \int_\CC \ro(z) 
       \oip{\expv(u)}{W(z)}{\expv(v)}
  \la_2(dz)
    \\
    &
    =
  \int_\CC \ro(z) 
       \ip{\expv(u)}{\bigl[e^{-\ip{z}{v}-\half\norM[text]{z}^2}\bigr]%
           \expv(z+v)}
  \la_2(dz)
    \\
    &
    =
  \int_\CC  e^{\ip{u}{v}}
           \ro(z)
           e^{-\half\norM[text]{z}^2}  
           e^{-\ip{z}{v} + \ip{u}{z}}
  \la_2(dz)
   \;.
 \end{aligN}
 Now 
 \begin{displaymatH}\displaystyle
  \ro(z) e^{-\half\norM[text]{z}^2} 
  = \frac{1}{\pi(1-s)} e^{-\half(\frac{1+s}{1-s} + 1)|z|^2}
  = \frac{1}{\pi(1-s)} e^{-\frac{1}{1-s}|z|^2}
  \;.
 \end{displaymatH}
 Further, writing $z=x+iy$, we calculate 
 \begin{aligN}
  - \ip{z}{v} + \ip{u}{z} 
  &
  = -\overline{z} v +  \overline{u}z
  \\
  & 
  = [-v + \overline{u}]x + i[v + \overline{u}]y
  \;.
 \end{aligN}
 Using this and the fact that for $w\in\CC$ and $a\in (0,\infty)$
 $\int_{-\infty}^\infty \exp(-ax^2 + wx) dx = \sqrt{\pi/a} \exp(w^2/4a)$
 we resume calculating
 \begin{aligN}\displaystyle
  \langle \expv(u) &
  | \Bigl\lgroup\int_\CC \ro(z) W(z) \la_2(dz)\Bigr\rgroup |
   \expv(v) \rangle  
   \\
   &
   =
   \frac{e^{\ip{u}{v}}}{(1-s)\pi}
   \int_{\infty}^\infty
       e^{-\frac{1}{1-s} x^2 + [-v + \overline{u}]x}
   dx
     \cdot
  \int_{\infty}^\infty
      e^{-\frac{1}{1-s} y^2  + i[v + \overline{u}]y}
    dy
   \\
   &
   =
  \frac{e^{\ip{u}{v}}}{(1-s)\pi}
  \cdot
  \sqrt{\pi(1-s)} \exp\Bigl\lgroup
                     \frac{1-s}{4} [-v + \overline{u}]^2
                      \Bigr\rgroup
  \cdot
  \sqrt{\pi(1-s)} \exp\Bigl\lgroup
                   - \frac{1-s}{4} [v + \overline{u}]^2
                      \Bigr\rgroup
  \\
  &
  =
   e^{\ip{u}{v}}
  \exp\Bigl\lgroup
         \frac{1-s}{4} \cdot  -4 \overline{u}v 
      \Bigr\rgroup
  = 
   e^{s\overline{u} v}
  \;
  .
 \end{aligN}
 Since $e_n=\frac{d^n}{dx^n} \expv(x)$, 
 \begin{aligN}
  \langle e_n & 
  | \Bigl\lgroup\int_\CC \ro(z) W(z) \la_2(dz)\Bigr\rgroup|
   e_m \rangle  
   \\
   &
   =
  \frac{1}{n!}\frac{1}{m!}
  \frac{d^n}{dx^n} \frac{d^m}{dy^m}
   \ip{\expv(x)}{%
          \Bigl\lgroup\int_\CC \ro(z) W(z) \la_2(dz)\Bigr\rgroup \expv(y)}  
   \\
   & 
   = 
  s^n \de_{n,m}
 \end{aligN}
 we see that 
 $ \int_\CC \ro(z) W(z) \la_2(dz)$ is in $\vN{E_n : n\in\NN}$, and
 that $\ip{e_n}{\int_\CC \ro(z) W(z) \la_2(dz) e_m} = s^n \de_{n,m}$
 i.e.\ $\int_\CC \ro(z) W(z) \la_2(dz)=s^N$.

\section{The uniform derivative of a absolutely summable operator-valued series\label{ap_uniformDerivative}}

Let $f(z)= \sum_k \al_k z^k$ be a series that is absolutely summable 
on an interval containing $0$,
and let $P_k$ be a bounded series of operators. 
Since $f(z)$ and its derivatives are all absolutely summable
($f(z)$ is a convergent power series 
on a disk around $0$ in $\CC$ and all its derivatives exist on that
disk and are hence absolutely convergent too), the series
\begin{aligN}
 \widetilde{f}(z) &: = \sum_{k=1}^\infty |\al_k| s^k P_k \\
 \widetilde{f}^{(n)}(z) &: 
   = \sum_{k=n}^\infty k\cdots (k-n+1)|\al_k| s^{(k-n)} P_k 
  ,\\  
\end{aligN}
and
\begin{aligN}\displaystyle
 A^{(0)}(s) &:= \sum_{k=1}^\infty \al_k s^k P_k 
   \\
 A^{(n)}(s) &:= \sum_{k=n}^\infty  k\cdot (k-1) \cdots (k-n+1) \al_k s^{k-n} P_k 
\end{aligN}
exist.

After defining
\begin{aligN}
 F^{(0)}(s) &:= \sum_{k=1}^\infty \al_k s^k P_k 
   \\
 F^{(n+1)}(s) &:= 
  \norM{}\text{-}\lim_{t\to s}\frac{F^{(n)}(t) - F^{(n)}(s)}{t-s}
   ,
\end{aligN}
we can prove by induction $A^{(n)}= F^{(n)}$ for all $n$.
In fact,
\begin{aligN}\displaystyle
  \Bigl\| & \frac{F^{(n)}(t) -  F^{(n)}(s)}{t-s} - A^{(n+1)}(s) \Bigr\|
  \\
  & 
   =
 \norM{\sum_{k=n+1}^\infty  k \cdots (k-n+1) \al_k  \Bigl[\frac{t^{k-n}-s^{k-n}}{t-s} - (k-n)s^{k-n-1}\Bigr]P_k}
  \\
  & 
  \leq
 \sup_n \norM{P_n} \cdot
  \sum_{k=n+1}^\infty 
      k \cdots (k-n+1)\cdot |\al_k| \cdot \Bigl|\frac{t^{k-n}-s^{k-n}}{t-s} - (k-n)s^{k-n-1}\Bigr|
  \\
  & = 
  \sup_n \norM{P_n} \cdot 
 \Bigl|\frac{ \widetilde{f}^{(n)}(t) -\widetilde{f}^{(n)}(s)}{t-s} 
              - \widetilde{f}^{(n+1)}(s) \Bigr|
   \to 0 
  \quad (t\to s).
\end{aligN}

\section{The count operator $d\mathcal{F}(P)$
\label{ap_countOperator}}
Given an orthogonal projection $P$ on $\mathcal{K}$, we can define a
self-adjoint operator $d\mathcal{F}(P)$ on $\mathcal{F}(\mathcal{K})$
via
\begin{equatioN}
 e^{i\lambda d\mathcal{F}(P)} : = \mathcal{F}(e^{i\lambda P})
 \quad
 (\lambda\in\RR)
 .
\end{equatioN}
Indeed, $P$ is self-adjoint and therefore $e^{i\lambda P}$ is a unitary
map $\mathcal{K}\to\mathcal{K}$: $\norM{e^{i\lambda P} f}=\norM{f}$ for all
$f\in\mathcal{K}$. As a consequence,
$\mathcal{F}(e^{i\lambda P})$ is unitary map 
$\mathcal{F}(\mathcal{K})\to\mathcal{F}(\mathcal{K})$, as
$\norM{f_1\otimes f_2 \otimes \ldots \otimes f_n} = \norM{f_1}\norm{f_2}\cdot
\norM{f_n}$ for all $f_1,\ldots, f_n\in\mathcal{K}$.
As one-parameter semi-group, $\lambda\mapsto\mathcal{F}(e^{i\lambda P})$
it has a generator $d\mathcal{F}(P)$ on $\mathcal{F}(\mathcal{K})$.

\section{Representations of $\Gamma(\mathcal{K})$ are unitarily equivalent\label{ap_repGamma(K)Equivalent}}

Suppose
$(\mathcal{W}_\mathcal{K}, \ph_\mathcal{K})$
and $(\mathcal{W}_\mathcal{K}^{'}, \ph^{'}_\mathcal{K})$ are
two representations of $\Gamma(\mathcal{K})$ over $\mathcal{H}$ and
$\mathcal{H}'$ respectively, then both of them contain
a cyclic vacuum representation of the CCR over $\mathcal{K}$, hence
there is a unitary map $U:\mathcal{H}' \to \mathcal{H}$ such that 
$W'(f)= U^{-1}W(f)U$ for all $f\in\mathcal{K}$. 
In other words, the linear map 
$\mathcal{B}(\mathcal{H})\to \mathcal{B}(\mathcal{H}')$,
$A\mapsto A'= U^{-1}A U $  sends 
$\{W(f) : f\in\mathcal{K} \} $ bijectively to 
$\{W'(f) : f\in\mathcal{K} \} $. Moreover, $A\mapsto A'$ is strongly continuous and therefore extends uniquely to the strong closures of
$\{W(f) : f\in\mathcal{K} \} $ and $\{W'(f) : f\in\mathcal{K} \} $ i.e.\
it is a strongly continuous, bijective map from
$\mathcal{W}_\mathcal{K}$ to $\mathcal{W}_\mathcal{K}^{'}$.

\section{A minimal dilation of the classical damped harmonic oscillator  \label{ap_minClassicDilationDampedHarmonicOsc}}

\proof
{\sl Existence:}
Then for all $t\ge0$ we have
\begin{aligN}
\ip{v}{R_tv}
   &=\int_{-\infty}^0e^{(\eta+i\om)x}e^{(\eta-i\om)(x-t)}2\gamma\,dx \\
   &=e^{(-\eta+i\om)t}\int_{-\infty}^0 e^{2\eta x} 2\gamma\,dx \\
   &=e^{(-\eta+i\om)t}\;.\\
\end{aligN}

Furthermore, the linear span of the vectors $R_tv$ is dense in 
$ L^2(\RR,dx) = L^2(\RR,2\gamma\,dx)$.
To see this, it suffices to show that $\{R_tv : t\in\RR\}^{\perp}=\{0\}$.
When we realize that the Fourier transformation extends to a unitary map  
$\F :  L^2(\RR,dx) \to  L^2(\RR,dx)$
	$$
  (\F g)(k) := L^2-\lim_{N\to\infty} \int_{-N}^{N} e^{-ikx}g(x)\,dx 
       \hbox{~ for almost all } k\in\RR 
	$$
that satisfies 
  $\F T_t\F^{-1}= M_{ e^{-it{\rm x}} }$, 
we see that we can just as well prove that 
	$$
  \{0\}=\{\F T_t\F^{-1}\F v: t\in\RR\}^{\perp}
       =\{e^{ -it{\rm x} }\; \F v : t\in\RR\}^{\perp} \, .
	$$
  To this end, let $f\in L^2$ with $f\perp e^{ it{\rm x} }\; \F v$ 
  for all $t\in\RR$.
  Then 
        $$
       0 = \int_{-\infty}^{\infty} f(k)\bar{( e^{ it{\rm x}} \F v)(k)}\,dk 
         = \int_{-\infty}^{\infty} f(k) e^{-itk} \bar{(\F v)(k)}\,dk  \quad
         \hbox{ for all } t\in\RR  \, .
	$$ 
  Apparently, the Fourier transform of $f\cdot\bar{\mathcal{F} v}$ 
  vanishes, so $f\cdot \bar{\mathcal{F} v} =0 $ \ae. 
  Since $(\mathcal{F} v)(k)=\frac{1}{\eta-i(\om+k)}\not = 0$ 
  for almost all $k\in\RR$,
  this means that $f=0$ in $L^2(\RR)$.

\textsl{Uniqueness:}
We claim that any minimal unitary dilation $(\K,v,U_t)$ of $(C_t)_{t\geq 0}$
induces --in the notation of theorem 3.2-- an Kolmogorov dilation 
$(\HH=\K,V(t)=U_t v)$ 
of the following, positive definite, kernel on $\RR$:
\begin{displaymatH}
   K(s,t)
   =\begin{cases}
    e^{(-\eta+i\om)(t-s)}&\text{if }t\geq s,\\
    e^{(-\eta-i\om)(s-t)}&\text{if }t\leq s.  
   \end{cases}
\end{displaymatH}
Before establishing this claim, let us first see how it implies the 
uniqueness: 
let $(\K,v,U_t)$ and $(\widetilde{\mathcal{K}},\widetilde{v},\widetilde{U}_t)$ be two minimal unitary dilations of 
$(C_t)_{t\geq 0}$, then --assuming the claim-- they are both 
Kolmogorov dilations of one and the same kernel. 
By the Kolmogorov Dilation theorem 3.2 there exists 
a unitary map $\tilde{U}: \K \to \widetilde{\mathcal{K}}$ 
such that for all $ t\in\RR$: ${ \tilde U}V(t) ={ \widetilde V}(t)$ 
i.e. ${\tilde U} U_t v = \widetilde{U}_t \widetilde{v} $ for all $t\in\RR$. Taking $t=0$ we have
$ {\tilde U}v=\widetilde{v}$ 
so that for all $t,t'\in\RR : 
{\tilde U}U_t(U_{t'}v)={\tilde U}U_{t+t'}v= \widetilde{U}_{t+t'}\widetilde{v}=\widetilde{U}_t\widetilde{U}_{t'}\widetilde{v}
                                         =\widetilde{U}_t{\tilde U}(U_{t'}v)$. 
Since the span of $\{U_tv:t\in\RR\}$ is dense, we may conclude that 
$\widetilde{U}_t={\tilde U}U_t{\tilde U}^{-1}$. So:  any two minimal unitary dilations
of $(C_t)_{t\geq 0}$ are unitarily equivalent.

We now establish the claim:
let $(\K,v,U_t)$  be an minimal dilation of $(C_t)_{t\geq 0}$.
Obviously,  $(s,t) \mapsto \ip{U_sv}{U_tv}$ is  
a positive definite kernel on $\RR$ with Kolmogorov dilation 
$(\HH=\K,V(t)=U_t v)$. Using the fact that $(U_t)_{t\in\RR}$ is 
a unitary one-parameter group, we see that for $t,s\in\RR$ :
\begin{displaymatH}
\ip{U_sv}{U_tv}=\ip{U_sv}{U_sU_{t-s}v}=\ip{v}{U_{t-s}v}=%
  \begin{cases}
       e^{(-\eta+i\om)(t-s)} & \text{if }t\geq s,\\
      \overline{\ip{v}{U_{s-t}v}}=e^{(-\eta-i\om)(s-t)} &\text{if }t\leq t  
  \end{cases}
\end{displaymatH}
So $K(s,t)=\ip{U_sv}{U_tv}$ is a positive definite kernel and 
$V: \RR \to \K$ , $V(t)=U_t v$ is an embedding such that 
$\ip{V(s)}{V(t)}=K(s,t)$ and $\overline{\bigvee V(\RR)}=\K$.
\qed

\chapter{The $2^{\text{nd}}$ quantization functor $\Gamma$
\label{ap_quantizationFunctor}}

\question{provide an introduction: the aim is to proof theorem about $\Gamma$,
streamline the argument: either [ first lemmas then proof of the theorem],
[go directly to proof and provide the lemmas as they are needed]. Preference
for the first: lemmas and their proof are rather long. End with a comparison
with real scalar field}

\section{Complete Positivity}
When working in representations, there is a simple
criterion for complete positivity:

\begin{theorem}[criterion complete positivity in representations]
Let $\mathcal{A}_1\subset\mathcal{B}(\mathcal{H}_1)$,
$\mathcal{A}_2\subset\mathcal{B}(\mathcal{H}_2)$ be Von Neumann algebras,
and let $\Gamma$ be a linear map $\mathcal{A}_1\to\mathcal{A}_2$.
Then $\Gamma$ is completely positive if and only if
for all $N\in\NN$, all $\xi_1,\ldots,\xi_n\in\mathcal{H}_2$,
 and all $A_1,\ldots,A_n\in\mathcal{A}_1$:
 \begin{displaymatH}
  \sum_{p,q=1}^N \ip{\xi_q}{\Gamma\bigl[A^{*}_q A_p \bigr] \xi_p}\; \geq 0
  . 
 \end{displaymatH}
\end{theorem}

\section{Strong operator topology}

\begin{lemma}
\label{lm_critStrongConv}
 Let $\mathcal{X}$ be a dense subset of a be a Hilbert space $\HH$.
 Then $F_\al \to F$ in $\B(\HH)$ strongly if and only if
 $\norM{(F_\al-F)\xi}\underset{\al}{\to} 0$ for all $\xi\in\mathcal{X}$
 and $\sup_\al \norM{F_\al}_{\B(\HH)} <\infty$.

 In particular, if $\HH$ is separable, then the strong operator topology
 is metrizable (and continuity coincides with sequential continuity)
 on operator norm bounded subsets of $\B(\HH)$.
\end{lemma}

\begin{pf}
 Replacing $F_\al$ by $G_\al:=F_\al-F$, we have 
 $\sup_\al \norM{G_\al}< \infty$ if and only if $\sup_\al\norM{F_\al}<\infty$.
 In other words, we may assume that $F=0$.

 \textsl{If:}
 suppose that $\norM{F_\al\xi} \to 0$ for all $\xi\in\mathcal{X}$, and
 $\sup_\al \norM{F_\al}_{\B(\HH)} <\infty$. 
 We now prove that $\norM{F_\al\eta} \to 0$ for all $\eta\in\HH$.
 To this end, let $\eta\in\HH$, and $\ep>0$. 
 Take a $\xi\in\mathcal{X}$ with $\norM{\eta-\xi}<\ep$, and
 take $\al(\ep)$ 
 such that $\norM{F_\al\xi}< \ep$ for $\al \succ \al(\ep)$.      
 Then for $\al\succ \al(\ep)$: 
 \begin{aligN}
  \norM{ F_\al\eta}
    \ats\leq 
   \norM{F_\al(\eta-\xi)} + \norM{F_\al\xi}
    \leq
   \bigl(\sup_\al \norM{F_\al}\bigr)\norM{\eta-\xi} + \ep
    \\
    \ats
    < 
   (\sup_\al \norM{F_\al} + 1) \ep 
 \end{aligN}
 Since $\ep$ can be chosen arbitrarily, it follows that   
 $\norM{F_\al\eta}\underset{\al}{\to}0$.

 \textsl{Only if:}
 Suppose $\norM{F_\al\eta}\to 0$ for all $\eta\in\HH$. 
 Then, $\sup_\al \norM{F_\al} < \infty$ 
 by the uniform boundedness principle, and trivialiter
 $\norM{F_\al\xi}\to 0$ for all $\xi\in\mathcal{X}$. 
\end{pf}


\section{Fock space}

\begin{lemma}
\label{lm_ipSep} 
 Suppose $w_1, w_2, \ldots, w_m\in \K$, $w_i\neq 0$ (all $i$).
 Then there exists a $u\in \K$ such that $\ip{u}{w_i}\neq 0$ (all $i$).

 As a consequence:
 if $f_1,\ldots, f_n$ are $n$ different vectors in $\K$,
 then there exists a $u\in\K$ such that $\ip{u}{f_1}, \ldots, \ip{u}{f_n}$
 are $n$ different (complex) numbers.
\end{lemma}

\begin{pf}
Let $H_i:=\{ f\in \K : \ip{f}{w_i} = 0 \}$, $i=1,\ldots,m$,   
be the ortho-complements of $\CC \cdot w_i$ in $\K$. 
All of these are closed (linear) 
subspaces of $\K$, and none of them contains an open ball (i.e.\ a subset 
$B(\ep,h):=\{g \in \K: \norM{g-h} < \ep \}$ for some $\ep>0$ and $h\in\K$).
Therefore their union cannot be all of $\K$.
\begin{pf}
 This is in fact an application of Baire's theorem 
 in finite dimensional complete metric spaces.
 Indeed, $U_i:= \K\setminus H_i$, $i=1,\ldots,m$ are open, dense subsets of
 $\K$. Their intersection is open too, and, since only a finite number of
 open sets are involved, the intersection is also dense in $\K$, 
 whence non-empty. That means that the union of the $H_i$ cannot be all of
 $\K$.
\end{pf}
Take $u\in \K$ such that $u \not \in H_i$ ($i=1,\ldots,m$). 
Then $\ip{u}{w_i} \neq 0$ for all $i$.

The corollary follows by taking as $w_i$ all the differences $f_k - f_l$ with
$k\neq l$.
\end{pf}

\begin{lemma}
Let $\K$ and $\HH$ be Hilbert spaces, 
and suppose that for each $f\in\K$ a vector $e(f)\in\HH$ is given 
such that $\ip{e(g)}{e(f)}_\HH = \exp(\ip{g}{f}_\K)$.
Then $\{e(f) : \: f\in\K \}$ is linearly independent.
\end{lemma}

\begin{pf}
 Let $f_1,\ldots, f_n$ be $n$ different vectors in $\K$, and suppose there
 exists scalars $\al_1,\ldots,\al_n$ such that 
 $\sum_{i=1}^n \al_i e(f_i) = 0$.

 Choose a $g\in \K$ such that $\ip{g}{f_1},\ldots,\ip{g}{f_n}$ are $n$ 
 different complex numbers (use lemma \ref{lm_ipSep}.
 Then:
 \begin{displaymatH}
   \sum_1^n \al_i \exp(t \ip{g}{f_i})
    = \ip{ e(tg)}{\sum_1^n \al_i e(f_i)}
    = 0
    \quad (t\in\RR)
    \;. 
 \end{displaymatH}
 Since the system of exponential functions,
 $\{ t\mapsto \exp(t z) : z\in \CC \}$, is linearly independent,
 we conclude that $\al_1=\ldots=\al_n=0$.
\end{pf}

Let $\Ep(\K):= \{ \expv(f): f\in \K \} $ 
be the set of exponential vectors  in $\K$. 
\begin{lemma}
 $\Ep(\K)$ is an independent total system of $\F(\K)$.
\end{lemma}

\begin{pf}
 The independence follows from the lemma above, for the totalness observe that 
 $\displaystyle \frac{d\phantom{t^n}}{dt^n} \expv(tf)
  = f \otimes \ldots \otimes f$ (elaborate).
\end{pf}

\section{(Cyclic) (vacuum) representations of the CCR 
         over a Hilbert space $\K$}

\begin{lemma}
 All cyclic vacuum representations are unitarily equivalent
\end{lemma}

The norm on $\B(\HH)$ can be expressed in term of the
vacuum state.

\begin{lemma}
 Expression of inner product, and hence the 
 norm in terms of vacuum state for cyclic representations

 $\ip{F_2\Om}{F_1\Om} = \om(F_2^{*}F_1)$
\end{lemma}

As a result of the above, a 
\textsl{cyclic vacuum} representation of the CCR 
over a Hilbert space $\K$
enables one to express operator norm, strong, and weak convergence 
in $\Gamma_0(\K)$ in terms of the vacuum state.

\begin{lemma}
\label{lm_convCritWeyl}
 Let $(W,\Om)$ be a cyclic vacuum representation of the CCR over $\K$.
 Then:
\begin{Properties}[.2em]

  \item $A_i \to A$ in  $\Gamma_0(\K)$ strongly if and only if 
        $\om\Bigl(F^{*}(A_i-A)^{*}(A_i-A)F\Bigr)\to 0$ 
        for all $F \in \Gamma_0(\K)$.
\end{Properties}
\end{lemma}

\begin{lemma}
 Let $(W,\Om)$ be a vacuum representation of the CCR over $\K$ 
 in $\U(\HH)$ (not necessarily cyclic),
 i.e.\ $W:\K \to \U(\HH)$ is a unitary representation of $(\K,+)$ and
 $\Om \in\HH$ is a unit vector such that
 $\ip{\Om}{W(f)\Om}= e^{-\half\norM{f}^2}$ for all $f\in\K$.

 Then $\{W(f): f\in\K\}$ is linearly independent in $\U(\HH)$.
\end{lemma}

\begin{pf}
 Let $e(f):= e^{\half\norM{f}^2}W(f)\Om$  ($f\in\K$). We will reduce the
 independence of $\{W(f): f\in\K\}$ to that of $\{e(f): f\in\K\}$.

 First, a calculation shows that the $e(f)$, $f\in\K$,
 satisfy $\ip{e(g)}{e(f)}= \exp(\ip{g}{f})$ ($f,g\in\K$) and 
 are therefore linearly independent.

 Now suppose that $f_1, \ldots, f_n$ are $n$ different vectors in $\K$ 
 such that $\sum_1^n \al_i W(f_i) = 0 $ for some scalars 
 $\al_1,\ldots,\al_n$.
 Then:
 \begin{aligN}
  0 \ats = \Bigl(\sum_1^n \al_i W(f_i) \Bigr) \Om
      =  \sum_1^n \bigl(\al_i e^{-\half\norM{f_i}^2}\bigr)
        \cdot \, e^{\half\norM{f_i}^2} W(f_i)\Om 
      \\
    \ats = \sum_1^n (\al_i e^{-\half\norM{f_i}^2})
        \cdot e(f_i)
    \;.
 \end{aligN}
 By the independence of the $e(f)$, ($f\in\K$), 
 $\al_i e^{-\half\norM{f_i}^2} = 0$, hence $\al_i=0$ for all $i$. 
\end{pf}

\begin{corollary}
 Let $(W,\Om)$ be a vacuum representation of the CCR over $\K$ 
 in $\U(\HH)$. 

 Let $\Gamma_0(\K)$ be the linear span of $\{W(f): f\in\K\}$ in $\B(\HH)$.
 Then every $A\in \Gamma_0(\K)$ has a unique decomposition 
 $A=\sum_1^n \al_i W(f_i)$.
\end{corollary}

\section{$\Gamma$-functor}

Given a contraction $C:\K_1\to\K_2$ and 
a representation of the CCR over $\K_1$, 
we can use the above corollary, 
to define a linear map $\Gamma(C): \Gamma_0(\K_1) \to \Gamma_0(\K_2)$
by
\begin{displaymatH}
 \Gamma(C): W(f) \mapsto e^{-\half(\norM{f}^2 - \norM{Cf}^2)}  W(Cf) 
  \quad
  (f\in \K_1) \;.
\end{displaymatH}

We set out to prove:

\begin{theorem}
\label{th_contFunctor}
 Let $(W_1,\Om_1)$ and $(W_2,\Om_2)$ 
 be \textsl{cyclic} vacuum representations 
 of the CCR over $\K_1$ and $\K_2$ respectively .	

 Then
 \begin{Properties}
  \item The map $\Gamma(C)$ is a completely positive map 
        $\Gamma_0(K_1) \to \Gamma_0(K_2)$.

  \item The map $\Gamma(C)$ is continuous with respect to the 
        operator norm topology, and it has therefore 
        a unique extension to the $C^*$-algebra 
        generated by $\Gamma_0(\K_1)$ and $\Gamma_0(\K_2)$.
        
  \item The map $\Gamma(C)$ is continuous with respect to the 
        strong operator topology, and it has therefore 
        a unique extension to the von Neumann-algebras $\Gamma(\K_1)$ and
        $\Gamma(\K_2)$ 
        generated by $\Gamma_0(\K_1)$ and $\Gamma_0(\K_2)$ respectively.
       
  \item By continuity, both extensions inherit the complete positivity.
 \end{Properties}
\end{theorem}

The proof of theorem \ref{th_contFunctor} rest on two  pillars.

First, in a \textsl{cyclic vacuum} representation one can 
express convergence in terms of the vacuum state 
(see lemma \ref{lm_convCritWeyl}).
Second, we can express the state of the system after the action of
$\Gamma(C)$ in terms of the state of the system before 
(see lemmas \ref{lm_GammaContTool}).

\medskip 

We now provide the details in a series of lemmata.
To lighten the burden of lengthy expressions in the calculus of Weyl operators,
we fix some notation, and derive some calculation rules which will thereafter
be used without further reference:

\question{Check that we use consistly
A FIXED CHOICE OF $\ip{\Om}{W(f)\Om}=\om(f)$
instead of$\ip{\Om}{W(f)\Om}=\om(W(f))$}

\begin{lemma}
\label{lm_calcWeyl}
 Let $(W,\Om)$ be a cyclic vacuum representation of the CCR over $\K$.
 Define
 \begin{aligN}
  \om(f): \ats =\exp(-\half\norM{f}^2)  \in [0,\infty) \quad (f\in\K) \;, 
   \\
  \sx{f}{g}:\ats =\exp(-i\Im\ip{f}{g}) \in \{ z\in\CC : |z|=1 \} 
   \quad (f,g\in\K) \;.
 \end{aligN}

 Then for $f,g,f_1,f_2,f_3,f_4\in\K$:
 \begin{Properties}[.4em]
  \item $\ip{\Om}{W(f)\Om}= \om(f)$;
  \item $W(f)W(g) = \sx{f}{g} W(f+g)$;
  \item $\sx{g}{f}= \sx{-f}{g}=\sx{f}{-g}=\overline{\sx{f}{g}}
                  =\sx{f}{g}^{-1}$;
  \item $\sx{f_1+f_2}{f_3+f_4}=\sx{f_1}{f_3}\; \sx{f_1}{f_4} \;
                               \sx{f_2}{f_3}\; \sx{f_2}{f_4} $;
  \item $\om(f_1+f_2)\sx{f_1}{f_2}=\om(f_1)\om(f_2)\exp(-\ip{f_1}{f_2})$;
  \item $\om(-f)=\om(f)$;
  \item $W(f)^*=W(-f)$, since $W$ is unitary.
  \item \label{it_omOfSum} 
     	$\om(f_1+f_2+f_3+f_4)$
        \\
        $ 
           = \om(f_1)\om(f_2)\om(f_3)\om(f_4)
        $
        \\
        $ \times 
        \exp(-\Re\ip{f_1}{f_2}) \exp(-\Re\ip{f_1}{f_3}) \exp(-\Re\ip{f_1}{f_4})
        $
        \\
        $
        \times
         \exp(-\Re\ip{f_2}{f_3})\exp(-\Re\ip{f_2}{f_4})
	    \exp(-\Re\ip{f_3}{f_4});
        $
  \item \label{it_sumWeyl}
        $W(f_1)\: W(f_2)\: W(f_3)\: W(f_4)$
        \\
        $ = 
         \sx{f_1}{f_2}\sx{f_1}{f_3}\sx{f_1}{f_4}
	 \sx{f_2}{f_3}\sx{f_2}{f_4}         
	 \sx{f_3}{f_4} W(f_1+f_2+f_3+f_4)$;
 \end{Properties}

 \medskip

 Conclusion: for $f_1,f_2,f_3,f_4\in\K$ we have $\oip{f}{g}{h}$
 \begin{aligN}
   \langle\Om | W(f_1)\: & \textstyle W(f_2)\: W(f_3)\: W(f_4) 
       | \Om \rangle
       = 
       \Pi_{i=1}^4 \om(f_i) \Pi_{1\leq i < j \leq 4} \exp(-\ip{f_i}{f_j})
            \\
          = \ats \;\;
          \om(f_1)\om(f_2)\om(f_3)\om(f_4) 
                   \\
          \ats
          \exp(-\ip{f_1}{f_2}\exp(-\ip{f_1}{f_3})\exp(-\ip{f_1}{f_4})
          \\
          \ats
          \exp(-\ip{f_2}{f_3})\exp(-\ip{f_2}{f_4})
          \\
          \ats
          \exp(-\ip{f_3}{f_4})
     \;.
 \end{aligN}
\end{lemma}

\begin{pf}
 
 (\romanref{it_omOfSum}) Observe  
 $\norM{\sum_1^4 f_i}^2= \ip{\sum_{i=1}^4 f_i}{\sum_{j=1}^4 f_j}
  = \sum_{i,j} \ip{f_i}{f_j} 
  = \sum_1^4 \norM{f_i}^2 
     + \sum_{1\leq i < j \leq 4} \bigl[\ip{f_i}{f_j} + \ip{f_j}{f_i}\bigr]
  = \sum_1^4 \norM{f_i} 
    + \sum_{1\leq i < j \leq 4} 2 \Re(\ip{f_i}{f_j})
  $.

 \medskip

 (\romanref{it_sumWeyl})
 Working from the right to the left we find 
 \begin{aligN}
  W(f_1)\ats \; W(f_2)\;\Bigl[ W(f_3) W(f_4)\Bigr] 
     =   W(f_1)\; W(f_2)\; \Bigl[ \sx{f_3}{f_4} W(f_3+f_4) \Bigr]
           \\
       \ats = \sx{f_3}{f_4} \;\; W(f_1) \Bigl[\sx{f_2}{f_3+f_4} W(f_2+ f_3+f_4)\Bigr]
         \\
       \ats = \sx{f_3}{f_4}  \sx{f_2}{f_3+f_4}\;\; 
                 \Bigl[\sx{f_1}{f_2+f_3+f_4} W(f_1+f_2+ f_3+f_4)\bigr]
         \\
       \ats =
        \underbrace{\sx{f_1}{f_2}\sx{f_1}{f_3}\sx{f_1}{f_4}}_{}
        \underbrace{\sx{f_2}{f_3}\sx{f_2}{f_4}}_{}
        \underbrace{\sx{f_3}{f_4}}_{}
        W(f_1+f_2+ f_3+f_4)
             \;.
 \end{aligN} 

 For the conclusion observe that $\sx{f}{g}\exp(-\Re\ip{f}{g})=\exp(-\ip{f}{g})$,
 and then combine (\romanref{it_omOfSum}) and (\romanref{it_sumWeyl}).
\end{pf}

We now return to our map $\Gamma(C):\Gamma_0(\K_1)\to\Gamma_0(\K_2)$.

To begin with, we calculate  $\Gamma(C) \bigl[ W_1(f') W_1(f) \bigr]$:
\begin{lemma}
\label{lem_prod_W}
 Take $f,f'\in\K_1$. Then:
 \begin{equation}
 \label{eq_prod_W}
 \Gamma(C) \bigl[ W_1(f') W_1(f) \bigr]
   = 
 {\displaystyle
        \frac{\om_1(f')\om(f)\exp(-\ip{f'}{f})}%
                        {\om_2(Cf')\om_2(Cf)\exp(-\ip{Cf'}{Cf})}%
       W_2(Cf')\;W_2(Cf)
      }
       \;.
 \end{equation}
 In particular,
 \begin{displaymatH}
 \Gamma(C) \bigl[ W_1(f')^{*} W_1(f) \bigr]
   = 
  L(f',f) \cdot W_2(Cf')\;W_2(Cf)
  \;,
 \end{displaymatH}
 where
 \begin{displaymatH}
  L(f',f) 
   = 
 \displaystyle
        \frac{\om_1(f')\om_1(f)}{\om_2(Cf')\om_2(Cf)}
        \exp(\ip{f'}{(I-C^{*}C)f})
  \;,
 \end{displaymatH}
 is a positive definite kernel.
\end{lemma}

\begin{pf}
 By the above calculation rules we have 
 $\om(f_1+f_2)\sx{f_1}{f_2}=\om(f_1)\om(f_2)\exp(-\ip{f_1}{f_2})$, 
 so that
 \begin{aligN}
  \Gamma(C) \bigl[ W_1(f') W_1(f) \bigr]
      \ats = 
       \sx{f'}{f}\Gamma(C)\bigl[ W_1(f'+f) \bigr]
      \\
      \ats
      = 
      {\displaystyle 
       \frac{\om_1(f'+f)}{\om_2(Cf'+Cf)}
       \sx{f'}{f}
       W_2(Cf'+Cf)
      }
      \\ 
      \ats
       =
      {\displaystyle
       \frac{\om_1(f'+f)}{\om_2(C(f'+f)}
       \frac{\sx{f'}{f}_1}{\sx{Cf'}{Cf}_2}
       W_2(Cf')\;W_2(Cf)
      }
      \\
      \ats
      =
      {\displaystyle
        \frac{\om_1(f')\om(f)\exp(-\ip{f'}{f})}%
                        {\om_2(Cf')\om_2(Cf)\exp(-\ip{Cf'}{Cf})}%
       W_2(Cf')\;W_2(Cf)
      }
       \;.
 \end{aligN}

 The second part of the lemma follows 
 by observing that $W(f')^{*}= W(-f)$ and that both
 \begin{displaymatH}
    (f',f)\mapsto{\displaystyle\frac{\om_1(f')\om(f)}{\om_2(Cf')\om_2(Cf)}}
 \end{displaymatH}
 and
 \begin{displaymatH}
   (f',f)\mapsto \exp(+\ip{f'}{(I-C^{*}C)f})
 \end{displaymatH}
 are positive definite kernels, and so their product is. 
\end{pf}

\section{Complete positivity of $\Gamma(C)$}
As a result the map $\Gamma(C)$ is completely positive:
\begin{lemma}
\label{lm_complPosGamma}
 Let $F_1,\ldots, F_N\in \Gamma_0(\K_1)$, and $\xi_1,\ldots,\xi_N \in \HH_2$.
 Then 
 \begin{displaymatH}
  \sum_{p,q=1}^N \ip{\xi_q}{\Gamma(C)\bigl[F^{*}_q F_p \bigr] \xi_p}\; \geq 0
   \;,  
 \end{displaymatH}
 i.e.
 $\Gamma(C):\Gamma_0(\K_1)\to\Gamma_0(\K_2)$ is completely positive.
\end{lemma}

\begin{proof}
 To facilitate the reasoning, we employ the following notational trick:
 we write $F_p=\sum_{k=n_p+1}^{n_{p+1}} \al_k W(f_k)$ 
 for suitable $0=:n_1< \ldots < n_{N+1}=:K$, and let
 and $\eta_k$, for $k\in\{n_p+1,\ldots, n_{p+1}\}$, be identical to $\xi_p$.

 Then
 \begin{aligN}
  \sum_{p,q}\ats \ip{\xi_q}{\Gamma(C)\bigl[F^{*}_q F_p \bigr] \xi_p}
    \\
    \ats =
  \sum_{p,q=1}^N \ip{\xi_q}%
             {%
       \Gamma(C)\Bigl[ \Bigl( \sum_{l=n_q+1}^{n_{q+1}} \al_l W(f_l) \Bigr)^{*}%
                       \Bigl( \sum_{k=n_p+1}^{n_{p+1}} \al_k W(f_k) \Bigr)%
                 \Bigr]%
              \xi_p%
             }%
    \\
    \ats = 
  \sum_{p,q=1}^N \sum_{l={n_q+1}}^{n_{q+1}}\sum_{k=n_p+1}^{n_{p+1}} 
                 \overline{\al_l} \al_k 
                   \cdot
                 L(f_l,f_k)
                   \cdot
                 \ip{\eta_l}{W(C f_l)^{*} \; W(C f_k) \eta_k}       
   \\
   \ats =
  \sum_{k,l=1}^K \overline{\al_{l}} \al_{k} 
                 \cdot
                 L(f_{l},f_{k})
                 \cdot
                 \ip{W(C f_{l})\eta_l}{W(C f_{k})\eta_k}       
  \;,
 \end{aligN}
 Now, the matrix 
 $M= \bigl( \ip{W(Cf_l)\eta_l}{W(Cf_k)\eta_k} \bigr)_{k,l=1}^K$ 
 is positive definite. Letting $N=N^{*}$ be the (positive) square root of $M$,
 we can write:
 \begin{displaymatH}
  \ip{W(Cf_l)\eta_l}{W(Cf_k)\eta_k}= \sum_{m=1}^K (NN^{*})_{kl}
   = \sum_m N_{km} N_{ml}^{*}  = \sum_m \overline{N_{lm}} N_{km}  
   \;,
 \end{displaymatH}
 and hence:
 \begin{aligN}
  \sum_{p,q}\ats \ip{\xi_q}{\Gamma(C)\bigl[F^{*}_q F_p \bigr] \xi_p}
    \\
    \ats =
  \sum_{k,l=1}^K \overline{\al_{l}} \al_{k} 
                 \cdot
                 L(f_{l},f_{k})
                 \cdot
                 \bigl( 
                        \sum_{m=1}^K \overline{N_{lm}} N_{km}
                 \bigr)
   \\
   \ats =
  \sum_m \Bigl[ \sum_{k,l} 
                       \overline{\bigl( \al_l N_{lm} \bigr)} 
                                 \cdot 
                       \bigl(\al_k N_{km} \bigr) 
                                 \cdot
                       L(f_{l},f_{k})
         \Bigr]
    \geq 0
    \;,
 \end{aligN}
 because $L(\cdot,\cdot)$ is a positive definite kernel.
\end{proof}

\begin{corollary}
\label{cor_complPos}
 Let $F_1,\ldots, F_N\in \Gamma_0(\K_1)$ 
 and $G_1,\ldots, G_N\in \Gamma_0(\K_2)$.
 Then 
 \begin{displaymatH}
  \sum_{p,q=1}^N G_q^{*} \; \Gamma(C)\bigl[ F_q^{*} F_p \bigr] \; G_p  \geq 0
  \;,
 \end{displaymatH}
 symbolically:
 \begin{displaymatH}
  \begin{matrix}
   G_1,\ldots,G_N   
  \end{matrix}
   \cdot
  \Gamma(C)
  \begin{matrix}
   F_1^{*}F_1 \ldots F_1^{*}F_N\\
     \ddots \\
   F_N^{*}F_1 \ldots F_N^{*}F_N
  \end{matrix}
   \cdot
  \begin{matrix}
   G_1\\
   \vdots\\
   G_N
  \end{matrix}
  \geq 
  0
 \end{displaymatH}
\end{corollary}

\begin{pf}
 Let $\xi\in \HH_2$. 
 Setting $\xi_n:=G_n \xi$, yields
 \begin{displaymatH}
  \ip{\xi}{\Bigl(%
            \sum_{p,q=1}^N G_q^{*}\;\Gamma(C)\bigl[ F_q^{*} F_p \bigr]\;G_p%
           \Bigr)%
           \xi}%
   =\sum_{p,q} \ip{\xi_q}{\Gamma(C)\bigl[ F_q^{*} F_p \bigr] \xi_p} \geq 0
   \;,
 \end{displaymatH}
 by the complete positivity of $\Gamma(C)$.
\end{pf}

An elementary consequence of the above is:
\begin{lemma}[Schwartz inequality for %
               $\Gamma(C):\Gamma_0(\K_1)\to\Gamma_0(\K_2)$]
\label{lm_SchwartzIneq}
 Let $F\in \Gamma_0(\K_1)$. 
 Then 
 $\bigl[\Gamma(C)(F)\bigr]^{*}\cdot \Gamma(C)(F) 
     \leq 
  \Gamma(C)\bigl[F^{*}F)\bigr]$.
\end{lemma}

\begin{pf}
 Take $N=2$, $F_1=I$, $F_2=F$, $G_1=\Gamma(C)(F)$, $G_2=-I$ 
 in \ref{cor_complPos}, and do not forget that 
 $\Gamma(C)\bigl[F^{*}\bigr] = \bigl[\Gamma(C)F\bigr]^{*}$.
\end{pf}

\section{Continuity of $\Gamma(C)$}

\begin{lemma}
\label{lm_om2INom1}
 Let $(W_1,\Om_1)$ and $(W_2,\Om_2)$ 
 be \textsl{cyclic} vacuum representations 
 of the CCR over $\K_1$ and $\K_2$ respectively.	
 Denote the corresponding vacuum states by $\om_1$ and $\om_2$ respectively.

Take $f,f'\in\K_1$, and $g',g\in\K_2$. Then:
\begin{aligN}
 \om_2 \ats
     \Biggl( W_2(g') \Bigl( 
                     \Gamma(C)\bigl[ W_1(f')W_1(f) \bigr] 
                     \Bigr) 
             W_2(g) 
     \Biggr)
    \\
   \ats 
    =  
   {\displaystyle\frac{\om_2(g')\om_2(g)}{\om_1(C^{*}g')\om_1(C^{*}g)}}
   \exp(-\ip{g'}{(I-CC^{*})g}_2)
   \\
   \ats
   \phantom{mmm}\times
   \om_1\Bigl( W_1(C^{*}g')\; W_1(f')\; W_1(f) W_1(C^{*}g) \Bigr)
   \;.
\end{aligN}
\end{lemma}

\begin{pf}
 Filling in the above expression and using the result of the above lemma,
 we find
 \nopagebreak
 \begin{aligN}
 \om_2 \ats
     \Biggl( 
             W_2(g') \Bigl( 
                     \Gamma(C)\bigl[ W_1(f')W_1(f) \bigr] 
                     \Bigr) 
             W_2(g) 
     \Biggr)
    \\
   \ats 
    =  
   {\displaystyle
    \Biggl(
    \frac{\om_1(f')\om_1(f)\exp(-\ip{f'}{f})}{\om_2(Cf')\om_2(Cf)\exp(-\ip{Cf'}{Cf})}
    \Biggr) 
   }
   \\
   \ats\phantom{mm}\times
   \om_2(g')\om_2(Cf')\om_2(Cf)\om_2(g)                 
    \\
    \ats
   \phantom{mm}\times
   e^{-\ip{g'}{Cf'})} e^{-\ip{g'}{Cf})} e^{-\ip{g'}{g}}
   e^{-\ip{Cf'}{Cf}}  e^{-\ip{Cf'}{g}}
   e^{-\ip{Cf}{g}}
   \\
   \ats
   =
   \om_1(f')\om_1(f)   \om_2(g')\om_2(g)
   \\
   \ats
   \phantom{mm}\times
   e^{-\ip{f'}{f}}
   e^{-\ip{g'}{Cf'}} e^{-\ip{g'}{Cf}}    e^{-\ip{g'}{g}}
   e^{-\ip{Cf'}{g}}
   e^{-\ip{Cf}{g}}
   \\ 
   \ats
    = 
  \Biggl(
  {\displaystyle
   \frac{\om_2(g')\om_2(g)}{\om_1(C^{*}g')\om_1(C^{*}g)}    
   \frac{\exp(-\ip{g'}{g}}{\exp(-\ip{C^{*}g'}{C^{*}g})}
  }
  \Biggr)
  \\
  \ats\phantom{mm}\times
  \om_1(C^{*}g')\om_1(f')\om_1(f)\om_1(C^{*}g)   
  \\
  \ats\phantom{mm}\times
   e^{-\ip{C^{*}g'}{f'}}   e^{-\ip{C^{*}g'}{f}}     e^{-\ip{C^{*}g'}{C^{*}g}}
   e^{-\ip{f'}{f}}    e^{-\ip{f'}{C^{*}g}}
   e^{-\ip{f}{C^{*}g}}
  \\
  =
  \ats 
  \Biggl(
  {\displaystyle
   \frac{\om_2(g')\om_2(g)}{\om_1(C^{*}g')\om_1(C^{*}g)}    
   \frac{\exp(-\ip{g'}{g})}{\exp(-\ip{Cg'}{Cg})}
  }
  \Biggr)
  \\
  \ats\phantom{mm}\times
  \om_1\Bigl( W_1(C^{*}g')\; W_1(f') \; W_1(f) \; W_1(C^{*}g) \Bigr)
  \;,
 \end{aligN}
 where in the one but last equality we have divided and multiplied by 
 $\om_1(C^{*}g')\om_1(C^{*}g)e^{-\ip{Cf'}{Cf}}$, and in the last
 equality we have used lemma \ref{lm_calcWeyl} for the identification 
 with $\om_1(\cdot)$.
\end{pf}

\begin{corollary}
\label{cor_om2INom1}
Take $f,f'\in\K_1$, and $g',g\in\K_2$. Then:
\begin{aligN}
 \Bigl< \Om_2 \ats 
         \Bigm|
             W_2^{*}(g') \Bigl( 
                     \Gamma(C)\bigl[ W_1^{*}(f')W_1(f) \bigr] 
                     \Bigr) 
             W_2(g) 
     \Bigm|
 \Om_2 \Bigr>
    \\
   \ats 
    =  
   K(g',g)
   \cdot
   \Bigl< \Om_1 
         \Bigm| 
           W_1^{*}(C^{*}g')\; W_1^{*}(f')\; W_1(f) W_1(C^{*}g) 
         \Bigm|
          \Om_1
    \Bigr>
   \;.
\end{aligN}
where
\begin{displaymatH}
   K(g',g) = 
   {\displaystyle\frac{\om_2(g')\om_2(g)}{\om_1(C^{*}g')\om_1(C^{*}g)}}
   \exp(+\ip{g'}{(I-CC^{*})g}_2)
  \qquad
  (g',g \in\K_2)
\end{displaymatH}
is a positive definite kernel.
\end{corollary}

\begin{pf}
 Since $W^{*}(h)=W(-h)$, we only need to replace $g'$ and $f'$ in lemma 
 \ref{lm_om2INom1} by $-g'$ and $-f'$ respectively. 
 Since $\om(-h)=\om(h)$ the factors 
 $\om(\cdot)$ do not change, however the factor $\exp(-\ip{g'}{(I-CC^{*})g})$ 
 becomes $\exp(+\ip{g'}{(I-CC^{*})g})$.

 As the positive definiteness concerns: both 
 \begin{displaymatH}
    (g',g)\mapsto
      {\displaystyle\frac{\om_2(g')\om_2(g)}{\om_1(C^{*}g')\om_1(C^{*}g)}}
 \end{displaymatH}
 and
 \begin{displaymatH}
   (g',g)\mapsto \exp(+\ip{g'}{(I-CC^{*})g})
 \end{displaymatH}
 are positive definite kernels,
 and so is there product. 
\end{pf}

The following tool is crucial for establishing the operator norm and strong
continuity of $\Gamma(C):\Gamma_0(\K_1)\to\Gamma_0(\K_2)$.

\begin{lemma}
\label{lm_GammaContTool} 
 Let $G\in\Gamma_0(\K_2)$ be fixed. 

 Then there exist $\widetilde{G}_1,\ldots, \widetilde{G}_m \in \Gamma_0(\K_1)$ 
 (only depending on $G$)
 such that 
\begin{equatioN}
 \om_2 
     \Biggl( G^{*} \Bigl( 
                     \Gamma(C)\bigl[ F^{*} F \bigr] 
                     \Bigr) 
              G 
     \Biggr)
    =  
   \dsum_{k=1}^m
   \om_1\Bigl( 
               {\widetilde{G}_k}^{*}\; \bigl[F^{*} F \bigr]\; \widetilde{G}_k 
        \Bigr)
  \qquad
  (F\in\Gamma_0(\K_1))
   \;.
  \tag{$*$}
\end{equatioN}
Alternatively formulated:

Let $\xi$ be in the dense linear subspace $\Gamma_0(\K_2)\Om_2$ of $\HH_2$.
Then there exist $\widetilde{\xi}_1,\ldots, \widetilde{\xi}_m$ in the
dense linear subspace $\Gamma_0(\K_1)\Om_1$ of $\HH_1$
(only depending on $\xi$)
such that
\begin{equatioN}
 \dsum_{k=1}^m
  \norM[big]{\widetilde{\xi}_k}_{\HH_1}^2
  =
 \norM[big]{\xi}_{\HH_2}^2   \;,
 \tag{$\%1$}
\end{equatioN}
and
\begin{equatioN}
 \norM[Big]{\Bigl(\Gamma(C)\bigl[F^{*}F\bigr]\bigr)\xi}_{\HH_2}^2
  =
 \dsum_{k=1}^m
  \norM[Big]{F \widetilde{\xi}_k}_{\HH_1}^2
  \qquad
  (F\in\Gamma_0(\K_1))
   \;,
  \tag{$\%2$}
\end{equatioN}
which implies by the Schwartz inequality that
\begin{equatioN}
 \norM[Big]{\bigl[\Gamma(C)F\bigr]\xi}_{\HH_2}^2
  \leq
 \dsum_{k=1}^m
  \norM[Big]{F \widetilde{\xi}_k}_{\HH_1}^2
  \qquad
  (F\in\Gamma_0(\K_1))
   \;.
  \tag{$\%3$}
\end{equatioN}

\end{lemma}

\begin{pf}
 \statement{Proof of $(*)$}

 Looking at corollary \ref{cor_om2INom1} while keeping in mind that
 $\om_1, \om_2$, and $\Gamma(C)$ are linear, we see that for \textsl{each} 
 $F=\sum_{i=1}^n \al_i W(f_i) \in\Gamma_0(\K_1)$
 \begin{aligN}
 \om_2 \ats
     \Biggl( W_2^{*}(g') \Bigl( 
                     \Gamma(C)\bigl[ F^{*}F \bigr] 
                     \Bigr) 
             W_2(g) 
     \Biggr)
    \\
   \ats 
    =  
   K(g',g)
   \cdot
   \om_1\Bigl( W_1^{*}(C^{*}g')\; F^{*}F\;  W_1(C^{*}g) \Bigr)
   \;.
 \end{aligN}

 Consequently, if we fix $G=\sum_{j=1}^m \be_j W_2(g_j) \in \Gamma_0(\K_2)$
 then
\begin{equatioN}
\begin{aligned}
\om_2 &
     \Biggl(\!\! G^{*} \Bigl( 
                     \Gamma(C)\bigl[ F^{*}F \bigr] 
                     \Bigr) 
             G 
     \!\Biggr)
    \\
    & =  
   \sum_{j,j'} \overline{\be_{j'}} \be_{j} K(g_{j'},g_j)
   \cdot
   \om_1\Bigl( {W_1(C^{*}g_{j'})}^{*}  \bigl[F^{*}F\bigr]  W_1(C^{*}g_j) \Bigr)
   \;.
\end{aligned}
   \tag{$\#$}
\end{equatioN}
 Since $(g',g)\mapsto K(g',g)$ is positive definite, 
 the $m\times m$ matrix $\bigl( K(g_{j'}, g_{j}) \bigl)_{j,j'=1}^m$ is
 positive definite: let $N=N^{*}= \bigl( N(g_{j'}, g_j) \bigr)_{j,j'=1}^m$
 be its (positive) square root. 
 Then 
 \begin{aligN}
  K(g_{j'}, g_j) 
    \ats = 
        {(N^2)}_{j',j} = (N^{*} N)_{j',j}= 
        \sum_{k=1}^m N^{*}_{j',k} N_{k,j}= 
        \sum_{k=1}^m \overline{N_{k,j'}} N_{k,j}
        \\
    \ats =
    \sum_{k=1}^m \overline{N(g_k,g_{j'})} N(g_k,g_j)
        \;.
 \end{aligN}
 Plugging this in $(\#)$  yields
\begin{displaymatH}
 \om_2 
     \Bigl( G^{*} \Bigl( 
                     \Gamma(C)\bigl[ F^{*}F \bigr] 
                     \Bigr) 
             G 
     \Bigr)
    =  
   \sum_{k=1}^m 
   \om_1\Bigl( \widetilde{G}^{*}_k
                  \bigl[F^{*}F\bigr]  
               \widetilde{G}_k
        \Bigr)
   \;,
 \end{displaymatH}
 where
 \begin{displaymatH}
   \widetilde{G}_k=\sum_{j}^m \be_j  N(g_k,g_j) \cdot W_1(C^{*}g_j)
   \;,
 \end{displaymatH}
 only depends on $G=\sum_{j}^m \be_j W_2(g_j)$ (via $C$ and $K(\cdot, \cdot)$).
 
 \statement{Proof of $(\%1-3)$}

 $(\%2)$ is a mere restatement of $(*)$, $\%1$ follows by taking $F=I$ 
 (realizing that $\Gamma(C)I=\Gamma(C) W(0)= I$), and $(\%3)$ results from
 $(\%2)$ by using the Schwartz inequality (see \ref{lm_SchwartzIneq}):
 \begin{aligN}
 \norM{\bigl[\Gamma(C)F\bigr]\xi}^2
  \ats
   = 
  \ip{\xi}{\bigl[\Gamma(C)F\bigr]^{*} \bigl[\Gamma(C)F\bigr]\xi} 
   \leq 
  \ip{\xi}{\Bigl(\Gamma(C)\bigl[F^{*}F]\Bigr)\xi}
  \\
  \ats
   =
 \norM{\bigl(\Gamma(C)\bigl[F^{*}F]\Bigr)\xi}^2
  \;.
 \end{aligN}
\end{pf}

\begin{corollary}
\label{cor_opNormContGamma}
 The map $\Gamma(C):\Gamma_0(\K_1)\to\Gamma_0(\K_2)$ is a contraction 
 with respect to the operator norm on $\B(\HH_1)$ and $\B(\HH_2)$ respectively,
 i.e.\ 
 \begin{displaymatH}
 \norM{\Gamma(C)F}_{\B(\HH_2)} \leq \norM{F}_{\B(\HH_1)}
 \qquad 
 (F\in\Gamma_0(\K_1)\,)
 \;.
 \end{displaymatH}
 In particular,  
 $\Gamma(C)$ can be uniquely extended to an
 operator norm continuous map between 
 $C^{*}$-algebras generated by $\Gamma_0(\K_1)$ and $\Gamma_0(\K_2)$ 
 respectively. The resulting extension is again completely positive.
\end{corollary}

\begin{pf}
 Let $F\in\Gamma_0(\K_1)$. We prove that 
 $\norM{\bigl[\Gamma(C)F\bigr]\xi}\leq \norM{F\xi}$ $(\xi\in\HH_2)$ in 
 two steps.

 \statement{(Step 1: $\norM{\bigl[\Gamma(C)F\bigr]\xi}\leq \norM{F}
                      \norM{\xi}$
            for $\xi$ in the dense linear subspace $\Gamma_0(\K_2)\Om_2$
            of $\HH_2$.)}

 Take $\xi \in\Gamma_0(\K_2)\Om_2$. 
 With $\widetilde{\xi}_1, \ldots,\widetilde{\xi}_m$ 
 as in lemma \ref{lm_GammaContTool} we have
 (by $(\%3)$ of the same lemma)
 \begin{aligN}
  \norM{\bigl[\Gamma(C)F\bigr]\xi}^2
   \ats 
    \leq
  \sum_{i=1}^m \norM{F\widetilde{\xi_i}}^2
    \leq
     \norM{F}^2 \cdot \sum_{i=1}^m \norM[text]{\widetilde{\xi_i}}^2
    = 
    \norM{F}^2 \norM{\xi}^2
    \;.
 \end{aligN}

 \statement{(Step 2: $\norM{\bigl[\Gamma(C)F\bigr]\xi}\leq \norM{F}
                      \norM{\xi}$
            for all $\xi \in \HH_2$)}

 Let $\xi\in\HH_2$ and let $\ep >0$. Take $\xi_0\in\Gamma_0(\K_2)\Om_2$ with
 $\norM{\xi-\xi_0}<\ep$.
 Then
 \begin{aligN}
  \norM{\bigl[\Gamma(C)F\bigr]\xi} 
   \ats
   \leq 
        \norM{\bigl[\Gamma(C)F\bigr](\xi-\xi_0)} 
         + 
        \norM{\bigl[\Gamma(C)F\bigr]\xi_0} 
   \\
   \ats
   \leq
       \norM{\Gamma(C)F}\norM{\xi-\xi_0} 
         +
        \norM{F}\norM{\xi_0} 
   \leq
       \norM{\Gamma(C)F}\ep
         +
        \norM{F}\Bigl( \norM{\xi_0-\xi} + \norM{\xi} \Bigr)
   \\
   \ats
   \leq
       \Bigl(\norM{\Gamma(C)F} + \norM{F}\Bigr) \ep
        +
       \norM{F}\norM{\xi}
       \;,
 \end{aligN}
 and since $\ep$ is arbitrarily, the proof is completed.
\end{pf}

Now for the strong continuity.
\begin{corollary}
 The map $\Gamma(C):\Gamma_0(\K_1)\to\Gamma_0(\K_2)$
 can be uniquely extended to a map between the von-Neumann algebras 
 $\Gamma(\K_1)$ and $\Gamma(\K_2)$ generated 
 by $\Gamma_0(\K_1)$ and $\Gamma_0(\K_2)$  respectively.
 The resulting extension is again strongly continuous and 
 completely positive.
\end{corollary}

\begin{pf}
First recall lemma \ref{lm_GammaContTool}:
\begin{equatioN}
  \xi\in\Gamma_0(\K_2)\Om_2 
    \implies
  \left\lgroup
  \begin{array}{l}
    \text{there exist }m\in\NN, \text{and } 
          \widetilde{\xi}_1,\ldots,\widetilde{\xi}_m
    \text{ such that } 
    \\[3pt]
    \quad\norM{\bigl[\Gamma(C)F\bigr]\xi}^2
      \leq
    \sum_{k=1}^m
         \norM[text]{F \widetilde{\xi}_k}^2 
    \qquad
    (F\in\Gamma_0(\K_1)
    \;.
  \end{array}
  \right.
  \tag{$\&$}
\end{equatioN}

 \statement{Definition of the extension of $\Gamma(C)$ to 
            $\Gamma(\K_1)$ and $\Gamma(\K_2)$}

 Let $\overline{F}\in\Gamma(\K_1)$. 
 We first define $\Gamma(C)\overline{F}$ on the dense linear subspace
 $\Gamma_0(\K_2)\Om_2$ of $\HH_2$ and extend it to all of $\HH_2$.

 Let $\xi\in\Gamma_0(\K_2)\Om_2$, 
 and let $\widetilde{\xi_1},\ldots,\widetilde{\xi}_m$ as in $(\&)$.
 Next, choose  a net  $(F_\al)_\al$ in 
 $\Gamma_0(\K_1)$ that strongly converges to $\overline{F}$.
 Then 
 \begin{displaymatH}
  \norM{\bigl[\Gamma(C)(F_\al - F_{\al'})\bigr]\xi}^2
      \leq
  \sum_{k=1}^m
      \norM[text]{(F_\al - F_{\al'})\widetilde{\xi}_k}^2
  \to 0 \quad(\al,\al'\to\infty)
    \;,
 \end{displaymatH}
 so $\bigl(\bigl[\Gamma(C)F_\al\bigr]\xi\bigr)_\al$ is a Cauchy net in $\HH_2$,
 having a limit $\eta$.

 If $(G_\la)_\la$ were another net in $\Gamma_0(\K_1)$ converging strongly to
 $\overline{F}$ then again by $(\&)$,
 \begin{displaymatH}
  \norM{\bigl[\Gamma(C)(F_\al - G_{\la})\bigr]\xi}^2
      \leq
  \sum_{k=1}^m	
      \norM[text]{F_\al\widetilde{\xi}_k - G_{\la}\widetilde{\xi}_k}^2
     \to 
    0
  \quad(\al,\la\to\infty)
    \;,
 \end{displaymatH}
 since both $(F_\al\widetilde{\xi}_k)_\al$ and 
 $(G_{\la}\widetilde{\xi}_k)_\la$ converge to 
 $\overline{F}\widetilde{\xi}_k$ for $k=1,\ldots, m$.
 Therefore, the net
 $\bigl([\Gamma(C)G_{\la}]\xi\bigl)_\la$ also converges to $\eta$.

 Furthermore, 
 since  $\sum_{k=1}^m \norM[text]{\widetilde{\xi}_k}^2 =\norM{\xi}^2$, 
 we see that
 \begin{aligN}
  \norM{\lim_\al\bigl[\Gamma(C)F_\al\bigr]\xi}^2 
   \ats
     =  
  \lim_\al\norM{\bigl[\Gamma(C)F_\al\bigr]\xi}^2  
     \leq 
  \sum_{k=1}^m \lim_\al\norM{F_\al\widetilde{\xi}_k}^2 
  \\
  \ats
   = 
  \sum_{k=1}^m \norM{\overline{F}\widetilde{\xi}_k}^2 
   \leq
   \norM{\overline{F}}^2 \cdot \sum_{k=1}^m \norM{\widetilde{\xi}_k}^2 
    =
 \norM{\overline{F}}^2 \cdot \norM{\xi}^2 
  \;,
 \end{aligN}

 Conclusion: we can define a map 
 $\Gamma(C)\overline{F}:\Gamma_0(\K_2)\Om_2\to\HH_2$ by setting
 \begin{displaymatH}
   \left\lgroup
     \begin{array}{l}
      \bigl[\Gamma(C)\overline{F}\bigr]\xi 
        =\lim_\al \bigl[\Gamma(C)F_\al\bigr]\xi
        \quad(\xi\in\Gamma_0(\K_2)\,)\;,
        \\[5pt]
        \quad
        \text{whenever } \bigl(F_\al\bigr)_\al 
        \text{ is a(ny) net in }\Gamma_0(\K_1)
        \text{ that strongly converges to }\overline{F} 
        \;,
   \end{array}
   \right.
 \end{displaymatH}
 and $\Gamma(C)\overline{F}$ can be uniquely extended to all of $\HH_2$, with
\begin{equatioN}
  \norM{\Gamma(C)\overline{F}}_{\B(\HH_2)} 
      \leq 
   \norM{\overline{F}}_{\B(\HH_1)}
  \qquad(\overline{F}\in\Gamma(\K_1)\;)
  \;,
 \tag{$\&1$} 
\end{equatioN}
 and
 \begin{equatioN}
  \xi\in\Gamma_0(\K_2)\Om_2 
    \implies
  \left\lgroup
  \begin{array}{l}
    \text{there exist }m\in\NN, \text{and } 
          \widetilde{\xi}_1,\ldots,\widetilde{\xi}_m
    \text{ such that } 
    \\[3pt]
    \quad\norM{\bigl[\Gamma(C)\overline{F}\bigr]\xi}^2
      \leq
    \sum_{k=1}^m
         \norM[text]{\overline{F} \widetilde{\xi}_k}^2 
    \qquad
    (\overline{F}\in\Gamma(\K_1) \;)
    \;.
  \end{array}
  \right.
  \tag{$\&2$}
\end{equatioN}

 \statement{Strong continuity of $\Gamma(C):\Gamma(\K_1)\to\Gamma(\K_2)$}

 Suppose $\overline{F}_\al\to 0$ strongly in $\B(\HH_1)$. Then
 \begin{equatioN}
  \norM[text]{\overline{F}_\al\widetilde{\xi}}^2\to 0 
  \quad(\widetilde{\xi}\in\Gamma_0(\K_1)\Om_1)
     \;, \; \text{and } 
   \textstyle\sup_\al \norM[text]{\overline{F}_\al}_{\B(\HH_1)}<\infty
     \;.
  \tag{$\&3$}
 \end{equatioN}
 
 To establish that $\Gamma(C)\overline{F}_\al \to 0$ strongly in $\B(\HH_2)$, 
 it suffices to check (see lemma \ref{lm_critStrongConv}) that 
 (i) $\sup_\al \norM[text]{\bigl[\Gamma(C)\overline{F}_\al\bigr]}_{\B(\HH_2} 
       <\infty$,
 and (ii) 
 $\norM[text]{\bigl[\Gamma(C)\overline{F}\bigr]\xi}^2\to 0$ for all 
 $\xi \in \Gamma_0(\K_2)\Om_2$.
 
 Now (i) follows from $(\&1)$ and $(\&3)$ above, and
 (ii) from $(\&2)$ and $(\&3)$.

 \statement{Complete positivity of $\Gamma(C):\Gamma(\K_1)\to\Gamma(\K_2)$}

 By lemma \ref{lm_complPosGamma}, we have 
 \begin{displaymatH}
  \sum_{p,q=1}^N \ip{\xi_q}{\Gamma(C)\bigl[F^{*}_q F_p \bigr] \xi_p}\; \geq 0
   \quad	 
   (F_1,\ldots,F_n\in\Gamma_0(\K_1)\;,\;\xi_1,\ldots,\xi_N\in\HH_2),  
 \end{displaymatH}
 and since $\Gamma(C)$ (and the operator multiplication) 
 is strongly continuous, 
 this inequality extends to $\overline{F}_1,\ldots,\overline{F}_n\in\Gamma(\K_1)$.
\end{pf}

\section{Properties of the $\Ga$-functor}
\label{ap_propertiesGammaFunctor}
\subsection{The functorial character of $\Ga$}
The map $\Ga$ from the category of Hilbert spaces with contractions to the
category of quantum probability spaces 
(von Neumann algebras with a normal state) with quantum operations
(completely positive, state and $\indicator$-preserving maps) is  
\Dindex{functorial} as expressed by the following theorem.

\begin{theorem}
 Let $C_1:\K_1\to\K_2$ and $C_2:\K_2\to\K_3$ be a contractions.

 Then $\Ga(C_1 C_2)=\Ga(C_1)\Ga(C_2)$.
\end{theorem}

\begin{pf}
 A one-line writing exercise shows that 
 \begin{displaymatH}
  \Ga(C_1)\Bigl( \Ga(C_2) F \Bigr) =    \Ga(C_1 C_2) F 
 \end{displaymatH}
 for $F=W(f)$ with $f\in\K_1$. 
 By linearity of $\Ga(C_1)\Ga(C_2)$ and $\Ga(C_1 C_2)$ this equality
 extends to $F\in\Ga_0(\K_1)$ and 
 by strong continuity also to $F\in\Ga(\K_1)$.
\end{pf}

\subsection{$\Ga(C)$ if $C=J$ is an isometry}

Recall that a linear, adjoint-, product-, (and unity-)preserving map between 
Von Neumann algebras is called a $*$ isomorphism.
\question{Remark: $J$ an isometry if and only if $J^*$ is a projection ?
Yes if $J$ is embedding of a closed subspace in enveloping Hilbert space. 
In this case $J^* J = \Identity$}

\begin{theorem}
Let $J:\K_1\to \K_2$ be an isometry.
Then $j:=\Ga(J):\Ga(\K_1) \to \Ga(\K_2)$ 
is an injective $*$-homomorphism, 
so a quantum random variable.
\end{theorem}

\begin{pf}
We first prove that $\Ga(J)$ is an $*$-homomorphism. 

Let $f',f\in\K_1$. 
Since $J$ is an isometry, 
we have $\ip{Jf'}{Jf}=\ip{f'}{f}$, 
and $\norM{Jf}^2=\norM{f}^2$, so $\om(f)=\om(Jf)$
(here we use the notation $\om(f)=\exp(-\half\norM{f}^2)$ of lemma
\ref{lm_calcWeyl} again).

Using this in formula (\ref{eq_prod_W}) 
(lemma \ref{lem_prod_W}, p.~\pageref{lem_prod_W}),
we see that 
\begin{displaymatH}
\Ga(J)\bigr[ W(f')\cdot W(f) \bigr] = W(Jf')\cdot W(Jf)=
   \bigl[\Ga(J)W(f') \bigr] \cdot \bigl[\Ga(J)W(f) \bigr]
  \;,
\end{displaymatH}
and 
\begin{displaymatH}
 \Ga(J)\bigr[  W(f)^*  \bigr] 
   =
  \Ga(J)\bigr[  W(-f)  \bigr] 
   = \frac{\om(-f)}{\om(-Jf)} W(-Jf)
   = W(Jf)^*
   = \bigl[\Ga(J)W(f) \bigr]^*
  \;,
\end{displaymatH}
so that by linearity $\Ga(J)$ is an $*$-homomorphism on $\Ga_0(\K_1)$, 
and by strongly continuous extension also on $\Ga(\K_1)$.

\medskip

For proving the isometric character, we turn to corollary \ref{cor_om2INom1}
with  $g'=g\in \K_2$,
which yields, using that $\Ga(J)$ is an $*$-homomorphism,
\begin{aligN}
 \Bigl<  W_2(g)\Om_2 
       &
      \Bigm| \bigl[\Gamma(J) W_1(f')\bigr]^* \cdot \bigl[\Ga(J) W_1(f)\bigr]
      \Bigm|
       W_2(g) \Om_2
 \Bigr>
    \\
   =  & \;  \frac{\om(g)^2}{\om(J^*g)^2} \cdot \exp(\ip{g}{(I-JJ^*)g}) 
    \\
    &
    \quad\times
 \Bigl< W_1(J^*g)\Om_1 
      \Bigm| W_1^*(f')\cdot W_1(f)
      \Bigm|
       W_1(J^*g) \Om_1
 \Bigr>
   \quad 
   (f,f'\in \K_1)
   \;.
\end{aligN}
Since $\Ga(J)$ (and the inner product) is linear,
we derive from the above that
\begin{aligN}
 \Bigl<  W_2(g)\Om_2 
       &
      \Bigm| \bigl[ \Gamma(J)F \bigr]^* \cdot \bigl[\Ga(J) F\bigr]
      \Bigm|
       W_2(g) \Om_2
 \Bigr>
    \\
   =  & \;  \frac{\om(g)^2}{\om(J^*g)^2} \cdot \exp(\ip{g}{(I-JJ^*)g}) 
    \\
    &
    \quad\times
 \Bigl< W_1(J^*g)\Om_1 
      \Bigm| F^*\cdot F)
      \Bigm|
       W_1(J^*g) \Om_1
 \Bigr>
   \quad 
   (F \in \Ga_0(\K_1)\: )
   \;,
\end{aligN}
and by continuity this extends equality extends to 
$\overline{F} \in \Ga(\K_1)$.
In other words,
\begin{aligN}
 \norM[Big]{ \bigl[
                   \Gamma(J) & \overline{F}
             \bigr]
            \Bigl( W_2(g) \Om_2 \Bigr)
           }^2
    =  
    \\
    &
   \displaystyle
   \underbrace{\frac{\om(g)^2}{\om(J^*g)^2} \exp(\ip{g}{(I-JJ^*)g})}_{\;> \;0}
   \;
   \norM[Big]{\overline{F} \Bigl( W_1(J^*g)\Om_1\Bigr)}^2
   \quad
   (\overline{F}\in\Ga(\K_1)\;, g\in\K_2)
   \;.
\end{aligN}
Therefore,
\begin{displaymath}
  \bigl[ \Gamma(J)  \overline{F} \bigr]  \Bigl( W_2(g) \Om_2 \Bigr) =0
 \Implies
  \overline{F} \Bigl( W_1(J^*g)\Om_1\Bigr) = 0
  \quad
 (\overline{F}\in\Ga(\K_1)\;, g\in\K_2)
  \;.
\end{displaymath}
Since $J^*(\K_2) = \K_1$ ($J^*J=\Identity_{\K_1}$) we see that 
if $\Ga(J)\overline{F} = 0 $ on $\HH_2$ 
(so on the dense linear subspace $\Ga_0(\K_2)\Om_2$
of $\HH_2$), 
then $\overline{F}=0$ 
on the dense linear subspace $\Ga_0(\K_1)\Om_1$ of $\HH_1$, so
$\overline{F}=0$ (by continuity).
In other words, the linear map $\Ga(J):\Ga(\K_1)\to\Ga(\K_2)$ has
kernel $\{0\}$ i.e.\ is injective.
\end{pf}

\subsection{$\Ga(C)$ if $C=J^*$ is a projection}

Let $j$  be a
injective $*$-homomorphism from $ (\mathcal{A}, \phi)$ into 
$(\widetilde{\mathcal{A}}, \widetilde{\phi})$. 
Recall that a map $P: (\widetilde{\mathcal{A}}, \widetilde{\phi})\to 
(\mathcal{A}, \phi)$ 
is called a conditional expectation with respect to $j$
if it is the (necessarily) unique quantum operation satisfying
$P\after j= \Identity_{\mathcal{A}}$.

On the level of Hilbert spaces, we have that if $J$ is an isometry
from $\mathcal{H}_1$ to $J(\mathcal{H}_1)\subset \mathcal{H}_2$,
then it adjoint $J^*=\adj{J}$ is a projection from $\mathcal{H}_2$ to 
$\mathcal{H}_1$ 
if we identify $\mathcal{H}_1 \isom J(\mathcal{H}_1)$.

\begin{theorem}
If $J^*$ is a projection ($J$ an isometry) $\K_0$ to $\K$
then $P:=\Ga(J^*)$ is a conditional expectation (by definition). 

In particular:
\begin{displaymath}
 \forall_{B_1,B_2 \in \Ga(\K_0} \forall_{A\in\Ga(\K)}	
  B_1 P(A) B_2 = P\bigl( j(B_1) A j(B_2) \bigr)
\end{displaymath}

\end{theorem}

\begin{pf}
We check the claim for $B_i=W(b_i)\in\Ga(\K_0)$ and $A=W(a)\in \Ga(\K)$.
By linearity and continuity the claim then extends to arbitrary 
$B_i\in\Ga(\K_0)$ and $A\in\Ga(\K)$.

Let $b_1,b_2 \in \K_0$ and $a\in \K$.
Since for $f,g,h$ in $\K_0$ ($\K$ respectively):
\begin{displaymatH}
 W(f)W(g)W(h)
  = 
 e^{-i\Im\ip{f}{g}} e^{-i\Im\ip{f}{h}} e^{-i\Im\ip{g}{h}} W(f+g+h)
 \,.
\end{displaymatH}
we have
\begin{equation}
\begin{aligned}
 W(b_1)\cdot &\Ga(J^*)W(a) \cdot  W(b_2) 
  \\
   &
  = 
  e^{-\half\bigl(\norM{a}^2 - \norM{J^* a}^2\bigr)}
  e^{-i\ip{b_1}{J^* a}} e^{-i\ip{b_1}{b_2}} e^{-i\ip{J^* a}{b_2}} W(b_1+a+b_2)
 \;.
\end{aligned}
\tag{$*$}
\end{equation}
On the other hand, $\Ga(J)W(b_i) = W(J b_i)$ since $J$ is an isometry,
and it follows that 
\begin{aligN}
 \Ga(J)W(b_1)\cdot & W(a) \cdot  \Ga(J)W(b_2) 
  \\
  \ats
  = 
  W(J b_1) W(a) W(J b_2)
  \\
  \ats
  e^{-i\Im\ip{Jb_1}{a}} e^{-i\Im\ip{jb_1}{jb_2}} e^{-i\Im\ip{a}{Jb_2}} 
  W(Jb_1+a+Jb_2)
 \;,
\end{aligN}
and therefore,
\begin{equation}
\begin{aligned}
 \Ga(J^*) & \Bigl[ \Ga(J)W(b_1)\cdot  W(a) \cdot  \Ga(J)W(b_2) \Bigr]
  \\
  &
  = 
  e^{-\half\bigl( \norM{Jb_1+a+Jb_2}^2 - \norM{J^*(Jb_1+a+Jb_2)}^2\bigr)}
  \\
  &
  \quad
  \times
  e^{-i\Im\ip{b_1}{J^*a}} e^{-i\Im\ip{b_1}{b_2}} e^{-i\Im\ip{J^*a}{b_2}} 
  \\
  &
  \quad
  \times
 W\bigl(J^*(Jb_1 + a + J b_2)\bigr)
 \;.
\end{aligned}
\tag{$*'$}
\end{equation}
Using that $J$ is isometry and that $\ip{Jf}{g}=\ip{f}{J^* g}$,
the term $\norM{Jb_1+a+Jb_2}^2$ - $\norM{J^*(Jb_1+a+Jb_2)}^2$
is seen to be $\norM{a}^2 - \norM{J^* a}^2$. 
Further $J^*(Jb_1 + a + J b_2)=b_1 + Ja+ b_2$, so that ($*$) equals
($*'$).
\end{pf}

\subsection{$\Ga(C)$ if $C=U$ is a unitary map}
Since $U:\K_1\to\K_2$ as well as $U^{-1}=\adj{U}:\K_2\to\K_1$ 
is an isometry, we have that 
$\Ga(U)$ is a $*$-isomorphism with inverse $\Ga(U)^{-1}=\Ga(\adj{U})$.  

In the particular case that $\K_1=\K_2=:\K$, $\Ga(U)$ is called 
an $*$-automorphism of $\Ga(\K)$.

\begin{example}
 The $*$-automorphism of $\B(\RR^n)$ are of the form $B\to UB\adj{U}$ 
 where $U:\RR^n\to\RR^n$ is itself a unitary map.
\end{example}
The above example illustrates that the time evolution of a quantum system
in $\mathcal{B}(\RR^n)$ implemented by a unitary map, is in fact 
a $*$-automorphism. Via this identification, time evolutions on
more general Von Neumann algebras can be defined.

\section{The usual approach to second quantization for the free (quadratic)
scalar field\label{ap_quantizationFreeScalarField}
\label{ap_usualSecondQuantization}}

The approach taken here largely follows \cite[p.~107-114]{Itzykson+Zuber}.
 
Suppose the Lagrangian density is given by
\begin{displaymatH}
  \mathcal{L}= \half(\partial \ph)^2 - V(\ph)
\end{displaymatH}
with $V$ a differentiable \textsl{real} function. 
The corresponding classical field equations
(of motion) read
\begin{displaymatH}\displaystyle
  \partial_\mu \frac{\partial\mathcal{L}}{\partial(\partial_\mu \ph)} 
    - 
  \frac{\partial\mathcal{L}}{\partial\ph} 
   =: 
  \Box\,\ph + \frac{dV(\ph)}{d\ph} 
   = 0
   \;,
\end{displaymatH}
where 
$\Box=
\frac{\partial^2}{\partial x^2} + \frac{\partial^2}{\partial y^2} 
+ \frac{\partial^2}{\partial z^2} - \frac{\partial^2}{\partial t^2}$.

\medskip

For simplicity we now restrict to the case of a quadratic potential 
(i.e.\ a harmonic oscillator potential), $V(\ph)=\half m^2\ph^2$, 
which has the Klein-Gordon equation
\begin{displaymatH}
  \Box\,\ph + m^2\ph = 0
   \;,
\end{displaymatH}
as classical field equation, and 
\begin{displaymatH}\displaystyle
   \pi=\frac{\partial \mathcal{L}}{\partial(\partial_0 \ph)} = \partial_0 \ph
      =\frac{\partial \ph}{\partial t}
\end{displaymatH}
as conjugate field (i.e.\ field impulse).

From the above, the Hamiltonian is given by
\begin{displaymatH}\displaystyle
  H=\int d^3 x \,
         \Bigl\{ \half\bigl[{\pi(\mathbf{x})}^2 
                  + \bigl(\nabla \phi(\mathbf{x})\bigr)^2 \bigr] 
                  +  m^2{\ph(\mathbf{x})}^2 
         \Bigr\}
  \:, 
\end{displaymatH}
which can be recognized as a continuum of coupled oscillators.
To elucidate this, assume that space has one dimension (coordinate $x$) 
and that the coordinate $x$ can only take discrete values which are integral
multiples of a an elementary length taken as unity. 
In that case the Hamiltonian 
\begin{equatioN}
\label{eq_real_scalar_field_H_cont}
  H=\int d x \,
         \Bigl\{ \half\bigl[{\pi(x)}^2 
                  + \left(\frac{d \phi}{d x}(x)\right)^2 \bigr] 
                  +  m^2{\ph(x)}^2 
         \Bigr\}
  \:, 
\end{equatioN}
is replaced by
\begin{equatioN}
\label{eq_real_scalar_field_H_discr}
  H= \half \sum_{n=-\infty}^\infty 
         \Bigl\{
           \pi_n^2 + 
           (\phi_{n+1}-\ph_n)^2 +  
           m^2\ph_n^2 
         \Bigr\}
\end{equatioN}
with commutators
\begin{displaymath}
   [\pi_n, \pi_{n'}] = [\ph_n, \ph_{n'}]= 0
     \:, \quad\text{and }\quad
   [\ph_n, \pi_{n'}] = i \delta_{n,n'}
   \:.
\end{displaymath}

A physical model having this Hamiltonian is a chain of mass points with the
following characteristics:
\begin{Enumerate}
  \item the displacement (along the $x$-axis) 
        of the $n^{\text{th}}$ mass point from its equilibrium is given
        by $\ph_n$, its impulse by $\pi_n$;
  \item mass point $n$ is attached with a spring to its equilibrium position,
        the restoring force being provided by the $m^2\ph_n^2$ term in the
        Hamiltonian;
  \item mass point $n$ is coupled to its two neighbors $n-1$ and
        $n+1$ which induces a additional force incorporated by the 
        $(\ph_n - \ph_{n\pm 1})^2$ contributions to the Hamiltonian.
\end{Enumerate}
Such a chain could model the vibrations of a one-dimensional crystal where 
$\ph_n$ is the displacement of the $n^{\text{th}}$ atom from its equilibrium,
and each atom is coupled to his equilibrium position and to its neighbors.
Like in the case of a string 
(which can also seen as a continuum limit of  a chain of oscillators), 
it is fruitful to look for normal eigenmodes for this model.

Instead of taking a Fourier series expansion of the discrete version of the 
system (\ref{eq_real_scalar_field_H_discr}) 
and then taking the continuum limit, 
we shall directly apply a Fourier transform to the continuous system 
(\ref{eq_real_scalar_field_H_cont}).

Using that 
\begin{displaymatH}\displaystyle
 \pi(x)^2  = \left(\int_{-\infty}^{\infty}
                     \frac{dk}{2\pi}\widehat{\pi}(k) e^{ikx}
              \right)^2 
            = \int \!\! \frac{dk}{2\pi}
               \int \!\! \frac{dk'}{2\pi}
                    \quad\widehat{\pi}(k) \,\widehat{\pi}(k')\, e^{i(k+k')x}
            \,,
\end{displaymatH}
and that formally
\begin{displaymatH}\displaystyle
 \int_{-\infty}^{\infty} \frac{dx}{2\pi} \; e^{i{k+k'}x} = \delta(k+k')
 \;, 
\end{displaymatH}
we see that
\begin{aligN}\displaystyle
 \int_{-\infty}^{\infty}\!\!dx\; {\pi(x)}^2
   & =
 \frac{1}{2\pi}\int \!\! dk\!\! \int \!\! dk' 
     \;\;\widehat{\pi}(k) \,\widehat{\pi}(k') 
       \cdot
      \left(\int \!\! \frac{dx}{2\pi}\; e^{i(k+k')x} \right)
   \\
   & =
 \frac{1}{2\pi}\int \!\! dk \;\;
       \widehat{\pi}(-k) \, \widehat{\pi}(k) 
     =
 \frac{1}{2\pi}\int \!\! dk \;\;
       \adj{\widehat{\pi}}(k) \, \widehat{\pi}(k)
   \;,
\end{aligN}
where the last equality uses that $\pi(x)$ is real.

Likewise $\int\!\! dx\; {\ph(x)}^2 =  
\int \!\! \frac{dk}{2\pi} \; \adj{\widehat{\ph}}(k)\, \widehat{\ph}(k)$,
whereas using 
\begin{aligN}\displaystyle
 \frac{d \ph(x)}{d x} 
  & = \frac{d}{d x}\int\!\! \frac{d k}{2\pi} \;\widehat{\ph}(k) e^{ikx}
    = \int\!\! \frac{d k}{2\pi} \; i k \widehat{\ph}(k) e^{ikx}
    \,,
\end{aligN}
we see that 
\begin{aligN}\displaystyle
 \int\!\! dx  \left(\frac{d \ph(x)}{d x}\right)^2 
   & = 
    \frac{1}{2\pi} 
     \int \!\! dk \int \!\! dk'
                    \quad ik \widehat{\ph}(k') \,ik'\widehat{\ph}(k)\, 
    \left(\int \frac{dx}{2\pi} \;e^{i(k+k')x} \right)
   \\
   &  = 
    \frac{1}{2\pi}    
     \int \!\! dk\; i^2 k (-k) \widehat{\ph}(-k) \widehat{\ph}(k) 
   \\
   & =
   \frac{1}{2\pi}   
    \int \!\! dk\;  k^2 \adj{\widehat{\ph}}(k) \widehat{\ph}(k) 
    \,.
\end{aligN}
Putting pieces together a Fourier transformed Hamiltonian obtains:
\begin{equatioN}
  H = \frac{1}{4\pi} \int\!\!dk
       \biggl\{ 
     \adj{\widehat{\pi}}(k) \widehat{\pi}(k) 
           +  
     \underbrace{[m^2 +k^2]}_{\om(k)} \adj{\widehat{\ph}}(k) \widehat{\ph}(k)  
       \biggr\}
     \;.
\end{equatioN}
The resemblance with a continuum of harmonic oscillators 
\begin{displaymatH}\displaystyle
   H(k) = \half\left\{\frac{p(k)^2}{m} + m\om(k)^2 q(k)^2\right\} 
   \quad\text{ with $m=1$}
\end{displaymatH}
begins to arise, except for the fact
that $\widehat{\ph}(k)$ and $\widehat{\pi}(k)$ are no \text{real} quantities,
hence cannot be quantized to self-adjoint operators. However,
$\adj{\widehat{\pi}}(k) \widehat{\pi}(k)$ and  
$\adj{\widehat{\ph}}(k) \widehat{\ph}(k)$ are, 
and one could quantize the (positive) square root of those.

Here we will adopt another approach by quantization of the corresponding
creation and annihilation operators (which are not required to be self-adjoint).

Let
\begin{align}
  a(k)       & := \frac{1}{\sqrt{4\pi\om(k)}}
               \left[
                     \om(k) \widehat{\ph}(k) + i\widehat{\pi}(k)
               \right]  \\
  \adj{a}(k) &  = \frac{1}{\sqrt{4\pi\om(k)}}
               \left[
                     \om(k) \adj{\widehat{\ph}}(k) - i\adj{\widehat{\pi}}(k)
               \right] \,. 
\end{align}
Then
\begin{align}
 \int\hspace{-.3em}dk \;
        \om(k)\Bigl[ \adj{a}(k)& a(k) +  a(k)\adj{a}(k) \Bigr]
    \\
    & = \frac{1}{4\pi}
        \int\hspace{-.3em}dk \;
         \biggl[\; \om(k)^2\Bigl(\adj{\widehat{\ph}}(k)\widehat{\ph}(k) + \widehat{\ph}(k)\adj{\widehat{\ph}}(k)\Bigr)
    \\ \notag
    & \phantom{\frac{1}{4\pi}\biggl[\om(k)^2}
           + \Bigl(\adj{\widehat{\pi}}(k)\widehat{\pi}(k) + \widehat{\pi}(k)\adj{\widehat{\pi}}(k)\Bigr)
         \;\biggr]
    \\ \notag
    & = \frac{1}{2\pi}
         \int\hspace{-.3em}dk \;
         \biggl[\; \om(k)^2 \adj{\widehat{\ph}}(k)\widehat{\ph}(k)
                   + 
                   \adj{\widehat{\pi}}(k)\widehat{\pi}(k)
         \biggr] 
           \;,
\end{align}
where we have used in the first equality 
that the ``mixed'' terms compensate each other
\begin{aligN}
 \int_{-\infty}^\infty  \adj{\widehat{\ph}}(k)\widehat{\pi}(k) dk
   \ats =
  \int_{-\infty}^\infty  \widehat{\ph}(-k)\widehat{\pi}(k) dk
     =
   \int_{-\infty}^\infty  \widehat{\ph}(k)\widehat{\pi}(-k) dk
   \\
   \ats =
   \int_{-\infty}^\infty  \widehat{\ph}(k)\adj{\widehat{\pi}}(k) dk
   \,, 
   \\
   \textstyle
   \int_{-\infty}^\infty  \adj{\widehat{\pi}}(k)\widehat{\ph}(k) dk
   \ats 
   = \ldots = 
  \int_{-\infty}^\infty  \widehat{\pi}(k)\adj{\widehat{\ph}}(k) dk
   \,, 
 \end{aligN}
and in the second equality that 
the quadratic terms in $\ph$ and $\pi$ respectively double  
\begin{aligN}
 \int_{-\infty}^\infty  \widehat{\ph}(k)\adj{\widehat{\ph}}(k) dk
   \ats =
 \int_{-\infty}^\infty  \widehat{\ph}(k)\widehat{\ph}(-k) dk 
   =
 \int_{-\infty}^\infty  \widehat{\ph}(-k)\widehat{\ph}(k) dk
  \\
  \ats 
   =
 \int_{-\infty}^\infty  \adj{\widehat{\ph}}(k)\widehat{\ph}(k) dk 
  \,,
  \\
 \textstyle
 \int_{-\infty}^\infty  \widehat{\pi}(k)\adj{\widehat{\pi}}(k) dk
  \ats
  =
 \int_{-\infty}^\infty  \adj{\widehat{\pi}(k)}\widehat{\pi}(k) dk
  \,.
\end{aligN}

Summarizing
\begin{equation}
 H = \half\int_{-\infty}^\infty \hspace{-1em} dk\quad
                \om(k) \Bigl[ \adj{a}(k)a(k)+ a(k)\adj{a}(k) \Bigr]
   \;.
\end{equation}

Quantization now proceeds by postulating
\begin{equation}
 [a(k),\adj{a}(k')]=  \delta(k-k')\,,
 \qquad
 [a(k),a(k')]=[\adj{a}(k),\adj{a}(k')]=0
 \,.
\end{equation}
Using the commutator product rule $[a_1,a_2a_3]= [a_1,a_2]a_3 + a_2[a_1,a_3]$,
\begin{aligN}
 [&H,a(k_0)]        &= -\,\om(k_0) a(k_0) \,,\\
 [&H,\adj{a}(k_0)]  &= +\,\om(k_0)a(k_0) \,,\\
\end{aligN}
so that, if $\ket{E}$ is an eigenstate of $H$ with energy $E$,
then
\begin{aligN}
 H \Bigl(a(k_0)\ket{E}\Bigr) 
      & = a(k_0) H\ket{E} + [H,a(k_0)]\ket{E}
        = E a(k_0)\ket{E} + -\om(k_0)a(k_0)\ket{E} 
      \\
      & = [E-\om(k_0)] \Bigl(a(k_0)\ket{E}\Bigr) 
      \\
 H \Bigl(\adj{a}(k_0)\ket{E}\Bigr) 
      & = \adj{a}(k_0) H\ket{E} + [H,\adj{a}(k_0)]\ket{E}
        = E \adj{a}(k_0)\ket{E} + \om(k_0)\adj{a(k_0)}\ket{E} 
      \\
      & = [E+\om(k_0)] \Bigl(\adj{a}(k_0)\ket{E}\Bigr) 
      \,
\end{aligN}
i.e.\ $a(k)$ and $\adj{a}(k)$ lower respectively raise the energy of the system
by an amount of $\om(k)$, which means that with respect to energy quanta,
they play the role of annihilation and creation operators respectively.

Here we see the relation between particles and fields: 
The field (\ref{eq_real_scalar_field_H_cont}) 
consists of a \textsf{continuum} of (uncoupled) 
harmonic oscillators, 
each with its own Hamiltonian 
 \begin{displaymatH}
    H(k)= \half\om(k) \Bigl[ \adj{a}(k)a(k)+ a(k)\adj{a}(k) \Bigr]
 \end{displaymatH}
and corresponding creation (annihilation) operator $\adj{a}(k)$ ($a(k)$);
each oscillator contributes an \textsf{integral} multiple of its 
eigen-energy $\om(k)$ to the total energy of the field.
Therefore the energy of the system can be considered as made up of 
energy \textsf{quanta} of different kinds (labeled by $\om(k)$), each with 
its own occupation number $n(k)$.

Eigenstates of the field are normally not taken against the field operator 
$\ph$ (position), or for that matter, the field impulse operator $\pi$, but
against the energy operator $H$.
The ground state $\ket{0}$ of $H$ is the state with all oscillators 
$H(k)= \om(k)[ \adj{a}(k)a(k)+ a(k)\adj{a}(k)]$
in their ground state i.e.\ with energy $\half\om(k)$ and occupation number
$n(k)=0$.
Working upwards, all other eigenstates (excitations)
are obtained by creating energy quanta of different (frequency) modes.   

However, since we are dealing with a continuum of oscillators (each with
non-zero ground-level energy $\om(k)/2$), 
the ground-state of $H$ has an infinite energy. 
Indeed, since
\begin{displaymatH}
 \adj{a}(k)a(k)\ket{0}= 0
   \,,\; 
  \text{and }\; 
  a(k)\adj{a}(k)\ket{0}=\ket{0}\,,
\end{displaymatH}
(note the difference between $0$, no state i.e.\ there is no field nor 
oscillators, and $\ket{0}$, the field exists and is in its ground state with
all oscillators in their ground-level energy)
we can evaluate
\begin{aligN}
  \oip{0}{H}{0}
    & = 
   \half\int_{-\infty}^\infty dk\quad
            \om(k) \oip{0}{a(k)\adj{a}(k)}{0} 
    \\
    & =
   \half\int_{-\infty}^\infty dk\quad
            \om(k) \oip{0}{[a(k),\adj{a}(k)]}{0} 
    \\ 
    & = \half\int_{-\infty}^\infty \hspace{-1em} dk
           \quad\om(k)\;\delta(0)\ip{0}{0}
      = \infty 
   \,,
\end{aligN}
although it would be more appropriate to say that the above is undefined by
the appearance of $\delta(0)$.
Since the ground-level energy of the field is not observable (unless the field
is destroyed), and only differences of energy levels are physically observable,
we can formally declare the ground-level energy of the field to be $0$,
and furthermore, take the ground-state to be normalized, $\ip{0}{0}=1$.

A manner to accomplish this is to split up $H$ as
\begin{align*}
 H & = \half\int_{-\infty}^\infty \hspace{-1em} dk\quad
                \om(k) \Bigl\{ 2\adj{a}(k)a(k) + [a(k),\adj{a}(k)] \Bigr\}
   \\
   & = \int_{-\infty}^\infty \hspace{-1em} dk\quad \om(k)\: \adj{a}(k)a(k)
          + 
      \half\int_{-\infty}^\infty \hspace{-1em} dk\quad \om(k)\: 
         [a(k),\adj{a}(k)]\ip{0}{0} \Bigr\}
   \\
   & = \int_{-\infty}^\infty \hspace{-1em} dk\quad \om(k)\: \adj{a}(k)a(k)
          + 
      \half\int_{-\infty}^\infty \hspace{-1em} dk
           \quad \om(k)\: \oip{0}{[a(k),\adj{a}(k)]}{0} 
    \\
    & = \int_{-\infty}^\infty \hspace{-1em} dk\quad \om(k)\: \adj{a}(k)a(k)
          + 
      \oip{0}{H}{0}
    \\
    & \overset{\text{declaration}}{=}
    \int_{-\infty}^\infty \hspace{-1em} dk\quad \om(k)\: \adj{a}(k)a(k)
      \,,
\end{align*}
where we have used in the third equality that $[a(k),\adj{a}(k)]$ 
is a scalar (``$c$-number'' in physical terminology).

The above way of handling infinite expected values in the ground-state, i.e.\ 
rewriting occurrences of $a(k)\adj{a}(k)$ to 
$\adj{a}(k)a(k) + [a(k),\adj{a}(k)]$ (the latter term of which is formally
$\delta(0)$) and then declaring the $[a(k),\adj{a}(k)]$ 
to vanish in the ground-state expectation, 
is referred to as putting the operator in ``normal ordering'' 
(also called Wick's ordering).

Having declared away the infinite ground state energy, 
problems arise again when forming excited states such as
$\adj{a}(k)\ket{0}$ (one energy quantum in mode $\om(k)$), since these states
are not normalizable:
\begin{displaymatH}
 \oip{0}{a(k)\adj{a}(k)}{0}= \oip{0}{[a(k),\adj{a}(k)]}{0}
                           = \delta(0)\ip{0}{0}
 \, 
\end{displaymatH}
where we have again used that $a(k)\ket{0}=0$.

Wave packets are therefore build using linear superposition: a state of the 
form 
\begin{displaymath}
 \adj{a}_f \ket{0}:= \int_\infty^\infty dk\; f(k) \adj{a}(k)\ket{0}
\end{displaymath}
is normalizable provided that
\begin{aligN}
  \infty 
   & >  \oip{0}{a_f \adj{a}_f}{0} 
     =
  \int_{-\infty}^\infty\hspace{-1em}dk \int_{-\infty}^\infty\hspace{-1em}dk'\quad 
   \overline{f(k')}f(k) \; 
    \langle 0 |\hspace{-1.3em}
     \underbrace{a(k')\adj{a}(k)}_{\adj{a}(k)a(k') + \delta(k-k')}
              \hspace{-1.3em} | 0 \rangle 
     \\
   & =
     \int_{-\infty}^\infty\hspace{-1em}dk \; |f(k)|^2
    \,,
\end{aligN}
i.e.\ the ``density kernel'' $f$ has to be square integrable.
Assuming $\adj{a_f}\ket{0}$ to be normalized (i.e.\ $\int\!\! dk\; |f(k)|^2 =1$),
we see that if $f(k)$ has a peak around $k_0$ with small spread, 
i.e.
\begin{displaymatH}
  \langle 0 | k a_f \adj{a_f}| 0 \rangle=\int\!\! dk\;  k |f(k)|^2 = k_0
  ,
\end{displaymatH}
and 
\begin{displaymatH}
  \langle 0 | (k-k_0)^2 \cdot a_f\adj{a_f}| 0 \rangle
   = 
  \int\!\! dk\; (k-k_0)^2 |f(k)|^2
\end{displaymatH}
is small, then $\adj{a_f}(k)\ket{0}$ can be interpreted 
as a wave packet around wave number $k=k_0$.
Using that
\begin{equation}
\label{eq_comm_Ak_Adaggerf}
 [a(k),\adj{a_f}]
  = 
 \int_{-\infty}^\infty \hspace{-1em}dk' \quad
  f(k') \underbrace{[a(k),\adj{a}(k')]}_{\delta(k-k')}
  = 
  f(k)
\end{equation}
we see that 
\begin{equation}
\label{eq_AkAdaggerk}
\begin{aligned}
  a(k)\; \adj{a_f} \ket{0}
  & =
  \adj{a_f}\underbrace{a(k)\ket{0}}_{0} 
   +   
  \underbrace{[a(k),\adj{a_f}]}_{f(k)}\ket{0}
  & =
  f(k) \ket{0}
  \,.
\end{aligned}
\end{equation}
If we use the latter in the first and third equality of
\begin{aligN}
 \oip{0}{a_f\; \adj{a}(k) a(k)\adj{a_f}}{0}
    & =
    f(k) \oip{0}{a_f \adj{a}(k)}{0}
     = 
    f(k) \overline{\oip{0}{a(k) \adj{a_f}}{0}}
     \\
    & = f(k) \overline{f(k)}\overline{\ip{0}{0}}  
      = |f(k)|^2
    \,,
\end{aligN}
we derive that the energy in state $\adj{a_f}\ket{0}$ is
\begin{equation}
\begin{aligned}
  \langle 0 | a_f H \adj{a_f} | 0 \rangle
    & =
  \int_{-\infty}^{\infty}\hspace{-1em}dk\quad \om(k) 
   \langle 0 | a_f \adj{a}(k)a(k) \adj{a_f} | 0 \rangle
   \\
   & = 
    \int_{-\infty}^{\infty}\hspace{-1em}dk\quad \om(k) |f(k)|^2
   \approx
   \om(k_0)
\end{aligned}
\end{equation}
as one would expect from a state that is concentrated around $k=k_0$.

In conclusion, to obtain meaningful expressions in quantum field theory 
the fundamental creation and annihilation operators 
$\adj{a}(k)$ and $a(k)$ have to be smeared (smoothed)
with square integrable functions in $k$-space. The resulting operators
$\adj{a_f}$ and $a_f$ are so-called operator-valued distributions.

Being a linear superposition of one-particle states, 
$a_f\ket{0}$ is to be interpreted as a one-particle state.
Multiple particle states are created by successively applying 
creation operator-valued distributions $\adj{a_f}$ to the ground state
$\ket{0}$. For example a (raw) $r$-particle state is given by
\begin{displaymatH}
  \ket{r} = \adj{a_{f_1}}\adj{a_{f_2}}\ldots \adj{a_{f_r}}\ket{0}
  \;.
\end{displaymatH}
$n$-particle states are orthogonal to $m$-particle states 
whenever $n \neq m$.
Indeed, from (\ref{eq_comm_Ak_Adaggerf}) we have
\begin{equation}
 [a_g,\adj{a_f}]
  = 
 \int_{-\infty}^\infty \hspace{-1em}dk \quad
  \overline{g(k)} \underbrace{[a(k),\adj{a_f}]}_{f(k)}
  = 
 \int_{-\infty}^\infty \hspace{-1em}dk \quad  \overline{g(k)}f(k)
  = \ip{g}{f}_{L^2}
  \,,
\end{equation}
and from there we derive successively that the vacuum is orthogonal to
$n \geq 1$-particle states:
\begin{displaymatH}
 \oip{0}{\adj{a_{f_1}}\adj{a_{f_2}}\ldots\adj{a_{f_n}}}{0}=
   \underbrace{\langle 0 | \adj{a_{f_1}}}_{0}
      \adj{a_{f_2}}\ldots\adj{a_{f_n}} | 0 \rangle = 0
  \,,
\end{displaymatH}
that $1$-particle states are orthogonal to $n\geq 2$-particle states
\begin{aligN}
 \langle 0 |a_{g_1} & \adj{a_{f_1}}\adj{a_{f_2}}\ldots\adj{a_{f_n}} | 0 \rangle
    =
 \langle 0 | \hspace{-1em}
 \underbrace{a_{g_1}\adj{a_{f_1}}}_{\adj{a_{f_1}}a_{g_1} + \; \ip{g_1}{f_1}_{L^2}}
             \hspace{-1em} 
 \adj{a_{f_2}}\ldots\adj{a_{f_n}} 
 | 0 \rangle
  \\[.3em]
  & = 
 \underbrace{\langle 0 | \adj{a_{f_1}} }_{0} 
         a_{g_1} \adj{a_{f_2}}\ldots\adj{a_{f_n}} | 0 \rangle
  +
 \ip{g_1}{f_1} \oip{0}{\adj{a_{f_2}}\ldots\adj{a_{f_n}}}{0}
  = 0
\end{aligN}
(the last term vanishes because the vacuum is orthogonal to a 
$(n-1)	$-particle state), and so on using that
\begin{aligN}
 [a_{g_1} & a_{g_2}\ldots a_{g_m}, \adj{a_{f_1}}]
  \\
  & =
 [a_{g_1}, \adj{a_{f_1}}] a_{g_2}\ldots a_{g_m} + 
  a_{g_1}[a_{g_2}, \adj{a_{f_1}}]a_{g_3}\ldots a_{g_m} 
  +  \ldots +
  a_{g_1}a_{g_2}\ldots a_{g_{n-1}}[a_{g_n}, \adj{a_{f_1}}]
  \\
  &  = 
 \ip{g_1}{f_1} a_{g_2}\ldots a_{g_m} +
  a_{g_1} \ip{g_2}{f_1} a_{g_3}\ldots a_{g_m} 
  +   
  a_{g_1} a_{g_2} \ip{g_3}{f_1} a_{g_4}\ldots a_{g_m} +   
    \ldots 
  \\
  & \quad + a_{g_1} a_{g_2} \ldots a_{g_{m-1}}\ip{g_{m}}{f_1}
  .
\end{aligN}

The number of quanta in a multiple particle state is counted by the number
operator 
\begin{equation}
 N:= \int_{-\infty}^{\infty} \hspace{-1em} dk \quad \adj{a}(k)a(k)
 \,,
\end{equation}
since
\begin{aligN}
 [a(k)&,\adj{a_{f_1}}\adj{a_{f_2}}\ldots\adj{a_{f_n}}]
  \\
  & = 
 [a(k),\adj{a_{f_1}}]\adj{a_{f_2}}\ldots\adj{a_{f_n}}
  + 
  \adj{a_{f_1}} [a(k),\adj{a_{f_2}}]\adj{a_{f_3}}\ldots\adj{a_{f_n}}+
  \\
  & \quad\ldots +  
  \adj{a_{f_1}}\ldots\adj{a_{f_{n-1}}}[a(k),\adj{a_{f_n}}]
  \\
  &  =
  f_1(k)\adj{a_{f_2}}\ldots\adj{a_{f_n}}
  + 
  \adj{a_{f_1}} f_2(k)\adj{a_{f_3}}\ldots\adj{a_{f_n}}+
  \\
  & \quad\ldots +  
  \adj{a_{f_1}}\ldots\adj{a_{f_{n-1}}} f_{n}(k)
  \,,
\end{aligN}
so that
\begin{align*}
 N& \adj{a_{f_1}}\adj{a_{f_2}}\ldots\adj{a_{f_n}}\ket{0}
  = \int_{-\infty}^{\infty}  \hspace{-1em} dk \quad \adj{a}(k) 
      \underbrace{a(k)}_{} 
      \underbrace{\adj{a_{f_1}}\adj{a_{f_2}}\ldots\adj{a_{f_n}}}_{}
      \ket{0} 
  \\
  & =
 \int_{-\infty}^\infty \hspace{-1em} dk\quad \adj{a}(k) 
      \underbrace{\adj{a_{f_1}}\adj{a_{f_2}}\ldots\adj{a_{f_n}}}_{} 
      \underbrace{a(k)\ket{0}}_{0} 
  + 
 \int_{-\infty}^\infty \hspace{-1em} dk\quad \adj{a}(k) 
      [\underbrace{a(k)}_{}, 
        \underbrace{\adj{a_{f_1}}\adj{a_{f_2}}\ldots\adj{a_{f_n}}}_{}
      ] 
      \ket{0}
  \\
  & = 
 \biggl(\int\!\! dk\quad 
   f_1(k)\adj{a}(k)\adj{a_{f_2}}\ldots\adj{a_{f_n}}
  + 
 \int\!\! dk\quad 
  \adj{a_{f_1}} f_2(k)\adj{a}(k) \adj{a_{f_3}}\ldots\adj{a_{f_n}}
  +
  \\
  & \quad\ldots +  
 \int\!\! dk\quad 
  \adj{a_{f_1}}\ldots\adj{a_{f_{n-1}}} f_{n}(k)\adj{a}(k)  
  \biggr)\ket{0}
  \,,
  = n \cdot  \adj{a_{f_1}}\ldots\adj{a_{f_{n-1}}} \adj{f_n}\ket{0}
\end{align*}


\bigskip

Using the $\adj{a_f}$ coherent states of the field can be introduced.

The field Hamiltonian is an (infinite) sum of independent (uncoupled)
harmonic oscillators. Each harmonic oscillator has its own coherent
states $\Coh(\la):=\exp\bigl(\la \adj{a}(k)\bigr)\ket{0}$, $\la\in\CC$, 
(where $k$ is the wave number identifying 
the oscillator).
Since the different modes (oscillators) are independent, 
simultaneous coherent states of the system (coherent states
of the field) can be formed by taking 
infinite products of individual harmonic oscillator
coherent states:
their form is 
\begin{displaymatH}
  \Coh(f): = K(f) \exp( \int\!\! dk\;  f(k) \adj{a}(k) )\ket{0}
  \,,
\end{displaymatH}
where $K(f)$ is a normalization factor and $f(k)$ is the weight.

We now calculate $\ip{\Coh(g)}{\Coh(f)}$.

Set $A:=\adj{[\int\!\! dk\;  g(k) \adj{a}(k)]}
= \int\!\! dk\;  \overline{g(k)} a(k)$ and
$B:= \int\!\! dk'\;  f(k') \adj{a}(k')$.
Then 
\begin{displaymatH}
 [A,B]=\int\!\! dk \int\!\! dk'\; 
        \overline{g(k)} f(k') \underbrace{[a(k),\adj{a}(k')]}_{\delta(k-k')}
      = \int\!\! dk \overline{g(k)} f(k)
      = \ip{g}{f}_{L^2}
      \,,
\end{displaymatH}
whence $A$ and $B$ commute with $[A,B]$, 
so that by the Baker-Hausdorff-Campbell formula
$e^A e^B = e^{A+B}  e^{\half[A,B]}$. 
Likewise, $ e^B e^A= e^{B+A} e^{\half[B,A]}= e^B e^A e^{-\half[A,B]}$,
so that $e^A e^B= e^B e^A e^{[A,B]}$.
As a result,
\begin{equation}
\label{eq_ipCoh}
\begin{aligned}
 \bigl[&\overline{K(g)} K(f)\bigr]^{-1}
 \ip{\Coh(g)}{\Coh(f)}
 \\
 &
   = 
 \bra{0}
 \adj{\Bigl(\exp\bigl(\int\!\! dk\;  g(k) \adj{a}(k)\bigr)\Bigr)}
 \:
 \Bigl(\exp\bigl(\int\!\! dk'\;  f(k') \adj{a}(k')\bigr)\Bigr)
 \ket{0}
 \\
 & 
 = \oip{0}{e^A e^B}{0} =  e^{[A,B]} \oip{0}{e^B e^A}{0} 
 =  e^{[A,B]} \ip{0}{0} 
 \\
 &
 = \exp\Bigl(\int\!\! dk\: \overline{g(k)} f(k)\Bigr)
 \;,
\end{aligned}
\end{equation}
where we have used in the before last equality that
$A =\int\!\! dk\;  \overline{g(k)} a(k)$ annihilates $\ket{0}$ so that
$e^A\ket{0}= (I + A + A^2/2! + A^3/3! + \ldots) \ket{0}= \ket{0}$, 
and by the same token $\bra{0} e^B = \bra{0}$.

In particular, $K(f)= \exp\Bigl(-\half\int\!\! dk\: \overline{f(k)} f(k)\Bigr)
= e^{-\half\norM{f}_{L^2}^2}$, and therefore
\begin{displaymath}
 \ip{\Coh(g)}{\Coh(f)} = e^{-\half\norM{f}^2-\half\norM{g}^2+ \ip{g}{f}}
     = e^{-i\Im\ip{g}{f} - \half\norM{g-f}^2} 
  \;. 
\end{displaymath}

From (\ref{eq_ipCoh}), we see that the coherent vectors are not orthogonal.
Nevertheless, they form a complete set, since
\begin{displaymatH}
 \frac{d}{dt} \Coh(tf) \mid_{t=0} = \adj{a_f} \ket{0}
  \,,
\end{displaymatH}
and more generally,
\begin{displaymatH}
 \frac{d^n}{dt^n} \Coh(tf) \mid_{t=0} = (\adj{a_f})^n \ket{0}
 \;, 
\end{displaymatH}
from which via a polarization formula,
\begin{displaymatH}
  a_1 a_2 \cdots a_n 
  = (2^n n!)^{-1} 
    \sum_{\ep_1, \ldots, \ep_n = \pm 1} (\ep_1 a_1 + \ldots + \ep_n a_n)^n
  \,, 
\end{displaymatH}
the state 
\begin{displaymatH}
 \adj{a_{f_1}}\adj{a_{f_2}}\cdots\adj{a_{f_n}} \ket{0}
\end{displaymatH}
can be obtained.


\chapter*{Thanks}


My thanks goes out to dr.\ Hans Maassen who introduced me 
to the wonder world of 
quantum probability theory and has supervised this thesis research.
It was a pleasure working with you, Hans, and I wish you many years to come 
full of discoveries and pleasure in this field.
\\[.3em]
I would like to thank professor Ted Janssen for his willingness to act as
second reader of the thesis. His comments have not only added to  
the quality of the thesis, but were also surprisingly fast 
(one weekend's time!).
\\[.3em]
Some people were not involved in the contents of the thesis, 
but nevertheless contributed to the habitat in which it could grow.
\\[0.3em]
The enthusiasm and interest of Luc Bouten has given me a precious memory. In 
particular, our strolls in the park, where we discussed our ups and downs
in science. Thanks, Luc!
\\[0.3em]
I would like to thank professor Van Rooij for his interest and 
the opportunity to do my PhD-study in mathematics in part-time which enabled
me to do the research for this thesis in parallel with the first two years of
my PhD-study.
\\[0.3em]
Last-but-not-least, my father, mother, and brother have been 
constant factors of support during the years. 
Although my father cannot see the completion of this thesis, I still
know he appreciates it. 
Attending my presentation of the work inspired my brother to write the poem 
on the first page of the thesis. Reading it opens not only my eyes for another
view on the work, but also on myself. Thanks, Rogier.
Finally, studying requires not only an intellectual effort, and in particular
I want to thank my mother for financial and mental support.
\\[2em]
Steven Teerenstra





\end{document}